\documentstyle[a4wide,epsf,epsfig,psfig,11pt,titlepage,twoside]{article}

\newcounter{nref}
\newcommand{\bbib}{%
  \renewcommand{\refname}{\large\bf References}%
  \begin{thebibliography}{99}%
  \setcounter{enumiv}{\thenref}}
\newcommand{\ebib}{%
  \end{thebibliography}%
  \setcounter{nref}{\arabic{enumiv}}}
\newcommand{\head}[3]{%
  \setcounter{nref}{0}
  \setcounter{figure}{0}
  \setcounter{table}{0}%
  \setcounter{footnote}{0}%
  \setcounter{equation}{0}%
  \section*{\LARGE\bf #1}%
  \stepcounter{section}%
  \addcontentsline{toc}{section}{#1}%
  \large\itshape%
  #2\\\vspace{0.1pt}\\%
  #3%
  \normalsize\upshape%
  \bigskip}

\newcommand{\talk}[2]                           
        {{      %
        \removelastskip
        \vglue 6pt plus 2pt minus 2pt
        \parskip=0pt
        \parindent=0pt
        \hrule height0pt                
        \noindent
        \vtop{\hsize=0.2\hsize \strut \rm #1 \hfil \quad} 
        \vtop{\hsize=0.79\hsize \strut \frenchspacing #2 \hfil \hskip 3pt}
        \hrule height0pt                
        }}

\newcommand{\contentitem}[3]
        {{      %
        \removelastskip
        \vglue 8pt plus 1pt minus 1pt
        \vbox   {\frenchspacing
                {\noindent \rm #1} \hfill \\
                \hglue 5mm {\em #2} \dotfill \quad {\rm #3}
                }
        \allowbreak
        }}

\newcommand{\TUM}{{Technische Universit\"at M\"unchen \\
                   Physik Department E15 \\
                   D--85747 Garching, Germany\\
                  }}
\newcommand{\TUMS}{{Technische Universit\"at M\"unchen \\
                   Physik Department E21 \\
                   D--85747 Garching, Germany\\
                  }}
\newcommand{\TUML}{{Technische Universit\"at M\"unchen \\
                   Physik Department T30 \\
                   D--85747 Garching, Germany\\
                  }}
\newcommand{\TUMK}{{Technische Universit\"at M\"unchen \\
                   Physik Department E12 \\
                   D--85747 Garching, Germany\\
                  }}
\newcommand{\MPA}{{Max-Planck-Institut f\"ur Astrophysik \\
                   Karl-Schwarzschild-Stra\ss e~1 \\
                   D--85740 Garching, Germany\\
                  }}
\newcommand{\MPP}{{Max-Planck-Institut f\"ur Physik \\
                   F\"ohringer Ring~6 \\
                   D--80805 M\"unchen, Germany\\
                  }}
\newcommand{\LMU}{{Ludwig-Maximilians-Universit\"at\\
                   Sektion Physik \\
                   Theresienstra\ss e~37 \\
                   D--80333 M\"unchen, Germany\\
                  }}
\newcommand{\USM}{{Ludwig-Maximilians-Universit\"at\\
                   Universit\"ats-Sternwarte \\
                   Scheinerstra\ss e~1\\
                   D--81679 M\"unchen, Germany\\
                  }}

\newcommand{\participant}[2]            
        {{      %
        \removelastskip
        \vglue 9pt plus 2pt minus 2pt
        \parskip=0pt
        \parindent=0pt
        \hrule height0pt                
        \noindent
        \vtop{\hsize=0.29\hsize \strut \rm #1 \hfil \quad} 
        \vtop{\hsize=0.70\hsize \strut  #2 \hfil \hskip 3pt}
        \hrule height0pt                
        }}

\def\alt{\mathrel{\mathop
  {\hbox{\lower0.5ex\hbox{$\sim$}\kern-0.8em\lower-0.7ex\hbox{$<$}}}}}
\def\agt{\mathrel{\mathop
  {\hbox{\lower0.5ex\hbox{$\sim$}\kern-0.8em\lower-0.7ex\hbox{$>$}}}}}

\begin{document}

\pagenumbering{roman}

\newpage {

\thispagestyle{empty}
\centerline {\normalsize \bf Sonderforschungsbereich 375 $\bullet$ 
Research in Particle-Astrophysics}
\centerline{\normalsize Technische Universit\"at M\"unchen (TUM) $\cdot$
Ludwig-Maximilians-Universit\"at M\"unchen (LMU)} 
\centerline{\normalsize Max-Planck-Institut f\"ur Physik (MPP) $\cdot$
Max-Planck-Institut f\"ur Astrophysik (MPA)}
\vskip 3truecm

\centerline{\Large\bf Proceedings of the Fourth SFB-375 Ringberg Workshop}
\vskip 12truemm
\centerline{\Huge\bf Neutrino Astrophysics}
\vskip 22truemm

\centerline{\normalsize\bf Ringberg Castle, Tegernsee, Germany}
\vskip 1truemm
\centerline{\normalsize\bf October 20--24, 1997}

\vskip 40truemm
\normalsize \noindent \begin{tabular}{ll}
Program Committee: &
Michael Altmann, Wolfgang Hillebrandt, Hans-Thomas Janka, \\
        &       Manfred Lindner, Lothar Oberauer, Georg Raffelt \\
\end{tabular}

\vskip 35 truemm
\centerline{\bf edited by }
\vskip 3truemm
\centerline{\bf Michael Altmann, Wolfgang Hillebrandt, 
        Hans-Thomas Janka and Georg Raffelt}
\vskip 7truemm
\centerline{\bf January 1998}
}

\newpage {

\thispagestyle{empty}

\begin{raggedright}
Proceedings of the Fourth SFB-375 Ringberg Workshop ``Neutrino Astrophysics''
\\

\vglue 8truecm

Sonderforschungsbereich 375 Research in Astro-Particle Physics \\
E-mail: {\tt depner@e15.physik.tu-muenchen.de} \\

\bigskip
Postal Address: \\

\smallskip
Technische Universit\"at M\"unchen \\
Physik Department E15 --- SFB--375 \\
James-Franck-Stra\ss e \\
D--85747 Garching \\
Germany\\

\vfill
\copyright\ Sonderforschungsbereich 375 and individual contributors \\

\bigskip 

Published by the Sonderforschungsbereich 375 \\
Technische Universit\"at M\"unchen, D--85747 Garching \\
January 1998 \\

\end{raggedright}


}\newpage{


\noindent {\Large\bf Preface}

\bigskip

\noindent This was the fourth workshop in our series of annual
``retreats'' of the {\it Sonderforschungsbereich Astroteilchenphysik}
(Special Research Center for Astroparticle Physics), or SFB for short,
to the Ringberg Castle above Lake Tegernsee in the foothills of the
Alps.  These meetings are meant to bring together the members of the
SFB which are dispersed between four institutions in the Munich area,
the Technical University Munich (TUM), the
Ludwig-Maximilians-University (LMU), and the Max-Planck-Institute for
Physics (MPP) and that for Astrophysics (MPA).  We always invite a
number of external speakers, including visitors at our institutions,
to complement the scientific program and to further the exchange of
ideas with the international community.

This year's topic was ``Neutrino Astrophysics'' which undoubtedly is
one of the central pillars of astroparticle physics.  We focused on
the astrophysical and observational aspects of this field,
deliberately leaving out theoretical particle physics and laboratory
experiments from the agenda.  Each day of the workshop was dedicated
to a specific sub-topic, ranging from solar, supernova and atmospheric
neutrinos over high-energy cosmic rays to the early universe.  A
session on future prospects served to conclude the workshop and
provide an outlook on the field in the next decade and beyond.  We
started every topical session with one or two introductory talks,
reviewing the status of theory and experiment and to providing some
background for the non-experts.

For the entire program we interpreted ``neutrino astrophysics'' in a
broad sense, including, for example, the physics of $\gamma$-ray
bursts or the recent observations of TeV $\gamma$-rays by the imaging
air-Cherenkov technique.  Some of the after-dinner-talks went
significantly beyond a narrow interpretation of the field!  From our
perspective the profile of neutrino astrophysics as defined by our
program worked very well, even better than we had hoped. We are proud
that the main complaint of the participants seemed to be that they did
not get enough mountain-hiking done because the sessions were too
interesting to miss.

Besides regular SFB resources this workshop was made possible by a
direct grant from the Max-Planck-Society and additional funds from the
Max-Planck-Institute for Astrophysics. Special thanks go to the SFB
secretary, Maria Depner, for her smooth and skillful management of all
practical matters related to the workshop.

We thank the participants for their high-level contributions and for
being extremely co-operative in submitting the ``extended abstracts''
of their talks on time and in a format that allowed us to produce
these proceedings in electronic form.  Anyone interested in a printed
version should write to the SFB secretary at the address given on the
previous page.  We hope that you will find this booklet a useful and
up-to-date resource for the exciting and fast-developing field of
neutrino astrophysics.

\bigskip\bigskip

\noindent{\it Michael Altmann, Wolfgang Hillebrandt,
Hans-Thomas Janka and Georg Raffelt}

\medskip

\noindent{\it Munich, January 1998}

\newpage
{\ }

}\newpage {

\section*{Contents}

\subsection*{History}

\contentitem{R.L.\ M\"o\ss bauer}
        {History of Neutrino Physics: Pauli's Letters}
        {3}
\contentitem{A. Dar}
        {What Killed the Dinosaurs?}
        {6}
%

\subsection*{Solar Neutrinos}

\contentitem{M. Stix}
        {Solar Models}
        {13}
\contentitem{H. Schlattl, A. Weiss}
        {Garching Solar Model: Present Status}
        {19}
\contentitem{M. Altmann}
        {Status of the Radiochemical Gallium Solar Neutrino Experiments}
        {22}
\contentitem{Y. Fukuda (for the Superkamiokande Collaboration)}
        {Solar Neutrino Observation with Superkamiokande}
        {26}
\contentitem{M.E. Moorhead (for the SNO Collaboration)}
        {The Sudbury Neutrino Observatory}
        {31}
\contentitem{L. Oberauer (for the Borexino Collaboration)}
        {BOREXINO}
        {33}
\contentitem{M. Junker (for the LUNA Collaboration)}
        {Measurements of Low Energy Nuclear Cross Sections}
        {36}
\contentitem{G. Fiorentini}
        {Solar Neutrinos: Where We Are and What Is Next?}
        {40}

\subsection*{Supernova Neutrinos}

\contentitem{W. Hillebrandt}
        {Phenomenology of Supernova Explosions}
        {47}
\contentitem{B. Leibundgut}
        {Supernova Rates}
        {51}
\contentitem{A.G. Lyne}
        {Pulsar Velocities and Their Implications}
        {54}
\contentitem{W. Keil}
        {Convection in Newly Born Neutron Stars}
        {57}
\contentitem{H.-Th. Janka}
        {Anisotropic Supernovae, Magnetic Fields, and Neutron Star Kicks}
        {60}
\contentitem{K. Sato, T. Totani, Y. Yoshii}
        {Spectrum of the Supernova Relic Neutrino Background and Evolution of
         Galaxies}
        {66}
\contentitem{G.G. Raffelt}
        {Supernova Neutrino Opacities}
        {73}
\contentitem{S.J. Hardy}
        {Quasilinear Diffusion of Neutrinos in Plasma}
        {75}
\contentitem{P. Elmfors}
        {Anisotropic Neutrino Propagation in a Magnetized Plasma}
        {79}
\contentitem{A.N. Ioannisian, G.G. Raffelt}
        {Cherenkov Radiation by Massless Neutrinos in a Magnetic Field}
        {83}
\contentitem{A. Kopf}
        {Photon Dispersion in a Supernova Core}
        {86}

\subsection*{Gamma-Ray Bursts}

\contentitem{D.H. Hartmann, D.L. Band}
        {Gamma-Ray Burst Observations}
        {91}
\contentitem{S.E. Woosley, A. MacFadyen}
        {Gamma-Ray Bursts: Models That Don't Work and Some That Might}
        {96}
\contentitem{M. Ruffert, H.-Th. Janka}
        {Models of Coalescing Neutron Stars with Different Masses and Impact
         Parameters}
        {101}
\contentitem{R.A. Sunyaev}
        {Physical Processes Near Black Holes}
        {106}

\subsection*{High-Energy Neutrinos}

\contentitem{K. Mannheim}
        {Astrophysical Sources of High -Energy Neutrinos}
        {109}
\contentitem{P. Gondolo}
        {Atmospheric Muons and Neutrinos Above 1$\,$TeV}
        {112}
\contentitem{D. Kie{\l}czewska}
        {Atmospheric Neutrinos in Super-Kamiokande}
        {116}
\contentitem{Ch. Wiebusch}
        {Neutrino Astronomy with AMANDA}
        {121}
\contentitem{M.E. Moorhead}
        {High Energy Neutrino Astronomy with ANTARES}
        {127}
\contentitem{R. Plaga}
        {Ground-Based Observation of Gamma-Rays (200$\,$GeV--100$\,$TeV)}
        {130}

\subsection*{Cosmology}

\contentitem{C.J. Hogan}
        {Helium Absorption and Cosmic Reionization}
        {139}
\contentitem{K. Jedamzik}
        {Non-Standard Big Bang Nucleosynthesis Scenarios}
        {141}
\contentitem{J.B. Rehm, K. Jedamzik}
        {Big Bang Nucleosynthesis With Small-Scale Matter-Antimatter Domains}
        {147}
\contentitem{M. Bartelmann}
        {Neutrinos and Structure Formation in the Universe}
        {149}

\subsection*{Future Prospects}

\contentitem{P. Meunier}
        {Neutrino Experiments with Cryogenic Detectors}
        {161}
\contentitem{P.F. Smith}
        {OMNIS---A Galactic Supernova Observatory}
        {165}
\contentitem{J. Valle}
        {Neutrinos in Astrophysics}
        {170}
\contentitem{L. Stodolsky}
        {Some Neutrino Events of the 21st Century}
        {178}

\subsection*{Appendix}

\contentitem{\hfill}{Workshop Program and List of Participants}
        {183}
\contentitem{\hfill}{Our Sonderforschungsbereich (SFB) 
        ``Astro-Teilchen-Physik'' and its Divisions}
        {193}

\newpage

{\ }

}


\newpage {


\thispagestyle{empty}
\pagenumbering{arabic}
\begin{flushright}
\Huge\bf
{\ }


History\\

\end{flushright}

\newpage

\thispagestyle{empty}

{\ }

\newpage


}\newpage{

\head{History of Neutrino Physics: Pauli's Letters}
{Rudolf L.\ M\"o\ss bauer}
{Physik Department E15, Technische Universit\"at M\"unchen, 85747 Garching,
 Germany}
       
\subsection*{Editors' Note}
Professor M\"o\ss bauer gave an evening lecture on the history of
neutrino physics during which he read Pauli's famous letters on the
neutrino hypothesis. Rather than writing a formal contribution to our
proceedings, Professor M\"o\ss bauer suggested that we reproduce these
letters.  (They were taken from Ref.~\cite{mossbauer.1} but can also be
found in Ref.~\cite{mossbauer.2}.) Because much of the humour of
Pauli's writing is lost in the translation we quote the original
German text. An English translation can be found, for example, in
Ref.~\cite{mossbauer.3}.

\begin{figure}[b]
\vskip0.4cm
\centerline{\epsfig{file=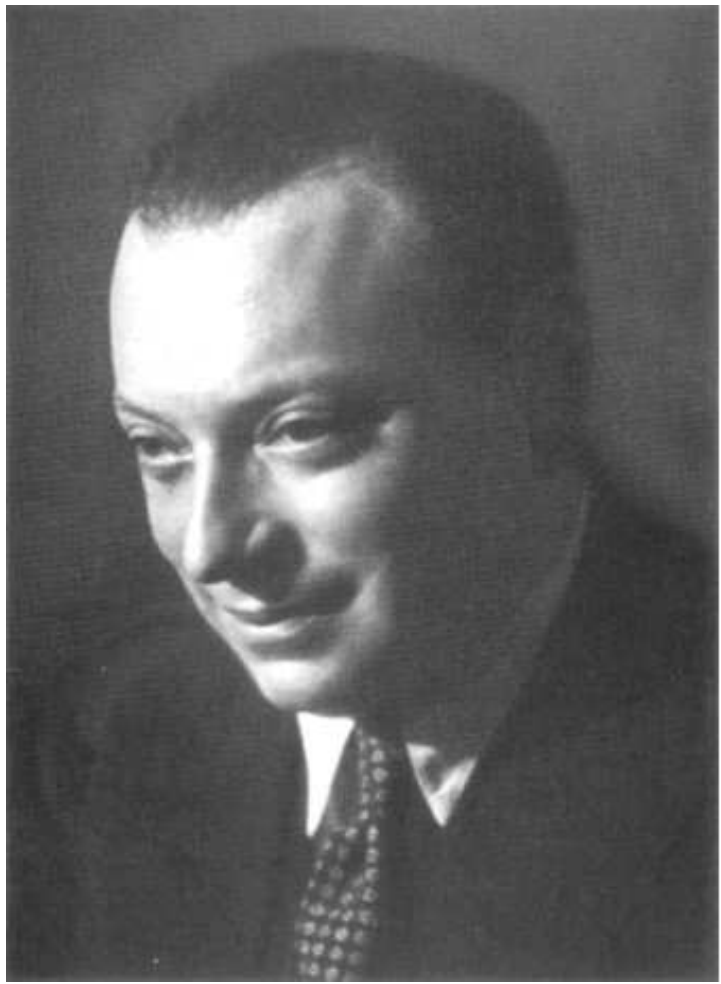,width=9.5cm}}
\caption{Wolfgang Pauli (1900--1958) in Vienna 1933.}
\end{figure}

\subsection*{Pauli's Letters}

\subsubsection*{Brief an Oskar Klein, Stockholm, vom 18.~2.~1929}

\noindent
Aber ich verstehe zu wenig von Experimentalphysik um diese Ansicht
beweisen zu k\"onnen und so ist Bohr in der f\"ur ihn angenehmen Lage,
unter Ausnutzung meiner allgemeinen Hilflosigkeit bei der Diskussion
von Experimenten sich selber und mir unter Berufung auf Cambridger
Autorit\"aten (\"ubrigens ohne Literaturangabe) da etwas beliebiges
vormachen zu k\"onnen.

\subsubsection*{Brief an Oskar Klein, Stockholm, 1929}

Ich selbst bin ziemlich sicher (Heisenberg nicht so unbedingt), da\ss\
$\gamma$-Strahlen die Ursache des kontinuierlichen Spektrums der
$\beta$-Strahlen sein m\"ussen und da\ss\  Bohr mit seinen diesbez\"uglichen
Betrachtungen \"uber eine Verletzung des Energiesatzes auf vollkommen
falscher F\"ahrte ist. Auch glaube ich, da\ss\  die w\"armemessenden
Experimentatoren irgendwie dabei mogeln und die $\gamma$-Strahlen ihnen
nur infolge ihrer Ungeschicklichkeit bisher entgangen sind.

\subsubsection*{Brief an die Gruppe der ``Radioaktiven'' 1930}

\noindent
Physikalisches Institut\\
der Eidg.~Technischen Hochschule\\
Z\"urich  \hfill Z\"urich, 4.~Dez.~1930

\medskip

\noindent
Liebe Radioaktive Damen und Herren!\\
Wie der \"Uberbringer dieser Zeilen, den ich huldvollst anzuh\"oren
bitte, Ihnen des n\"aheren aus\-ein\-an\-der\-setzen wird, bin ich
angesichts der falschen Statistik der N- und Li~6-Kerne, sowie
des kontinuierlichen $\beta$-Spektrums auf einen verzweifelten
Ausweg verfallen, um den Wechselsatz der 
Statistik\footnote{Heute Pauli'sches 
Ausschlie\ss ungsprinzip} 
und
den Energiesatz zu retten. N\"amlich die M\"oglichkeit, es k\"onnten
elektrisch neutrale Teilchen, die ich Neutronen\footnote{Heute Neutrinos} 
nennen will, in den Kernen
existieren, welche den Spin 1/2 haben und das Ausschlie\ss ungsprinzip
befolgen und sich von Lichtquanten au\ss erdem noch dadurch unterscheiden,
da\ss\ sie nicht mit Lichtgeschwindigkeit laufen. ---
Das kontinuierliche $\beta$-Spektrum w\"are dann verst\"andlich unter der
Annahme, da\ss\  beim $\beta$-Zerfall mit dem Elektron jeweils noch ein
Neutron emittiert wird, derart, da\ss\  die Summe der Energien von
Neutron und Elektron konstant ist.

Nun handelt es sich weiter darum, welche Kr\"afte auf die Neutronen wirken.
Das wahr\-schein\-lich\-ste Modell f\"ur das Neutron scheint mir aus
wellenmechanischen Gr\"unden dieses zu sein, da\ss\  das ruhende Neutron
ein magnetischer Dipol von einem gewissen Moment $\mu$ ist. Die Experimente
verlangen wohl, da\ss\  die ionisierende Wirkung eines solchen Neutrons
nicht gr\"o\ss er sein kann als die eines $\gamma$-Strahls, und dann
darf $\mu$ wohl nicht gr\"o\ss er sein als $e\cdot (10^{-13}$~cm).
Ich traue mich vorl\"aufig aber nicht, etwas \"uber diese Idee zu publizieren,
und wende mich erst vertrauensvoll an Euch, liebe Radioaktive, mit der 
Frage, wie es um den experimentellen Nachweis eines solchen Neutrons 
st\"ande, wenn dieses ein ebensolches oder etwa 10mal gr\"o\ss eres
Durchdringungsverm\"ogen besitzen w\"urde wie ein $\gamma$-Strahl. 
$\ldots$

Also, liebe Radioaktive, pr\"ufet, und richtet.---Leider kann ich nicht
pers\"onlich in T\"ubingen erscheinen, da ich infolge eines in der Nacht
vom 6.~zum 7.~Dez. in Z\"urich stattfindenden Balles hier
unabk\"ommlich bin. $\ldots$
Euer untert\"anigster Diener
\hfill W.~Pauli

\subsubsection*{\bf Brief an Oskar Klein, Stockholm, vom 12.~12.~1930}

Ich kann mich vorl\"aufig nicht entschlie\ss en, an ein Versagen des
Energiesatzes ernstlich zu glauben und zwar aus folgenden Gr\"unden
(von denen ich nat\"urlich zugebe, da\ss\ sie nicht absolut zwingend
sind).  Erstens scheint es mir da\ss\ der Erhaltungssatz f\"ur
Energie-Impuls dem f\"ur die Ladung doch sehr weitgehend analog ist
und kann keinen theoretischen Grund daf\"ur sehen, warum letzterer
noch gelten sollte (wie wir es ja empirisch \"uber den $\beta$-Zerfall
wissen) wenn ersterer versagt. Zweitens m\"u\ss te bei einer
Verletzung des Energiesatzes auch mit dem Gewicht etwas sehr
merkw\"urdiges passieren.

\subsubsection*{Telegramm von Reines und Cowan vom 14.~6.~1957 an 
Wolfgang Pauli}

We are happy to inform you that we have definitely detected neutrinos from
fission fragments by observing inverse beta decay of protons. Observed
cross section agrees well with expected $6\cdot 10^{-44}$ cm$^2$.

}\newpage{


\head{What Killed The Dinosaurs?} 
{Arnon Dar$^{1,2}$}
{$^1$Max-Planck-Institut f\"ur Physik
(Werner-Heisenberg-Institut)\\ 
F\"ohringer Ring 6, 80805 M\"unchen, Germany.\\ 
$^2$Department of Physics and Space Research Institute,\\
Technion, Israel Institute of Technology, Haifa 32000, Israel.}

\noindent
The early history of life during the Precambrian until its end
570~million years (My) ago is poorly known. Since then the diversity
of both marine and continental life has increased exponentially.
Analysis of fossil records shows that this diversification was
interrupted by five massive extinctions and some smaller extinction
peaks~\cite{Dar.1}. The largest extinction occurred about 251~My ago
at the end of the Permian period.  The global species extinction
ranged then between 80\% to 95\%, much more than, for instance, the
Cretaceous-Tertiary extinction 64~My ago which killed the dinosaurs
and claimed $\sim 47\%$ of existing genera~\cite{Dar.2}. In spite of
intensive studies it is still not known what caused the mass
extinctions. Many extinction mechanisms have been proposed but no
single mechanism seems to provide a satisfactory explanation of the
complex geological records on mass extinctions~\cite{Dar.3}. These
include terrestrial mechanisms such as intense volcanism, which
coincided only with two major extinctions~\cite{Dar.4} or drastic
changes in sea level, climate and environment that occurred too often,
and astrophysical mechanisms, such as a meteoritic impact that
explains the iridium anomaly which was found at the
Cretaceous/Tertiary boundary~\cite{Dar.5} but has not been found in
any of the other extinctions, supernova explosions~\cite{Dar.6} and
gamma ray bursts~\cite{Dar.7} which do not occur close enough at a
sufficiently high rate to explain the observed rate of mass
extinctions.

The geological records, however, seem to indicate that an accidental
combination of drastic events~\cite{Dar.3} occurred around the times
of the major extinctions.  For instance, the dinosaur extinction
coincides in time with a large meteoritic impact, with a most
intensive volcanic eruption and with a drastic change in sea level and
climate. The origin of these correlations is still unclear.
Meteoritic impacts alone or volcano eruptions alone or sea regression
alone could not have caused all the major mass extinctions.  An impact
of a 10~km wide meteorite with a typical velocity of $30~\rm
km~s^{-1}$ was invoked~\cite{Dar.5} in order to explain the
Cretaceous-Tertiary (K/T) mass extinction 64~My ago, which killed the
dinosaurs, and the iridium anomaly observed at the K/T boundary. But
neither an iridium anomaly nor a large meteoritic crater have been
dated back to 251~My ago, the time of the Permian/Triassic (P/T) mass
extinction, which was the largest known extinction in the history of
life~\cite{Dar.3} where the global species extinction ranged between
80\% to 95\%.  The gigantic Deccan volcanism in India that occurred
around the K/T boundary~\cite{Dar.4} and the gigantic Siberian basalts
flood that occurred around the P/T boundary have ejected approximately
$ 2\times 10^6~\rm km^3$ of lava~\cite{Dar.4}.  They were more than a
thousand times larger than any other known eruption on Earth, making
it unlikely that the other major mass extinctions, which are of a
similar magnitude, were produced by volcanic eruptions. Although there
is no one-to-one correspondence between major mass extinctions, large
volcanic eruptions, large meteoritic impacts, and drastic
environmental changes, there are clear time correlations between them.
We propose that near encounters of Earth with ``visiting planets''
from the outer solar system are responsible for most of the mass
extinctions on planet Earth and can explain both the above
correlations and the detailed geological records on mass
extinctions~\cite{Dar.71}.

Recent observations with the Hubble Space Telescope of the Helix
Nebula (Fig.~1) the nearest planetary nebula, have
discovered~\cite{Dar.8} that the central star is surrounded by a
circumstellar ring of about 3500 gigantic comet-like objects
(``Cometary Knots'') with typical masses about $10^{-5}M_\odot$,
comparable to our solar system planets ($M_{\rm Earth}\approx 3\times
10^{-6}M_\odot$ and $M_{\rm Jupiter}\approx 9.6\times 10^{-4}M_\odot$).
It is not clear whether they contain a solid body or uncollapsed gas.
They are observed at distances comparable to our own Oort cloud of
comets but they seem to be distributed in a planar ring rather than in
a spherical cloud like the Oort cloud. It is possible that these
Cometary Knots have been formed together with the central star since
star formation commonly involves formation of a thin planar disk of
material possessing too high an angular momentum to be drawn into the
nascent star and a much thicker outer ring of material extending out
to several hundred AU. Evidence for this material has been provided by
infrared photometry of young stars and also by direct imaging.

It is possible that such Cometary Knots and the recently discovered
gigantic asteroids~\cite{Dar.9} in the outer solar system between the
Kuiper belt and the Oort cloud are the high mass end of the vastly
more numerous low mass comets. The massive objects are more confined
to the ecliptic plane because of their relatively large masses, and
form a circumstellar ring, while the very light ones are scattered by
gravitational collisions into a spherical Oort cloud. Gravitational
interactions in the ring can change their parking orbits into orbits
which may bring them into the inner solar system. In fact, various
``anomalies'' in the solar planetary system~\cite{Dar.10} could have
resulted from collisions or near encounters with such visitors in the
early solar system. These include~\cite{Dar.11} the formation of the
moon, the large eccentricity and inclination of some planetary orbits,
the retrograde orbits of 6 moons of Jupiter, Uranus and Neptune and
the tilted spin planes of the Sun, the planets and moons relative to
their orbiting~plane.

Strong gravitational tidal forces can cause frictional heating of
planetary interiors leading to strong volcanic activity, as seen, for
instance, on Jupiter's moon Io, the most volcanically active object
known in the solar system. The moon, the Sun and the known planets are
too far away to induce volcanic eruptions on Earth, but relatively
recent ``visits'' of planet-like objects near Earth could have
generated gigantic tidal waves, large volcanic eruptions, drastic
changes in climate and sea level, and impacts of meteorites which were
diverted into a collision course with Earth by the passage through the
astroids and Kuiper belts~\cite{Dar.12}. Thus, visiting planets may
provide a common origin for the diverse mass extinction patterns as
documented in the geological records.
 
Although exact calculations of surface tidal effects are a formidable
scientific effort that was begun by Newton and has continued and
improved since then by many of the great mathematicians and
physicists, an approximate estimate of the flexing ($h$) of Earth
(radius~$R_E$) by a passing planet (distance $d$) can be easily
obtained by neglecting the rotation of Earth and the speed of the
passing planet (mass $M_p$) and by assuming quasi hydrostatic
equilibrium:
\begin{equation} 
h\approx {3\over 4}{M_p\over M_E}\left({R_E\over
d}\right)^3R_E. 
\end{equation} 
The maximal crustal tide due to the moon is 27~cm.  However, a
visiting planet with a typical mass like that of the Cometary
Knots~\cite{Dar.8} which passes near Earth at a distance comparable to
the Earth-Moon distance produces gigantic oceanic and crustal tidal
waves which are a few hundred times higher than those induced by the
moon.  Oceanic tidal waves, more than 1~km high, can flood vast areas
of continental land and devastate sea life and land life near
continental coasts. The spread of ocean waters by the giant tidal wave
over vast areas of land and near the polar caps will enhance
glaciation and sea regression.

Flexing the Earth by $h\sim 100$~m will deposit in it $\sim \alpha
GM_E^2h/R_E^2\approx 10^{34}$~ergs, where $\alpha\sim 0.1$ is a
geometrical factor. It is approximately the heat release within Earth
during $10^6$~y by radioactive decays. The flexing of Earth and the
release of such a large energy in a very short time upon contraction
might have triggered the gigantic volcanic eruptions that produced the
Siberian basalts flood at the time of the P/T extinction and the
Deccan basalts flood at the time of the K/T extinction.

A reliable estimate of the masses and the flux of the visiting
planets/planetesimals is not possible yet. However, we have fixed them
from the assumption that the unaccounted energy source of Jupiter and
its tilted spin plane relative to its orbital plane are both due to
accretion of visiting planets/moons.  From the $3.13^\circ$ tilt of
Jupiter's spin and the accretion rate, we inferred that $N_J\approx
16$ planets of average mass $M_p\approx 0.5M_E$ have crashed into
Jupiter during its $\sim 4.57$~Gy lifetime. Similar estimates for
other planets, although yielding the correct order of magnitude, are
less reliable because the inferred number of accreted planets is too
small. The $7^\circ$ tilt of the spin of the Sun could have been
produced by the impact of $\sim 3\times 10^4$ such planets (ignoring
possible angular momentum loss by the solar wind).  This means that
the Sun has accreted $\sim5\%$ of its mass after its formation, at a
rate of $\sim 7$ planets per My.  In each capture episode, $\sim
6\times 10^{42}$~erg of gravitational energy is released in the Sun's
convective layer. It produces optical and x-ray flashes at a rate
$\sim 7~\times 10^{-6}~L_\odot^{-1}{\rm y}^{-1}$ for Sun-like stars.
It also causes a significant luminosity rise for an extended time
which may have induced climatic and sea level changes on Earth, and
extinctions of species which could not have adapted to large
environmental changes. The predicted rate is consistent with the
observed rate of large changes in O$^{18}$ concentration in sea water
sediments which record large changes in sea water level and total
glacier volume.

Using our inferred planet flux from Jupiter and its collimation by the
Sun, we obtained that a ``visiting rate'' of once every $t_v=100$~My
for planets with $M_p\sim 0.5M_E $ which fall towards the Sun implies
a passing distance of approximately 170,000~km from Earth which
produces crustal tidal waves of $h\geq 125$~m and water tidal waves of
$\sim 1~\rm km$ height.  The combination of tidal waves, volcanic
eruptions, meteoritic impacts and environmental and climatological
changes can explain quite naturally the biological and time patterns
of mass extinctions. For instance, the giant tidal waves devastate
life in the upper oceans layers and on low lands near coastal lines.
They cover large land areas with sea water, spread marine life to dry
on land after water withdrawal, and sweep land life into the sea. They
flood sweet water lakes and rivers with salt water and erode the
continental shores where most sea bed marine life is concentrated.
Amphibians, birds and inland species can probably survive the ocean
tide.  This may explain their survival after the K/T extinction.
Survival at high altitude inland sites may explain the survival of
some inland dinosaurs beyond the K/T border. Volcanic eruptions block
sunlight, deplete the ozone layer, and poison the atmosphere and the
sea with acid rain. Drastic sea level, climatic and environmental
changes inflict further delayed blows to marine and continental life.
But high-land life in fresh water rivers which are fed by springs,
that is not so sensitive to temperature and climatic conditions, has
better chances to survive the tidal waves, the volcano poisoning of
sea water, and the drastic sea level, environmental and climatic
changes.

Altogether, visiting planets offer a simple and testable solution to
the puzzling correlations between mass extinctions, meteoritic
impacts, volcanic eruptions, sea regression and climatic changes, as
documented in the geological records. Perhaps the best test will come
from the MACHO sample of $\sim$100,000 light curves of variable stars:
Planets/moons crashing onto stars produce a big flash of light and
soft x-rays from the hot spot at the impact point on main sequence
(rotating) stars, a nova-like thermonuclear explosion on the surface
of a white dwarf (rare), or a strong gamma ray flash from the surface
of a neutron star (very rare).

\begin{figure}[ht]
  \centerline{\epsffile{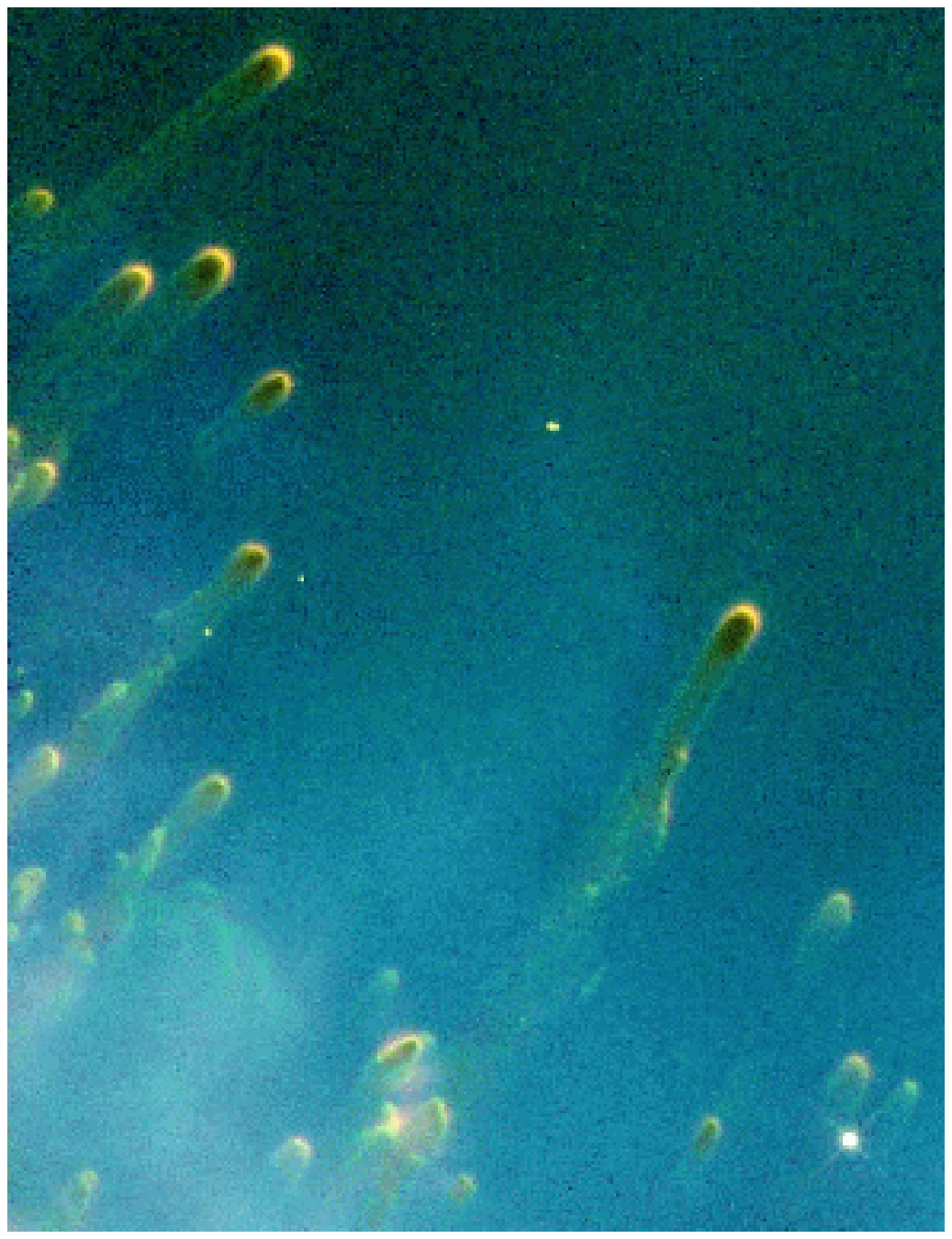}}
  \caption{A small section of a distant ring of gigantic 
   comet-like objects around the central star 
   in the Helix nebula, the nearest planetary nebula, as seen recently
   by the Hubble Space Telescope [9]. These objects have a
   typical mass of $5M_E$. Their gaseous heads have a typical size
   of $10^{15}$~cm. It is not clear whether they contain a solid body
   or uncollapsed gas.}
  \label{Dar.fig1}
\end{figure}

\subsection*{Acknowledgements}
It is a pleasure to thank the organizers of the workshop for their
kind invitation and for an excellent and enjoyable meeting.  Very
exciting and useful discussions with P. Gondolo, G. Raffelt and L.
Stodolsky during and after lunches at MPP, M\"unchen, are greatfully
acknowledged.

}


\newpage {


\thispagestyle{empty}

\begin{flushright}
\Huge\bf
{\ }


Solar\\
\bigskip
Neutrinos

\end{flushright}

\newpage

\thispagestyle{empty}

{\ }

 }\newpage{


\head{Solar Models}
     {M.\ Stix}
     {Kiepenheuer-Institut f\"ur Sonnenphysik, Freiburg, Germany}

\subsection*{Introduction}

A number of solar model calculations has been presented in recent
years, especially in view of the predicted flux of neutrinos from the
Sun~\cite{stix.4, stix.5, stix.6, stix.9, stix.12, stix.20,
stix.27}. In the present contribution I shall discuss mostly the {\it
Standard Solar Model}.  This model, including some refinements, is
confirmed by seismological tests. Non-standard models, on the other
hand, face severe difficulties.  I close with a remark on the Sun's
magnetic field in the neutrino context.

\subsection*{The Standard Model}

The standard solar model is a gas sphere in hydrostatic equilibrium.
This may seem trivial. However, some of the earliest non-standard
models were based on the assumption of a very strong internal magnetic
field, or of rapid rotation of the solar core; both of these
assumptions violate the spherical symmetry as well as the hydrostatic
equilibrium.

The input to the standard model should be to the best of our present
knowledge. Let me begin with the mass, luminosity, and radius,
$$ m_\odot = (1.989 \pm 0.001) \cdot 10^{30} {\rm kg}\ ,\quad L_\odot
   = (3.845 \pm 0.008) \cdot 10^{26} {\rm W}\ , \quad r_\odot = (6.960
   \pm 0.001) \cdot 10^{8} {\rm m} \ .$$ 
The value given for the luminosity is from the ACRIM experiment on the
Solar Maximum Mission, with the error as quoted in \cite{stix.15}.
The radius is for the optical depth $\tau = 2/3$ where the temperature
is equal to the effective temperature $(L_\odot/4\pi\sigma
r_\odot^2)^{1/4}$.

The age of the Sun is found by measuring the decay of long-lived
radioactive isotopes in meteorites,
$$t_\odot = (4.57 \pm 0.05) \cdot 10^9 {\rm years}\ . $$
Wasserburg, in an appendix to \cite{stix.5}, gives an even smaller
error; the main source of uncertainty is the not well-known state of
the formation of the Sun at the time of the melting and
crystallization of the meteoritic material.

\subsubsection*{a)~Chemical Composition}
The composition of the solar surface (and the entire perfectly mixed
outer convection zone) is derived from solar spectroscopy and from
meteoritic data. Recent results \cite{stix.16} show a slightly
decreased content of the heavy elements in comparison to earlier
results, e.g.~\cite{stix.1}. For oxygen, the most abundant element
after H and He, the change is $- 0.06$ in the usual logarithmic scale,
or $\approx - 13\%$.  Altogether, the ratio of the mass fractions $Z$
and $X$ of the heavy elements and hydrogen is 0.0245, instead of the
earlier value 0.0267.

For helium the most accurate determination is based on
helioseismology. This is because both the sound velocity and the
acoustic cut-off frequency depend on the adiabatic
exponent~$\Gamma_1$:
$$c^2 = {P\Gamma_1 \over \rho} \ ,\qquad
  \omega_A = {g\Gamma_1 \over 2c} \ , \qquad
  \Gamma_1 = \biggl({d \ln T \over d \ln P}\biggr)_S \ .$$
In the depth range where an abundant element is partly ionized
$\Gamma_1$ is greatly affected by the energy of ionization. Thus the
He abundance, in particular by its effect in the zone of partial HeII
ionization, has some influence on the eigenfrequencies of the Sun's
pressure (p) modes of oscillation. The inversion of observed
frequencies yields a mass fraction $Y_S = 0.242 \pm 0.003$
(\cite{stix.22}, other authors find similar results).  It should be
noted that the original He abundance, $Y_0$, is adjusted so that the
luminosity of the present model, at age $t_\odot$, equals $L_\odot$.
The result of this procedure is $Y_0 = 0.27 \dots 0.28$, depending on
other input to the model. The difference to the seismically determined
$Y_S$ is appropriate in view of helium settling in the radiative solar
interior, see below.

\subsubsection*{b) Nuclear Reactions} 
Concerning the nuclear reactions I shall concentrate on the pp chains,
which provide $\approx 99\%$ of the energy. The reaction rates, and
therefore the branching between the three chains leading to helium,
depend on temperature (see below), and on the ``astrophysical
$S$-factor'' $S(E)$. This factor depends weakly on the center-of-mass
energy $E$ and measures the cross section after separation of $1/E$
times the penetration probability through the Coulomb barrier. For the
most important reactions Parker \cite{stix.21} reviews results of
$S$-factors (at zero energy):

\bigskip
\begin {tabular}{llr}
\ \ p(p,e$^+\nu$)d \qquad & $S_{11}(0) = 3.89\cdot 10^{-25}$ MeV$\cdot$b& $\pm 1\%$\\
\smallskip
  $^3$He($^3$He,2p)$^4$He \qquad\qquad   & $S_{33}(0) = 5.0$ MeV$\cdot$b& $\pm 6\%$ \\
\smallskip
  $^3$He($^4$He,$\gamma )^7$Be \qquad & $S_{34}(0) = 533$ eV$\cdot$b& $\pm 4\%$\\
\smallskip
$^7$Be(p,$\gamma )^8$B \qquad & $S_{17}(0) = 22.2$ eV$\cdot$b&\qquad $\pm 14\%$ 
\end{tabular}
\medskip

\noindent
It is difficult to assess the errors. $S_{11}$ is so small that it can
only be calculated. The other $S$-factors are measured in the
laboratory but must be extrapolated to zero energy, and a correction
must be applied for electron screening at low energy. In particular
the $14\%$ error of $S_{17}$ has been criticized, e.g.~\cite{stix.17},
as it is derived from diverse experiments with partially contradicting
results in the range 17.9--27.7 eV$\cdot$b. Perhaps better results
will soon become available as the measurements are extended to lower
energy, cf.~the contribution of M.~Junker to these proceedings.

\subsubsection*{c) Equation of State and Opacity} 
It is appropriate to discuss these two important ingredients to the
solar model together, since both depend on the knowledge of the number
densities of the diverse particles, and especially on the ionization
equilibria and the electron density. To first order the Sun consists
of a perfect gas; but significant corrections, at the percent level,
arise e.g. from the electrostatic interaction of the particles
(especially at the depth where abundant species are partially ionized)
and from partial electron degeneracy (in the core),
cf.~\cite{stix.23}.  Modern standard models are usually calculated
with a tabulated equation of state and opacity. The most recent tables
from the Lawrence Livermore Laboratory \cite{stix.18} still exhibit
unexplained discrepancies of up to 20\% as compared to various other
calculations; in the energy-generating region of the Sun the
uncertainty probably is much less, 2.5\% according to \cite{stix.4}.

Generally, the recent opacities \cite{stix.18} are somewhat increased
in comparison to earlier results, e.g. from Los Alamos, mainly because
more elements have been included in the calculations. The increase
renders the radiative transport of energy less effective. The solar
model responds with a slightly raised central temperature
(cf. discussion below), and a slightly lowered base of the outer
convection zone. As for the depth of the convection zone, there are
two other input modifications that may lead to an increase: convective
overshoot (see below) and element diffusion.

\subsubsection*{d) Element Diffusion} 
Driven by the gradients of pressure, temperature, and composition,
helium and the heavy elements diffuse toward the solar center, while
hydrogen diffuses upward. The process is slow, with a characteristic
time exceeding the Sun's age by a factor 100 or more.  Nevertheless
diffusion should be included into the standard solar model since it
produces significant effects, especially in view of the details that
can be seen by helioseismology. The following table summarizes some
results \cite{stix.5}:

\bigskip \begin {tabular}{lllllll}

Diffusion & \ $Y_0$&\ $Y_S$&$r_b/r_\odot$ &$T_c (10^6)$ K & snu(Cl) & snu(Ga)\\
\hline
---             & 0.268  & 0.268 & 0.726  & \ \ 15.56     &\ \ 7.0  &\ \ 126 \\
He              & 0.270  & 0.239 & 0.710  & \ \ 15.70     &\ \ 8.1  &\ \ 130 \\
He \& heavy el. & 0.278  & 0.247 & 0.712  & \ \ 15.84     &\ \ 9.3  &\ \ 137    
\end{tabular}\bigskip

\noindent The obvious effect of a decreased surface mass fraction 
$Y_S$ of He is essentially in agreement with the results of other
authors, and with the helioseismological result (see above); the
increase of $Y_S$ obtained if heavy element diffusion is included is
due to the larger initial helium content, $Y_0$, which is required in
this case for the luminosity adjustment.  Also, the central
temperature rises: Helium diffusion changes the mean molecular weight,
which must be compensated by a higher temperature in order to maintain
the hydrostatic equilibrium; heavy element diffusion increases the
opacity, which must be compensated by a steeper temperature gradient
(and hence larger $T_c$) to maintain the radiative transport of
energy. Together with $T_c$, the neutrino rates predicted for the
chlorine and gallium experiments are increased, as listed in columns
6~and~7 of the table. It should be noted that the $T_c$ effect found
in other calculations is somewhat smaller, cf. the contribution of
Schlattl and Weiss to these proceedings. Also, the effect of helium
diffusion on the depth of the convection zone (column 4) is in
contrast to the result of Cox et~al.~\cite{stix.12}.

The depth of the convection zone has been determined from
helioseismology by several authors. Below the convection zone the
temperature gradient is determined by the requirement of radiative
energy transport; within the convection zone the gradient is nearly
adiabatic because of the large heat capacity and hence large
effectivity of the convective transport. Thus, a prominent transition
occurs in $dT/dr$ and, accordingly, in $dc^2/dr$. As the sound speed
$c(r)$ can be obtained from the p-mode frequencies, the transition can
be located. Christensen-Dalsgaard et~al.~\cite{stix.10} find
$r_b/r_\odot = 0.713 \pm 0.003$, a more recent study \cite{stix.7}
yields an even smaller uncertainty, $0.713 \pm 0.001$, with the claim
that systematic errors are included! --- The temperature at the base
of the convection zone slightly depends on the helium abundance; for
the value of $Y_S$ quoted above a good estimate \cite{stix.10} is
$(2.22 \pm 0.05) \cdot 10^6$~K.

\subsection*{Neutrinos}

The standard model of the Sun predicts the flux of neutrinos, as a
function of neutrino energy, that should be observed on Earth. So far
5 detectors have been used for such observations: One using chlorine,
two using gallium and two using water. The detectors as well as the
observational results are reviewed elsewhere in these
proceedings. Briefly the results are as follows: (1) All the detectors
measure a significant neutrino flux, and so confirm that nuclear
energy generation actually takes place in the Sun. (2) The results of
the two gallium experiments agree with each other, and the results of
the two water experiments agree with each other.  (3) All experiments
measure a flux that lies significantly below the prediction of the
standard solar model.  (4) The deficit depends on neutrino
energy. This last point is of particular interest for the solar model
builder, because it is for this reason that it appears to be
impossible to repair all the deficits simultaneously by modifications
of the solar model.

The significance of the neutrino deficit and its energy dependence has
been illustrated by a calculation of 1000 standard models
\cite{stix.3} based on input parameters having normal distributions
with appropriate means and standard deviations. For the diverse
neutrino experiments these 1000 models predict results that have a
certain spread but are clearly in conflict with the actual
measurements. If the predicted flux of energetic neutrinos originating
from the beta decay of $^8$B is replaced by the value obtained in the
water experiments ($\approx 42\%$ of the predicted), then the
predictions for the chlorine and gallium experiments become smaller
but are still significantly above the measurements. In other words,
the existing experiments cannot simultaneously be reconciled with the
standard solar model.

In two recent calculations \cite{stix.13, stix.14} all the input
parameters have been pushed to the extreme in order to minimize the
neutrino flux; nevertheless the prediction is still significantly too
high.

\subsection*{Non-Standard Models}

The aim of most non-standard solar models is to lower the temperature
in the energy-generating central region and thereby change the
branching ratios of the three pp chains, in particular in order to
suppress a part of the high-energy $^8$B neutrino flux. Castellani
et~al.~\cite{stix.9} discuss in detail the influence of the diverse
input parameters on the central temperature of the Sun. For example, a
45\% increase of $S_{11}$, or a 50\% decrease of $Z/X$, or a 29\%
decrease of the opacity would result in a 4\% smaller central
temperature. For the neutrino fluxes resulting from $^7$Be and $^8$B
the temperature dependence is \cite{stix.2}
$$\Phi_\nu(^7{\rm Be}) \propto T_c^8 \ ,\qquad 
  \Phi_\nu(^8{\rm B}) \propto T_c^{18}\ .$$
Hence, a 4\% cooler solar core reduces the predicted $^8$B neutrino
flux to 0.48 of the standard value, and the flux of $^7$Be neutrinos
to 0.72.  This may help to remove the discrepancy for the water
experiments, and to reduce the discrepancy for the chlorine
experiment, but it provides almost no help for the gallium
experiments. Of the 137 snu (above table, last line) there are only 16
from the decay of $^8$B, but 38 from the electron capture of
$^7$Be. Thus a large discrepancy remains; more specifically, the
gallium experiments appear to leave no room for the predicted $^7$Be
neutrinos. This is the major difficulty of the non-standard solar
models.

Perhaps the neutrino discrepancy will finally be resolved by a
combination of various effects. A slight decrease of the central solar
temperature $T_c$ may be one of these effects, although it is entirely
unclear at present how such a decrease of $T_c$ could be
achieved. Heavy element diffusion {\it increases} $T_c$, as we have
seen. Mixing of the solar core apparently helps, but the mixed-core
model seems to fail the seismological test \cite{stix.28}.  Other
handles, such as the opacity or the equation of state, permit only
variations of $T_c$ that are too small for a substantial effect.

The conclusion is that most of the discrepancy should rather be
resolved by {\it non-standard neutrinos}. The energy-dependent
conversion of electron neutrinos into other neutrino flavours by the
Mikeyev-Smirnov-Wolfenstein effect is a possibility; for a recent
review see \cite{stix.17}.

\subsection*{Convective Overshoot and Magnetism}

Non-local versions of the mixing-length theory allow for overshooting
flows at the base of the solar convection zone \cite{stix.24,
stix.25}. The overshoot layers calculated so far have a nearly
adiabatic temperature gradient.  The thickness of such a layer should
not exceed a few thousand kilometers, according to helioseismological
results \cite{stix.11}. On the other hand, some convective overshoot
will certainly occur and must be included into a calculation that is
designed to exactly reproduce the depth of the convection zone. The
following table, adapted from \cite{stix.25}, gives a few examples of
standard models, calculated without element diffusion.  The input
consists of the opacity, the type of convection formalism (MLT:
mixing-length theory, CM: Canuto \& Mazzitelli), equation of state
(SS: Stix \& Skaley \cite{stix.26}, MHD: Mihalas
et~al.~\cite{stix.19}), initial helium mass fraction $Y_0$,
mixing-length parameter $\alpha$. The results are the relative radius
$r_b/r_\odot$ where convection ceases, the central temperature $T_c$,
and the predicted neutrino rates for the Cl and Ga experiments. There
is very little variation of these neutrino rates.

\bigskip 

\begin{tabular}{rllllllllll}
     \hline
Nr& Opacity   & MLT & EOS  &\quad $Y_0$  & \quad $\alpha$&$r_b/r_\odot$
  &\ \ $T_{c6}$ &   snu (Cl)   & snu (Ga)   \\
     \hline
 1 &Alamos &local &SS     &0.2733 &2.403 &0.7361 &15.56 &\ \ 7.51 &\ 128.3 \\
 2 &Alamos &local &MHD    &0.2731 &2.164 &0.7325 &15.55 &\ \ 7.49 &\ 128.2 \\
 3 &Alamos &nonl. &SS     &0.2733 &2.409 &0.7210 &15.56 &\ \ 7.51 &\ 128.4 \\
 4 &Alamos &nonl. &MHD    &0.2732 &2.168 &0.7201 &15.55 &\ \ 7.49 &\ 128.2 \\
 5 &Opal92 &local &SS     &0.2769 &1.847 &0.7231 &15.56 &\ \ 7.68 &\ 129.4 \\
 6 &Opal92 &local &MHD    &0.2768 &1.677 &0.7194 &15.56 &\ \ 7.66 &\ 129.3 \\
 7 &Opal92 &nonl. &SS     &0.2770 &1.850 &0.7114 &15.57 &\ \ 7.68 &\ 129.4 \\
 8 &Opal92 &nonl. &MHD    &0.2768 &1.679 &0.7096 &15.56 &\ \ 7.66 &\ 129.3 \\
 9 &Opal95 &local &Opal95 &0.2774 &1.660 &0.7220 &15.59 &\ \ 7.81 &\ 130.0 \\
10 &Opal95 &nonl. &Opal95 &0.2774 &1.661 &0.7116 &15.59 &\ \ 7.81 &\ 130.0 \\
11 &Opal95 &CM    &Opal95 &0.2775 &1.033 &0.7218 &15.58 &\ \ 7.80 &\ 129.9 \\
12 &Opal92 &nonl. &MHD    &0.2768 &1.680 &0.7059 &15.56 &\ \ 7.66 &\ 129.3 \\
     \hline
\end{tabular}
\bigskip

\noindent
The overshoot layer appears especially attractive because it is the
only place that can accommodate a toroidal magnetic field of $\approx
10^5$~G, suitable for the 11-year sunspot cycle. A similar field
strength is inferred from the latitude of field emergence at the Sun's
surface, as well as from the tilt angle of bipolar spot groups with
respect to the east-west direction~\cite{stix.8}. With $10^5$~G, and a
layer thickness of, say, 5000 km, a neutrino should have a magnetic
moment of $\approx 10^{-10} \mu_{\rm Bohr}$ in order to suffer a
noticeable spin precession. This would be in conflict with supernova
SN1987A observations which indicate an upper limit of $10^{-12}
\mu_{\rm Bohr}$ for the magnetic moment of the neutrino. The reasoning
may be even stronger as only a fraction of the overshoot layer
normally will be filled with magnetic flux. Matter-enhanced (resonant)
spin flip should also not occur since it requires an electron density
exceeding that at the base of the convection zone.  Therefore, there
appears to be no reason to speculate about a solar-cycle dependence of
the neutrino flux from the Sun.


}\newpage {

\head{Garching Solar Model: Present Status}
     {H. Schlattl, A. Weiss}
     {Max-Planck-Institut f\"ur Astrophysik, 
      Karl Schwarzschild-Str.~1, 85748 Garching, Germany}

\noindent
The Garching solar model code is designed to calculate high precision
solar models.  It allows to control the numerical accuracy and has the
best available input physics implemented~\cite{schlattl.1}. It
uses the OPAL-equation of state \cite{schlattl.2} and for the
opacities those of~\cite{schlattl.3} complemented by~\cite{schlattl.4} in
the low-temperature regions. Pre-main sequence evolution is also taken
into account. The microscopic diffusion of hydrogen, helium, the
isotopes participating in the CNO-cycle and some additional metals
(Ne, Mg, Si) is incorporated following the description
of~\cite{schlattl.5} for the diffusion constants. The nuclear reaction
rates were taken from~\cite{schlattl.6}.

In this work we want to emphasize the sensitivity of the structure of
a solar model on the interpolation technique used for the opacity
tables. As the run of opacities ($\kappa$) with temperature ($T$)
and/or density ($\rho$) may show very rapidly changing gradients,
choosing a suitable interpolation procedure is not trivial. Our
program uses two-dimensional bi-rational cubic
splines~\cite{schlattl.8} to interpolate in the $T$-$R$-grid of the
opacity tables ($R=\rho/T_6^3,\; T_6 = T/10^6 {\rm K}$). Apart from
the general problem to choose suitable outer boundary conditions for
the calculation of the spline functions one can introduce an
additional parameter to avoid artificial unphysical oscillations, as
they are typical in the case of cubic splines and rapidly varying
slopes.  Increasing this \emph{damping parameter} leads to almost
linear interpolation just between two grid points and to a very
rapidly changing gradient at the grid points themselves. The higher
the damping parameter the more extended gets the linear region (a
value of $0$ corresponds to standard cubic splines).
\begin{figure}[h]
\centerline{\epsfxsize=.9\textwidth\epsffile{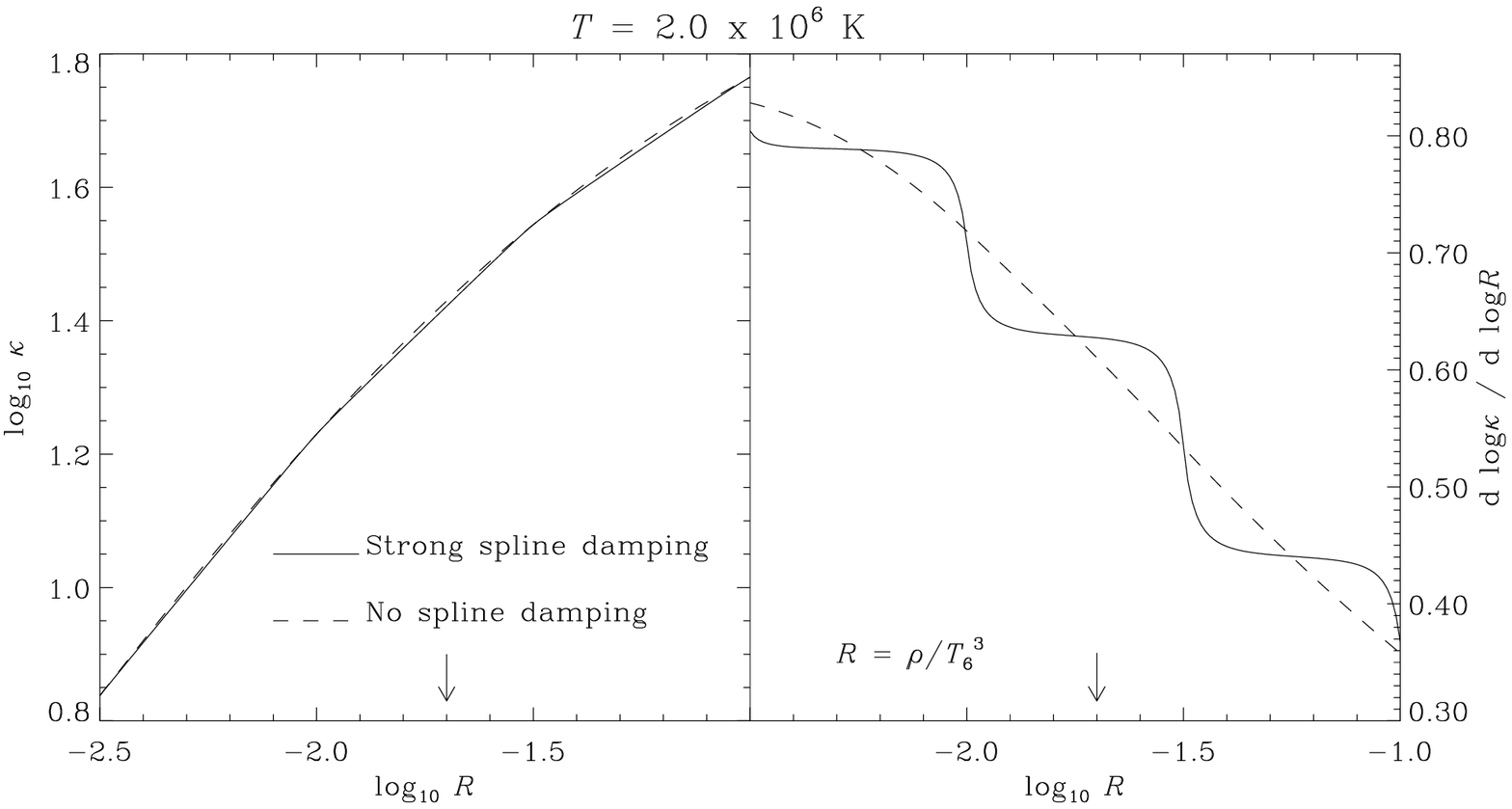}}
\caption{The run of $\kappa$ and ${\rm d log}\kappa / {\rm d log}R$ 
with $R$ at fixed temperature of 2 million Kelvin using pure cubic
splines (dashed line) and damped cubic splines (full line). The arrows
mark the value of $R$ at this temperature in a typical solar model.}
\label{schlattl.fig1}
\end{figure}
The disadvantage of this damping is that in regions where the opacity
has a very slowly changing gradient too high values for the damping
parameter lead to an interpolated run of opacity with $T$ and/or $R$
which shows step-like gradient variations.  This is illustrated in
Fig.~\ref{schlattl.fig1} where the full line shows the interpolated
$\kappa(R)$ using a high damping parameter and the dashed line the run
of $\kappa$ without damping. In this case a simple cubic spline seems
to describe a better fit. The greatest deviations of $\kappa$ between
two tabulated values is only about 3\%, which is smaller than the
quoted uncertainty for the opacities of about 10\%.  We would like to
note here that neither of the chosen interpolation schemes can claim
to reproduce the true values, as interpolation is always an
estimation.

To illustrate the influence of the opacity interpolation, two solar
models were calculated, GARSOM3 with no and GARSOM2 with strong spline
damping.  The run of sound speed of both models is compared with a
seismic mode from~\cite{schlattl.7} in Fig.~\ref{schlattl.fig2}. The
deviation of GARSOM2 (dashed line) from the seismic model just below
the convective zone is about twice as large as compared with GARSOM3
(full line). At $r = 0.67\; R_\odot$ the temperature is approximately
2 million Kelvin, ${\rm log_{10}} R \approx -1.7$. Regarding
Fig.~\ref{schlattl.fig1} one notices that at this radius the model is
in $T$-$R$-regions of the opacity tables where the different
interpolations schemes differ most.

\begin{figure}[ht]
\centerline{\epsfxsize=.7\textwidth\epsffile{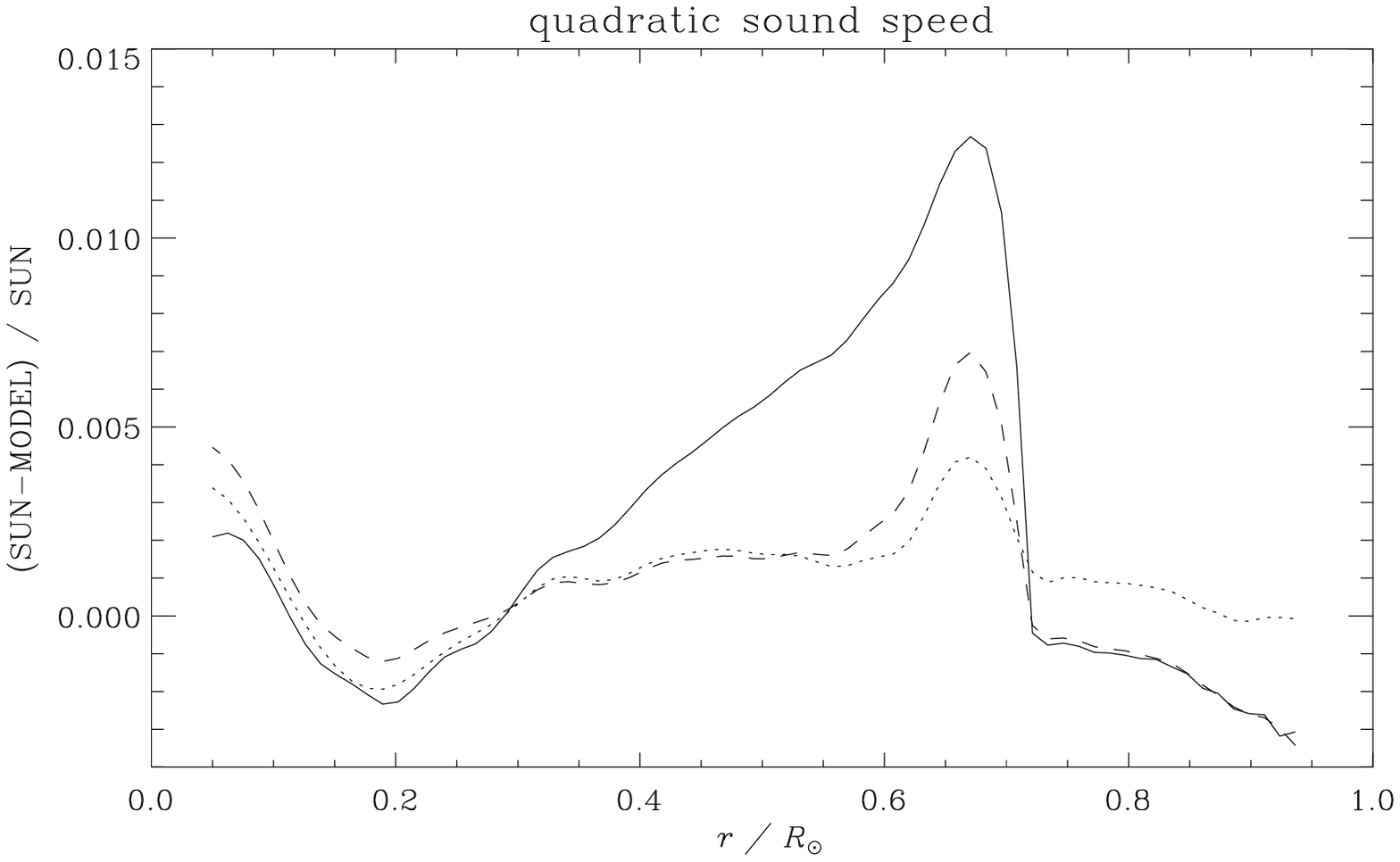}}
\caption{The run of quadratic sound speed of various models compared 
to seismic model from [7]. Full line: GARSOM2, dashed line: GARSOM3,
dotted line: reference model from which the seismic model is
inferred.}  \label{schlattl.fig2}
\end{figure}

As the possible error in the tabulated opacities (10\%) is bigger than
the difference due to interpolation (3\%), neither of the models can
be favoured from the theoretical point of view. Although we must admit
that $\kappa(R,T)$ inferred from cubic splines without damping does
look more reliable, we can not really rule out GARSOM2. It is
therefore necessary to improve the grid density in the tables and the
input physics in the opacities.

To show that GARSOM3 is compatible with solar models from other
groups, we have plotted in Fig.~\ref{schlattl.fig2} also the run of
sound speed of the reference model which was used to infer the seismic
model of \cite{schlattl.7}. The remaining discrepancies between the
reference model and GARSOM3 may be due to slightly different ages or
nuclear reaction rates.

\newpage

\subsection*{Acknowledgments}

We would like to thank J{\o}rgen Christensen-Dalsgaard who helped to
detect the sensitivity of opacity to interpolation. Additionally we
acknowledge Sarbani Basu who provided us her seismic model.

 }\newpage{

\newcommand{\G}{{\sc Gal\-lex}}
\newcommand{\Gno}{{\sc Gno}}
\newcommand{\Sag}{{\sc Sage}}
\newcommand{\as}{$\rm {^{71}As}$}
\newcommand{\ger}{$\rm {^{71}Ge}$}
\newcommand{\gal}{$\rm {^{71}Ga}$}
\newcommand{\chr}{$\rm {^{51}Cr}$}
\newcommand{\ber}{$\rm {^{7}Be}$}
\newcommand{\gaeinfang}{$\rm {^{71}Ga} + \nu_e \to {^{71}Ge} + e^-\,$}
\newcommand{\gezerfall}{$\rm {^{71}Ge} + e^- \to {^{71}Ga} + \nu_e\,$}
\newcommand{\gacl}{$\rm GaCl_3$}

\head{Status of the Radiochemical Gallium Solar Neutrino Experiments}
     {Michael Altmann}
     {Physik Department E15, Technische Universit\"at M\"unchen, D--85747
Garching \\
        and Sonderforschungsbereich 375 Astro-Particle Physics}

\noindent
With the successful completion of \G \ after six years of operation and the
smooth transition to \Gno \ a milestone in radiochemical solar neutrino
recording has been reached.
The results from \G , $77 \pm 6 \pm 5\,$SNU, 
and \Sag , $74 \,\,^{+11}_{-10} \,\, ^{+5}_{-7}\,$SNU,   
both being significantly below all solar model predictions, confirm  the
long standing solar neutrino puzzle and constitute an indication for
non-standard neutrino properties.
This conclusion is validated by the results of \chr \ neutrino
source experiments which have been performed by both collaborations 
and  \as \ doping tests done in \G .

\subsection*{GALLEX and SAGE: Radiochemical Solar Neutrino Recording}

The radiochemical gallium detectors, \G \ and \Sag , 
have been measuring the integral solar neutrino
flux exploiting the capture reaction \gaeinfang . 
The energy threshold being only $233\,$keV, this reaction allows to
detect the pp-neutrinos from the initial solar fusion step which
contribute about $90\,$\% to the total solar neutrino flux.

In a typical run the target,
consisting of $30\,$tons of gallium in the form of
$101\,$t \gacl \ solution for \G \ and $55\,$t of metallic gallium for \Sag ,
is exposed to the solar neutrino flux for 3-4 weeks.
In the following I will mainly focus on \G , 
as for \G \ and \Sag\ the experimental 
procedure -- apart from the chemical extraction
of the neutrino produced \ger \
and the stable germanium carrier which is added at the beginning of each run --
is rather similar.
Both experiments use the signature provided by the
Auger electrons and X-rays associated with the decay \gezerfall \ for
identification of \ger \ during a several months counting time.
Referring to \cite{altmann.gal92,altmann.hen92} 
for a detailed description of the detector 
setup and experimental procedure I concisely summarize 
the \G \ experimental program in table \ref{altmann.tab1}.

\begin{table}[hbt]
\begin{center}
\caption{\small \label{altmann.tab1}
GALLEX experimental program. It comprises 4 periods of solar neutrino
observations (Gallex 1 -- Gallex 4), two chromium neutrino source experiments
(Source I and Source II) and the arsenic test.
}
\smallskip
\noindent
\begin{tabular}{|c|c|c|c|}\hline 
\small date & \small exposure period &\small number of runs &\small result \cr
\hline
\small 5/91--4/92 & 
\small Gallex I  & \small 15 solar $+$ 5 blank & 
      \small    $ 81 \pm 17 \pm 9\, $SNU \cr
\small  8/92--6/94 & 
\small  Gallex II  & \small  24 solar  $+$ 22 blank & 
        \small          $75 \pm 10 \,^{+4}_{-5}\,$SNU \cr
\small  6/94--10/94 & 
\small  Source I ($\rm {^{51}Cr}$) & \small  11 source runs& 
        \small          $R = 1.01 \,\, ^{+0.11}_{-0.10} $ \cr
\small  10/94--10/95 & 
\small  Gallex III  & \small  14 solar  $+$ 4 blank & 
        \small  $54 \pm 11 \pm 3\, $SNU \cr
\small  10/95--9/96 & 
\small  Source II ($\rm {^{51}Cr}$) & \small  7 source runs & 
        \small  $R = 0.84 \,\, ^{+0.12}_{-0.11} $ \cr
\small  9/96--1/97 & 
\small  Gallex IV  & \small  12 solar  $+$ 5 blank & 
        \small          $118 \pm 19 \pm 8\,$SNU \cr
\small   1/97--3/97 & 
\small  $\rm {^{71}As}$-test & \small  4 arsenic runs & 
        \small          $Y = 1.00 \pm 0.03 $ \cr
\hline
\end{tabular}
\end{center}
\end{table}

\noindent
The combined result of all 65 \G \ solar runs  is 
$76.4 \pm 6.3 \,\, ^{+4.5}_{-4.9}\,$SNU. We note, however, 
that for the Gallex-IV period pulse shape information is used only for K-peak
signals.
The present result from \Sag , $74 \,\,^{+11}_{-10} \,\, ^{+5}_{-7}\,$SNU   
\cite{altmann.sag97},  is in perfect agreement with \G .

The overall results from both experiments 
are only about 60\% of the predictions from solar model
calculations, which constitutes an indication for non-standard neutrino
properties, even without considering the results from
other solar neutrino experiments.

\subsection*{The $^{51}$Cr neutrino source experiments}
In order to validate this conclusion and check the reliability and efficiency
of their detectors, both collaborations have prepared  intense 
($\approx 2\,$MCi (\G ) and $517\,$kCi (\Sag )) \chr \ neutrino
sources by neutron irradiating isotopically enriched Cr \cite{altmann.cri96}. 
\chr \ decays by EC and emits neutrinos of energy $750\,$keV (decay to g.s.,
90\%) and $430\,$keV (decay to $320\,$keV excited level, 10\%), nicely
accommodating the neutrino energies of the solar $\rm {^7Be}$ branch.

For \G , the source has been inserted in the central tube of the target
tank. Altogether 18 extractions have been performed with the source in place,
divided into two series of measurements with the source being re-activated
in between.
Figure \ref{altmann.fig1} shows  the individual run results.

\begin{figure}[hbt]
\vskip 10truecm
\vbox   {                               
\hbox to \hsize{                        
\hskip -0.6 truecm                      
\vbox {\hsize=\hsize
            \hfil\break
            \noindent
            \hskip 0truemm
                \includegraphics{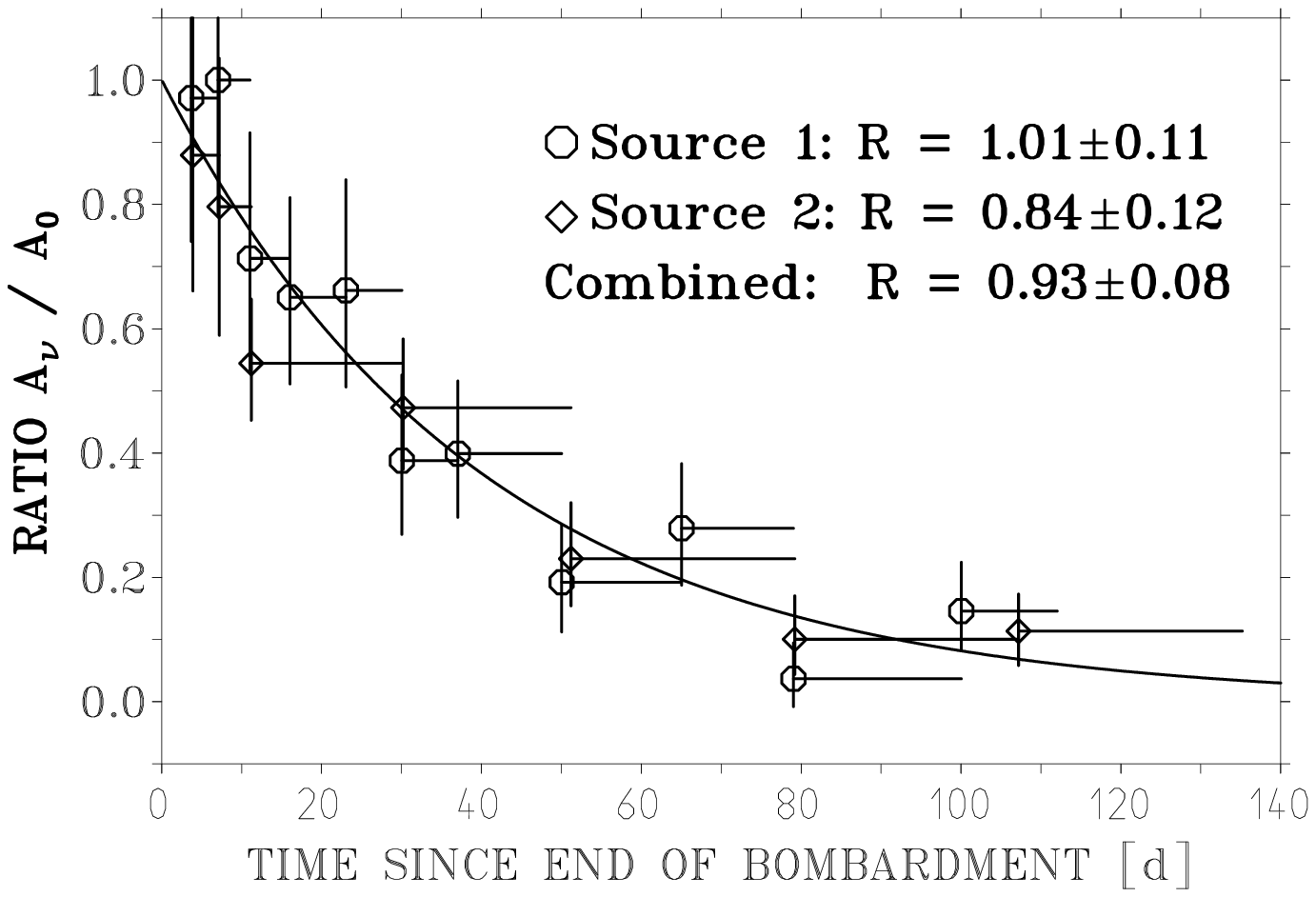}
           } \hss
\hfil          }                
\vskip -1.5truecm
        }
\caption        {\label{altmann.fig1} 
Individual run results of the GALLEX chromium neutrino source experiments, 
normalized to the
known source activity, as a function of time after the end of neutron
irradiation of the source. 
The horizontal bars indicate the durations of the exposures, the
exponential curve the expectation from the decay of the source. 
Source I and Source II runs are represented by  circles and
diamonds, respectively.
        }
\end{figure}

\smallskip
\noindent
An analysis of Source I and Source II 
yields $R = 1.01 \,\, ^{+0.11}_{-0.10}$  and 
$R = 0.84 \,\, ^{+0.12}_{-0.11}$, respectively, where $R$ is
the source activity as deduced from the neutrino measurement normalized to the
known true source activity.
The combined analysis of both series results in $R = 0.93 \pm 0.08$, 
clearly demonstrating the absence of large systematic errors
which could account for the observed 40\% solar neutrino deficit.

A similar experiment 
has also been performed by the \Sag \ collaboration.
They got a quantitative recovery of $R = 0.95\pm
0.12$ \cite{altmann.abd96}.

The fact that both experiments, though employing different chemistry,
demonstrated in these experiments full efficiency, proves the trustworthiness of
the radiochemical method and shows that the 40\% solar neutrino
deficit observed in the radiochemical gallium experiments 
cannot be ascribed to unknown systematic errors. 
In particular, the \chr\ neutrino energies nicely accommodating those from the
solar \ber -branch, the full efficiency of the gallium experiments to \ber
-neutrinos is demonstrated.

\subsection*{$^{71}$As experiments}

However, though the \chr \ source in \G \ 
did outperform the sun by more than a factor
$ 15$ after insertion into the target tank, the experiments still are low
statistics, involving only several dozens of neutrino produced \ger \ atoms.
Therefore, at the very end of \G \ the collaboration has 
performed a large-scale test
of potential effects of hot chemistry, which might lead to a different chemical
behaviour of \ger \ produced in a nuclear reaction compared to 
the stable Ge carrier isotope: The in-situ production of \ger \ by
$\beta-$decay of \as . A known quantity of \as \ (O($10^5$) atoms) has been
added to the tank (t-sample), 
where it decayed with $T_{1/2} = 2.9\,$d to \ger .
Four runs have been made under different operating conditions 
(mixing, carrier addition, standing time), cf.\ table \ref{altmann.tab2}.
For every spike a reference sample (e-sample) was kept aside, making possible
to calculate the ratio of t- and e-sample which does not suffer from 
most of the systematic uncertainties associated with \ger -counting.

\begin{table}[hbt]
\begin{center}
\caption {\label{altmann.tab2}
        Experimental conditions and results (ratio t-sample/e-sample) 
        of the \as \ runs.
        }
\smallskip
\begin{tabular}{|c|c|c|r|c|}\hline
\small run & \small mixing conditions & \small Ge-carrier & \small standing &
        \small result \cr
          & \small $\rm [h \times m^3/h]$ &\small addition  & \small time & 
                (tank / external) \cr
\hline
\small A1 & \small $\rm 22 \times 5.5$
        & \small with As        & \small $19.9\,$d      
                & \small $1.01 \pm 0.03$ \cr 
          & \small $\rm + 0.17 \times 60$ & & & \cr
\small A2 & \small $\rm 6 \times 5.5$   & \small no Ge-carrier
        & \small $19.9\,$d      & \small $1.00 \pm 0.03$ \cr 
\small B3-1 & \small $\rm 24 \times 5.5$ & \small after As      
                & \small $2.0\,$d       & \small $1.01 \pm 0.03$ \cr 
\small B3-2 & ---       & \small after As       & 
        \small $22.0\,$d        & \small $1.00 \pm 0.03$ \cr 
\hline
\end{tabular}
\end{center}
\end{table}

\noindent
In all cases a quantitative recovery of 100\% was achieved. 
This demonstrates on a 3\%-level the
absence of withholding effects even under unfavourable
conditions like carrier-free operation\footnote{The stable germanium 
carrier is not only used for determination of the extraction yield, but also
plays the role of an 'insurance' to saturate potential trace impurities which
might capture $^{71}$Ge in non-volatile complexes.}.

\subsection*{Gallium Neutrino Observatory}
With the \as -tests \G \ has completed its large-scale experimental program.
However, solar neutrino measurements with a gallium target at Gran Sasso 
will be re-commenced in spring 1998 in the frame of the Gallium Neutrino
Observatory (GNO) \cite{altmann.gno96} 
which is designed for long-term operation
covering at least one solar cycle. 
In its first phase GNO will use the $30\,$ton target of \G . 
However, for a
second phase it is planned to increase the target mass to $60\,$tons and 
later to $100\,$tons.
In addition, as it is equally important to decrease the systematic uncertainty,
effort is made to improve \ger \ counting, both by improving the
presently used proportional counters, 
and by investigating novel techniques like semiconductor
devices and cryogenic detectors \cite{altmann.alt97}.

\subsection*{Acknowledgements}
Our contribution to \G \ and \Gno \ is supported by grants from 
the german BMBF, the SFB-375, and the Beschleunigerlaboratorium Garching.

%

}\newpage {

\head{Solar Neutrino Observation with Superkamiokande}
     {Yoshiyuki Fukuda  (for the Superkamiokande Collaboration)}
     {Institute Cosmic Ray Research, University of Tokyo, Japan}

\subsection*{Introduction}

Superkamiokande, which is a second generation solar neutrino
experiment, is an imaging water Cherenkov detector with 50,000 tons of
pure water in the main tank. The detector is located 1000 m
underground (2700 m water equivalent) in Kamioka Zinc mine in the Gifu
prefecture of Japan, at 36.4$^{\circ}$N, 137.3$^{\circ}$E and
25.8$^{\circ}$N geomagnetic latitude.

\begin{figure}[ht]
\centerline{\epsfxsize=0.50\textwidth\epsffile{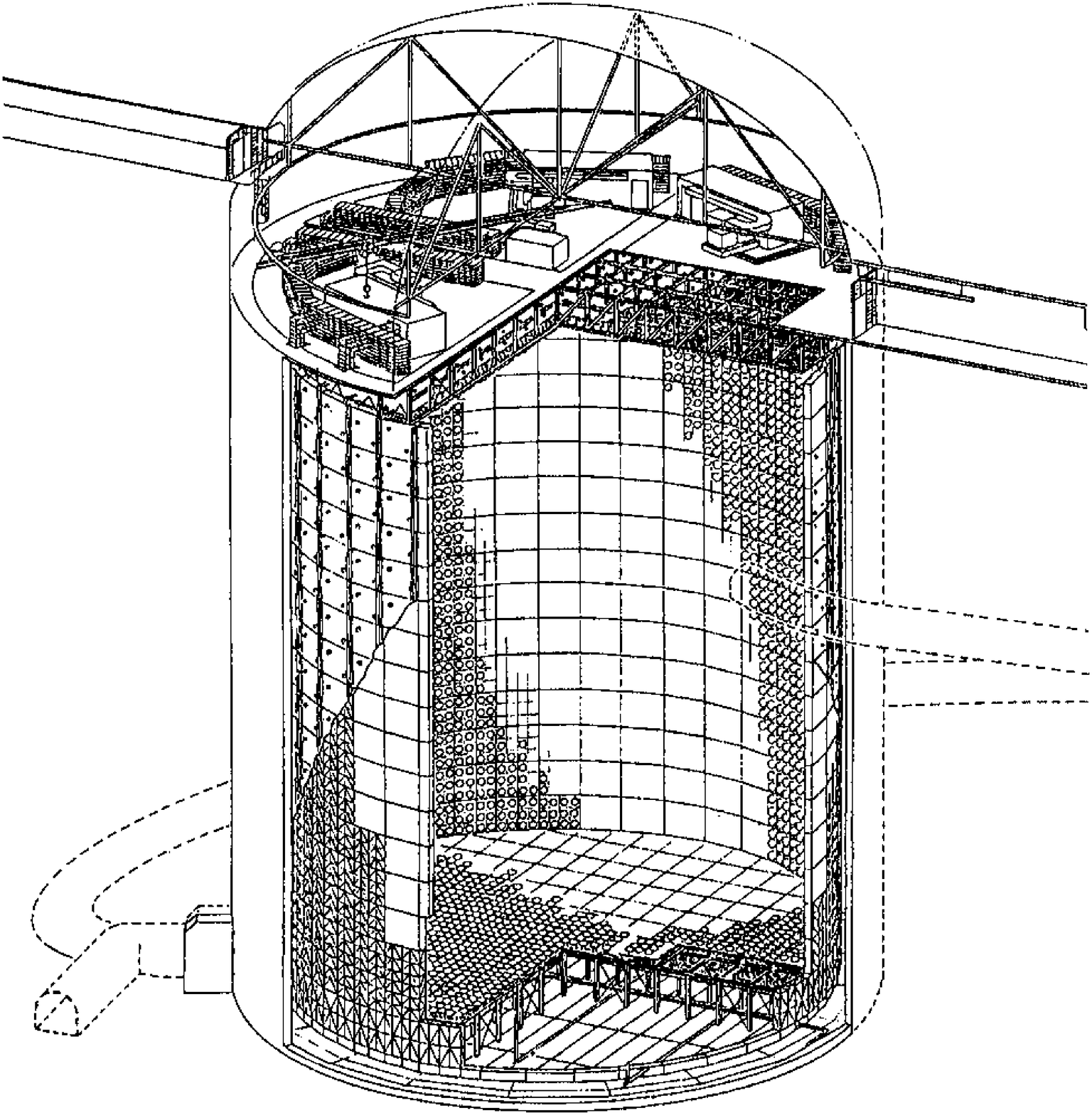}}
\caption{Schematic view of the Superkamiokande detector.}
\label{detector}
\end{figure}

The detector consists of a main inner counter and an outer
anti-counter. A schematic view of the detector is shown in
Fig.~\ref{detector}.  The inner counter is contained in a cylindrical
stainless-steel tank and has a volume of 39.3 m in diameter $\times$
42.0 m in height, containing 50,000 metric tons of water. A total
11,146 photo multipliers (PMTs) with 20 inch $\phi$ photo cathode area
cover 40\% of the entire inner surface of the tank. The fiducial mass
for the solar neutrino measurement is 22,500 tons, with boundaries 2.0
m from the inner surface. On the other hand, a 4$\pi$ solid-angle
anti-counter surrounding the inner counter is also a water
\v{C}erenkov counter of total mass 13,000 metric tons with 1850 PMTs
to detect any signals coming from outside of the detector and to
shield against gamma-rays and neutrons.

\subsection*{Calibrations}

The timing calibration for all PMTs is done by the Xe lamp. We usually
take those data for every 3 months and the maps of T-Q response for
all PMTs are produced. This table is used for correct timing as a
function of observed charge in real data analysis.

The variation of the water transparency is obtained by stopping muon
spectrum (Michael spectrum). This is very important for obtaining the
energy scale to be stable as a function of time.  The energy is almost
proportional to the number of hitted PMTs, however, the number maybe
variable to the attenuation length of the water.  Corrected number of
hitted PMTs is very stable to the variation of water
transparency. This is also confirmed by the $\gamma$-ray source which
is emitted by the reaction of Ni(n,$\gamma$)Ni$^{*}$. The absolute
values of water transparency at various points of wavelength are
measured by DYE-laser calibration. Recent water transparency itself is
very stable.

The performances of the detector, such as the absolute energy scale,
the energy resolution, the vertex resolution and the angular
resolution are mainly calibrated by LINAC system. The LINAC can
generate an electron beam with 5 MeV to 16 MeV. This beam is induced
via beam pipe and bent by magnetic coil into the several positions of
the tank. The typical values for each resolution is 16\%, 70cm and 22
degree for 10 MeV electrons. The absolute energy for electron beam is
calibrated by Ge solid state detector at each calibration time.

Monte Carlo simulation was tuned by the water transparency and other
parameters to reproduce the energy scale in various tank positions
within 1\% difference between MC and real data.

\subsection*{Solar neutrino analysis}

Superkamiokande has started from April 1996 and now processed about
400 days data for solar neutrinos measurement.  Analysis procedures
are (1) noise reduction, (2) vertex reconstruction and tight noise
reduction, (3) spallation products cut, (4) fiducial volume (22.5
kton) cut and (5) gamma ray cut. Some of analysis techniques are
similar to Kamiokande's ones, however, most of them have been newly
developed. In first step, we eliminate $\mu$ and decayed electron and
several electronics noise. The second step reconstructs the vertex
using the timing of hitted PMTs within selected 50 n second window.
Main backgrounds in the residual data are spallation products. In
order to eliminate those events, we used the likelihood method using
the time difference and the distance between induced muon and those
spallation events as a function of muon energy. We can reject most of
spallation products with 20\% dead time.  In the last reduction, we
reject external $\gamma$-rays coming from outside of the
detector. Most of those events are sitting at very close to the edge
of fiducial volume and have an opposite direction with respect to the
detector wall.  Event rate of the final sample is obtained by 175
events per day per 22.5 kton for 6.5 to 20 MeV.

\begin{figure}[ht]
\centerline     {
  \epsfxsize=0.45\textwidth \epsffile{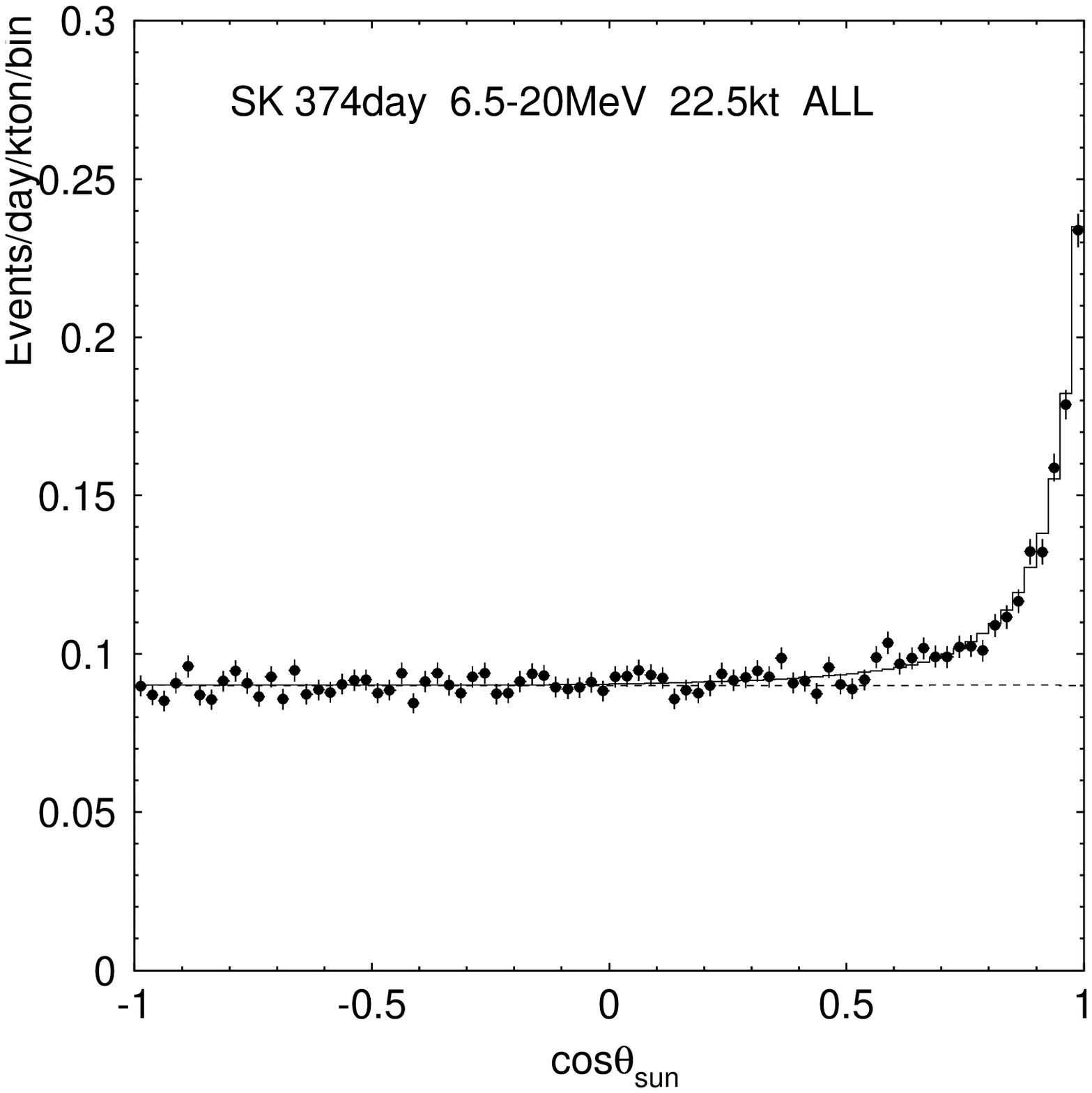}
  \epsfxsize=0.42\textwidth \epsffile{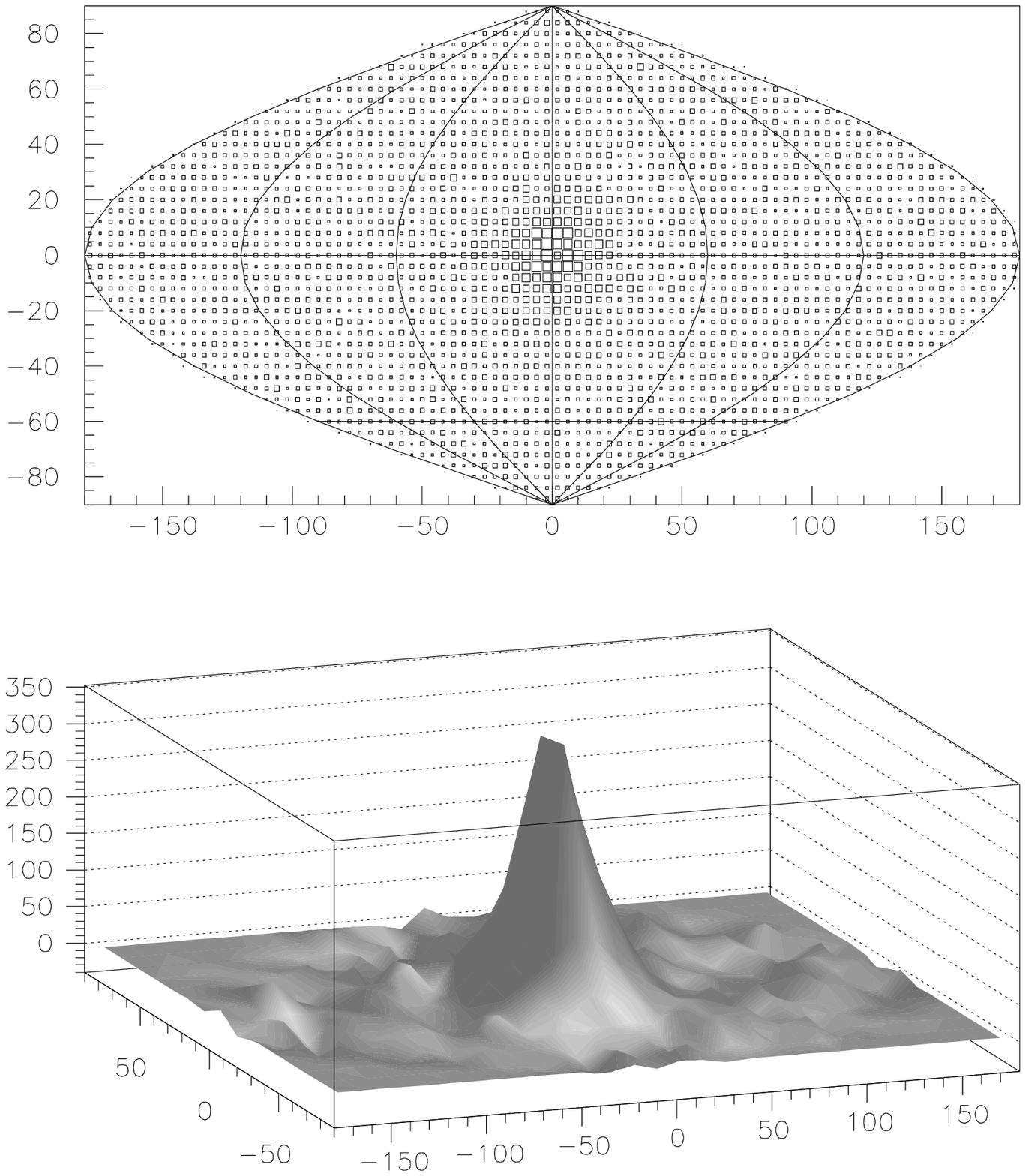}
                }
\caption{(a)~Angular distribution to the solar direction and 
(b)~heliograph for 374.2 days data.}
\label{soldir}
\end{figure}

From 1st May 1996 to 22 Oct 1997, we obtained 374.2 days data from
Superkamiokande measurement~\cite{fukuda.fl}.  Figure~\ref{soldir}
shows the angular distribution to the solar direction and the
heliograph for obtained final sample. In Fig.~\ref{soldir}(a), best
fit line which is expected by MC is also shown. Extracted number of
solar neutrinos is obtained by this fit as $4951 ^{+117.9}_{-111.3}$
for 374.2 days data. Observed $^{8}$B solar neutrino flux is given by
\begin{displaymath} 
\phi(^{8}B) = 2.37^{+0.06}_{-0.05}({\rm stat.})^{+0.09}_{-0.07}
 ({\rm syst.}) \times  10^{6}\,{\rm cm^{-2} sec^{-1}}
\end{displaymath}
or by taking ratio to the BP95 flux~\cite{fukuda.bp95};
\begin{displaymath}
\frac{\rm Data}{\rm SSM_{BP95}} 
= 0.358^{+0.009}_{-0.008}({\rm stat.})^{+0.014}_{-0.010}({\rm syst.}).
\end{displaymath}

\begin{figure}[ht]
\centerline     {
  \epsfxsize=0.47\textwidth \epsffile{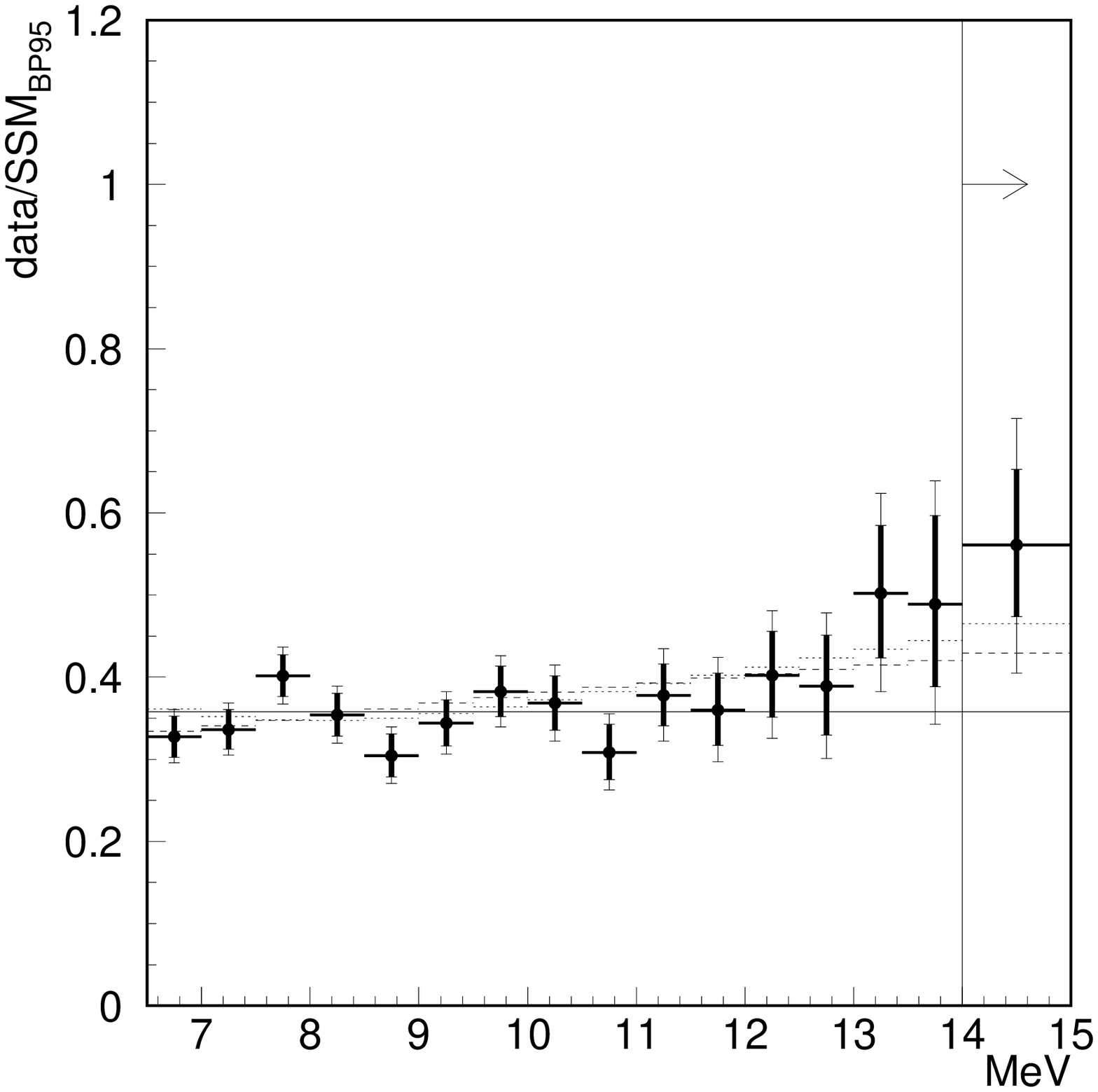}
  \epsfxsize=0.47\textwidth \epsffile{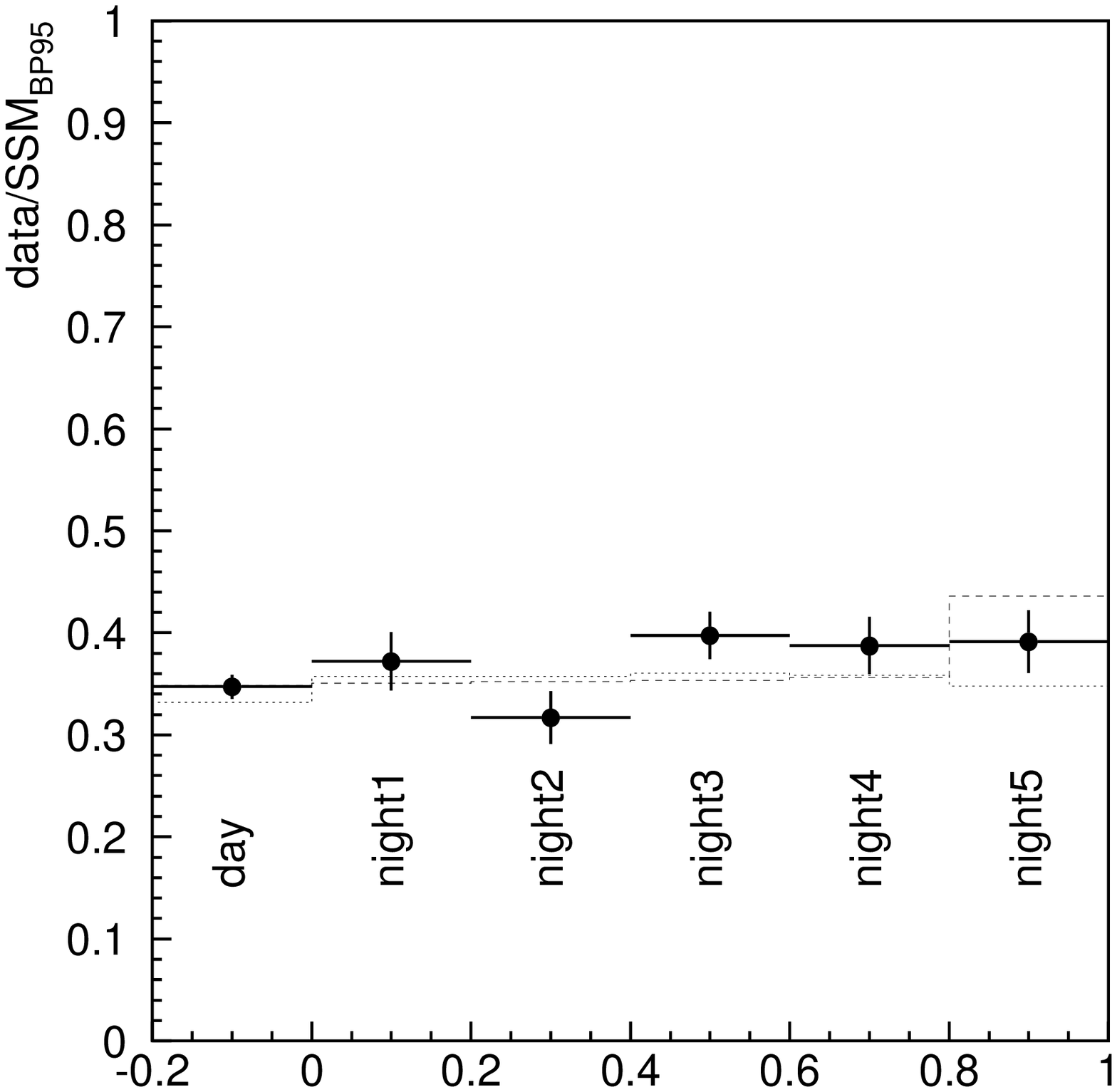}    
                }
\centerline     {  
  \epsfxsize=0.47\textwidth \epsffile{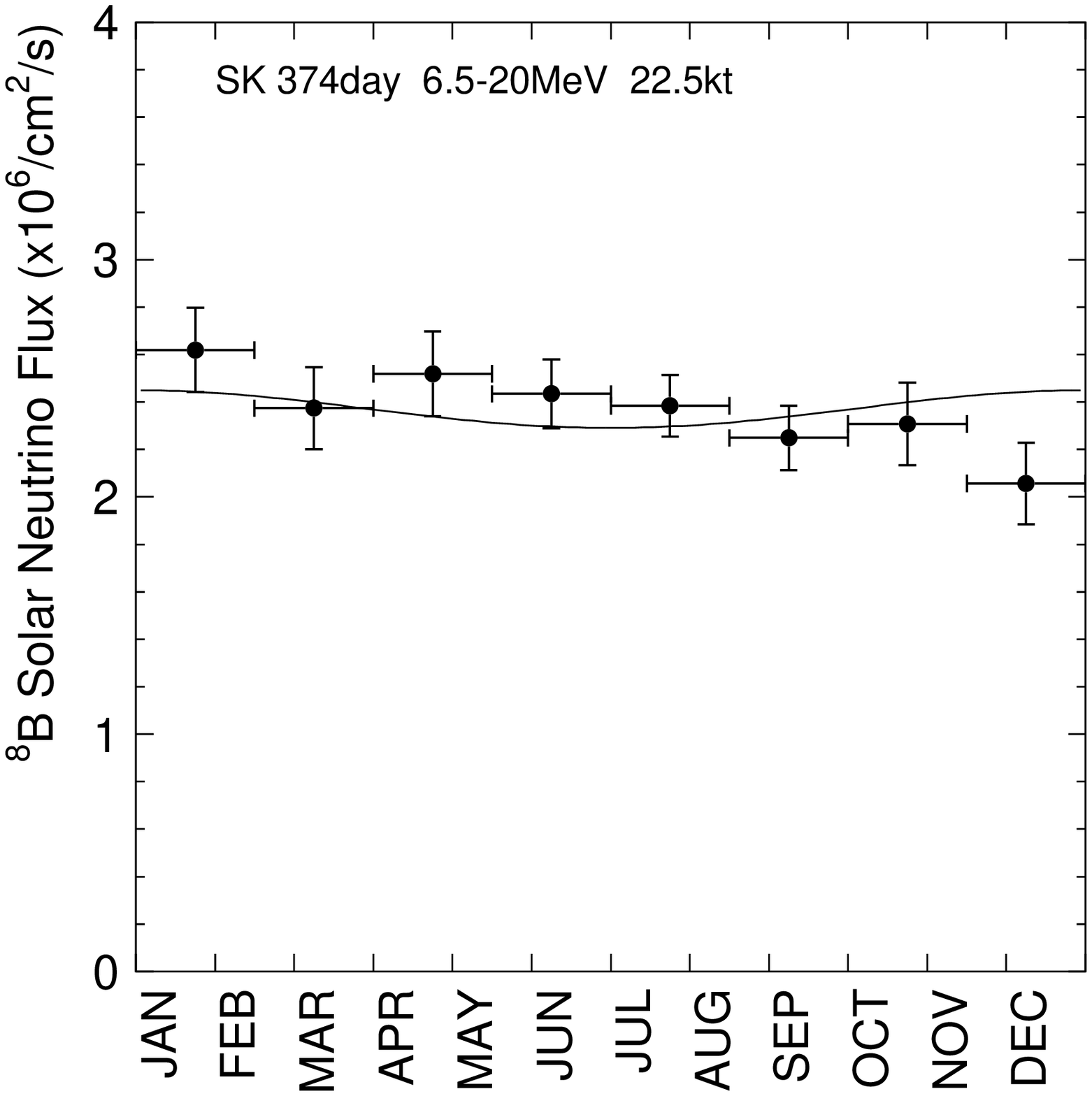}
                }
\caption{(a)~Energy spectrum, (b)~Day/Night flux variation and 
(c)~seasonal variation of obtained solar neutrino signal.}
\label{spectrum}
\end{figure}

The systematic error related to the energy scale has been reduced by
LINAC calibration and is obtained as 2.3\%. Observed $^{8}$B neutrino
flux is significantly deficit to the expectation of SSM(BP95) and it
is consistent with the result from Kamiokande~\cite{fukuda.km}.

The energy spectrum of observed neutrinos is shown in
Fig.~\ref{spectrum}(a). In this figure, expected spectra of MSW small
angle parameter and just-so parameter are also shown. At first sight,
small angle solution has better fit than flat (no oscillation) and
just-so solution, but it is not significant within an experimental
error. The day and night flux difference is obtained by;
\begin{displaymath}
\frac{D-N}{D+N} = -0.031 \pm 0.024 ({\rm stat.}) \pm 0.014 
({\rm syst.}).
\end{displaymath}
If night data are divided into five bins, those differences are shown
in Fig.~\ref{spectrum}(b). In this figure, typical day/night flux
variation of the large angle and the small angle solutions are also
shown. There is no significant difference in day/night fluxes in
present observation. Also Fig.~\ref{spectrum}(c) shows the seasonal
variation of solar neutrino fluxes.  Each season is pile up among
different years. Solid line corresponds to the expected variation from
an eccentricity of the Sun orbit. Within experimental error, there is
no seasonal variation in present analysis. Those results are also same
ones from Kamiokande.

\subsection*{Two flavor neutrino oscillation}
 
For astrophysical solution, it is generally difficult to explain the
solar neutrino problem with the modification of SSM including the
observations from helioseismology. On the other hands, the elementary
particle solution using MSW neutrino oscillation~\cite{fukuda.msw}
seems to be an excellent for explanation of the solar neutrino
problem, because it can distort the spectra of solar neutrinos. Also
MSW oscillation can give a effect in the day/night fluxes
variation. As obtained by Fig.~\ref{spectrum}, our observed spectrum
can be seen slightly as distorted one, but not seen in variance
between day/night fluxes. Using these results, we can obtained the
excluded region at 95\% C.L. in MS diagram as shown in 
Fig.~\ref{MSW}(a).

\begin{figure}[t]
\center\leavevmode
\hbox{\epsfxsize=0.45\textwidth\epsffile{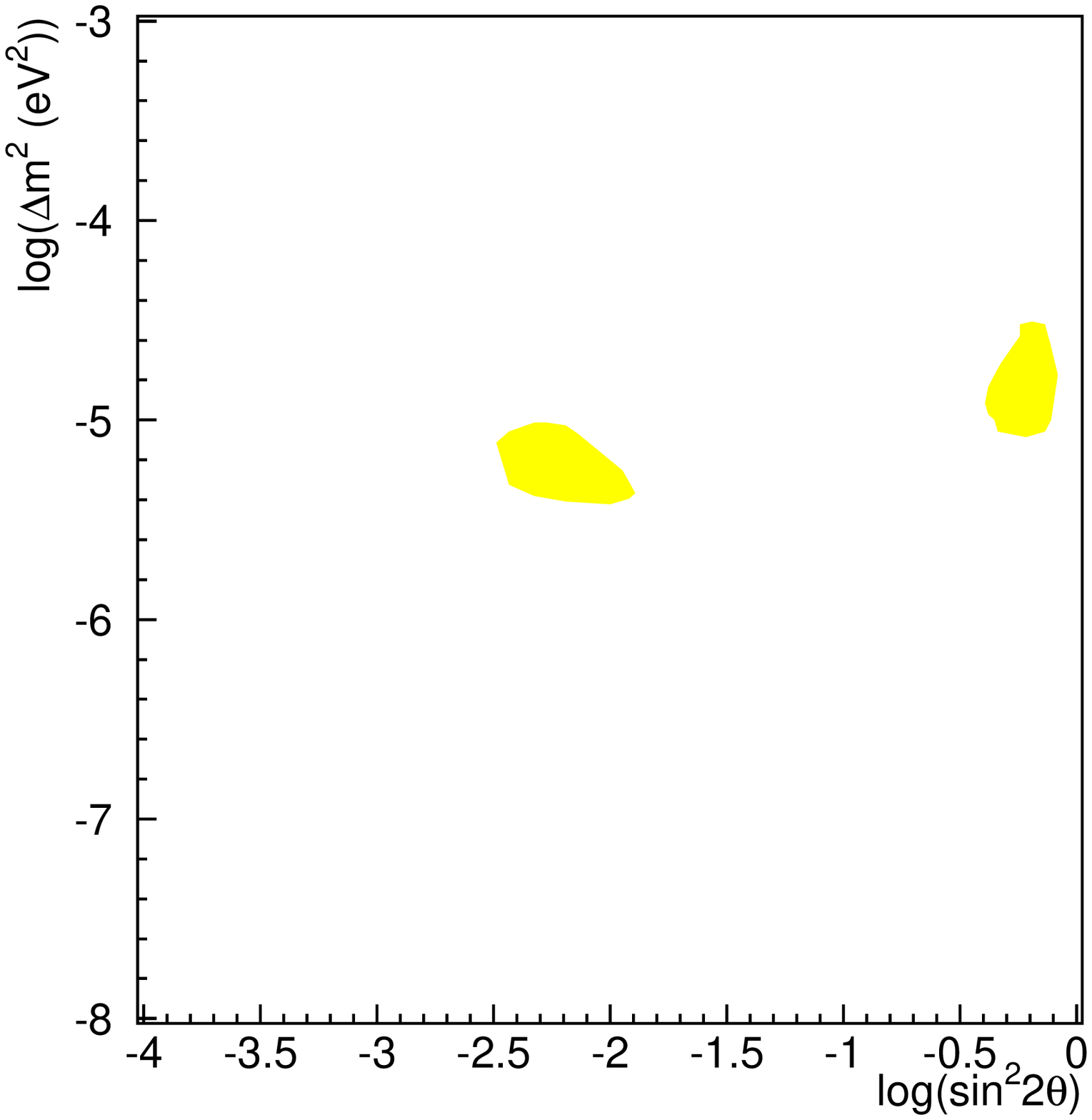}%
\hskip-0.45\textwidth
\epsfxsize=0.45\textwidth \epsffile{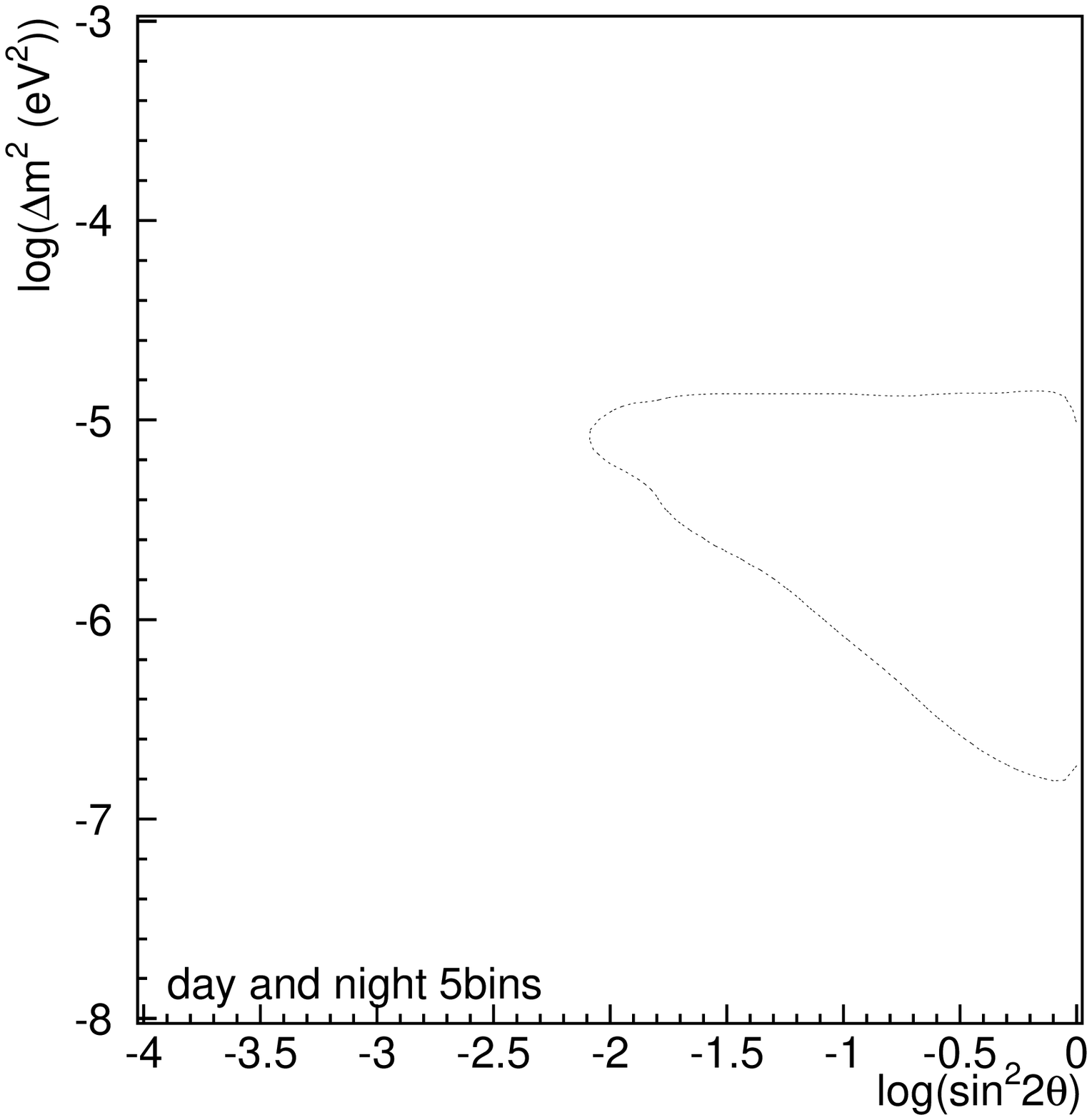}%
\hskip-0.45\textwidth
\epsfxsize=0.45\textwidth \epsffile{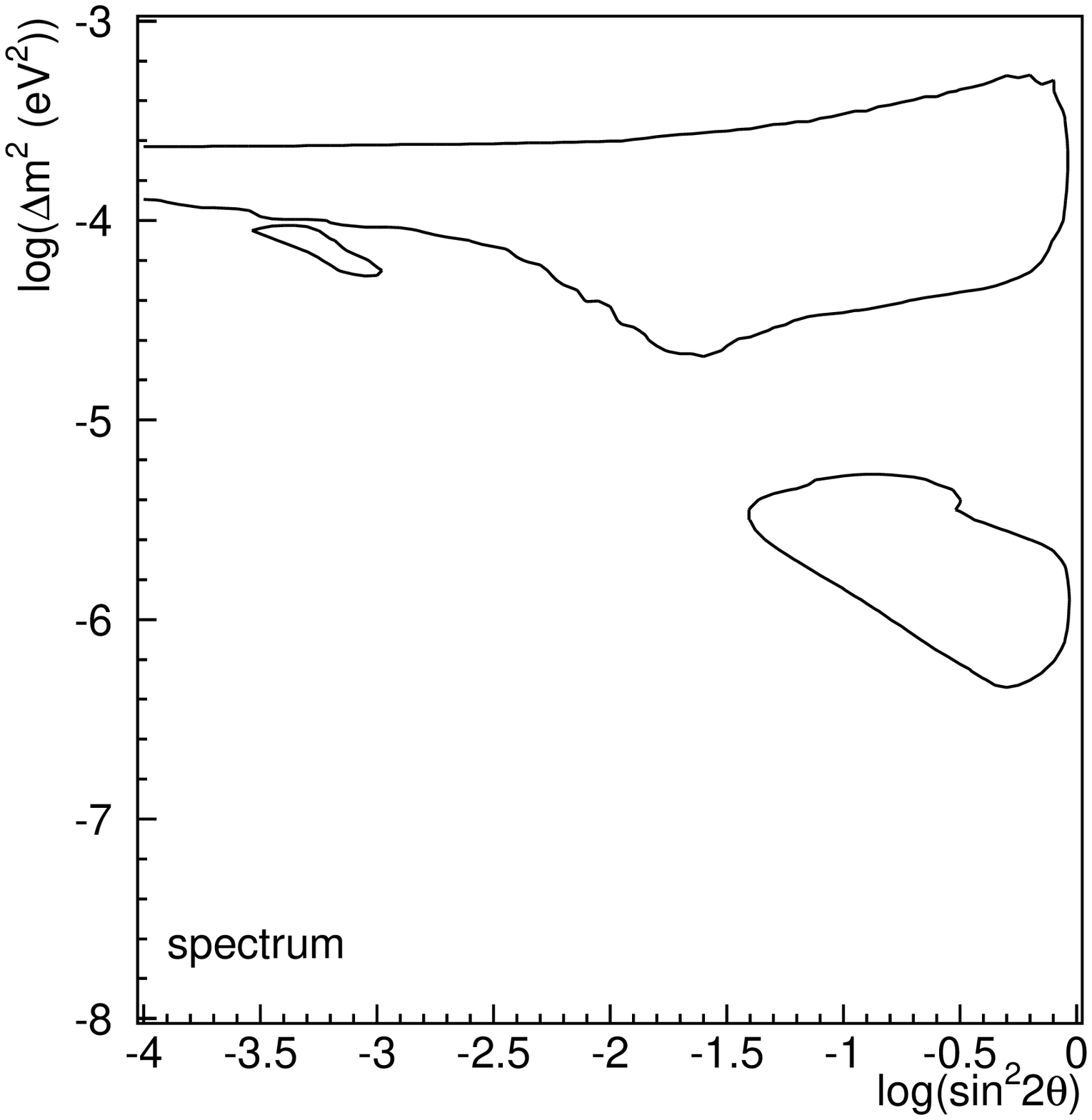}
\hskip0.5cm
\epsfxsize=0.45\textwidth \epsffile{fukuda_msw.eps}%
\hskip-0.45\textwidth
\epsfxsize=0.45\textwidth \epsffile{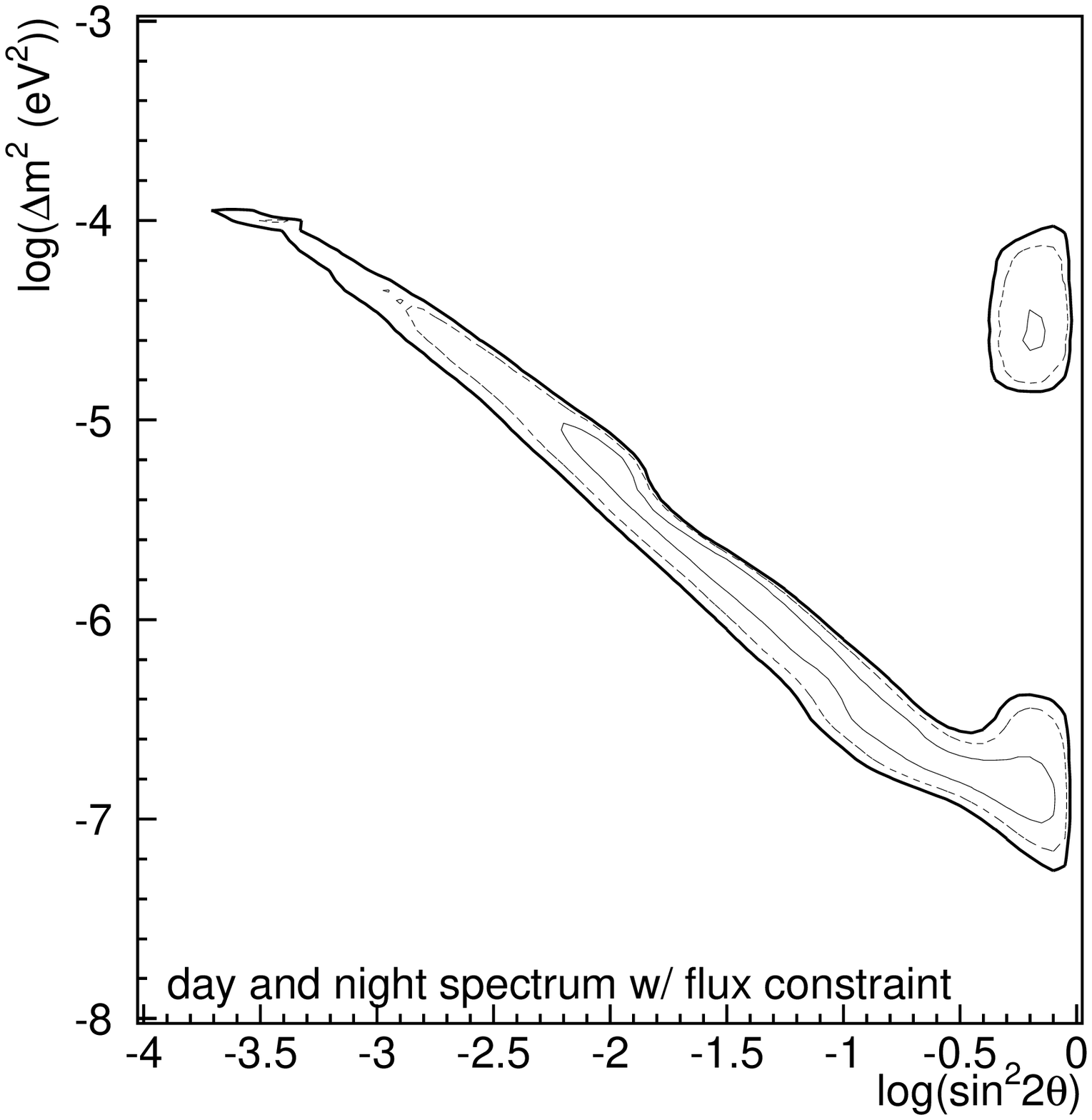}}
\caption{(a)~95\% C.L. excluded region in MSW parameters. Inside of 
solid and dotted line are excluded by the observed energy spectrum and
the day/night flux difference, respectively. (b)~Allowed region in MSW
parameters with measured $^{8}$B neutrino flux. Thick solid line,
dashed line and thin solid line correspond to 95\%, 90\% and 68\%
C.L., respectively.  Shaded region is allowed at 95\% C.L.\ by Hata.}
\label{MSW}
\end{figure}

If we take the constraint of measured $^{8}$B solar neutrino flux,
allowed region of 95\%, 90\% and 68\% C.L. as shown in
Fig.~\ref{MSW}(b).  From these figures, our results and the allowed
region given by Ref.~\cite{fukuda.hl} are consistent in MSW
oscillation analysis.

\subsection*{Summary}
In summary, solar neutrino observation in Superkamiokande has started
since May 1996 and has already taken 374.2 days data. The observed
$^{8}$B solar neutrino flux is about 37\% of the prediction from
SSM(BP95) and it is almost consistent with the result from
Kamiokande. Using LINAC calibration system, we can reduce the
systematic errors related to the energy scale. Obtained energy
spectrum is likely distorted and it is indicated that the new physical
solution will be solved the solar neutrino problem. New challenge to
lower threshold analysis ($\geq$ 5 MeV) has been started. Present
radioactive ($^{222}$Rn) level is about 3 mBq/m$^{3}$, however, we
will able to reduce that level to factor 1/5. Even though the present
analysis, we have succeeded to extract solar neutrino signals from
5--6.5~MeV region. Those data will give a strong indication to the
solution for the solar neutrino problem within a few years.

}\newpage{

\newcommand{\dto}{D$_2$O}
\newcommand{\hto}{H$_2$O}
\newcommand{\mohe}{$^3$He}
\newcommand{\mgcl}{MgCl$_2$}

\head{The Sudbury Neutrino Observatory}
     {M.E.~Moorhead (on behalf of the SNO Collaboration)}
     {Particle and Nuclear Physics Laboratory,
      Keble Road, Oxford OX1 3RH, UK}

\noindent 
The Sudbury Neutrino Observatory (SNO) \cite{MEMS.SNO} is a 1,000 ton
heavy water (\dto) Cherenkov detector in Sudbury, Ontario (Canada)
which will start taking data in `98. The $\nu$ reactions which occur
in \dto\ and the extremely low background environment of the detector
will allow the following measurements: i) the $\nu_e$ and $\nu_x$
(flavour independent) fluxes, and their {\em ratio}, for $^8$B solar
$\nu$'s, ii) the {\em energy spectrum} of $^8$B $\nu_e$'s above 5 MeV,
iii) time dependence in the $^8$B flux, and iv) detailed studies of
the $\nu$ burst from a galactic supernova, including a search for
$\nu_{\mu}$ and $\nu_{\tau}$ masses above 20~eV. For the $^8$B solar
$\nu$ measurements, the $\nu_e / \nu_x$ flux ratio and the $\nu_e$
energy spectrum constitute two separate tests of $\nu$ oscillations
which are both independent of solar model flux calculations
\cite{MEMS.Chen}. If the currently favoured MSW solution
\cite{MEMS.Hata} of the solar neutrino problem is correct then the
$\nu_e / \nu_x$ flux ratio should provide conclusive proof of
oscillations with one year's data yielding a 17 sigma departure from
unity (statistical error only).

The detector is situated two kilometers underground in a dedicated
laboratory that has been excavated in the Creighton nickel mine of
INCO Corporation. This laboratory comprises facilities for changing
into clean-room clothing, a lunch room, a car wash for bringing
equipment into the clean area, a utility room where the \hto\ and
\dto\ systems are located, a control room and a $30~{\rm m}\times
23~{\rm m}$ barrel shaped cavity for the detector itself. The walls of
the cavity have been coated with concrete and low-activity Urylon, a
water proof radon barrier.  Inside this cavity is located a 12~m
diameter spherical acrylic vessel (AV), recently completed, for
containing the \dto\ neutrino target. Almost completely surrounding
this AV, stands a 17~m geodesic sphere supporting 9,500
20-cm-Hamamatsu PMTs, each of which is equipped with a light
reflecting concentrator to increase its effective photocathode area by
a factor 1.7. Beginning in March `98, the detector will be filled with
7,000 tons of high purity \hto\ outside the acrylic vessel (to act as
shielding for high energy gamma rays coming from the rock and the
PMTs) and 1,000 tons of \dto\ inside the AV. The filling will take
between 3--4 months and hence data taking will begin in mid-98.

The event rates for solar $\nu$'s, assuming the full SSM $^8$B flux
\cite{MEMS.Bahcall}, and for a supernova (SN) at the center of our galaxy are
given in Table 1. Apart from the neutral current (NC) reaction, all of the
events are detected by the array of 9,500 PMTs via the Cherenkov radiation
emitted by a single electron (or positron) of energy $\geq 5$ MeV, the
detector's  threshold. The NC reaction produces a free neutron in the \dto\
which can be detected, after thermalization, by observing a subsequent neutron
capture reaction. There are three capture reactions of interest depending on
what additives are placed in the \dto: 

i) Pure \dto: In the case of no additive there is a 30\% probability of
capture on deuterium, which produces a 6.25 MeV gamma. This
gamma converts to electrons by Compton scattering and pair production,
and the resulting Cherenkov light is detected by the PMTs.
Five hundred events a year are expected above the detector's
5 MeV threshold.

ii) MgCl$_2$: Dissolving 2 tons of MgCl$_2$ in the D$_2$O gives an
83\% chance of neutron capture on $^{35}$Cl which produces an 8.5 MeV gamma 
cascade. The higher efficiency and Q-value of this capture (c.f.\
capture on deuterium in the pure \dto\ case) increases the number of 
detected events by a factor of 5 to 2,500 per year.

iii) $^3$He Counters: An array of $^3$He proportional counters
\cite{MEMS.NCD} (5~cm diameter tubes of 800~m total length) placed
vertically in the D$_2$O in a square grid of 1m spacing, gives a 42\%
chance of neutron capture on $^3$He.  The energy and rise-time of the
signals are used to separate $n$-capture (2,000 per year) from
internal alpha and beta backgrounds.
                                   
The dominant background for all these NC detection methods will probably come
from photodisintegration of deuterium which produces free neutrons that are
indistinguishable from NC neutrons. Thus, the detector components have been
carefully selected for extremely low levels of thorium and uranium chain 
contamination so that the photodisintegration rate is small compared  with the
NC rate. This small residual photodisintegration rate must be measured, in
order to subtract its contribution to the neutron capture signal. Several
methods have been developed for this purpose: i) radiochemical extraction and
counting of $^{228}$Th, $^{226}$Ra, $^{224}$Ra, $^{222}$Rn and $^{212}$Pb, ii)
analysis of low energy signals seen by the PMT array, iii) delayed coincidences
between signals seen by the PMT array, and iv) prompt and delayed coincidences 
between signals seen by the PMTs and signals in  the \mohe\ proportional
counters.

\begin{table*}[t]
\setlength{\tabcolsep}{1.5pc}
\newlength{\digitwidth} \settowidth{\digitwidth}{\rm 0}
\catcode`?=\active \def?{\kern\digitwidth}
\caption{Neutrino event rates in SNO (including detection efficiency). SSM
refers to the event rate per year for $^8$B solar neutrinos assuming
the standard solar model \protect\cite{MEMS.Bahcall} and the small angle MSW
solution \protect\cite{MEMS.Hata}. SN refers to a type II supernova at a
distance of 10 kpc (the center of the galaxy).} 
\label{tab:effluents}
\medskip
\begin{tabular*}{\textwidth}{@{}l@{\extracolsep{\fill}}lrr}
\hline
Neutrino reaction && \multicolumn{1}{r}{SSM} 
                  & \multicolumn{1}{r}{SN} \\
\hline
Charged Current (CC):&$\nu_e + d \rightarrow p + p + e^-   $ & 3000 & 80 \\
Neutral Current (NC):&$\nu_x + d \rightarrow p + n + \nu_x $ & 2500 & 300 \\
Electron Scattering (ES):&$\nu_{e,x} + e^- \rightarrow  \nu_{e,x} + e^- $ & 
400 & 20 \\
Anti-neutrino CC in \dto:&$\bar\nu_e + d \rightarrow n + n + e^+ $ & 0 & 70 \\
Anti-neutrino CC in \hto:&$\bar\nu_e + p \rightarrow n + e^+ $ & 0 & 350 \\
\hline
\end{tabular*}
\end{table*}

}\newpage {


\head{BOREXINO}
     {L.\ Oberauer (for the Borexino Collaboration)}
     {Technische Universit\"at M\"unchen, Physik Department E15, 85747
Garching, Germany \\
and      Sonderforschungsbereich 375 Teilchen-Astrophysik}

\subsection*{Physics Goals and Neutrino Detection with Borexino}

The aim of the solar neutrino experiment Borexino is to measure in
real time the solar neutrino flux with low energy threshold at high
statistics, and energy resolving via pure leptonic neutrino electron
scattering $\nu + e \to \nu + e$.

Motivation for Borexino comes from the long standing solar neutrino
puzzle. Data analysis of the existing experiments leads to the
assumption of severe suppression of the solar $^7{\rm Be}$-branch,
which probably cannot be explained by modifications of the standard
astrophysical model of the sun.  The monoenergetic $^7{\rm
Be}$-neutrinos give rise to a compton like recoil spectrum in
Borexino. Its edge will be at 660 keV, significantly higher than the
aimed energy threshold of ca.~250 keV.  Thus $^7{\rm Be}$-neutrino
detection is very efficient in Borexino.

Assuming validity of the standard model a counting rate for $^7{\rm
Be}$-neutrinos, which would consist in this case purely as $\nu_e$, of
roughly 55/day in Borexino is expected.

In scenarios of total neutrino flavour conversion (i.e.~for neutrino
mass differences $ \Delta m^2 \approx 10^{-6}{-}10^{-5} \,\, {\rm
eV}^2 $) a reduced flux of approximately 12/day would be measured due
to the lower cross section of $\nu_{\mu ,\tau}$ scattering, which
occurs only via neutral current interaction.

In case of vacuum oscillations (i.e.~for neutrino mass differences $
\Delta m^2 \approx 10^{-10} \,\, {\rm eV}^2 $) Borexino would see a
distinct time dependent periodical neutrino signal due to the seasonal
eccentricity of the earth's orbit around the sun.

For neutrino mass differences in the range of $ \Delta m^2 \approx
10^{-7} \,\, {\rm eV}^2 $ and for large mixing Borexino should see a
`day/night' effect due to electron neutrino regeneration
during the path through the earth.

Borexino also can serve for additional projects in neutrino physics.
Search for a magnetic moment can be performed by means of terrestrial
neutrino sources by investigating the electron recoil shape at low
momentum transfer.  Via the inverse beta-decay $\bar\nu_e + p \to
e^+ + n$ Borexino can look for signals from geophysical neutrinos as
well as for neutrinos emitted by european nuclear power plants.  The
latter would serve as a long baseline neutrino oscillation experiment
probing the so-called large mixing angle solution for the solar
neutrino problem.

\subsection*{The Detector and Background Considerations}

The detector is shielded successively from outer radioactivity by
means of an onion-like structure. Here the adjacent inner layer serves
as shielding and has to provide an increased purity in terms of
internal radioactivity.

Borexino is placed in hall C of the underground laboratory at Gran
Sasso, Italy. An overburden of ca.~3500~m.w.e. suppresses the cosmic
muon flux to roughly $1/{\rm m}^2$.  The outer part of the
detector consists of a steel tank with 18~m in height and diameter.
Inside this `external' tank a stainless steel sphere will support 2200
phototubes on the inside and 200 tubes at the outside.  Most of the
tubes inside the sphere will be equipped with light guides in order to
increase the geometrical coverage and hence the energy resolution.
Between external tank and sphere high purity water will serve as
shielding against external gamma rays and as active Cherenkov counter
against cosmic muons.  The steel sphere will be filled with a
tranparent, high purity buffer liquid which itself holds a nylon
sphere, filled with organic scintillator. The active scintillator mass
will be around 300~t.  By means of time of flight measurements event
position can be reconstructed and a fiducial volume for solar neutrino
interaction defined. The latter should be about 100~t, establishing a
counting rate of 55 neutrinos per day according to the standard solar
model. The outer part of the scintillator sphere serves as active
shielding.

The demands on purity in terms of radioactivity in Borexino,
especially for the scintillator itself, are very severe. In order to
be able to extract a clear signal from background events also in case
of total flavour conversion, an intrinsic concentration in Uranium and
Thorium of ca. $10^{-16}$ should not be exceeded significantly.  The
amount of $^{14}{\rm C}/ ^{12}{\rm C}$ must not be higher than
$\approx 10^{-18}$.  In order to test scintillating materials a large
Counting Test Facility (CTF) has been built up in hall C of the
underground laboratory at Gran Sasso, which resembles to a small
prototyp (ca.~5~t of scintillator) of Borexino.  From beginning of
1995 until summer 1997 several tests about the feasibilty of Borexino
including procedures to maintain the purity of the scintillator has
been performed, which showed very encouraging results: $ ^{14}{\rm
C}/{}^{12}{\rm C} =1.85 \cdot 10^{-18} $, $ ^{238}{\rm U} = (3.5 \pm
1.3) \cdot 10^{-16} \,\, {\rm g/g} $, $ ^{232}{\rm Th} = (4.4 \pm 1.5)
\cdot 10^{-16} \,\, {\rm g/g}$.  A complete discussion of the CTF
results including experimental techniques for further background
suppression is given in~\cite{oberauer1} and~\cite{oberauer2}.
Details about the experimental setup of the CTF can be found in
\cite{oberauer3}.

Highly developed neutron activation analysis of scintillation
samples performed in Munich is now sensitive in the same regime.
For uranium an upper limit of
${}^{238}{\rm U} < 2 \cdot 10^{-16} \,\,{\rm g/g}$ (90\% CL)
has been obtained.
In addition concentration values or limits have been measured by this
method for a various amount of isotopes, including man-made nuclei.
For details, see \cite{oberauer4}.

Background studies for Borexino include also the interaction of cosmic
muons. The direct detector response on muons has been determined by a
coincidence measurement between the CTF and a muon telescope on top of
it. The time distribution of such events can be used to discriminate
between muon events and neutrino candidates at a level of 98\%.
However, to reach the sensitivity needed to reach the goals in
Borexino, the leak rate for muons must not exceed a level of $\approx
10^{-4}$.  Our design of the muon veto system therefore is threefold:
The outer region between external tank and steel sphere acts as
Cherenkov counter, a special configuration of the tubes inside the
sphere will act as an additional muon identification system, and
finally the offline study of event topology like the time structure
will help to suppress this kind of background sufficiently.

Cosmogenic generation of radioactive nuclei has been sudied this fall
at the 180 GeV muon beam at SPS in CERN.  Most dangerous source of
events will come from $^{11}{\rm C}$-production in the scintillator
and surrounding buffer liquid.  However, the energy spectrum of these
events is between 1 MeV and 2 MeV since the decay mode is positron
decay at 1 MeV endpoint energy. Thus the detection of $^7{\rm
Be}$-neutrinos is not affected, however that of pep-neutrinos.

\subsection*{Prospects}

Borexino is an international collaboration of about 60 scientists.
Approved funding already comes from INFN (Italy) and from BMBF and DFG
(Germany).  A substantial part should also be covered by NSF (USA) in
the near future.  Work on the external tank of Borexino is almost
completed.  We expect to finish with the inner steel sphere in 1999.
Simultaneously the CTF will be upgraded. Finally it will serve as test
facility for Borexino scintillator procurement in batch mode.  Given
full funding also for our american collaborators first data taking may
be expected at the end of the year 2000.


}\newpage{

\head{Measurements of Low Energy Nuclear Cross Sections}
     {M.~Junker (for the LUNA Collaboration)}
     {Laboratori Nazionali Gran Sasso, Assergi (AQ), Italy\\ 
      and Institut f\"ur Experimentalphysik III, Ruhr-Universit\"at 
      Bochum, Germany}

\noindent
The nuclear reactions of the pp-chain play a key role in the
understanding of energy production, nucleosynthesis and neutrino
emission of the elements in stars and especially in our sun
\cite{junker.1}. A comparison of the observed solar neutrino fluxes
measured by the experiments GALLEX/SAGE, HOME\-STAKE and KAMIOKANDE
provides to date no unique picture of the microscopic processes in the
sun \cite{junker.2}. A solution of this so called ``solar neutrino
puzzle" can possibly be found in the areas of neutrino physics, solar
physics (models) or nuclear physics. In view of the important
conclusions on non-standard physics, which might be derived from the
results of the present and future solar neutrino experiments, it is
essential to determine the neutrino source power of the sun more
reliably.

Due to the Coulomb Barrier involved in the nuclear fusion reactions,
the cross section of a nuclear reaction drops nearly exponentially at
energies which are lower than the Coulomb Barrier, leading to a
low-energy limit of the feasible cross section measurements in a
laboratory at the earth surface.  Since this energy limit is far above
the thermal energy region of the sun, the high energy data have to be
extrapolated down to the energy region of interest transforming the
exponentially dropping cross section $\sigma(E)$ to the astrophysical
$S$-factor
\mbox{$S(E)= \sigma(E)\, E\, {\exp(2\,\pi\, \eta)}$},
with the Sommerfeld parameter given by 
\mbox{$2\, \pi\, \eta=31.29\, 
Z_1\,Z_2(\mu/E)^{1/2}$~\cite{junker.1}}.  
The quantities $Z_1$ and $Z_2$ are the nuclear charges of the
interacting particles in the entrance channel, $\mu$ is the reduced
mass (in units of amu), and $E$ is the center-of-mass energy (in units
of keV). In case of a non resonant reaction $S(E)$ may then be
parameterized by the polynomial
\mbox{$ S(E)  = S(0) + S'(0)E + 0.5\,S''(0)E^2.$}

As usual in physics, extrapolation of data into the ``unknown" can
lead onto ``icy ground".  Although experimental techniques have
improved over the years to extend cross section measurements to lower
energies, it has not yet been possible to perform measurements within
the thermal energy region in stars.

For nuclear reactions studied in the laboratory, the target nuclei and
the projectiles are usually in the form of neutral atoms/molecules and
ions, respectively. The electron clouds surrounding the interacting
nuclides act as a screening potential: the projectile effectively sees
a reduced Coulomb barrier.
This leads to a higher cross section, $\sigma_{\rm s}(E)$,
than would be the case for bare nuclei, $\sigma_{\rm b}(E)$, with an
exponential enhancement factor 
$f_{\rm lab}(E) = \sigma_{\rm s}(E)/ \sigma_{\rm b}(E) \simeq
\exp(\pi\eta\, U_{\rm e}/E)$ \cite{junker.ass87,junker.bra90}
where $U_{\rm e}$ is the electron-screening potential energy
(e.g.~$U_{\rm e} \simeq Z_1\cdot Z_2 \cdot e^2/R_{\rm a}$
approximately, with $R_{\rm a}$ an atomic radius). 
For a stellar plasma the value of $\sigma_{\rm b}(E)$
must be known because the screening in the plasma can be quite
different from that in laboratory studies $\cite{junker.ric95}$, and
$\sigma_{\rm b}(E)$ must be explicitly included in each
situation. Thus, a good understanding of electron-screening effects is
needed to arrive at reliable $\sigma_{\rm b}(E)$ data at low energies.
Low-energy studies of several fusion reactions involving light
nuclides showed $\cite{junker.gre95,junker.pra94,junker.lan96}$ indeed
the exponential enhancement of the cross section at low energies.  The
observed enhancement (i.e.~the value of $U_{\rm e}$) was in all
cases close to or higher than the adiabatic limit derived from
atomic-physics models. An exception are the $^{\rm 3}$He+$^{\rm 3}$He\
data of Krauss et al.~\cite{junker.kra87}, which show apparently no
electron screening down to $E$=25~keV, although the effects of
electron screening should have enhanced the data at 25 keV by about a
factor 1.2 for the adiabatic limit $U_{\rm e}$=240~eV. Thus, improved
low-energy data are particularly desirable for this reaction.
\looseness=1

The low-energy studies of thermonuclear reactions in a laboratory at
the earth's surface are hampered predominantly by the effects of
cosmic rays in the detectors.  Passive shielding around the detectors
provides a reduction of gammas and neutrons from the environment, but
it produces at the same time an increase of gammas and neutrons due to
the cosmic-ray interactions in the shielding itself. A 4$\pi$ active
shielding can only partially reduce the problem of cosmic-ray
activation. The best solution is to install an accelerator
facility~\cite{junker.gre94} in a laboratory deep
underground~\cite{junker.5}. The worldwide first underground
accelerator facility has been installed at the Laboratori Nazionali
del Gran Sasso (LNGS) in Italy, based on a 50~kV accelerator. This
pilot project is called LUNA and has been supported since 1992 by
INFN, BMBF, DAAD-VIGONI and NSF/NATO.

The major aim of the LUNA project is to measure the cross section of
$^3$He($^3$He,2p)$^4$He which is one of the major sources of
uncertainties for the calculation of the neutrino source power.  It
has been studied previously \cite{junker.kra87} down to about $E_{\rm
cm}$=25 keV, but there remains the possibility of a narrow resonance
at lower energies that could enhance the rate of path I at the expense
of the alternative paths of the pp-chain that produce the
high-energies neutrinos ($E_{\rm\nu} > 0.8$~MeV).  The LUNA--facility
allows to study this important reaction over the full range of the
solar Gamow Peak, where the cross section is as low as 8 pbarn at
$E_{\rm cm}$=25 keV and about 20~fbarn at $E_{\rm cm}$=17~keV.

Figure~\ref{junker.fig1} shows the results obtained in the energy
region between $E_{\rm cm}$=25 keV and $E_{\rm cm}$=20.7~keV.  The
lowest counting rate was of 3~events per day at $E_{\rm cm}$=20.7~keV.
At this energy about 1000~Cb of $^{\rm 3}$He$^+$ have been accumulated
on the target.  The data obtained at higher energies (450 kV
accelerator in Bochum) with the LUNA setup \cite{junker.gre94} are
included for completeness.  Previous data obtained \cite{junker.kra87}
at $E_{\rm cm}$=25 are also shown in
figure~\ref{junker.fig1}.  The LUNA data have been obtained at
energies within the solar Gamow Peak, i.e.~below the 21 keV center of
this peak, and represent the first measurement of an important fusion
cross section in the thermal energy region.

\begin{figure}[ht]
\centerline{\epsfxsize=0.8\textwidth\epsffile{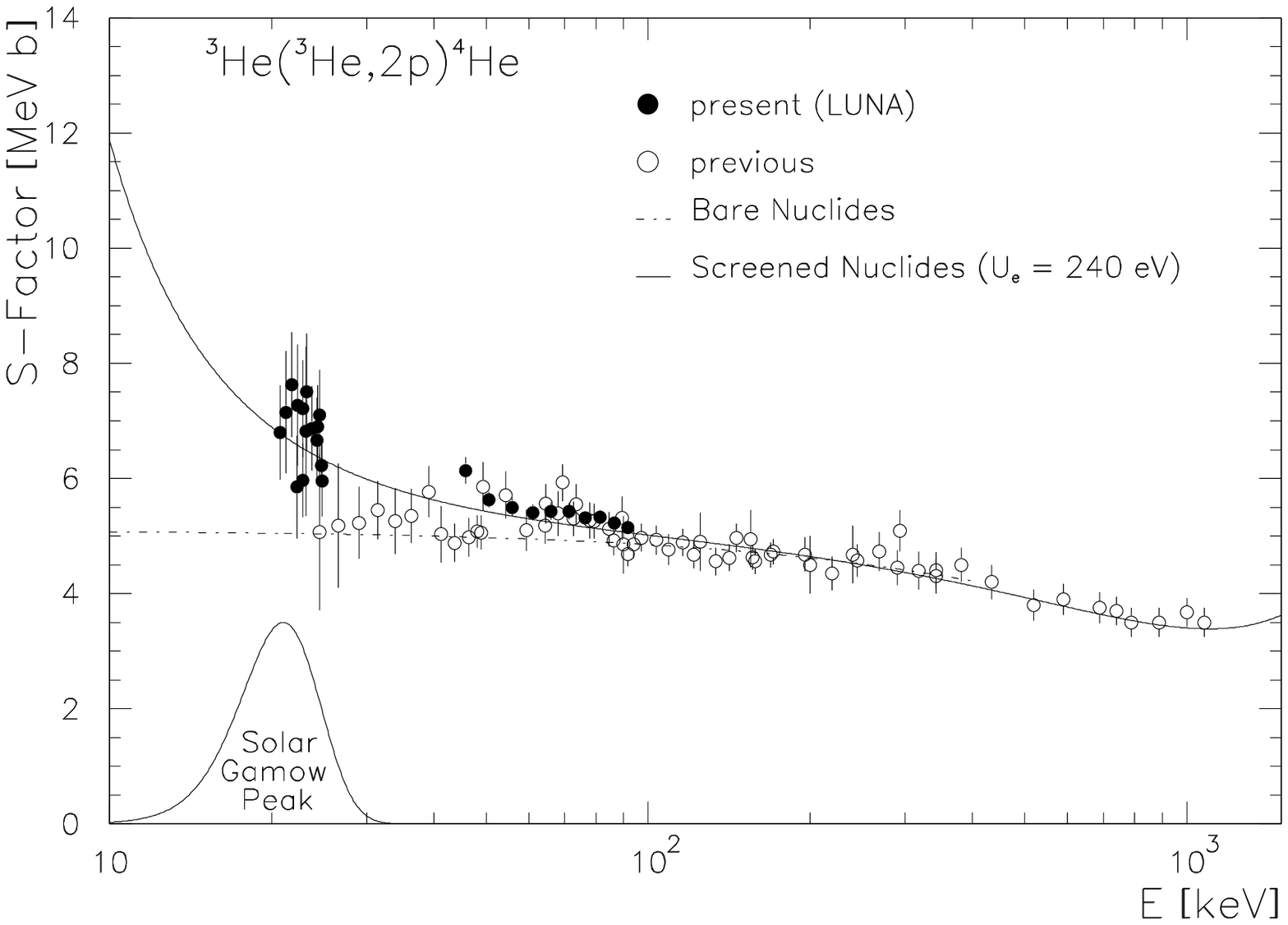}}
\caption{\label{junker.fig1}
The $S(E)$ factor of $^{\rm 3}$He$^{\rm 3}$He\ from 
\protect\cite{junker.kra87} and the present work. The dashed and
solid curves represent $S_{\rm b}(E)$ and $S_{\rm s}(E)$ respectively.
The solar Gamow Peak is shown in arbitrary~units.}
\end{figure}

The astrophysical $S$-factor has been extrapolated to zero energy
\cite{junker.lun97}:
\[ 
S(0)=5.40 \pm 0.05{\rm (stat)}\pm 0.30 {\rm (sys)}\pm 0.30{\rm
(screen)}. \nonumber
\]
The first error contains the statistical error (one standard
deviation) including counting statistics and apparative variations of
pressure, beam power and temperature measurement. The second error is
the systematical error (one standard deviation) including
uncertainties in pressure, beam energy and power, efficiency and
energy loss data. The 10\% error of the energy loss data transforms to
an uncertainty of 0.1 to 0.2~keV in the effective energy. This in turn
leads to an error of 1 to 3.5~\% for the astrophysical
$S$-factor. This is due to the exponential decrease of the cross
section at low energies \cite{junker.lun97}.  The third component of
the error results from the lack of understanding of electron
screening: Fitting all existing data sets with $S(0)$, $S'(0)$,
$S''(0)$ and $U_{\rm e}$ as free parameter gives an Electron Screening
Potential $U_{\rm e}$ of 323~eV with $S(0)=5.30$~MeVb. Fixing $S(0)$,
$S'(0)$ and $S''(0)$ at energies higher than 100 keV (where the
electron screening effect is neglectible) gives $S(0)=5.1$~MeVb. In
turn fitting $U_{\rm e}$ in a second step including also the low
energy data gives $U_{\rm e}=432$~eV. A fit to the data assuming no
electron screening gives $S(0)=5.7$~MeVb. The adiabatic limit for
$U_{\rm e}$ based on theoretical atomic physics is 240~eV.  The first
two methods give $U_{\rm e}$ values higher than the adiabatic limit
(240~eV), consistent with observations in other fusion reactions.  The
difference between observed and predicted $U_{\rm e}$ values is not
understood at present giving a considerable uncertainty on $S(0)$.

Based on the success of the LUNA project and on the fruitful
international collaboration (in parti\-cular the support provided by
LNGS), further experiments with the 50 kV machine for a better
understanding of the electron screening effect are planned. The
proposed reactions are D(p,$\gamma$)$^3$He, D($^{\rm 3}$He,p)$^4$He,
$^7$Li(p,$\alpha$)$^4$He and
$^{11}$B(p,$\alpha$)$^7$Be($\alpha$)$\alpha$. In addition, some
interesting aspects in nucleosynthesis of the early universe are
related to the low energy cross section of the first reaction
\cite{junker.lun2}. 

While studying these reactions an enlarged LUNA II collaboration will
install a 200~kV accelerator facility at Gran Sasso to study the
pp--reactions $^3$He($\alpha$,$\gamma$)$^7$Be and
$^7$Be(p,$\gamma$)$^8$B and the key reaction of the CNO--cycles,
$^{14}$N(p,$\gamma$)$^{15}$O, at energies far below the present
limits.  All these reactions are critical to the solar neutrino
puzzle.  The reaction rate of $^{14}$N(p,$\gamma$)$^{15}$O is also one
of the ingredients needed to determine the theo\-re\-ti\-cal scenario
used to constrain both the age and the distance of the oldest stellar
system in our galaxy, namely the Globular Clusters. In such a way,
more stringent limits to the cosmology will be obtained.

In addition, the new 200 kV accelerator will afford the study of many
(p,$\gamma$) reactions of the NeNa and MgAl cycles below an incident
proton energy $E_{\rm p}$=200 keV.  Experimental data about these
channels, very important for the understanding of nucleosynthesis
processes, are up today still missing or very uncertain.  For example
the NeNa cycle may play a role in understanding the almost pure
$^{22}$Ne abundance found in meteorites samples, while the MgAl cycle
may provide the mechanisms for production of $^{26}$Al, the decay of
which gives rise to the $^{26}$Mg/$^{27}$Al anomaly found in some
meteorites.  All the involved (p,$\gamma$) cross sections of these
cycles are scarcely known at low energies: a continuation of the
underground experimental program of the LUNA experiment could be
devoted in the future to the measurement of these channels for proton
incident energies below 200 keV, where many unmeasured resonances are
present \cite{junker.14}. For example the strength of the low lying
resonances in the reaction $^{25}$Mg(p,{$\gamma$})$^{26}$Al, which is
crucial for the production of $^{26}$Al, could be experimentally
determined for the first time. Also the other reactions of these
cycles can be investigated using stable targets and accelerator and
detector systems of the LUNA experiment.

}\newpage{


\head{Solar Neutrinos: Where We Are and What Is Next?}
     {G.~Fiorentini }
     {Dipartimento di Fisica dell'Universit\'a di Ferrara and
      Istituto Nazionale di Fisica Nucleare, Sezione di Ferrara,
      I-44100 Ferrara, Italy}

\subsection*{What has been Measured?}

All five experiments report a deficit of solar neutrinos with respect
to the predictions of Standard Solar Models (SSMs), see
Ref.~\cite{Fiorentini.1}.

\subsection*{What Have We Learnt on Solar Neutrinos, 
Independently of SSMs?}

Actually, the solar neutrino puzzle (SNP) is not just the discrepancy
between experimental results and the predictions of standard solar
models.  Rather, experimental results look inconsistent among each
other with the only assumption that the {\em present} total neutrino
flux can be deduced from the {\em present} solar luminosity, unless
something happens to neutrinos during the trip from Sun to Earth, see
Fig.~\ref{Fiorentini.fig1} and 
Refs.~\cite{Fiorentini.1,Fiorentini.2,Fiorentini.3}.
 
\subsection*{What Has Been Calculated?}

Accurate predictions of solar neutrino fluxes are anyhow extremely
important.  If neutrino masses (differences) are as small as suggested
by several proposed solutions to the SNP, then the only way to measure
neutrino masses is through the interpretation of future solar neutrino
experiments, which requires accurate theoretical predictions of solar
properties.
        
Refined solar models are thus necessary.  All these have to be
confronted with the powerful helioseismic constraints, see
Ref.~\cite{Fiorentini.4}, particularly for a quantitative (and
conservative) determination of the accuracy of solar properties as
deduced from helioseismology.
   
Recent SSM calculations, using accurate equations of state, recent
opacity calculations and including microscopic diffusion, look in
agreement with heliosesimology, see Figs.~\ref{Fiorentini.fig2}
and~\ref{Fiorentini.fig3} and Ref.~\cite{Fiorentini.4}.  Alternative
solar models should be as successful as these are~\cite{Fiorentini.5,
Fiorentini.6}.

Actually, one can exploit helioseismology within a different
strategy. One can relax some assumptions on the most controversial
ingredients of solar models (e.g.~opacity and metal abundance) and
determine them by requiring that helioseismic constraints are
satisifed.  These helioseismically constrained solar models (HCSM) all
yield the same central temperature within about one
percent~\cite{Fiorentini.7}. The main uncertainties for the
determination of solar neutrino fluxes arise now from nuclear physics
measurements.  After the succesful LUNA experiments at
LNGS~\cite{Fiorentini.8}, the main uncertainties are now from the
$^3$He+$^4$He and $^7$Be+p reactions.
      
\subsection*{What Is Missing?}

In a prophetical paper of 1946~\cite{Fiorentini.9} Bruno Pontecorvo
wrote: ``direct proof of the {\em existence} of the neutrino $\ldots$
must be based on experiments the interpretation of which does not
require the law of conservation of energy, i.e.~on experiments in
which some characteristic process produced by free neutrinos $\ldots$
is observed.''
      
The situation now looks very similar, just change {\em existence} with
{\em nonstandard properties}, in that the strongest argument for a
particle physics solution to the SNP arises from energy conservation
(the luminosity constraint) and actually we need a direct footprint of
some neutrino property, not predicted within the minimal standard
model of electroweak interactions.
     
The four most popular particle physics solutions (small and large
angle MSW effect, just so oscillations and universal oscillations) all
predict specific signatures which are being or will be tested by the
new generation of experiments (Superkamiokande, Borexino, SNO,
$\ldots$), see~Table~\ref{Fiorentini.tab1}.
 
The hypothesis of universal oscillation proposed in
Ref.~\cite{Fiorentini.10} has just been falsified by the recent
negative result of CHOOZ~\cite{Fiorentini.11}. This nice and small (in
comparison with the gigantic solar neutrino devices) experiment at a
nuclear reactor is cleaning some of the fog in the air.

Let us wait and wish that (at least) one of the fingerprints of
neutrino oscillations is detected by the new experiments.

\begin{table}
\caption{The proposed solutions, their fingerprints and the 
experiments
looking at them.}
\hbox{\hss
\begin{tabular}{lccccc}
&&&&\\
\hline
\hline
Proposed solutions & \multicolumn{5}{c}{Signatures}\\
\hline
 & Oscillation & spectral & Day-night &  Seasonal & CC/NC \\
 & at reactor  & deformation & variation & modulation & events \\
 \hline
 MSW small angle & NO & TINY & TINY & NO & YES \\
 MSW large angle & NO & TINY & YES & NO & YES \\
 JUST-SO         & NO & YES & NO & YES & YES \\
Universal oscil. & YES & NO & NO & NO & YES \\
\hline
Experiment        & CHOOZ & SUPERKAM. & SUPERKAM. & BOREXINO & SNO \\
Data             & now & now & now & 2000 & 2000 \\
\hline
\hline
\end{tabular}
\hss}
\label{Fiorentini.tab1}
\end{table}

\newpage

\begin{figure}[htp]
\centering
\epsfig{file=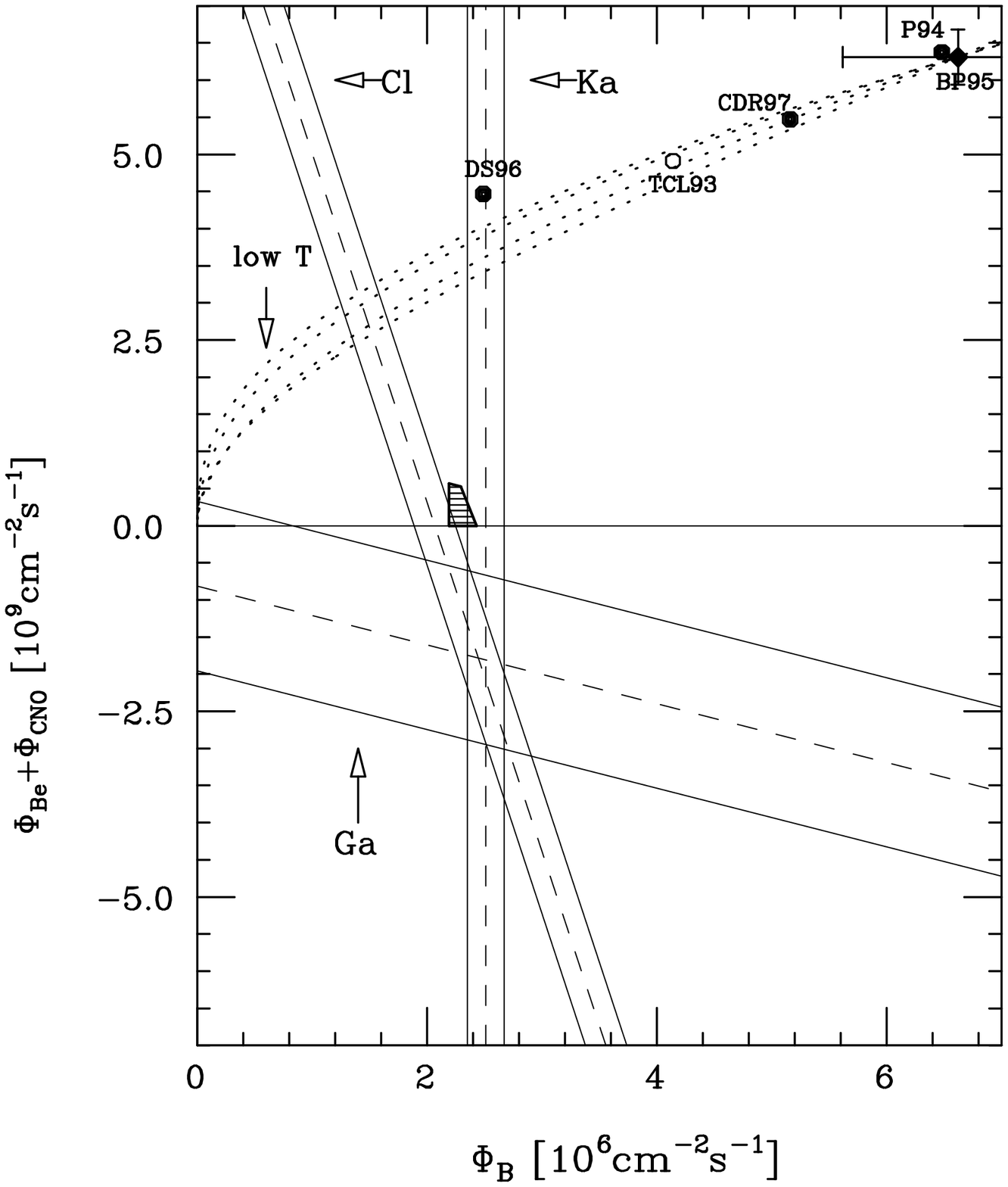,width=12cm}
\caption[aaa]{
The $^8$B and $^7$Be+CNO neutrino fluxes, consistent with the
luminosity constraint and experimental results for standard neutrinos.
The dashed (solid) lines correspond to the central ($\pm 1\sigma$)
experimental values for chlorine~\cite{Fiorentini.13},
gallium~\cite{Fiorentini.14} and $\nu$-$e$ scattering
experiments~\cite{Fiorentini.15}.  The dashed area corresponds to the
region within $2\sigma$ from each experimental result.  The
predictions of solar models including element diffusion (full
circles)
\cite{Fiorentini.12,Fiorentini.16,Fiorentini.17,Fiorentini.18}
and neglecting diffusion (open circles)~\cite{Fiorentini.19} are also
shown.  The dotted lines indicate the behaviour of nonstandard solar
models with low central temperature~\cite{Fiorentini.1}.}
\label{Fiorentini.fig1}
\end{figure}

{\ }

\newpage

\begin{figure}[ht]
\vbox{
\centering
\epsfig{file=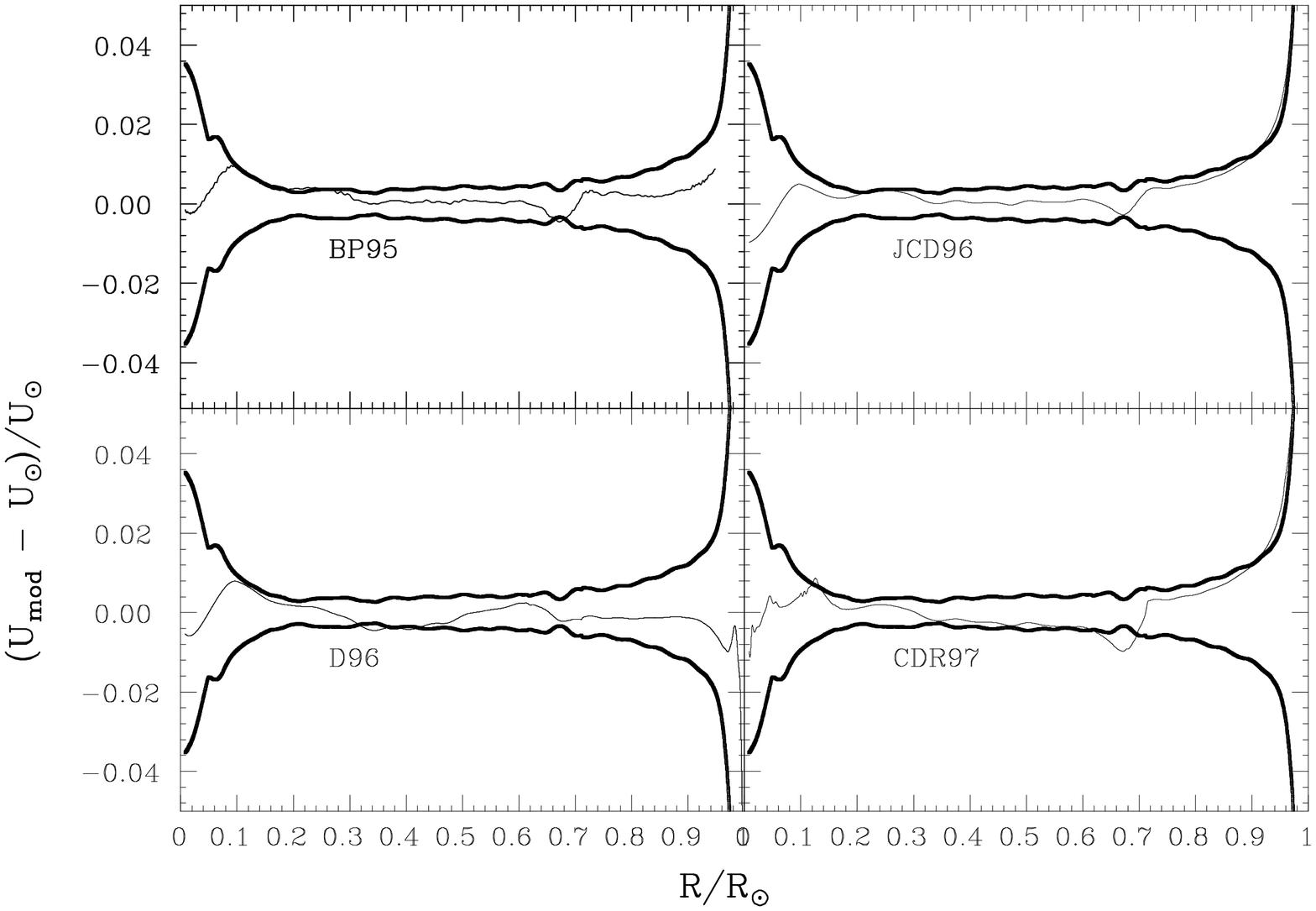,height=12cm,angle=90}
\vskip-1cm
\caption[bbb]{The difference between $U$ as predicted by selected 
solar models, $U_{\rm mod}$, and the helioseismic determination,
$U_{\odot}$, normalized to this latter (thin line).  The allowed area
is that within the thick lines.  BP95 is the model with metal and He
diffusion of Ref.~\cite{Fiorentini.12}; JCD96 is the``model S'' of
Ref.~\cite{Fiorentini.20}; D96 is the ``model 0'' of
Ref.~\cite{Fiorentini.21}; CDR97 is the ``best'' model with He and
heavier elements diffusion of Ref.~\cite{Fiorentini.17}.}
\label{Fiorentini.fig2}

\vskip1cm
\centering
\epsfig{file=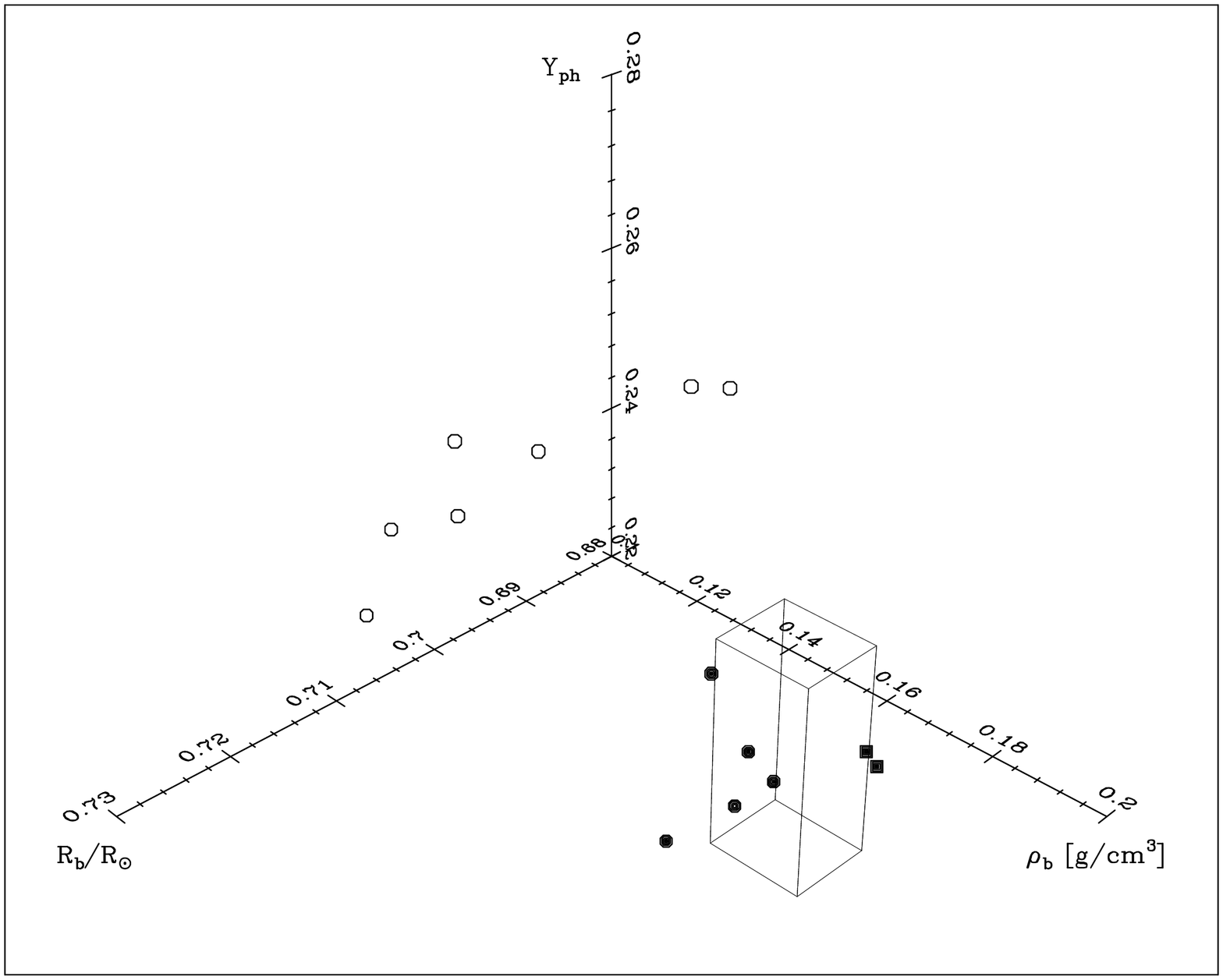,height=10cm, angle=90}
\caption[ccc]{Helioseismic determinations and solar model 
predictions of properties of the convective envelope.  The box defines
the region allowed by helioseismology.  Open circles denote models
without diffusion, squares models with He diffusion, full circles
models with He and heavier elements diffusion~\cite{Fiorentini.4}.}
\label{Fiorentini.fig3}
}
\end{figure}

\newpage

}

\newpage {


\thispagestyle{empty}

\begin{flushright}
\Huge\bf
{\ }


Supernova\\
\bigskip
Neutrinos

\end{flushright}

\newpage

\thispagestyle{empty}

{\ }

\newpage

}\newpage{


\head{Phenomenology of Supernova Explosions}
     {Wolfgang Hillebrandt}
     {Max-Planck-Institut f\"ur Astrophysik, D-85748 Garching, 
      Germany}

\subsection*{Abstract}

An attempt is made to match observed properties of supernovae, such as
spectra, lightcurves etc.\ with theoretical ideas and predictions.
It is demonstrated that neither do observational data constrain the
models in a unique way nor does theory allow for an unambiguous
interpretation of the data. Possible improvements on the theoretical
side are briefly discussed.

\subsection*{Some Observational Facts}

In general, supernovae are classified according to their maximum light
spectra.  Those showing Balmer lines of H are called Type II's, and
all the others are of Type I. Those Type I's which show a strong Si
absorption feature at maximum light are named Type Ia, and the others
are Ib's or Ic's, depending on whether or not they have also He I
features in their spectra~\cite{hillebrandt.1}.  At later times,
several months after the explosion, when the supernova ejecta become
optically thin, Type II spectra are dominated by emission lines of H,
O, and Ca, whereas Type Ia's have no O, but Fe and Co. Type Ib,c's, on
the other hand side, show emission lines of O and Ca, just like the
Type II's \cite{hillebrandt.1}.  However, the spectral
classification is not always as clear. For example, SN 1987K started
out as a Type II with H lines, but changed into a Type Ib,c like
spectrum after 6 months~\cite{hillebrandt.2}.

As far as the light curves are concerned, Type II's seem to be more
complicated than Type I's. Type II-L are characterized by a peak
lasting for about 100 days, followed by a ``linear'' decay in the blue
magnitude vs. time diagram. In contrast, Type II-P's have a somewhat
wider peak, followed by a ``plateau'' phase and an occasionally rather
complicated tail decaying not like a single exponential. Typically,
Type II-L's are brighter than Type II-P's in the blue. In addition,
there are objects like SN 1987A which possess a very complicated light
curve, with an early narrow peak, a first minimum, followed by a broad
hump after a few months, and a final nearly exponential decay with
indications of some flattening at late times. SN 1987A-like objects
are much dimmer than all other Type II's \cite{hillebrandt.1}.
              
In contrast, all Type Ia light curves are quite similar, making them
good candidates for standard candles to measure the cosmic distance
scale.  Moreover, since they are the brightest among all supernovae,
they can be observed even at high redshifts and, in fact, a Type Ia
supernova at a redshift of about 1 has recently been observed
\cite{hillebrandt.3}.  Although their absolute peak luminosity may
vary by about one magnitude, an observed correlation between the
luminosity and light curve shape (the brighter ones have broader light
curves) allows one to correct for the differences. So recently
observations of Type Ia's have become a powerful tool in attempts to
determine cosmological parameters \cite{hillebrandt.3}. Of course,
one has to make sure that the supernovae one is observing are indeed
of Type Ia, and not of Type Ib,c which are intrinsically fainter, but
this can be achieved if a spectrum near maximum light is available.
 
Other observational information on supernovae is, in general, less
solid. With the exception of SN 1987A, we do not have direct
information on the properties of the progenitors, the energetics, or
the masses of the ejecta and of the (compact) remnants, if there are
any. Indirect information can be obtained from the light curves and
the spectra leading, however, to an ``inverse'' problem as far as
theoretical interpretations are concerned.
 
\subsection*{Theoretical Classification}
 
The theoretical classification of supernovae is usually done according
to the suspected progenitor stars and the explosion mechanism, and it
dates back to a classic paper of Fred Hoyle and Willy Fowler in 1960
\cite{hillebrandt.4}. Based on very few observational facts
available at that time, they postulated that Type II supernovae are
the consequence of an implosion of non-degenerate stars, whereas Type
I's are the result of the ignition of nuclear fuel in degenerate
stars, and today this is still believed to be true in general, if
``Type I's'' are substituted by ``Type Ia's''.

\begin{figure}[ht]
  \centerline{\epsfxsize=0.9\textwidth\epsffile{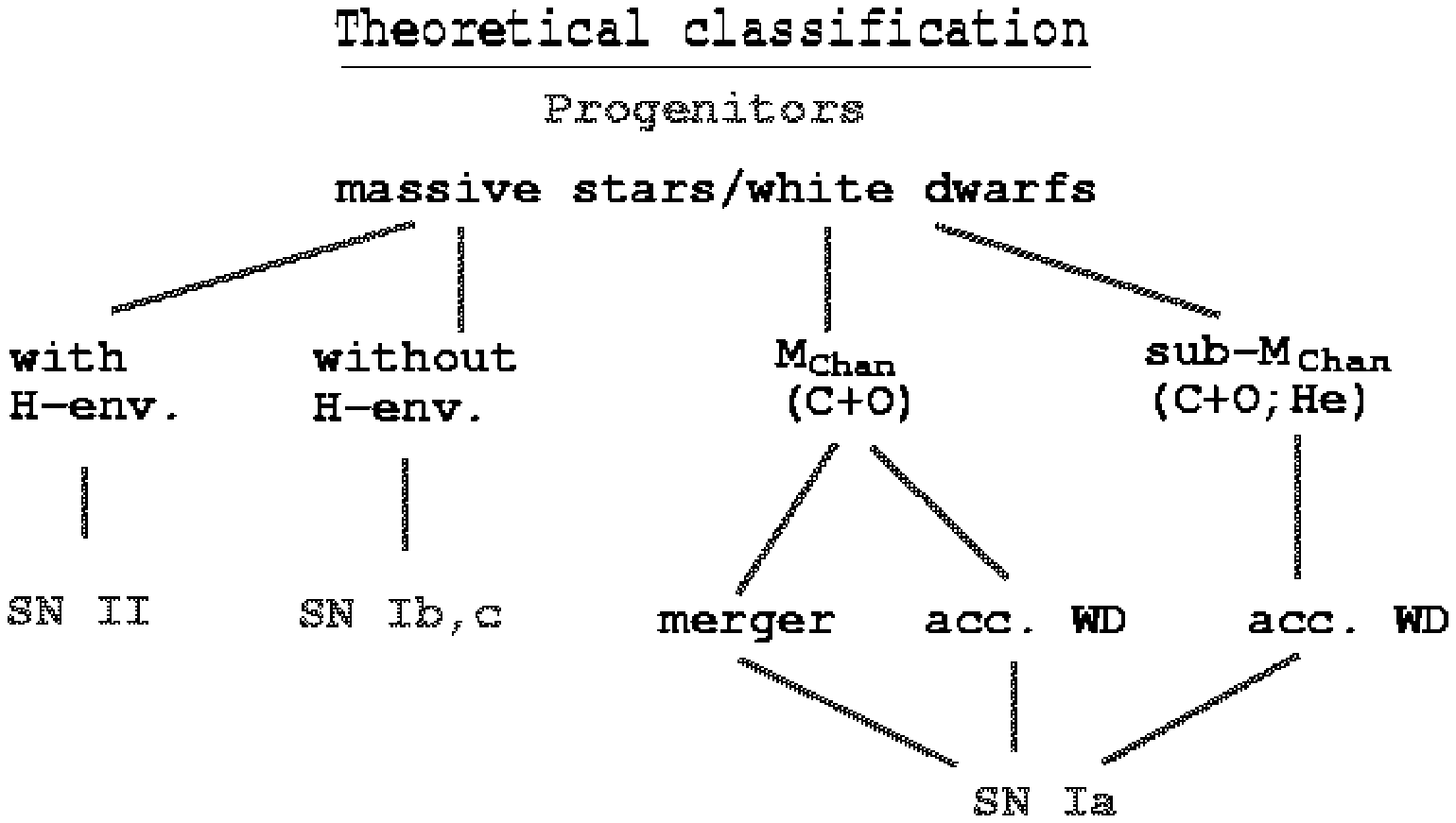}}
        \vglue -2.2truecm
  \caption{Theoretical classification scheme of supernovae according
  to their progenitors.}  \label{hillebrandt.fig1}
\end{figure}

\begin{figure}[ht]
  \centerline{\epsfxsize=0.9\textwidth\epsffile{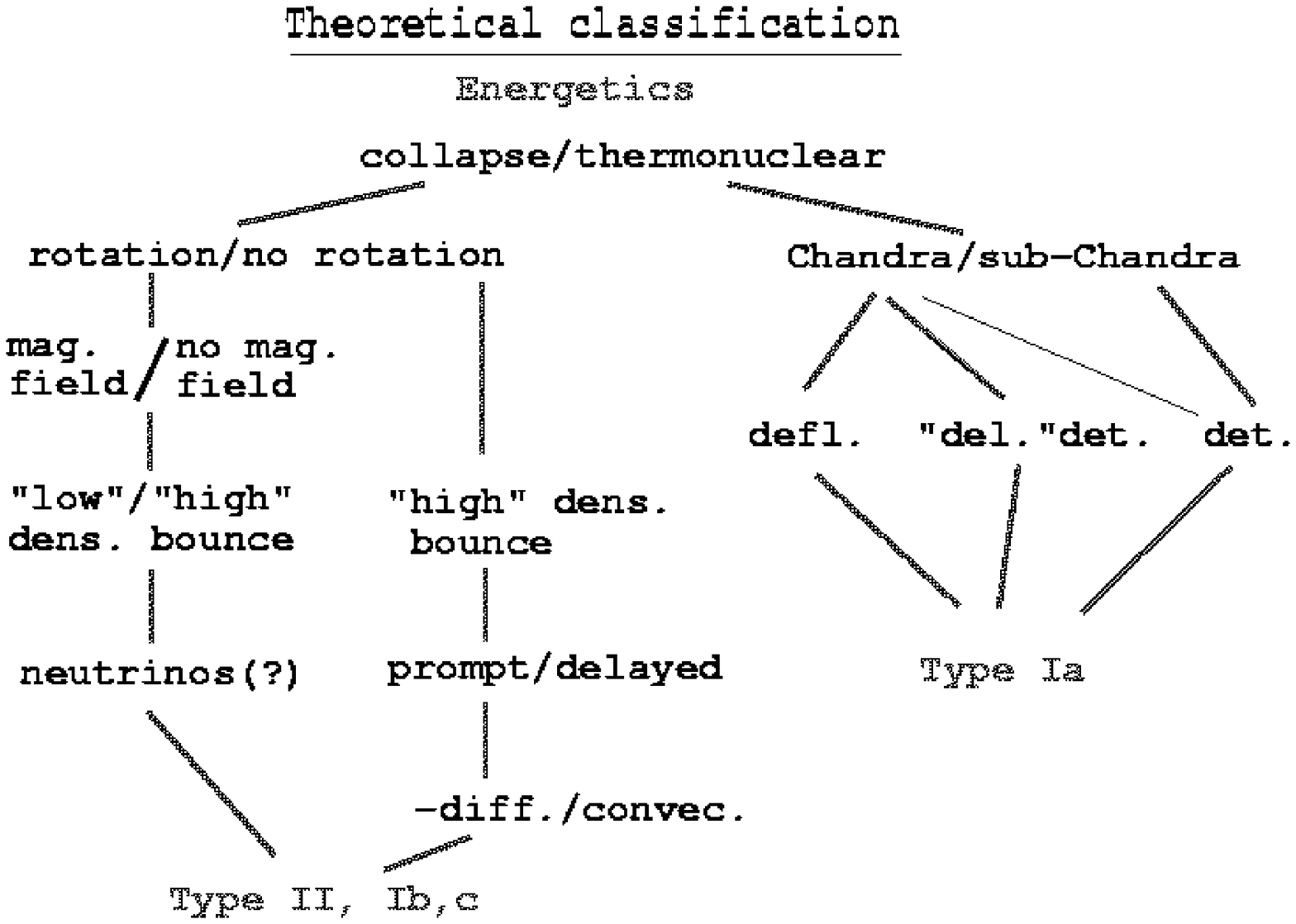}}
  \caption{Theoretical classification scheme of supernovae according
  to their energy sources.}  \label{hillebrandt.fig2}
\end{figure}
 
To be more specific, the present theoretical classification schemes
based on progenitor properties and the energetics of the explosion are
given in Figs.~\ref{hillebrandt.fig1} and~\ref{hillebrandt.fig2},
respectively.  It is now generally believed that Type II and Ib.c
supernovae stem from collapsing massive stars with and without
hydrogen envelopes, respectively, and that SN Ia originate from
explosions of white dwarfs, although the way to explosion (accretion
vs. merging of two white dwarfs) as well as the mass of the white
dwarf (Chandrasekhar vs.\ sub-Chandrasekhar mass) just prior to the
explosion are still heavily disputed (see Ref.~\cite{hillebrandt.5}
and references therein).
 
The classification with respect to the energetics is even more
complex, and a variety of possible models is available with no unique
and well accepted answer. For example, in the case of thermonuclear
explosions, it is not even clear what the mode of propagation of the
burning front is, and observations supply poor constraints only. A
(fast) deflagration wave in a Chandrasekhar-mass white dwarf can
explain the spectra and light curves of Type Ia's, but so can
deflagrations changing into detonations at low densities with or
without pulsations (so-called delayed detonations), or even pure
detonations in stars with low enough densities are not completely
ruled out. As far as the physics of core collapse supernovae is
concerned, the situation is not significantly better. Whether or not
rotation is important, whether or not magnetic fields play any role,
whether neutrinos are diffusively or convectively transported from the
cooling proto-neutron star to the stellar mantle, and many other
questions are largely unanswered.

\subsection*{Merging Theory and Observations}
 
Of course, a simple way to try to improve the rather unsatisfactory
situation outlined in the previous section is to call for more
observations. Since computing realistic theoretical light curves and
synthetic spectra for a given supernova model has become feasible
recently (although computing reliable spectra for Type Ia's still
requires novel techniques) high quality data might help to rule out
certain possibilities.  But as long as most theoretical models contain
a large number of more or less free parameters one may doubt the
success of this approach.
 
A few examples may serve as illustrations. As long as for a
thermonuclear explosion the propagation velocity of the burning front
and the degree of mixing can still be treated as free functions, it is
easy to fit a given light curve and spectrum of a Type Ia supernova
(see, e.g.\ Ref.~\cite{hillebrandt.1}), but one does not prove that
the model is correct.  Similarly, in the core collapse scenario,
neutrinos can transfer momentum to the mantle of a star and cause a
supernova explosion, provided their luminosity is sufficiently
high. If this luminosity results from low opacities or from convection
does not matter. Also, only thousands of neutrino events from a single
supernova seen in a neutrino detector could possibly tell the
difference.  Finally, a similar momentum transfer can, in principle,
be mediated by appropriate combinations of rotation and magnetic
fields. Again, a lucky accident of a galactic supernova may supply
enough data, including detections of gravity waves, to eliminate
several models, but only for one special event and not for the entire
class.
 
Therefore, it appears to be more promising to search for ways to
reduce the freedom one still has in building theoretical models. An
obvious possibility is to replace parameter studies by first principle
calculations, whenever this is possible. Again, a few examples are
given to demonstrate that there is plenty of room for improvements. A
first and obvious example is again neutrino transport in core collapse
supernovae. One will note that very primitive (and likely incorrect)
approximations for neutrino interactions with dense nuclear matter are
commonly used in all hydrodynamic simulations, and only very recently
first attempts have been made to compute those cross sections on the
basis of microscopic theories~\cite{hillebrandt.6,hillebrandt.7}.
Similarly, well-developed methods to calculate the properties of dense
nuclear matter are available, but the equations of state used in
numerical studies are still computed ignoring nucleon-nucleon
correlations, despite the fact that the stiffness of the equation of
state is very important ingredient.  As far as thermonuclear models
are concerned, numerical methods to handle turbulent combustion are
presently developed for combustion in engines and for reactor safety,
and there is no principle problem in applying them also to
supernovae. Moreover, simple microscopic models can be calibrated to
laboratory combustion experiments, which again will help to reduce the
number of parameters.
 
In conclusion, not underrating the importance of new and better
observational data in supernova research, I still think that most of
the progress in the near future has to come and can come from theory.

}\newpage {


\head{Supernova Rates}
     {Bruno Leibundgut}
     {European Southern Observatory\\
      Karl-Schwarzschild-Strasse 2,
      D-85748 Garching,
      Germany}

\noindent
Apart from stellar winds supernovae are the only mechanism which
releases chemically pro\-cessed material from stars into the
interstellar and intergalactic gas. They are almost exclusively
responsible for the chemical enrichment of galaxies and the universe
as a whole. As one of the end stages of stellar evolution and one of
the production channels of pulsars and black holes supernovae are also
placed at a major link of stellar and non-stellar matter. They further
inject kinetic energy into the interstellar gas which may be important
for gas heating, cosmic rays, and possibly the kinematics of galaxies.

Knowing the frequency of supernovae means to know the rate with which
these processes occur. Having further knowledge of the temporal
evolution of this frequency can provide the chemical history of
galaxies. Supernova statistics are thus a main ingredient in the study
of formation and transformation of matter in the universe.

Recent reviews of supernova rates has been published by van den Bergh
\& Tammann (1991), Tammann (1994), and Strom (1994).  The visual
searches have been summarized by van den Bergh and McClure (1994).  A
series of papers by Cappellaro et al.\ (1993a, b, 1997) have explored
the derivation of the supernova rate in detail.  SN statistics
starting from the available SN catalogs and exploiting various
indirect routes were presented by Tammann, L\" offler, \& Schr\" oder
(1994).

Supernova statistics make use of minimal information about the
explosions themselves. The main ingredients are the type of the
explosion and a crude description of the parent population in which
the explosion occurred. This is normally restricted to the
morphological type of the host galaxy. The last piece of information
is the time when the explosion took place.  The frequency of
supernovae $\nu_{\rm SN}$ 
is described as the ratio of known SNe $N_{\rm SN}$
over the total galaxy luminosity $L_{\rm gal}$ per unit time $t$
\[ \nu_{\rm SN} = \frac{N_{\rm SN}}{L_{\rm gal} \cdot t}.  \]
The SN rate is normally expressed in Supernova Units (SNu) which
corresponds to one supernova per $10^{10}$ (blue) solar luminosities
per 100 years.

Although in principle easy to determine, the derivation of supernova
rates suffers from a number of problems.  Supernovae are very rare
objects.  The number of supernovae in galaxies with known parameters
(morphological type, luminosity, color) is rather small. There are
about 60 to 80 supernovae per year, but only about 10 bright
supernovae in nearby galaxies (cf.~Timmes \& Woosley 1997).  The
surveyed galaxy sample has to be well described and defined. Many
supernovae occur in galaxies which are not part of the current
catalogs, thus being lost for the derivation of the rates.  Another
problem is the control times, i.e.~the time the galaxies have actually
been surveyed for supernovae. Large corrections can be incurred when
the galaxies are not observed frequently.

The small number statistics becomes even more apparent when we
consider that the known supernovae have to be split into a number of
subsamples to become meaningful. Typically there is a subdivision into
three SN types and about five galaxy morphologies, which immediately
creates 15 bins to group the supernovae. With typical numbers of about
200 SNe (e.g.~van den Bergh \& McClure 1994, Cappellaro et al.\ 1997)
in the sample the rates are not immune against small number
statistics.

Over 1200 supernovae been discovered to date.  Many of these objects
are not classified and can not be used for the statistics. However,
about 80\% of all supernovae since 1989 have a type associated. The
distribution is roughly 55\% SNe~Ia (thermonuclear explosions of white
dwarfs) and 41\% SNe~II and SNe~Ib/c (core-collapse in massive stars).
Since SNe~Ia are on average about 2 magnitudes brighter than other
supernovae, they can be detected over a larger volume and thus are
present at a higher percentage. Excluding all the supernovae from
searches specifically targeted at very distant supernovae (Perlmutter
et al.\ 1997, Leibundgut \& Spyromilio 1997), which mostly are
Type~Ia, we arrive at a near equipartition between thermonuclear and
core-collapse SNe (48\% SNe~Ia and 49.5\% SNe~II and SNe~Ib/c since
1989). With the same argument as above (SNe~Ia more luminous) we can
immediately conclude that core-collapse SNe are more frequent than
thermonuclear SNe in the nearby universe. This result is further
amplified by dust obscuration which affects core-collapse SNe stronger
than thermonuclear SNe.

Several corrections enter the determination of supernova rates.  The
best known is the Shawn effect (Shawn 1979) which describes the effect
of supernovae lost in the glare of centers of galaxies. Visual and CCD
searches are less affected than older photographic searches which lose
the contrast in bright regions. Absorption can obscure supernovae in
the discs of galaxies and if, as currently believed, core-collapse
supernovae stem from massive stars and are thus more affected by dust,
the derived SN statistics are skewed. Several proposals how to correct
for dust have been put forward (e.g.~van den Bergh \& Tammann 1991).
The luminosity of the various subtypes of supernovae differs by
factors of about 10 and the differences in light curve shapes
introduces varying control times for the SN classes. These have to be
factored in when the SN rate is derived.

There are two distinct approaches to derive SN statistics. The `search
technique' uses only supernovae which have been discovered in
controlled searches. The most recent and complete description of this
method is given by Cappellaro et al.\ (1997). The advantage here is
that the control times and the galaxy sample are defined very
well. The drawback is the small number of supernovae in the
sample. Cappellaro et al.\ (1997) had to discard about half of the
discovered supernovae, because they appeared in galaxies which are not
part of the catalogs. They were left with 110 supernovae from five
searches which have lasted for almost 20 years and amounted to a total
of 42500 years $\times 10^{10} L_{\odot}$ control time (sometimes
referred to as `galaxy years').

The `catalog technique' makes use of all supernovae discovered in a
known galaxy sample (Tammann et al.\ 1994). This maximizes the number
of supernovae as all supernovae discovered in galaxies of the sample
are included in the statistics, but strong assumptions on the control
time have to be made. The latter quantity can be derived very badly as
serendipitous SN discoveries enter the catalogs.

Despite all these problems, there is a fair agreement for the rates of
thermonuclear supernovae (Type~Ia) in star-forming (spiral)
galaxies. The distribution is independent of the detailed
morphological type of the galaxy at about 0.2 SNu. A large discrepancy
exists for old stellar systems (elliptical galaxies) for which Tammann
et al.\ find a rate increasing to a level three times higher, while
Cappellaro et al.\  actually find indication of slightly lower SN~Ia
rate in ellipticals.  There is only a very small number of supernovae
known in elliptical galaxies, however, and the statistics are very
uncertain.

Core-collapse supernovae (Types II and Ib/c) have not been observed in
elliptical galaxies.  Their rate increases with the fraction of young
stars in galaxies, i.e.~the morphological type. The most prolific
supernova producers are Sc and Sd galaxies with about 1.2 SNu.  There
is a discrepancy of about a factor of two in the absolute rates
between Tammann et al.\ and Cappellaro et al.\ for all galaxy
types. The relative rates, however, agree quite well. The SN rates in
irregular galaxies are largely undetermined due to the small numbers
of known SNe.

The Galactic SN rate is about $20\pm 8$ per millennium. This number
depends on the morphological type and the total blue luminosity of the
Galaxy. There are 6 historical supernovae on record (van den Bergh \&
Tammann 1991) which implies that roughly 2/3 of all Galactic
supernovae have been hidden.  Supernovae in the Local Group have been
observed in M31 (S~And: SN~1885A) and in the Large Magellanic Cloud
(SN~1987A). The prediction for the Local Group is slightly above 2
(excluding the Galaxy), which appears consistent. The small
observational baseline, however, excludes any firm conclusion from
such a comparison.

}\newpage{


\def\<{\langle}
\def\>{\rangle}

\head{Pulsar Velocities and Their Implications}
     {A.G.~Lyne}
     {University of Manchester, Jodrell Bank, Macclesfield,
     Cheshire SK11 9DL, UK}

\noindent
In several respects, the Galactic distribution of pulsars mimics the
distribution of population I species, such as young stars, HII regions
and supernova remnants.  In particular, the galactocentric radial
distributions are very similar.  However, the widths of the
distributions in distance from the Galactic plane, Z, differ by nearly
a factor of 10.  Gunn and Ostriker (1970)~\cite{aglyne.1} realised
that this difference could result from high velocities which may be
given to pulsars at birth, possibly arising in the violence of their
formation in supernova collapse.  It was several years later before
the first measurement of the proper motion of a pulsar gave direct
evidence for this hypothesis~\cite{aglyne.2}.  Transverse velocities
are now known for more than 100 pulsars and their high values are
generally well established.

In this review, I describe briefly the techniques used in the
measurement of pulsar velocities, summarise the main features of the
distribution of the velocities and discuss their implications for the
physics of the formation events and for their spatial distribution.

There are two main techniques which have been used for the
determination of pulsar velocities, in both cases providing the
transverse components of the velocity only.  These techniques
respectively involve the measurement of proper motion and of the
velocity of the interstellar scintillation pattern across the Earth.
Proper motions are determined from high-resolution radio
interferometry or from timing measurements.  Like optical techniques,
the former involves measurement of the position of a pulsar relative
to a nearby source over a period of time.  With modern instruments,
position accuracy of 10 milliarcseconds (mas) is readily achievable,
so that only a few years are required to give proper motion errors of
only a few mas/yr~\cite{aglyne.3,aglyne.4,aglyne.5,aglyne.6,aglyne.7}.
Timing measurements can give similar precision in principle, but are
usually limited by irregularities in the rotation rate of the pulsar,
known as ``timing noise,'' which is predominant in young pulsars.
However, timing is particularly useful for millisecond pulsars which
have very stable rotation and whose narrow pulses allow very 
high~accuracy.

Estimates of pulsar transverse velocities $V_t$ are then obtained from
the proper motion $\mu$ (mas/yr) and the distance $D$ (kpc):
$V_t=4.74\,\mu D$ km/s.  The distance $D$ is obtained from the
dispersion measure and an electron density model~\cite{aglyne.8} which
is based upon a number of independent distance measurements.  Nearly
100 such measurements of transverse velocity are available.

\begin{figure}[ht]
        \vglue 5truemm
  \centerline{\epsfxsize=0.6\textwidth\epsffile{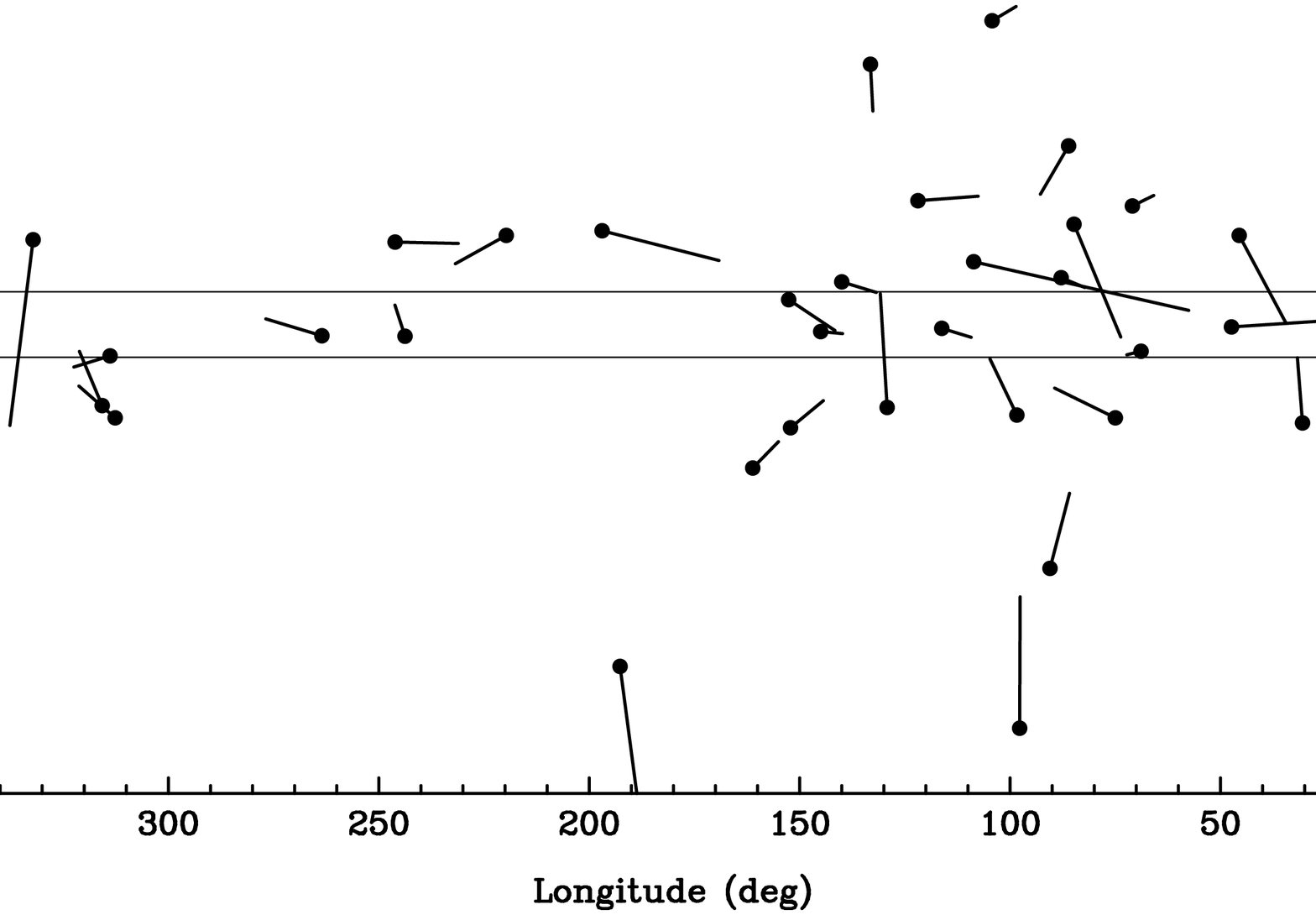}} 
  \vglue-4.8truecm 
  \caption{The
  motions of pulsars relative to the galactic plane.  Pulsars are
  represented as filled circles in galactic longitude and Z-distance,
  the tail representing its approximate motion in the past 1 Myr.}
  \label{aglyne.fig1} \vglue 1truemm
\end{figure}

The velocities of pulsars estimated from observations of the speed of
the interstellar scintillation patterns~\cite{aglyne.9,aglyne.10} are
considered rather less reliable than the more direct proper motion
measurements, mainly because it has recently been
shown~\cite{aglyne.11} that the values are systematically low by an
average factor of 2 because of a localisation of the scattering medium
close to the galactic plane.  There are about 71 pulsars for which
scintillation speeds are available~\cite{aglyne.10}, although most of
these pulsars have rather better proper motion measurements, leaving
27 which have only scintillation measurements.  Combining both
techniques gives a sample of about 100 values of $V_t$, of which about
8 are upper limits.

From this sample, the mean transverse velocity is found to be
$\<V_t\>=300\pm30$ km/s.  With a small number of exceptions, the
velocity vectors shown in Fig.~\ref{aglyne.fig1} demonstrate a clear
movement away from the galactic plane, consistent with pulsars being
formed from massive, young, population I stars close to the plane and
receiving velocities of a few hundred km/s at the same time.  How
large are the typical space velocities?  Unfortunately, there is a
selection effect visible in the data in which high velocity pulsars
quickly move away from the plane of the Galaxy and become
undetectable, leaving an excess of low-velocity pulsars.  This effect
can be minimised by considering the velocity distribution of only the
youngest pulsars, which cannot have moved far since birth.  The 29
pulsars with characteristic age of less than 3 Myr give
$\<V_t\>=350\pm70$ km/s.  Monte Carlo simulations show that the mean
space velocity required to give such an observed transverse velocity
distribution is~$\<V_s\>=450\pm100$~km/s~\cite{aglyne.12}.

There are two other pieces of evidence which support such high
velocities: the increase of the mean distance of pulsars from the
plane as a function of characteristic age and the movement of a small
number of pulsars from the centres of the putative supernova remnants
associated with their birth~\cite{aglyne.12}.

Similar analyses for the 13 millisecond pulsars with proper motion
measurements give $\<V_t\>=88\pm19$ km/s and $\<V_s\>=130\pm30$ km/s,
showing them also to be a rather high-velocity population.  Not
surprisingly, no migration from the galactic plane is seen, because of
their large ages.  These velocities probably represent the
low-velocity end of the ``kick'' spectrum---higher velocity kicks are
likely to disrupt any binary system, preventing any millisecond pulsar
formation by accretion spin-up in a binary system.

Clearly, the high velocities explain why only a few percent of pulsars
are in binary systems, compared with the high proportion of the likely
progenitor stars in binaries.  However, the high velocities pose
something of a problem for the retention of pulsars in globular
clusters: the velocities mostly exceed the cluster escape velocities
and yet there is a substantial pulsar population which has remained
gravitationally bound~\cite{aglyne.13,aglyne.14}.  It is also worth
noting that about half of the pulsar population will escape the
gravitational potential well of the Galaxy, ending up in intergalactic
space, the remainder being in a large galactic halo.

There has been much speculation on the origin of the high velocities,
the three main possibilities being the disruption of binary systems,
an electrodynamic ``rocket'' effect in the early life of the pulsars
and asymmetry in the supernova explosions.  The first two have great
difficulty in explaining the size of the velocities and the most
likely origin seems to lie in the supernova explosion itself. The high
momentum asymmetry of the neutron star after the explosion must be
matched by that of either the ejecta or neutrinos.  The origin of the
asymmetry is not clear, although both convection and the magnetic
field have been cited as the possible determining agents.  It is worth
noting that up to now, the magnetic properties of the neutron stars do
not seem to implicate the magnetic field in the production of 
the~high~velocities.

}\newpage{


\def \la {\mathrel{\vcenter
     {\offinterlineskip \hbox{$<$}\hbox{$\sim$}}}}
\def \ga {\mathrel{\vcenter
     {\offinterlineskip \hbox{$>$}\hbox{$\sim$}}}}
\def\ave#1{\langle #1 \rangle}
\def\eck#1{\left\lbrack #1 \right\rbrack}
\def\rund#1{\left( #1 \right)}

\head{Convection in Newly Born Neutron Stars}
     {W.~Keil}
     {Max-Planck-Institut f\"ur Astrophysik\\
      Karl-Schwarzschild-Str.~1, D-85740 Garching, Germany}

\subsection*{Introduction}

Observations of SN 1987A instigated recent work on multi-dimen\-sional
numerical modeling of Type-II supernova explosions. Two-dimensional
(2D) simulations~\cite{wkeil.1,wkeil.2,wkeil.3,wkeil.4} indicate that
overturn instabilities and mixing in the $\nu$-heated region between
PNS and supernova shock may indeed help $\nu$-driven explosions to
develop for conditions which do not allow explosions in spherical
symmetry. Nevertheless, a critical investigation of the sensitivity of
the explosion to the value of the $\nu$ flux from the neutron
star~\cite{wkeil.3,wkeil.4} cannot confirm claims that a ``convective
engine'' ensures the robustness of the explosion for a wide range of
conditions. Instead, the explosion turns out to be extremely sensitive
to the $\nu$ fluxes from the inner core. Explosions can occur in 2D as
well as in 1D, provided the $\nu$ luminosity is large
enough. Similarly, for too small core $\nu$ fluxes, strong convective
overturn in the $\nu$-heated region cannot develop and the explosion
fizzles. These findings are supported by recent simulations of
Mezzacappa and collaborators~\cite{wkeil.5}. In addition, all
currently successful numerical models of supernova explosions produce
nucleosynthesis yields that are in clear contradiction with
observational abundance constraints. These facts indicate severe
problems of the numerical modeling of the explosion.  The first 2D
simulations of the evolution of the nascent neutron
star~\cite{wkeil.6} suggest that long-lasting quasi-Ledoux convection
{\it in} the PNS can significantly raise the $\nu$ luminosities and
may have a positive impact on the supernova explosion and conditions
for explosive nucleosynthesis.

\subsection*{Results of Two-Dimensional Hydrodynamical Simulations}

Our simulations were performed with an elaborate description of the
nuclear equation of state~\cite{wkeil.7} and the $\nu$ transport and
were started with a PNS model of Bruenn~\cite{wkeil.8} at $\sim 25$~ms
after core bounce. These calculations showed that driven by negative
lepton fraction and entropy gradients convective activity can
encompass the whole PNS within $\sim 1\,{\rm s}$ and can continue for
at least as long as the deleptonization of the PNS takes place.
Because of the importance of the diffusive $\nu$ transport this
convection is not an ideal, adiabatic Ledoux convection. Also specific
thermodynamical properties of the nuclear equation of state have to be
taken into account. In differentially rotating PNSs stabilizing
angular momentum distributions seem to suppress convection near the
rotation axis.


\begin{figure}[ht]
 \begin{tabular}{cc} 
 \hbox{\epsfxsize=12cm\epsffile{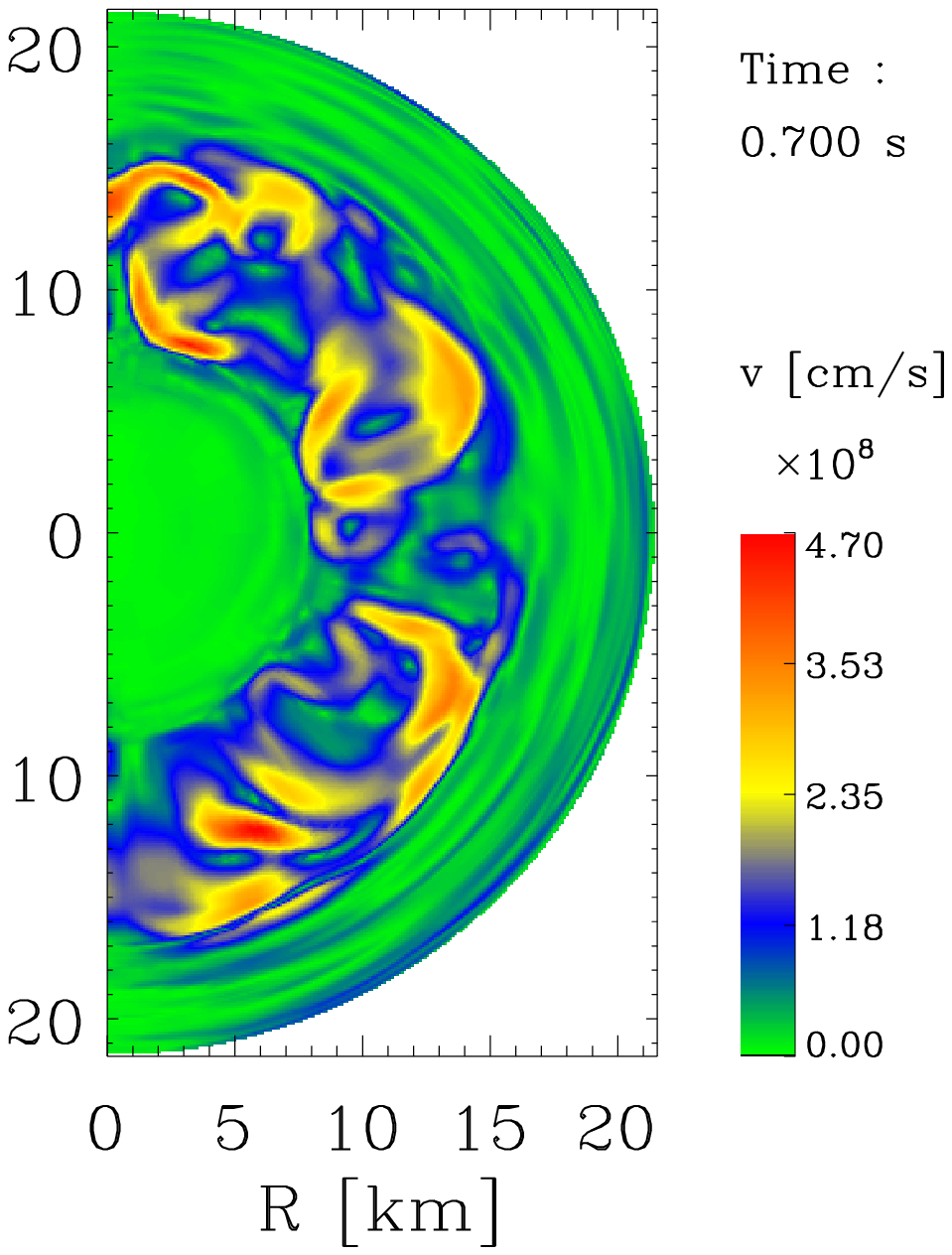}}
 \put(-160.0,10.0){{\Large \bf a}} & 
 \hbox{\hskip-4.5truecm
 \epsfxsize=12cm \epsffile{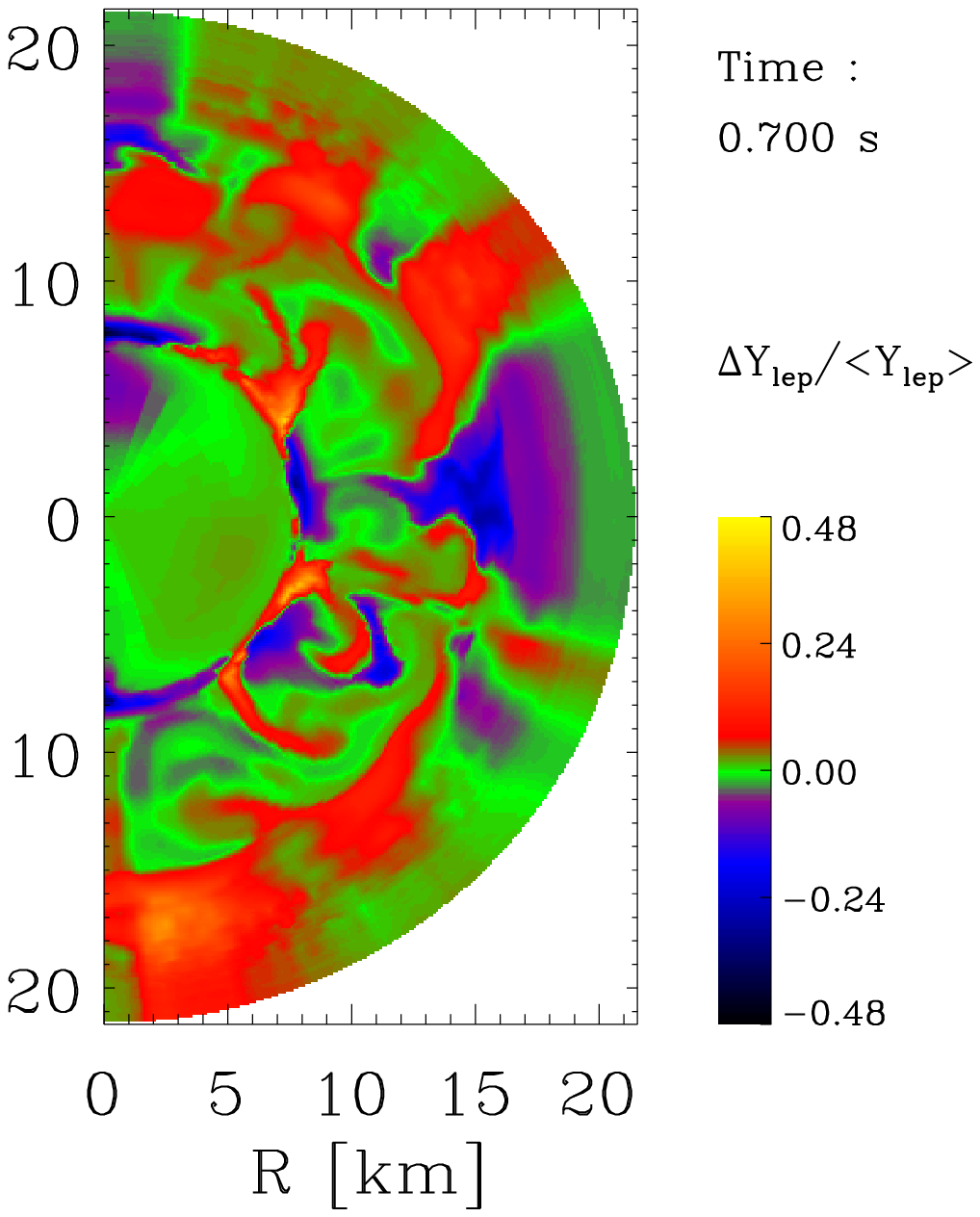}
 \put(-160.0,10.0){{\Large \bf b}} 
 \hskip-5.0truecm}
 \end{tabular} 
 \caption{Panel {\bf a} shows the absolute values
 of the convective velocity for a non-rotating PNS model $0.7~{\rm s}$
 after the start of the simulation in units of $10^8\,{\rm cm/s}$.
 The computation was performed in an angular wedge of $180^{\circ}$
 assuming axisymmetry.  Panel {\bf b} displays the relative deviations
 of the lepton fraction $Y_{\rm lep}$ from the angular means
 $\ave{Y_{\rm lep}}$ at each radius for $t = 0.7~{\rm s}$. The maximum
 deviations are of the order of 50\%. Lepton-rich matter rises while
 deleptonized material sinks in.} \label{wkeil.fig1}
\smallskip
\end{figure}


The convective pattern is extremely non-stationary and has most
activity on large scales with radial coherence lengths of several km
up to $\sim 10\,{\rm km}$ and convective ``cells'' of
20$^{\circ}$--30$^{\circ}$ angular diameter, at some times even
45$^{\circ}$ (Fig.~\ref{wkeil.fig1}a). The maximum convective
velocities are $\sim 5\times 10^8\,{\rm cm/s}$, but peak values of
$\sim 10^9\,{\rm cm/s}$ can be reached. Relative deviations of the
lepton fraction $Y_{\rm lep}$ from the angular mean can be several
10\% in rising or sinking buoyant elements (Fig.~\ref{wkeil.fig1}b),
and for the entropy ($S$) they can reach 5\% or more. Rising flows
always have larger $Y_{\rm lep}$ {\it and} $S$ than their
surroundings.


\begin{figure}[ht]
 \begin{tabular}{cc}
 \hskip -0.5 truecm
 \epsfxsize=7.5cm\epsffile{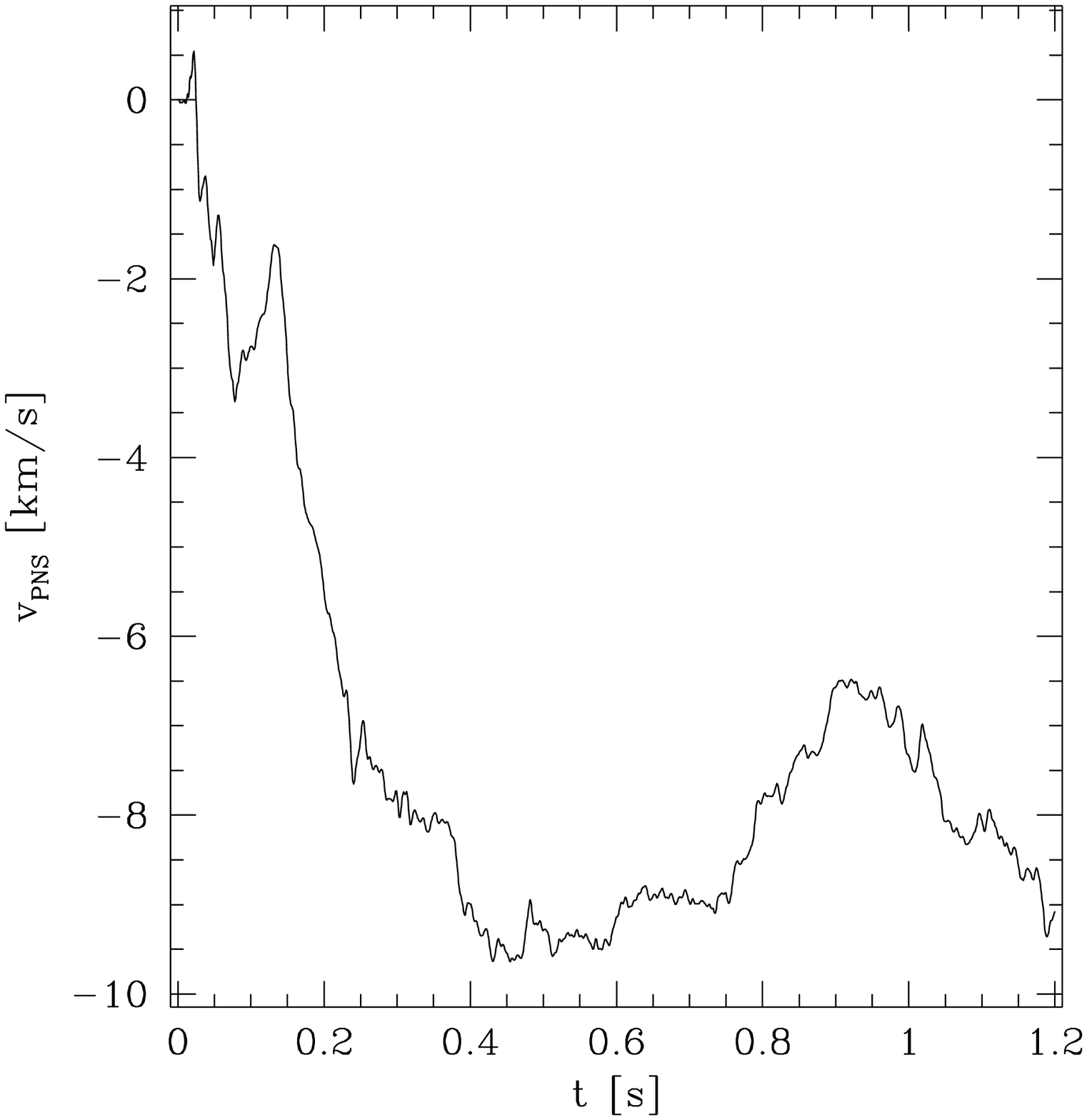}
 \put(0.0,0.0){{\Large \bf a}} &
 \epsfxsize=7.5cm \epsffile{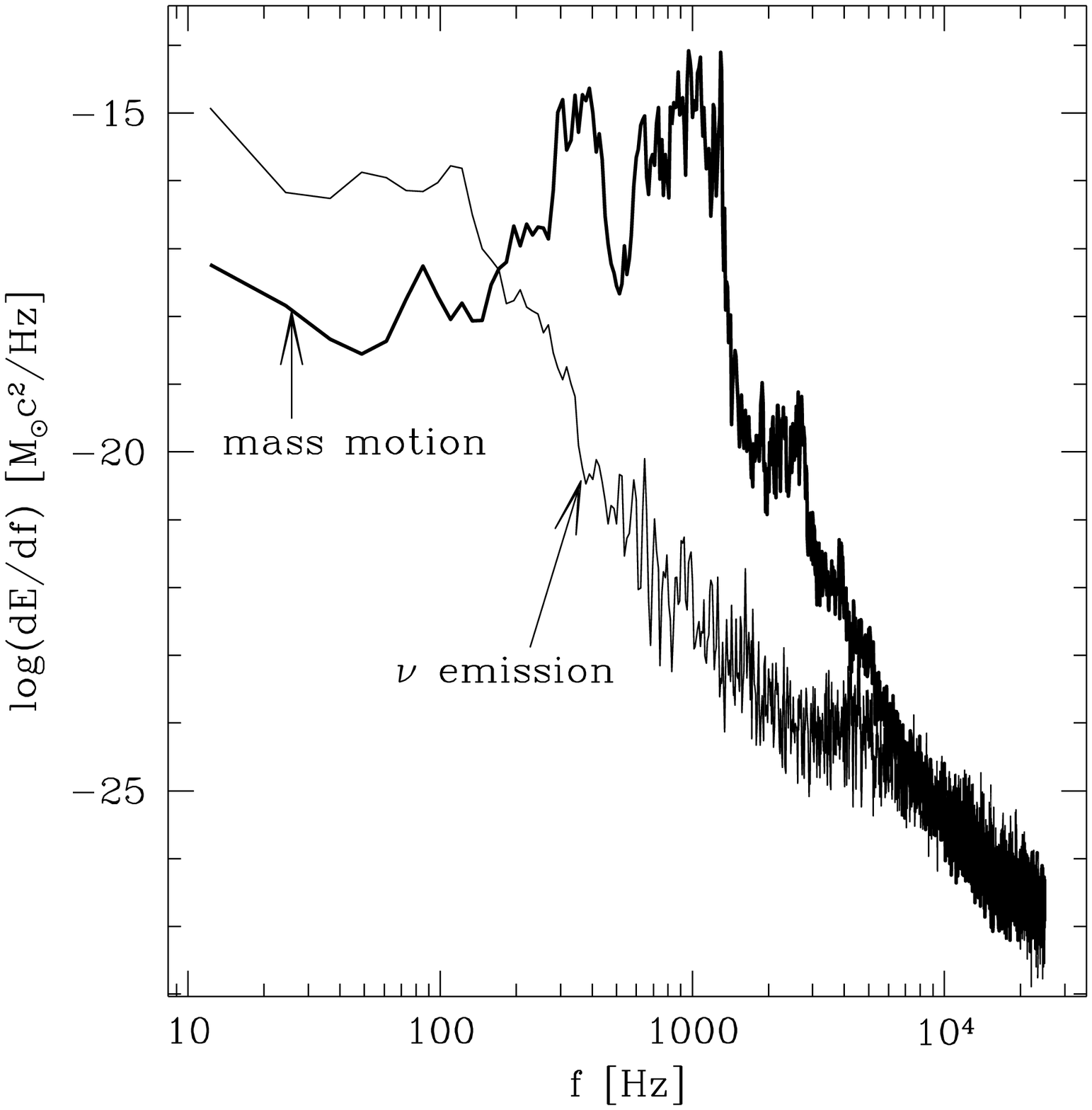} 
 \put(0.0,0.0){{\Large \bf b}}
 \end{tabular}
 \caption{
  Panel {\bf a} shows the recoil velocity of our PNS model in km/s 
  as a function of time resulting from stochastical asymmetries in the
  $\nu$ flux.  
  Panel~{\bf b} displays the spectral energy density of the emitted
  gravitational radiation caused by mass motions and anisotropic $\nu$
  emission.}
 \label{wkeil.fig2}
\end{figure}


Since convective activity continues for more than 1 s in a large
region of the PNS, convective $\nu$ transport shortens the cooling and
deleptonization timescales of the PNS compared to results of 1D
simulations and enhances the calculated $\nu$ luminosities of the star
by up to 65\%.  The latter can have important influence on the
interpretation of the measured $\nu$ signal from SN 1987A and might
change limits of various quantities in nuclear and elementary particle
physics which have been derived from these measurements. Moreover, the
enhancement of the $\nu$ fluxes can also have important consequences
for the explosion mechanism of the supernova, because it aids
$\nu$-driven explosions.  The faster deleptonization modifies the
luminosity ratio of $\nu_e$ and $\bar{\nu}_e$ in such a way that the
electron fraction $Y_e^{\rm ej}$ in the $\nu$-heated SN ejecta will be
raised during the first few 100 ms after core bounce, but it will be
lowered for $t \ga 1$~s compared to 1D simulations.  This might help
to solve the severe problems of the nucleosynthesis in current models
of Type-II supernovae which disregard long-lasting convective
activities in the PNS. Finally, convection in the PNS causes
stochastical asymmetries of the $\nu$ flux. In our calculation the
resulting recoil accelerates the PNS to $\sim$ 9~km/s during 1.2~s
(Fig.~\ref{wkeil.fig2}a). This is too small to explain measured proper
motions of pulsars of a few 100~km/s. Both convective mass motions and
anisotropic $\nu$ emission are a source of gravitational waves for
more than 1 second (Fig.~\ref{wkeil.fig2}b). Most of the gravitational
radiation is emitted at frequencies of 300--1400~Hz.


\subsection*{Acknowledgements} 
We thank S.W.~Bruenn for kindly providing us with the post-collapse
model used as initial model in our simulations. Support by the ``SFB
375-95 f\"ur Astro-Teilchenphysik'' of the German Science Foundation
is acknowledged.

}\newpage{


\def\gtrsim{\,\hbox{\hbox{$ > $}\kern -0.8em \lower
1.0ex\hbox{$\sim$}}\,}
\def\lesssim{\,\hbox{\hbox{$ < $}\kern -0.8em \lower
1.0ex\hbox{$\sim$}}\,}
\def\ave#1{\langle #1 \rangle}
\def\eck#1{\left\lbrack #1 \right\rbrack}
\def\rund#1{\left( #1 \right)}

\head{Anisotropic Supernovae, Magnetic Fields,\\ 
      and Neutron Star Kicks}
     {H.-Th.\ Janka}
     {Max-Planck-Institut f\"ur Astrophysik\\
      Karl-Schwarzschild-Str.~1, D-85740 Garching, Germany}

\subsection*{Abstract}

Hydrodynamic instabilities during the supernova explosion and neutrino
cooling phase lead to stochastic acceleration of the nascent neutron
star and are hardly able to account for recoil velocities $\gtrsim
100\,$km/s.  It is argued that an internal magnetic field of $\sim
10^{14}\,{\rm G}$ can define a preferred direction of the neutrino
emission with an anisotropy of several per cent which is sufficiently
large to produce kicks even in excess of 1000$\,$km/s and thus can
explain the fastest motions of observed pulsars.

\subsection*{Introduction}

Pulsars are observed with large mean velocities of several 100~km/s
(see A.~Lyne, this conference, for references to observations), and a
few neutron stars are measured with velocities of more than
1000~km/s. Similar velocities can also be deduced from associations of
neutron stars with nearby supernova remnants. Binary evolution
arguments strongly suggest that neutron stars experience a recoil
already during their birth in the supernova explosion of a massive
progenitor star (e.g.,~\cite{janka.-9,janka.-8}). The reason for this
acceleration is not finally understood and several mechanisms were
proposed or investigated, among them anisotropic core collapse and
mass ejection during the supernova explosion (e.g.,~\cite{janka.-7}),
large-scale hydrodynamic instabilities during the
explosion~(e.g.,~\cite{janka.-6,janka.-5}), or anisotropic neutrino
emission of the cooling proto-neutron star~\cite{janka.-4}, for
example associated with convective processes
(e.g.,~\cite{janka.-6,janka.-5,janka.-3}), magnetic fields
(e.g.,~\cite{janka.-2,janka.3}), or neutrino oscillations
(e.g.,~\cite{janka.-1,janka.0}). A 1\% anisotropy of the neutrino
emission during the neutrino cooling phase of the newly formed neutron
star would be sufficient to accelerate the remnant to about
300$\,$km/s.

While some of the proposed kick mechanisms are highly speculative and
involve one or even more components of new or non-standard physics
(e.g., neutrino masses or magnetic moments, superstrong magnetic
fields), other mechanisms are unsatisfactory in the sense that they
rely on certain initial conditions in the collapsing stellar core or
in the nascent neutron star which are very specific and cannot be
easily justified from stellar evolution calculations (e.g., special
magnetic field configurations or density distributions to produce
global anisotropies).  More ``natural'' anisotropies which develop in
self-consistent simulations from hydrodynamic instabilities without
the use of specially chosen initial conditions, on the other hand,
turned out to be sufficient to account for small velocities of a few
10~km/s but seem hardly able to explain the very fast motions
of pulsars.

\subsection*{Hydrodynamic Instabilities and Neutron Star Convection}

For example, anisotropic mass motions due to convection in nascent
neutron stars lead to gravitational wave production and anisotropic
emission of neutrinos (see W.~Keil, this conference). The angular
variations of the neutrino flux determined by 2D simulations are of
the order of 5--10\%~\cite{janka.1,janka.2}. With the typical size
(10--30 degrees angular diameter) and short coherence times (a few
milliseconds) of the convective structures, however, the global
anisotropy of the neutrino emission from the cooling proto-neutron
star is certainly less than 1\% (more likely only 0.1\%) and observed
high kick velocities in excess of 300~km/s can definitely not be
explained. In the simulation described by Keil, the determined recoil
velocity is disappointingly low, only about 10~km/s! There is no
reason to expect this case to be an especially unfavorable one. Even
worse, in three dimensions the convective elements tend to become
smaller than in 2D because of fragmentation and turbulent energy
transport from larger to smaller scales which is forbidden in 2D due
to angular momentum conservation. This makes large recoil velocities
of neutron stars by the described mechanism even more unlikely.

Convective motions in the neutron star have important consequences for
the magnetic field structure. Specific kinetic energies of the order
of ${1\over 2}v^2\sim 10^{17}\,$erg/g could lead to equipartition
fields of more than $B\sim 10^{16}\rho_{14}^{1/2}\,$G.  Such strong
fields increase the neutrino opacities~\cite{janka.3,janka.4} and
could affect the anisotropy of the neutrino emission.  In the dense
medium of the neutron star, where the electrons are degenerate, the
influence of the magnetic field on the charged-current reactions
becomes essential when the characteristic energy of an electron on a
Landau level reaches the size of the electron chemical potential,
i.e.~when $B\gtrsim B_{\rm c}\cdot(\mu_e/m_ec^2) \sim 5\times
10^{15}\ldots2\times 10^{16}\,$G where $B_{\rm c} = m_e^2c^3/(\hbar e)
= 4.4\cdot 10^{13}\,{\rm G}$. The presence of such strong fields
changes the phase space distribution of the electrons and thus the
cross section for $\nu_e$ absorption on neutrons~\cite{janka.4}.
Whether this results in sizable neutron star kicks or not will depend
on the strength and structure of the magnetic fields inside the
neutron star. Turbulent convective motions will certainly produce a
much more irregular magnetic field distribution than the idealized
mirror asymmetry assumed by Bisnovatyi-Kogan~\cite{janka.3} who
suggested that an initial, strong toroidal field component is
amplified by a wound-up poloidal field in one hemisphere, whereas both
toroidal field components (the initial one and the one produced by
winding the poloidal field) are superposed destructively in the other
hemisphere.  A more irregular field evolution will imply a much
smaller global anisotropy of the neutrino emission than estimated in
reference~\cite{janka.3}.

Rapid rotation of the collapsed stellar core has very important
effects on the development and strength of
convection. Two-dimensional, self-consistent hydrodynamic
simulations~\cite{janka.1,janka.2} reveal that a positive gradient of
the specific angular momentum in the nascent neutron star stabilizes
hydrodynamically unstable entropy or lepton number stratifications.
This positive gradient of the specific angular momentum suppresses
convective motions near the rotation axis and allows strong convection
to occur only close to the equatorial plane. Therefore, the neutrino
emission is convectively enhanced around the equatorial plane while it
is essentially unchanged near the poles. For the same reason, the
magnetic field structure may retain an initial dipole component and
irregular fields must be expected to be produced by convective mass
motions only in a belt around the equator.

\subsection*{Parity Violation and Cumulative Asymmetry}

According to a recent suggestion by Horowitz and Li~\cite{janka.5}, a
global anisotropy of the neutrino emission of the nascent neutron star
could originate from the cumulative effect of a large number of
neutrino scatterings off polarized nucleons in the magnetized stellar
medium. Although the polarization corrections to a single
neutrino-lepton interaction are small, typically of the order of only
$10^{-5}$ of the cross section~\cite{janka.5,janka.6}, Horowitz and Li
recognized that a macroscopic asymmetry can build up because the
anisotropy of the neutrino flux increases with the average number of
scatterings per neutrino, i.e.~with the optical depth of the
medium. The polarization term in the cross section represents a
slightly different chance for a neutrino to be scattered into the
direction of the external magnetic field than into the opposite
direction, or means a tiny enhancement of the scattering probability
of neutrinos moving into (or opposite to) the field direction.
Formally, the neutrino flux is not only described by a diffusive
propagation mode (proportional to the gradient of the neutrino energy
density), but an additional ``advective'' vector component in the
magnetic field direction (see~\cite{janka.7,janka.8}).  The ratio
between advective and diffusive component grows approximately linearly
with the scattering optical depth $\tau$.  Physically, this can be
understood~\cite{janka.5} by recalling that a diffusing particle has
travelled a mean distance $d = \sqrt{N}\lambda$ after $N$ steps with
(constant) scattering mean free path $\lambda$.  If the particle has a
tiny chance $P$ to be scattered into the field direction, the distance
it has propagated along the field after $N$ scatterings is $l =
PN\lambda$. Therefore the relative anisotropy increases as $l/d
\propto P\sqrt{N}\propto \tau$ (for $P\sqrt{N}\ll 1$).  The second
proportionality holds because the optical depth is defined as $\tau =
d/\lambda$, thus $\tau = \sqrt{N}$.

The discussed process would be highly interesting if the magnetic
fields required to create a few per cent anisotropy of the neutrino
emission of a neutron star were much smaller than the very strong
fields ($\gtrsim$ several $10^{15}\,{\rm G}$) where the neutrino
opacity is affected by the change of the electron phase space
distribution.  Horowitz and Li~\cite{janka.5} have only considered
very simple situations of scattering media to show the fundamental
characteristics of the cumulative process and have disregarded
essential complications of neutrino transport in neutron stars, e.g.,
neutrino absorption, transport of different neutrino types with
opposite signs of the polarization terms, and the feedback of the
transport on the energy distribution in the neutron star. The question
must be asked whether the proposed cumulative process leads to a
global anisotropy of the neutrino emission of the nascent neutron star
and how large this anisotropy can be for a given magnetic field.

\begin{figure}[t]
 \centerline{\epsfxsize=0.49\textwidth\epsffile{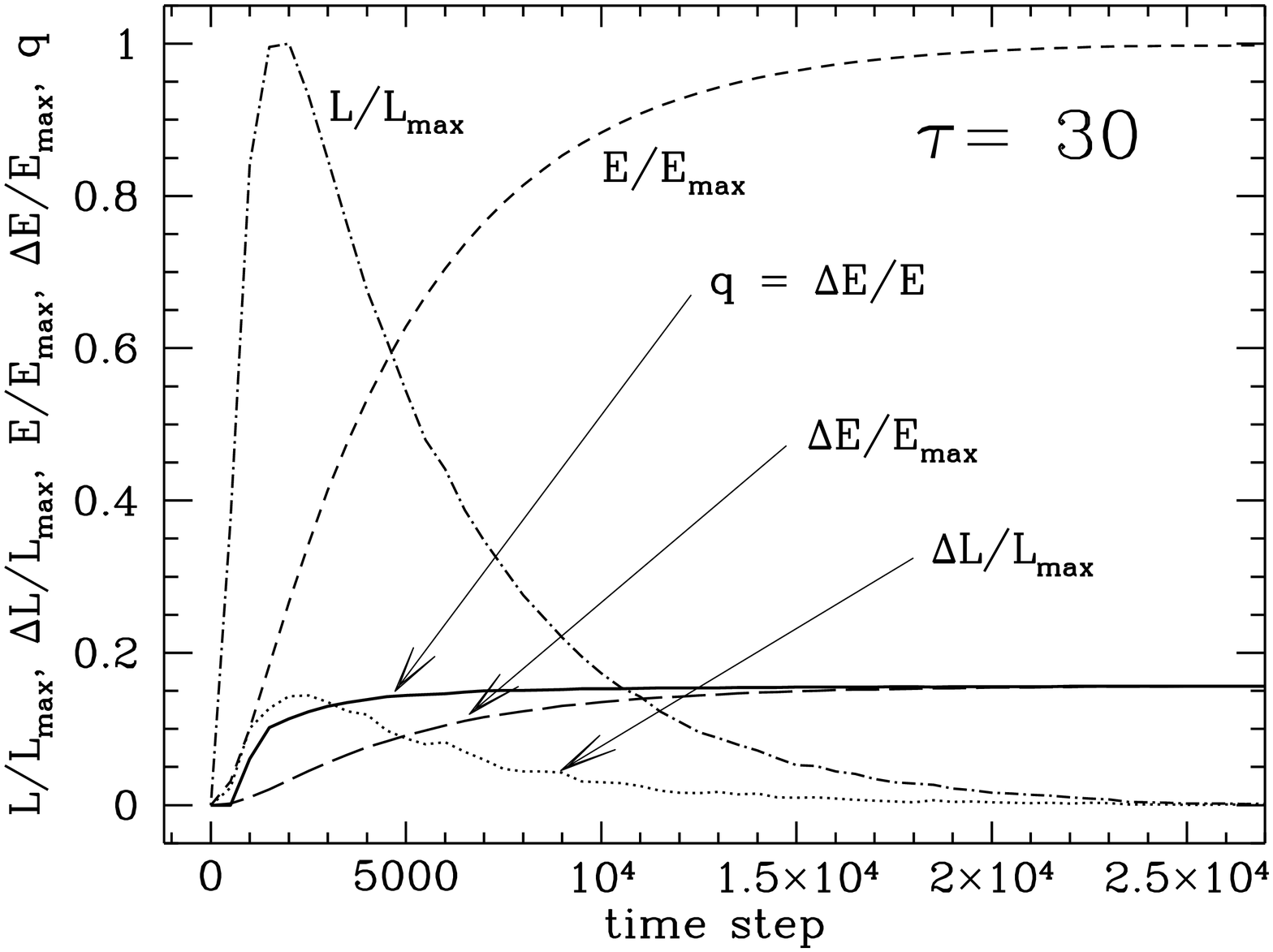}
             \epsfxsize=0.49\textwidth\epsffile{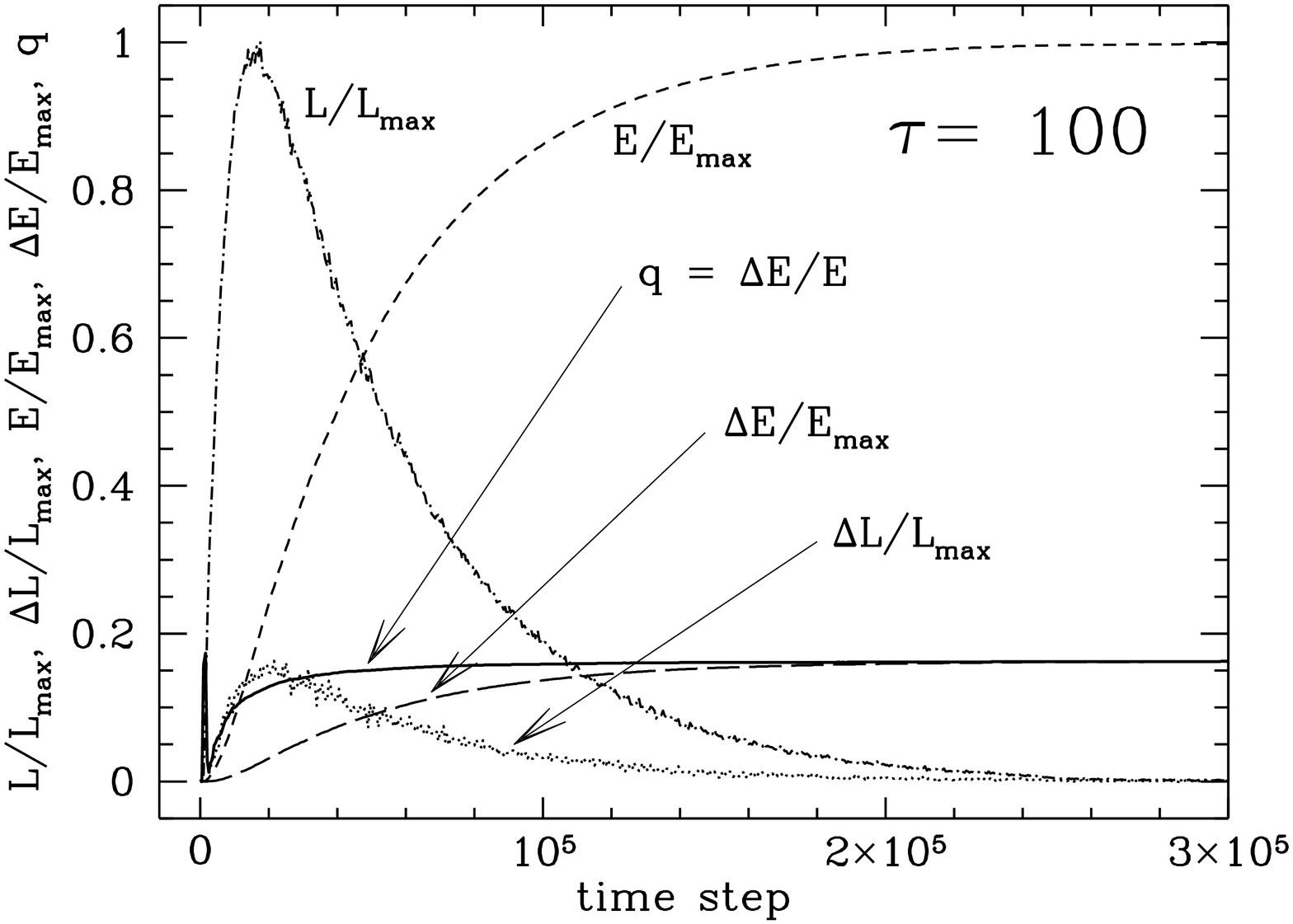}}
             \caption{\small Neutrino luminosity and emission
             anisotropy as functions of time step for two Monte Carlo
             simulations of the neutrino cooling of a one-dimensional
             toy model of a neutron star.  $L = L_+ + L_-$ is the
             total neutrino luminosity, $\Delta L = L_+-L_-$ $\Delta L
             = L_+-L_-$ the luminosity difference between the two
             hemispheres (both normalized to the maximum luminosity
             $L_{\rm max}$), $E = E_+ + E_-$ the energy radiated in
             neutrinos, and $\Delta E = E_+ - E_-$ the energy
             difference (both normalized to the final energy $E_{\rm
             max}$). The integral asymmetry of the neutrino emission
             is given by the parameter $q = \Delta E/E$ which
             approaches nearly 16\% towards the end of the
             simulations. The three-dimensional asymmetry in the
             integrated neutrino luminosity, $q_{\rm 3D}$, would be
             smaller by a factor 1/2 (for small anisotropies) compared
             to the displayed one-dimensional results.  The Monte
             Carlo simulations were performed for models with total
             optical depths of $\tau = 30$ (left) and $\tau = 100$
             (right), where the opacity polarizations were rescaled
             such that the results should be identical if the
             anisotropy grows linearly with the optical depth. The
             very good agreement between both calculations (left and
             right) thus confirms the scaling of the cumulative
             anisotropy with the optical depth of the star.}
\label{janka.fig1}
\end{figure}
 
The neutron star situation was more realistically investigated by Lai
and Qian~\cite{janka.8} who estimated a maximum recoil velocity of
about 200$\,$km/s for average magnetic fields of $\sim 10^{14}\,{\rm
G}$, provided the magnetic field in a nascent neutron star possesses a
strong dipole component. Their analysis, however, is based on the
assumption that neutrinos and antineutrinos, in particular $\nu_e$ and
$\bar\nu_e$, have the same opacity for interactions with the stellar
medium. Therefore their kicks are only produced by the deleptonization
neutrinos which carry away only a minor fraction ($\sim$ 10\%) of the
gravitational binding energy of the neutron star which is released by
neutrino emission. This assumption, however, leads to a large
underestimation of the recoil acceleration~\cite{janka.9}.

Monte Carlo simulations~\cite{janka.9} which include neutrino
production and absorption, different opacities and polarizations for
different neutrino and antineutrino flavors ($\nu_e$, $\bar\nu_e$, and
heavy lepton neutrinos $\nu_x$ and $\bar\nu_x$), and the evolution of
the stellar background in a simplified but reasonable, approximative
description\footnote{Monte Carlo simulations become extremely
expensive and CPU-time consuming for high optical depths. Instead of
considering a neutron star with neutrino optical depth of
10000--100000, a toy model with optical depth $\tau = 100$ and
correspondingly rescaled polarization terms of the neutrino
interactions was investigated.  The scaling of the results was tested
by comparative computations for $\tau = 30$ and for $\tau = 300$.},
suggest that a polarization term in the neutrino-nucleon scattering
opacity of $\kappa_{\rm s,pol}/\kappa_{\rm s}\sim 10^{-5}$ can lead to
a global asymmetry $q$ of the integrated neutrino luminosity of about
16\% for the one-dimensional (plane) case (Fig.~\ref{janka.fig1}).
Taking into account projection effects by multiplying this result with
1/2, the corresponding three-dimensional asymmetry $q_{\rm 3D}$ is
estimated to about 8\%, i.e.\ $q_{\rm 3D}\sim 0.08\cdot
\eck{\rund{\kappa_{\rm s,pol}/\kappa_{\rm s}}/ 10^{-5}}$. This can be
related to the anisotropy parameter $\alpha_{\rm 3D}$ for the momentum
transfer to the neutron star: $\alpha_{\rm 3D} = {\textstyle {2\over
3}} q_{\rm 3D}\sim 0.05\cdot \eck{\rund{\kappa_{\rm s,pol}/\kappa_{\rm
s}}/10^{-5}}$.  The effective polarization of neutrino-nucleon
scatterings depends on the magnetic field strength $B$, the plasma
temperature $T$, and the abundances $Y_n$ and $Y_p$ of neutrons and
protons, respectively:
\begin{equation}
{\kappa_{\rm s,pol}\over \kappa_{\rm s}}\,\sim\,
\rund{6.4\times 10^{-5} - 13.3\times 10^{-5}{Y_p\over Y_n}}\cdot
\rund{1 + 1.134\,{Y_p\over Y_n}}^{\! -1}\cdot
{B_{14}\over T_{10}} \; ,
\label{eq-1}
\end{equation}
where $B_{14}$ is measured in $10^{14}\,{\rm G}$ and $T_{10}$ in
$10\,{\rm MeV}$. Using representative values for temperature and
composition during the phase when most of the gravitational binding
energy of the neutron star is radiated away in neutrinos, $T =
10\,...\,30\,{\rm MeV}$, $Y_n = 0.8\,...\,0.9$, and $Y_p = 1-Y_n =
0.2\,...\,0.1$, a space-time average of $\kappa_{\rm
s,pol}/\kappa_{\rm s}\sim 10^{-5}$ can be expected for a mean interior
magnetic field $\ave{B_z}$ of the neutron star of $\sim 10^{14}\,{\rm
G}$.  This leads to an estimated recoil velocity of
\begin{eqnarray}
v_{\rm ns} & \cong & 1800\rund{{\alpha_{\rm 3D}\over 0.05}}
\rund{{1.4\,M_{\odot}\over M_{\rm ns}}}
\rund{{E_{\nu}\over 3\times 10^{53}\,{\rm erg}}}\,{\rm km\,s^{-1}}
\nonumber \\
& \sim &
1800\rund{{\ave{B_z}\over 10^{14}\,{\rm G}}}
\rund{{1.4\,M_{\odot}\over M_{\rm ns}}}
\rund{{E_{\nu}\over 3\times 10^{53}\,{\rm erg}}}\,{\rm km\,s^{-1}}\; .
\label{eq-2}
\end{eqnarray}

The reason for a nearly 10 times larger anisotropy compared to the
result found by Lai and Qian~\cite{janka.8} is twofold.  During
deleptonization and neutronization the neutron star plasma becomes
increasingly asymmetric, i.e.~the abundance of protons decreases and
the plasma becomes neutron-rich. On the one hand, the unequal
concentrations of neutrons and protons cause different opacities for
$\nu_e$ and $\bar\nu_e$, on the other hand the growing abundance of
neutrons implies an increase of the total polarization of the
nucleonic medium because the contributions from neutrons and protons
have opposite signs and partly cancel. For a typical post-collapse
composition, $Y_p\approx 0.35$ and $Y_n \approx 0.65$, this
cancellation is severe and reduces the polarization to roughly 1/8 of
the value for a neutronized medium with $Y_p \approx 0.1$ and $Y_n
\approx 0.9$. The latter composition is representative of the late
Kelvin-Helmholtz cooling phase after the deleptonization of the
proto-neutron star. Most (60--90\%) of the neutrino energy is radiated
during this phase which is therefore decisive for a ``rocket engine''
effect by asymmetric neutrino emission.

Of course, the estimated velocity in Eq.~(\ref{eq-2}) is an upper
limit which could be realized only if the interior magnetic field were
uniform in the $z$-direction throughout the neutron star.  Convective
processes that reach deep into the star will certainly destroy the
ordered structure of the field. In case of rapid rotation of the newly
formed neutron star (rotation periods of a few milliseconds), however,
convection can develop only near the equatorial plane but is
suppressed near the rotation axis where an ordered field structure
could persist in a large fraction of the neutron star volume.

\subsection*{Conclusions}

If parity violation of weak interactions plays the crucial role to
kick pulsars and if convection is important during the early phases of
a neutron star's life, one would therefore conclude that the observed
velocities of fast pulsars between several $100\,$km/s and more than
$1000\,$km/s require interior magnetic fields from a few
$10^{13}\,{\rm G}$ to $\sim 10^{14}\,{\rm G}$, and that the largest
velocities might indicate rapid rotation of the newly formed neutron
star.

\subsection*{Acknowledgements}

This work was supported by the ``Sonderforschungsbereich 375-95 f\"ur
Astro-Teilchenphysik'' der Deutschen Forschungsgemeinschaft.
Stimulating discussions with S.~Hardy, G.~Raffelt, and S.~Yamada are
acknowledged.

}\newpage{


\head{Spectrum of the Supernova Relic Neutrino\\ Background and 
        Evolution of Galaxies} 
{Katsuhiko Sato$\,^{1,2}$, Tomonori Totani$\,^{1}$
 and Yuzuru Yoshii$\,^{2,3}$}
{$^1\,$Department of Physics, School of Science, 
        the University of Tokyo \\
        7-3-1 Hongo, Bunkyo-ku, Tokyo 113, Japan\\ \hbox{\ }\\ 
$^2\,$Research Center for the Early Universe, School of Science, 
        the University of Tokyo\\
        7-3-1 Hongo, Bunkyo-ku, Tokyo 113, Japan\\ \hbox{\ }\\
$^3\,$Institute of Astronomy, Faculty of Science, 
        the University of Tokyo \\
        2-21-1 Osawa, Mitaka, Tokyo 181, Japan}

\subsection*{Abstract}

Neutrinos emitted from the supernovae which have ever occurred in the
universe should make a diffuse neutrino background at present.  This
supernova relic neutrino background (SRN) is one of the targets of the
Superkamiokande (SK) detector which will be constructed in this year
and the SRN, if at all detected, would provide a new tool to probe the
history of supernova explosions in the universe. The SRN spectrum is
calculated by using a realistic, time-dependent supernova rate derived
from a standard model of galaxy evolution based on the population
synthesis method.  The expected event rate at the SK is also
calculated.  The SRN spectrum we show here is the most realistic at
present, because the largest uncertainty in previous theoretical
predictions has come from unrealistic assumptions of the supernova
rate so far made.  Our major results include: (1) the supernova rate
is much higher in the early phase of evolution of galaxies and there
appears a hump in the SRN spectrum in the low-energy region of
$\alt 5$ MeV, (2) the SRN flux depends on the Hubble constant
($H_0$) in a way approximately proportional to $H_0^2$ and only weakly
on the density parameter of the universe ($\Omega_0$) and a
cosmological constant ($\lambda_0$), (3) the uncertainty in the star
formation history of spiral galaxies affects the resulting SRN flux by
about a factor of 3, and (4) the plausible event rate at the SK is 1.2
yr$^{-1}$ in the observable energy range of $15$--$40$ MeV.  Such a
low event rate is due mainly to a quite low supernova rate at present
which is averaged over the morphological types of galaxies.  The most
optimistic rate in our model is found to be 4.7 yr$^{-1}$ in the same
energy range, and if more events are detected, we will have to
reconsider our current understanding of collapse-driven supernovae and
evolution of galaxies.

\subsection*{Summary and Conclusions}

We have calculated the SRN spectrum and also the expected event rate
at the SK detector by using a realistic, time-dependent supernova rate
from a reasonable model of galaxy evolution based on the population
synthesis method. This model of galaxy evolution has been used to
investigate the geometry of the universe from comparison with the
number counts of faint galaxies, and the predicted number counts are
well consistent with the recent observations. In our analysis the
absolute SRN flux in the model is determined from the luminosity
function and the mass-to-luminosity ratio of nearby galaxies, but not
from the present supernova rate of our Galaxy which is subject to a
large margin of uncertainty.

In the model we used, the supernova rate is much higher in the early 
phase of elliptical galaxies, and more than half of total supernovae 
explode during the initial 1 Gyr from the formation of galaxies.  
Thereafter, the supernova rate in spiral galaxies becomes dominant 
and the total number of supernovae until present is consistent with 
the nucleosynthesis requirement.  A constant rate of supernova 
explosions, which was assumed in most of previous papers, is not 
justified from a viewpoint of galaxy evolution, and therefore the
SRN spectrum we show here is the most realistic at the present time.

Supernovae produced at a very high rate in the early phase of
ellipticals are responsible for the appearance of a hump in the
low-energy part ($\alt$ 5 MeV) of the SRN spectrum, because the
energy of neutrinos from those supernovae is degraded by the
cosmological redshift effect.  This energy degradation makes early
neutrinos unobservable and the high supernova rate causes almost no
enhancement in the event rate in the observable energy range of 15--40
MeV at the SK.

We used a low-density, flat universe with a non-zero $\Lambda$ as a
standard cosmological model ($\Omega_0$=0.2, $\lambda_0=0.2$, and
$h=0.8$), and investigated the SRN spectrum for other cosmological
models relative to the standard model. The SRN flux strongly depends
on $H_0$ as roughly proportional to $H_0^2$. The SRN flux changes with
$\Omega_0$ and $\Lambda$ in a way that smaller $\Omega_0$ and/or
larger $\Lambda$ suppresses the flux in the high-energy region
($\agt$~10MeV), however, these changes are rather small. The shape
of the SRN spectrum is almost insensitive to either of the above three
parameters.

We investigated the uncertainty of the SRN spectrum from the model of
galaxy evolution.  The simple, one-zone and $n = 1$ (S1) model, which
we use as a standard evolution model of spiral galaxies, gives an
intermediate SRN flux among the four different models of spiral
galaxies considered in this paper (for explanation of the models see
Refs.~\cite{Sato.1,Sato.2}). The difference between the highest flux
from the I1 model and the lowest flux from the S2 model is within a
factor of 3.  In the case of elliptical galaxies, however, the
different SFRs hardly change the SRN spectrum in either shape or
intensity.  The effect of changing $z_{F}$ on the SRN is almost
negligible above 15 MeV, and from $z_F=3$ to 5 the event rate at the
SK changes only by a factor of 1.3.

The total SRN flux in the whole energy range becomes 44
cm$^{-2}$s$^{-1}$ in our `standard' model ($\Omega_0=0.2$,
$\lambda_0=0.8$, $h=0.8$, $z_F=5$, the S1 model for spirals, the
standard SFR model for ellipticals, and ${\cal L}_B=1.93\times 10^{8}h
\; L_{B\odot}$Mpc$^{-3}$), which is a moderate value compared with the
previous estimates, however, the SRN flux in the observable energy
range of 15--40 MeV is as small as 0.83 cm$^{-2}$s$^{-1}$
corresponding to a low end of the SRN flux estimated before.  This
comes from the very small present supernova rate which is averaged
over morphological types of galaxies with the weight of the
mass-to-luminosity ratio. This small flux directly leads to a low
event rate, and makes the SRN detection at the SK more difficult.

The event rates at the SK detector are calculated for the various
models considered in this paper (table~\ref{sato.tab1}), 
and in the case of the
`standard' model, the expected rate is as small as 1.2 yr$^{-1}$ in
the observable range of 15--40 MeV.  The event rate depends mainly on
the evolution model of spiral galaxies, the Hubble constant, and the
luminosity density of galaxies, while only weakly on the geometry of
the universe (i.e., on $\Omega_0$ and $\lambda_0$) and the epoch of
galaxy formation.  These properties are in contrast with those of the
number counts of faint galaxies, and the SRN observation may be
complementary to the number counting of faint galaxies, if the SRN is
detected.  From the dependence of the flux on the Hubble constant,
the SRN signal would provide a new test for measuring the Hubble
constant.  However, because of presumably low event rates, such a test
may not be possible in practice.

\begin{table}[ht]
\caption{\label{sato.tab1}Expected Event Rates at the Superkamiokande 
Detector.}
\vglue 2truemm
\centerline
{\footnotesize
\begin{tabular}{lccccccc} \hline\hline
 & & & & & & \multicolumn{2}{c}{Event Rate (yr$^{-1}$)} \\ \cline{7-8}
 $\Omega_0$ & $\lambda_0$ & h & $z_F$ & S-type evolution$\;^a$ & 
 ${\cal L}_B\;^b$ & 15--40 MeV$\;^c$ & 20--30 MeV$\;^c$ \\ \hline
 1.0 & 0.0 & 0.5 & 5 & S1 & 1.93 & 0.63 & 0.28 \\
 0.2 & 0.0 & 0.8 & 5 & S1 & 1.93 & 1.4 & 0.62 \\
 0.2 & 0.8 & 0.5/0.8/1.0 & 5 & S1 & 1.93 & 0.40/1.2/1.9 & 0.17/0.51/0.83 \\
 0.2 & 0.8 & 0.8 & 3/4/5 & S1 & 1.93 & 1.4/1.3/1.2 & 0.66/0.54/0.51 \\
 0.2 & 0.8 & 0.8 & 5 & S1/S2/I1/I2 & 1.93 & 1.2/0.78/1.8/1.5 & 
                                            0.51/0.34/0.79/0.65 \\
 0.2 & 0.8 & 1.0 & 3 & I1 & 1.93+0.8 & 4.7 & 2.1 \\ 
 \hline\hline 
\end{tabular}
}
\vglue 2truemm

$^a\;$ The standard SFR model is used in common to represent the E/S0-type 
evolution.

$^b\;$ The luminosity density of galaxies in units of
$10^8 h \times L_{B\odot}$ Mpc$^{-3}$ where\newline
\hbox{\hskip1.0em$h\equiv H_0/100$\,km$\,$s$^{-1}$Mpc$^{-1}$.}
        
$^c\;$ Kinetic energy of recoil positrons.
        
Note.---We refer the `standard' model to the case with 
($\Omega_0$, $\lambda_0$, $h$, $z_F$, S-type evolution, ${\cal L}_B$)=
(0.2, 0.8, 0.8, 5, S1, 1.93), and the `most optimistic' model to the case 
given in the last row with 
(0.2, 0.8, 1.0, 3, I1, 1.93+0.8).

\vglue 2truemm
\end{table}

Changing the model parameters within the allowable range in a way of
increasing the event rate ($h\rightarrow 1$, $z_F\rightarrow 3$,
S1$\rightarrow$ I1, and ${\cal L}_B \rightarrow (1.93+0.8)\times
10^{8}h \; L_{B\odot}$Mpc$^{-3}$), we predict the `most optimistic'
event rate at the SK, which is 4.7 yr$^{-1}$ in the energy range of
15--40 MeV.  The very small rate of expected SRN events at the SK
implies the difficulty of gaining some insight on the star formation
history in galaxies from the SRN observation in the SK experiment.
More events exceeding 4.7 yr$^{-1}$, if detected, may enforce a
serious revision in our current understanding of collapse-driven
supernovae and evolution of galaxies.

\subsection*{Acknowledgements}

We would like to thank Y. Totsuka for valuable discussion and useful
information about the Superkamiokande detector.  We would also like to
thank P.O.~Lagage for the updated data of the reactor $\bar\nu_e$
flux.  We also appreciate a critical reading of this manuscript by
L.M.~Krauss.  This work has been supported in part by the Grant-in-Aid
for COE Reaearch (07CE2002) of the Ministry of Education, Science, and
Culture in Japan.


\begin{figure}[htp]
\begin{center}
\epsfig{file=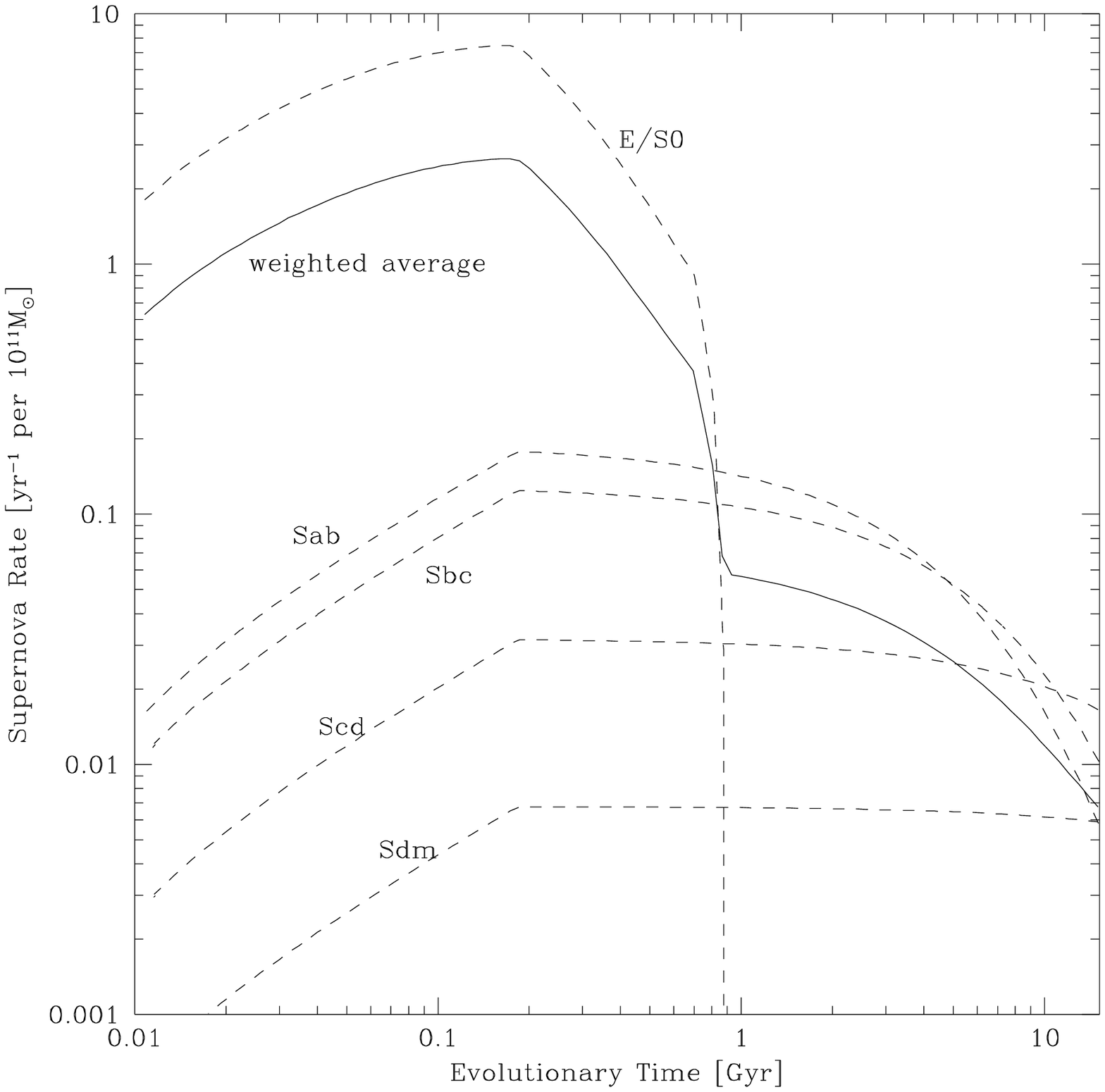,width=14cm}
\end{center}
\label{fig:snr-type}
\caption{ Time variation of supernova rates in a galaxy with the mass
of $10^{11}M_{\odot}$ for various galaxy types derived from the
standard model of galaxy evolution based upon the population synthesis
method.  The time is elapsed from the epoch of galaxy formation. The
dashed lines show the rates for the individual galaxy types, and the
thick line for the weighted average over the types.  The S1 model is
used to represent the evolution of spiral galaxies, and the standard
SFR model for elliptical galaxies.}
\end{figure}

\begin{figure}[htp]
\begin{center}
\epsfig{file=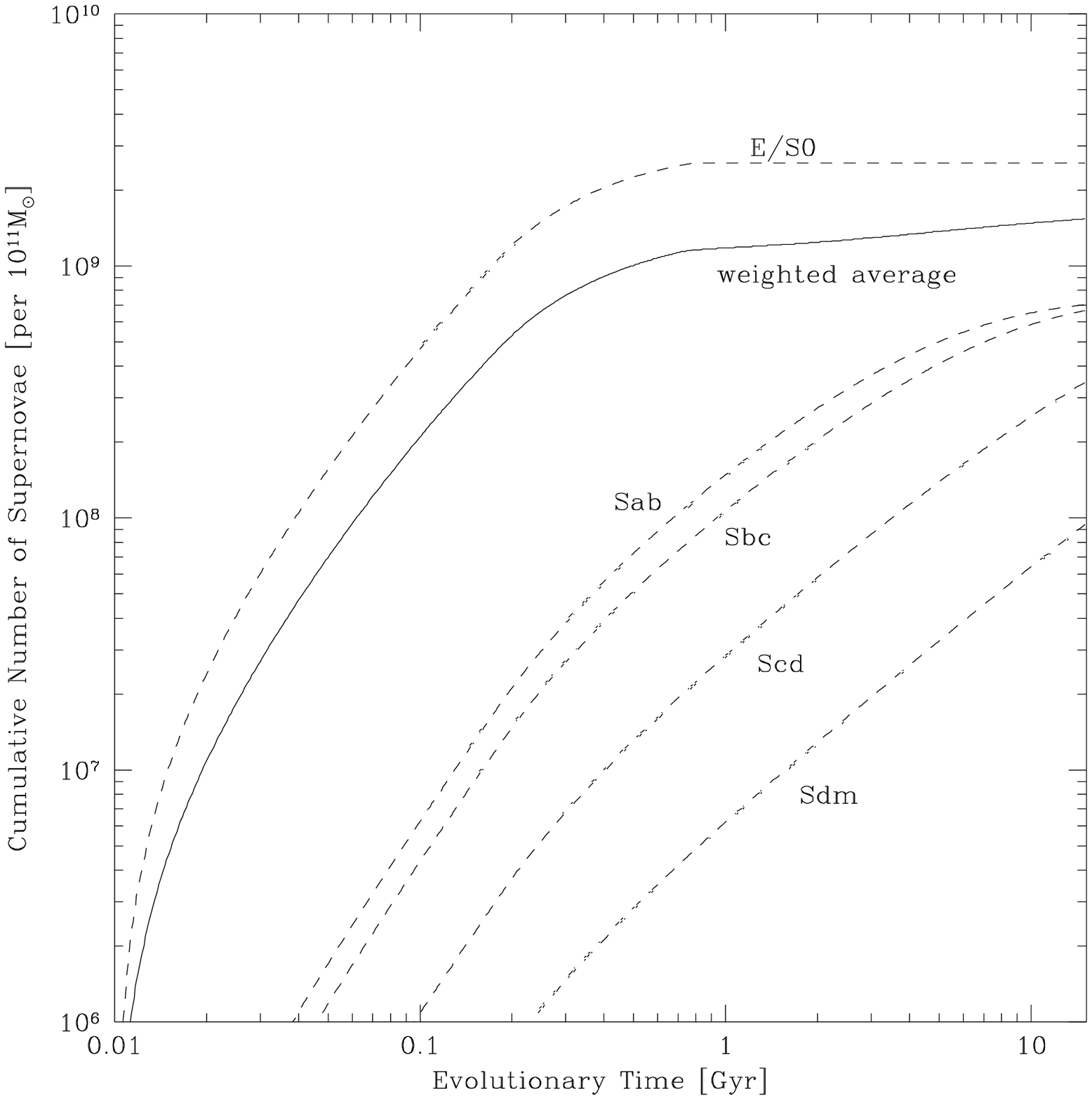,width=14cm}
\end{center}
\label{fig:total-SN}
\caption{Cumulative numbers (i.e., the time integration of Fig.~1) of
supernovae in a galaxy with the mass of 10$^{11}M_\odot$ as a function
of time elapsed from the epoch of galaxy formation.  The dashed lines
correspond to the cases for the individual galaxy types, and the thick
line to the weighted average over the types.  The S1 model is used to
represent the evolution of spiral galaxies, and the standard SFR model
for elliptical galaxies.  }
\end{figure}

\begin{figure}[htp]
\begin{center}
\epsfig{file=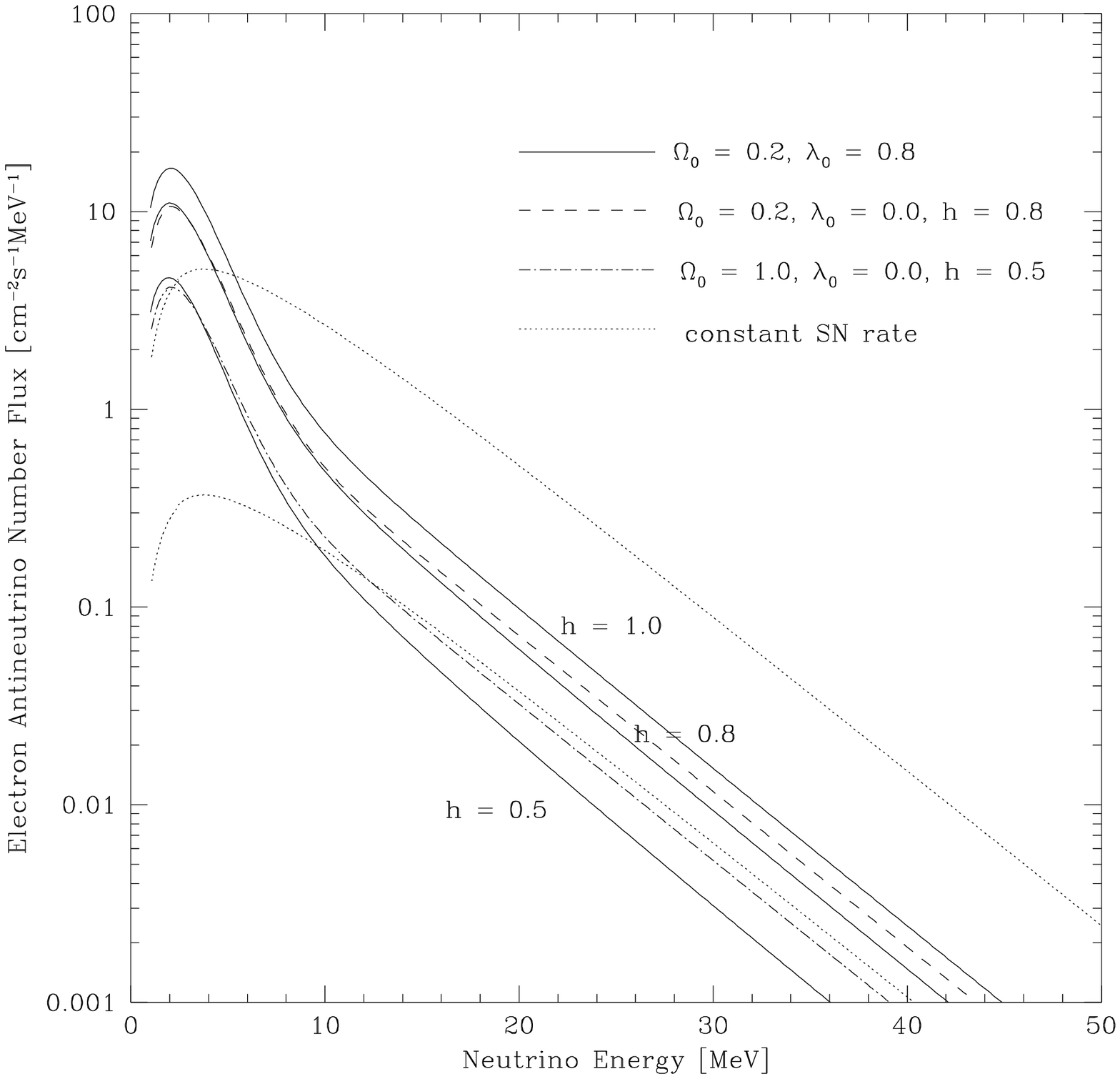,width=14cm}
\end{center}
\label{fig:flux-cp}
\caption{Energy spectra of the supernova relic neutrino background
(SRN) for various cosmological models. All lines represent the
differential number flux of electron antineutrinos ($\bar\nu_e$'s).
The thick lines show the SRN spectrum for a low-density, flat universe
with $\Omega_0 = 0.2$, $\lambda_0 = 0.8$, and three different values
of the Hubble constant; $h=0.5$ (bottom line), $h=0.8$ (middle line),
and $h=1.0$ (top line). The dashed line shows the SRN spectrum for a
low-density, open universe with $\Omega_0=0.2$, $\lambda_0=0.0$, and
$h=0.8$, and the dot-dashed line for an Einstein-de Sitter universe
with $\Omega_0=1.0$, $\lambda_0=0.0$, and $h=0.5$.  The S1 model is
used to represent the evolution of spiral galaxies, and the standard
SFR model for elliptical galaxies.  The epoch of galaxy formation is
set to be $z_F=5$.  For an illustrative purpose of comparison, two SRN
spectra based on a model with a constant supernova rate are also shown
by the dotted lines. The supernova rates of the two curves are chosen
as follows. In the case of the upper curve, the time-integrated number
of supernovae is the same with that of our standard model (thick,
middle line), and in the case of the lower curve, the present rate of
supernovae is the same with that of the standard model.}
\end{figure}

\begin{figure}[htp]
\begin{center}
\epsfig{file=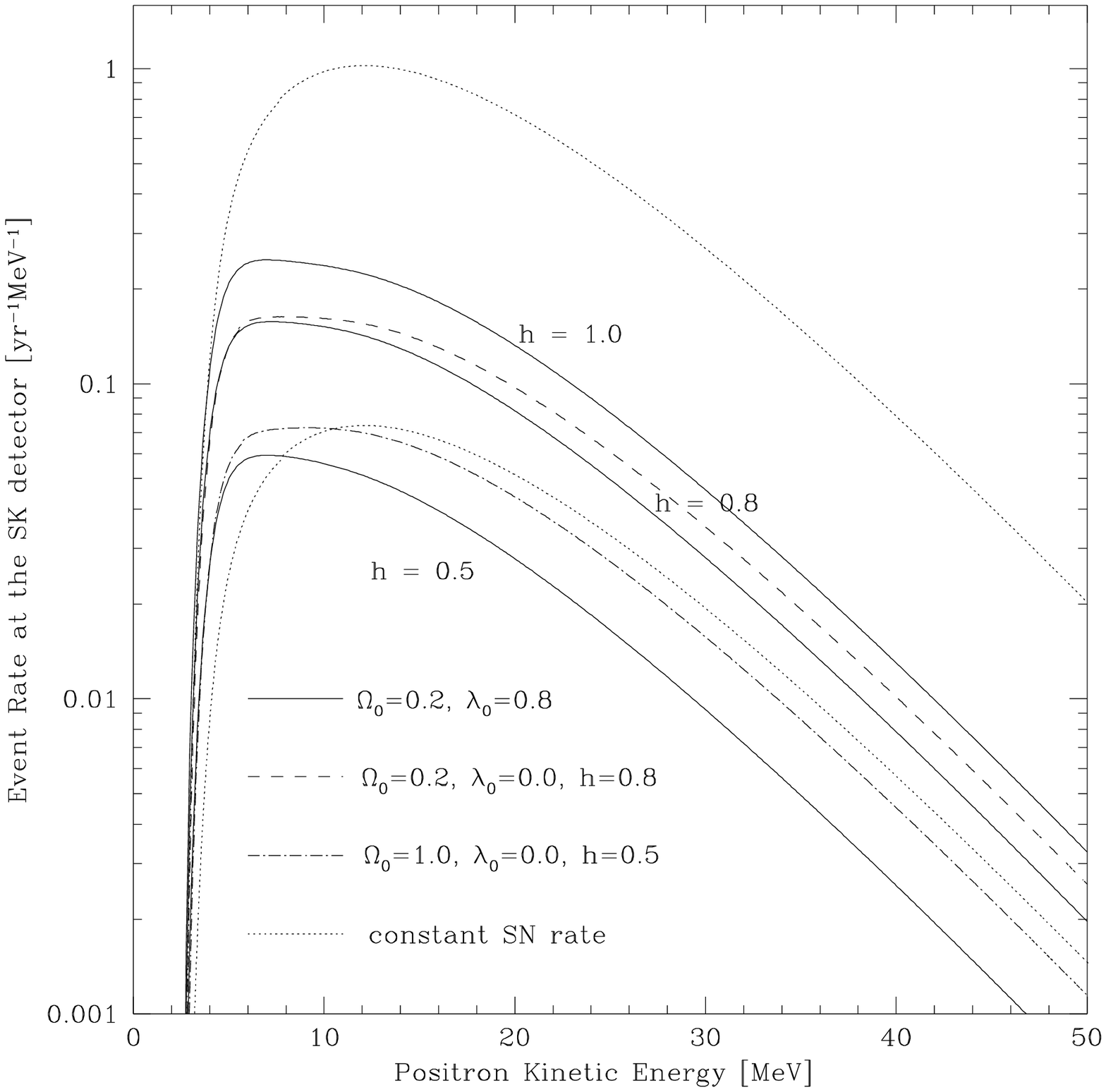,width=14cm}
\end{center}
\label{fig:sk-cp}
\caption{Expected event rates at the Superkamiokande (SK) detector for
various cosmological models, as a function of kinetic energy of
positrons produced by the reaction $\bar\nu_e p \rightarrow e^+ n$.
This figure corresponds to Figure 3 of the SRN spectrum. The thick
lines show the event rate for a low-density, flat universe with
$\Omega_0 = 0.2$, $\lambda_0 = 0.8$, and three different values of the
Hubble constant; $h=0.5$ (bottom line), $h=0.8$ (middle line), and
$h=1.0$ (top line).  The dashed line shows the event rate for a
low-density, open universe with $\Omega_0=0.2$, $\lambda_0=0.0$, and
$h=0.8$, and the dot-dashed line for an Einstein-de Sitter universe
with $\Omega_0=1.0$, $\lambda_0=0.0$, and $h=0.5$.  The S1 model is
used to represent the evolution of spiral galaxies, and the standard
SFR model for elliptical galaxies.  The epoch of galaxy formation is
set to be $z_F=5$.  For an illustrative purpose of comparison, two
event rate curves based on a model with a constant supernova rate are
also shown by the two dotted lines. The used rates of supernovae are
the same with those used in Fig.~3.}
\end{figure}
 
\vfill
\eject

}\newpage{


\head{Supernova Neutrino Opacities}
     {G.G.~Raffelt}
     {Max-Planck-Institut f\"ur Physik (Werner-Heisenberg-Institut)\\
      F\"ohringer Ring 6, 80805 M\"unchen, Germany}

The collapsed cores of supernovae are so dense and hot that neutrinos
are trapped. Therefore, the transport of energy and lepton number is
governed by the neutrino transport coefficients, i.e.\ by their
effective scattering and absorption cross sections.  While electrons
cannot be entirely ignored, it is the interaction with nucleons which
is thought to dominate the neutrino opacities.  The cross sections can
be dramatically modified by medium effects; strong magnetic fields
which can also be important will not be considered here.

Final-state blocking by degenerate nucleons is one obvious
modification of the vacuum cross sections~\cite{Raffelt.SS79}.  Even
this trivial effect depends sensitively on the nucleon dispersion
relation; a reduced effective nucleon mass increases the degree of
degeneracy. Moreover, depending on the equation of state the
composition may deviate from a naive proton-neutron mixture in that
there may be a significant fraction of hyperons. Studies of opacities
with self-consistent compositions and dispersion relations reveal
large modifications~\cite{Raffelt.RP97}.

The neutrino interaction with nucleons is dominated by the axial
vector current, i.e.\ it is a spin-dependent phenomenon. At
supernuclear densities one thus expects a dominant role for spin-spin
correlations~\cite{Raffelt.Sawyer89}; the use of uncorrelated
single-particle states for an opacity calculation amounts to an
uncontrolled approximation.

\begin{figure}[b]
\centerline{\epsfxsize=7cm\epsffile{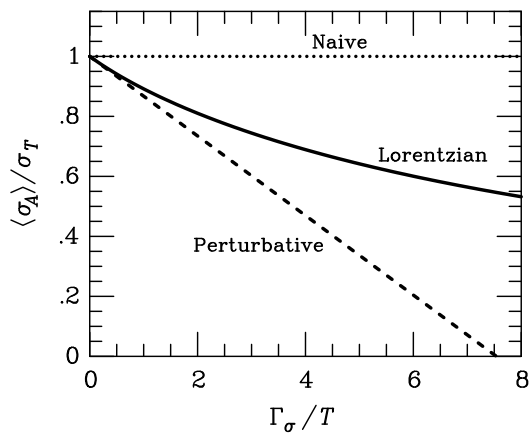}}
\caption{Variation of the thermally averaged axial-current neutrino
scattering cross section on nucleons as a function of the nucleon
spin-fluctuation rate. The curve marked ``Lorentzian'' is obtained
from a ``resummed'' dynamical structure
function~\protect\cite{Raffelt.JKRS}.}
\end{figure}

A less intuitive effect is the cross-section reduction by the nucleon
spin fluctuations which are described by the dynamical dependence of
the spin-density structure function
\cite{Raffelt.RS95,Raffelt.Sawyer95,Raffelt.JKRS}.  In Fig.~1 the
reduction of the average neutrino-nucleon axial-vector cross section
is shown as a function of the assumed nucleon spin fluctuation
rate~$\Gamma_\sigma$. A perturbative calculation yields
$\Gamma_\sigma/T\approx20$--50 for a SN core, but the true value is
not known. Still, spin fluctuations (dynamical correlations) certainly
have a significant impact on the neutrino opacities.

A controlled determination of the neutrino opacities requires a
calculation of the full dynamical spin and isospin density structure
function of the correlated nuclear medium at finite temperature, a
task which has not been accomplished in spite of the recent progress
in shell-model Monte Carlo calculations of nuclear structure
\cite{Raffelt.Langanke97}. Currently the most reliable information of
SN neutrino opacities arises from the SN~1987A neutrino pulse
duration~\cite{Raffelt.KJR}. It indicates that the true opacities
cannot be too different from the ``naive'' ones, but this conclusion
is based on the few late events.  The observation of a high-statistics
neutrino light-curve from a future galactic SN would go a long way
towards an empirical calibration of the neutrino opacities in hot
nuclear matter!

The fluctuating nucleon spins allow for a more effective transfer of
energy between the medium and the neutrinos than is possible by
elastic collisions alone. This implies that $\mu$ and $\tau$ neutrinos
may lose more energy between their ``energy sphere'' and the surface
of the SN core than had been thought previously, causing their spectra
to become more similar to $\bar\nu_e$
\cite{Raffelt.JKRS,Raffelt.HR97}.  If the magnitude of this
differential effect is as large as estimated in
Refs.~\cite{Raffelt.JKRS,Raffelt.HR97} a number of interesting
consequences would obtain, notably for scenarios of SN neutrino
oscillations which depend on the spectral differences between
different flavors.

\subsection*{Acknowledgements}

Partial support by the European Union contract CHRX-CT93-0120 and by
the Deutsche Forschungsgemeinschaft under grant No. SFB-375 is
acknowledged.

}\newpage{


\head{Quasilinear Diffusion of Neutrinos in Plasma}
     {S.J.~Hardy}
     {Max-Planck-Institut f\"ur Astrophysik,\\
      Karl-Schwarzschild-Str.~1, 85748 Garching, Germany}

\subsection*{Introduction}
It has been recognized for some years that a neutrino propagating in a
plasma acquires an induced charge \cite{hardy.1}.  This charge is due
to the forward scattering interactions between the neutrino and the
electrons in the plasma. As these scatterings are electroweak
interactions, the induced charge is rather small, for example,
$10^{-31} \, \mbox{C}$, for the dense plasma near the core of a type
II supernova (SN).  Though weak, this charge allows neutrinos in a
plasma to undergo processes which would usually be restricted to
charged particles, such as Cherenkov emission or absorption of a
photon.

The electromagnetic interactions of a neutrino in a plasma have been
considered in a variety of physical and astrophysical scenarios (for a
review, see \cite{hardy.2}). Some recent work by Bingham et
al. \cite{hardy.3} proposed a ``neutrino beam instability'' where an
intense flux of beamed neutrinos propagating through a plasma leads to
the production of an exponentially growing number of photons in the
plasma through stimulated Cherenkov emission. This form of
instability, caused by electron or photon beams, is well known in
plasma physics. The proposed application of the neutrino process was
as a reheating mechanism for the plasma behind the stalled shock of a
type II SN. While such an instability is possible in principle, it has
recently been shown \cite{hardy.4} that this does not occur in type II
SNe.

More recently, Tsytovich et al.~\cite{hardy.5} have proposed an
alternative mechanism whereby the neutrinos from a type II SN may
diffuse slightly in momentum space by propagating through a
pre-existing saturated thermal distribution of photons. The diffusion
mechanism, known as quasilinear diffusion, is based on the averaged
effect of the individual interactions that occur between the photons
and the neutrinos. Again, the analogous effect involving electrons is
well known is plasma physics \cite{hardy.6}. In their initial
calculation, Tsytovich et al.  obtained a timescale for angular
diffusion of the neutrinos from the core of the SN of $\tau_{ang}
\approx 10^{-4} \, \mbox{s}$, this corresponds to a scattering length
of approximately $30 \, \mbox{km}$, independent of energy, which would
be of great interest for neutrino transport near the shock of a type
II SN. The calculation reported here represents a more rigorous
calculation of this process and application of the results to a model
calculation of the plasma properties of a type II SN. It is concluded
that this process is only likely to be of importance to low energy
neutrinos (below $10\, \mbox{keV}$) and is unlikely to have any
bearing on the explosion of a type II SN.

\subsection*{Quasilinear Diffusion Rate}

Given the nature of the plasma behind the shock of a type II SN, it is
reasonable to assume a high level of plasma turbulence. Within the
weak turbulence approximation, this turbulence is represented by a
distribution of longitudinal photons in the plasma. The strongest
level of plasma turbulence allowed would be where the energy density
associated with the longitudinal photons is equal to the energy
density associated with the thermal motion of the plasma. Assuming
this maximum turbulence gives a natural limit on the strength of the
plasma-neutrino interaction.

One may describe the photons in the plasma and the neutrinos from the
SN core through distributions functions, $N({\bf k})$, and $f({\bf
q})$, respectively. Here, ${\bf k}=(k,\theta,\phi')$ denotes the
photon momentum, and ${\bf q}=(q,\alpha,\phi)$ the neutrino momentum,
both given in spherical polar coordinates. The neutrino distribution
evolves through the absorption and emission of longitudinal photons, a
process which occurs with probability per unit time denoted by $w({\bf
q},{\bf k})$. Given these definitions the equation governing the time
evolution of the neutrino distribution is (for~the electron
equivalent, see \cite{hardy.6})
\begin{equation}
{d f({\bf q}) \over dt}  = 
{1 \over \sin\alpha}{\partial \over \partial \alpha} \left[ \sin\alpha
 D_{\alpha\alpha}({\bf q}) {\partial \over \partial \alpha}  f({\bf q}) \right], \label{eq:nql}
\end{equation}
where
\begin{equation}
D_{\alpha \alpha}({\bf q}) = \int k^2 {d k \over 2 \pi} \int {d
(\cos\theta) \over 2 \pi} \int { d \phi' \over 2 \pi} (\Delta\alpha)^2 w({\bf
q},{\bf k}) N({\bf k}), \label{eq:daa}
\end{equation}
and with
\begin{equation}
\Delta\alpha = {k\over p}\left[ \cos(\phi - \phi') \cos\alpha \sin\theta -
\sin\alpha \cos\theta \right],
\end{equation}
and
\begin{equation}
w_L({\bf q},{\bf k}) = {G_F^2 c_V^2 \over 4 \alpha} \omega k^2 \left(1- {\omega^2 \over k^2}\right)^2
\delta(w - {\bf k}\cdot {\bf v}). \label{eq:wl}
\end{equation}

In equation (\ref{eq:nql}), only the largest component of the
quasilinear equation has been kept. Equation (\ref{eq:daa}) may be
evaluated for a saturated thermal distribution of waves and one
obtains
\begin{equation}
D_{\alpha \alpha}(q) = {G_F^2 c_v^2 n_e \over 16 \pi m_e c^2} {\omega_p W_L
\over (qc)^2},
\end{equation}
where $W_L$ is the energy density in the waves, and $\omega_p$ is the
plasma frequency.

To obtain a simple estimate of the importance of this effect, a few
simplifying assumptions about the nature of the neutrino distribution
are made. It is assumed that the distribution is separable, and that
the angular distribution is given such that
\begin{equation}
f(q,\alpha) = f(q) \Phi(\alpha),
\end{equation}
with
\begin{equation}
\Phi(\alpha) = {2 \over \alpha_0^2} \exp \left\{ {-\alpha^2 \over 2
\alpha_0^2} \right\}.
\end{equation}

Now, assuming that the angle subtended by the core of the SN on the
sky is small, $\alpha \ll 1$, one has 
\begin{equation}
{d f(q,\alpha) \over dt} = {f(q,\alpha) \over \alpha_0^2} D_{\alpha
\alpha}(q) \left[ 2 - {\alpha^2 \over \alpha_0^2} \right].
\end{equation}
Thus,
\begin{equation}
{1 \over\tau_{ang}} \approx {D_{\alpha \alpha}(q) \over \alpha_0^2},
\end{equation}
where $\tau_{ang}$ represents the timescale on which the angular scale
of the neutrino distribution grows through one e-folding.

Assuming a saturated thermal distribution, and $W_L \approx N_e m_e
v_{Te}^2$ and that $v_{Te} \approx c/2$ leads~to
\begin{equation}
{1 \over \tau_{ang}} = {G_F^2 c_v^2 \over 16 \pi} {N_e^2 \over p^2
c^2} {\omega_p \over \alpha_0^2}, 
\end{equation}
which may be parameterized,
\begin{equation}
{1 \over \tau_{ang}} = 10\, \mbox{s}^{-1} \left( {N_e \over {10^{38}
\mbox{ m}^{-3}}} \right)^{5/2} \left( {pc \over {1 \mbox{ MeV}}}
\right)^{-2} \left({\alpha_0 \over 10^{-2}}\right)^{-2}. 
\end{equation}

Figure \ref{hardy.fig1} shows this relation plotted for a neutrino
spectrum with temperature of 1 MeV propagating through the interior
regions of a model of a 25 solar mass star which was followed through
core collapse and bounce by S. Bruenn. Note that the small-angle
approximation made in producing the angular diffusion timescale break
down close to the SN core, thus the upper values of this plot are
overestimated. Near the shock, at around $10^7\ \mbox{cm}$, the
scattering timescale is of the order of tens of milliseconds, which is
several orders of magnitude too slow to be of importance. However,
given the inverse square dependence on the energy of the neutrino, low
energy neutrinos could be scattered quite effectively. Unfortunately,
this does not represent a reheating mechanism for the post shock
plasma, as a thermal neutrino distribution of 3 MeV (a realistic
temperature) has 99\% of its energy in neutrinos above 2.5MeV, and
only a tiny fraction at energies below 10keV, where the neutrinos are
scattered.

\begin{figure}[b]
\centerline{\epsfxsize=0.6\textwidth\epsffile{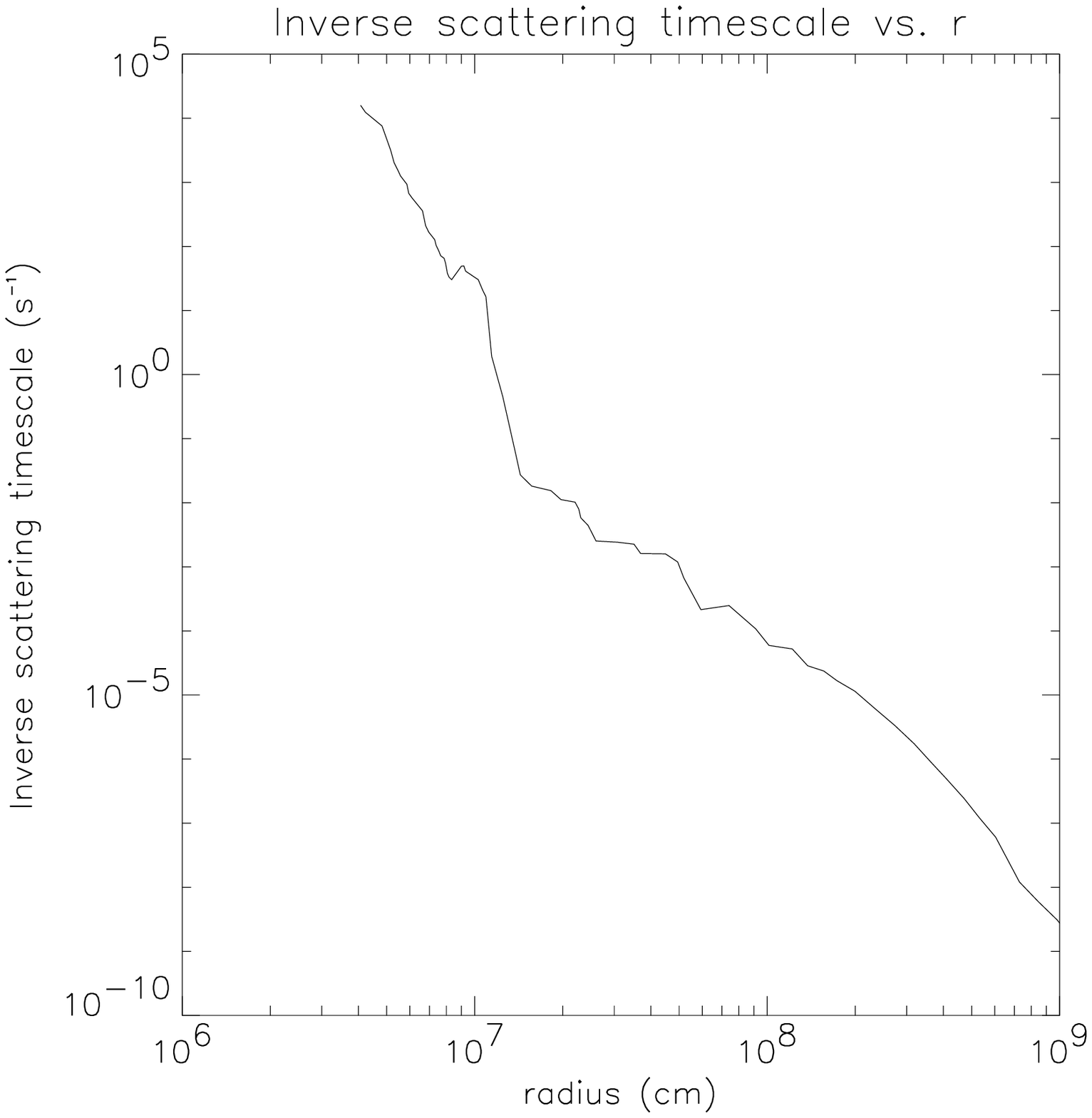}}
\caption{Inverse timescale for angular scattering for 1 MeV neutrino.
Calculation of collapse and bounce by S. Bruenn, from a 15 solar mass
progenitor, t= 0.225.}
\label{hardy.fig1}
\end{figure}

In conclusion it is noted that, while an intrinsically interesting
physical effect, the scattering of neutrinos off plasma turbulence in
a type II supernova is not strong enough to modify the bulk of the
distribution of neutrinos in the plasma surrounding the SN core.

}\newpage{


\def\Gabs{\Gamma_{\rm abs}}
\def\Gcre{\Gamma_{\rm cre}}
\def\EE#1{\mbox{$\times 10^{#1}$}}
\def\sign{\rm sign}
\def\eq#1{\mbox{Eq.~(\ref{#1})}}
\def\>{\rangle}
\def\<{\langle}
\def\slask{\hspace{0.em}\not\hspace{-0.25em}}
\def\tr{{\rm tr\,}}
\def\abs#1{\left|#1\right|}
\def\nn{\nonumber\\}

\head{Anisotropic Neutrino Propagation in a\\ Magnetized Plasma}
     {Per~Elmfors}
     {Stockholm University, Fysikum, Box 6730, S-113 85 Stockholm, 
        Sweden}

\subsection*{Introduction}
The idea that a large-scale magnetic field can be responsible for a
significant neutrino emission asymmetry in a supernova explosion goes
back to Chugai \cite{Chugai84} in 1984.  Since then there have been a
number of suggestions of how to implement this idea in more
detail. Lately, in particular after the Ringberg workshop, several
papers appeared dealing with how to calculate neutrino cross sections
in a magnetic field more accurately and with more realistic neutron
star parameters \cite{BezchastnovH96,BeneshH97}.  The various
approaches are either to use the exact Landau levels for the charged
particles which is necessary for very strong fields, or to include the
field effects only through the polarization of the medium. The
disadvantage with the full Landau level approach is that the number of
Landau levels that have to be included grows quadratically with the
Fermi energy. If the field is $B\sim 1$~(MeV)$^2\simeq
1.6\EE{14}$~Gauss and Fermi energy $E_F=\sqrt{\mu_e^2-m_e^2}\sim
100$~MeV approximately $E_F^2/2eB\simeq 15000$ Landau levels are
filled, and with all the transition matrix elements that have to be
calculated the problem becomes a numerical challenge. For weak fields,
where the strength should be compared with $E_F^2$ rather than
$m_e^2$, it seems logical to rely on a weak-field approximation.

To explain the observed peculiar velocities of neutron stars by
asymmetric neutrino emission an asymmetry of a few percent is
needed. It is found here, as in other recent papers \cite{BeneshH97},
that the asymmetry in the damping rate is itself much too small for
fields of the order of $10^{14}$--$10^{15}$~Gauss.  On the other hand,
it is certain that the final asymmetry cannot be esitmated simply by
the asymmetry in the damping rate. One step towards a more complete
analysis was taken in \cite{HorowitzL97} where a cumulative parity
violation was suggested to enhance the asymmetry. For a reliable
estimate it would be desirable to run a full magneto-hydrodynamics
simulation.

In this presentation I shall discuss the asymmetry from the electron
polarization in neutrino--electron scattering and URCA processes. In
addition one will have to add the asymmetry from proton and neutron
polarization in URCA processes which may be as important as the
electron polarization. The reason is that polarization of the
degenerate electron gas is suppressed by the large chemical potential
even though the electron magnetic moment is much larger than the
nuclear magnetic moment. Furthermore, I am here using free propagators
for the nucleons which may have to be improved to account for rapid
spin-flip processes~\cite{RaffeltS91}.

\subsection*{Neutrino Damping and the Imaginary Part of the 
Self-Energy}

The simplest formalism for computing damping coefficients in thermal
field theory is, in my opinion, from the imaginary part of the
self-energy \cite{Weldon83,KobesS85}.  The relation to be used for a
neutrino with momentum $Q$ is
\begin{equation}\label{GimS}
        \Gamma(Q)=-\frac{\<\overline{\nu}|{\rm Im} \Sigma(Q)|\nu\>}
        {\<\overline{\nu}|\nu\>}~~.
\end{equation}
The damping coefficient consists of two parts $\Gamma=\Gabs+\Gcre$
corresponding to processes that decrease and increase the number of
neutrinos in the state $|\nu(Q)\>$. The precise meaning of $\Gamma$ as
defined in \eq{GimS} becomes clear from the master equation for the
neutrino distribution function \cite{Weldon83}
\begin{equation}\label{master}
        \frac{d f_\nu(Q)}{dt}=-\Gabs f_\nu(Q)+\Gcre (1-f_\nu(Q))~~,
\end{equation}
from which one can show that the neutrino equilibration rate is
exactly $\Gamma$. For bosons the equilibration rate would have been
$\Gamma=\Gabs-\Gcre$ which also comes out automatically from the
imaginary part of the self-energy.  For URCA processes $\Gabs$ is
exactly the damping from absorption of $\nu$ while for $\nu$-$e$
processes $\Gabs$ includes both annihilation with $\overline{\nu}$ and
scattering with $e$, i.e. anything that removes a neutrino from
$|\nu(Q)\>$.

\begin{figure}[h]
\begin{picture}(150,80)(0,370)
   \centerline{\epsfxsize=0.8\textwidth\epsffile{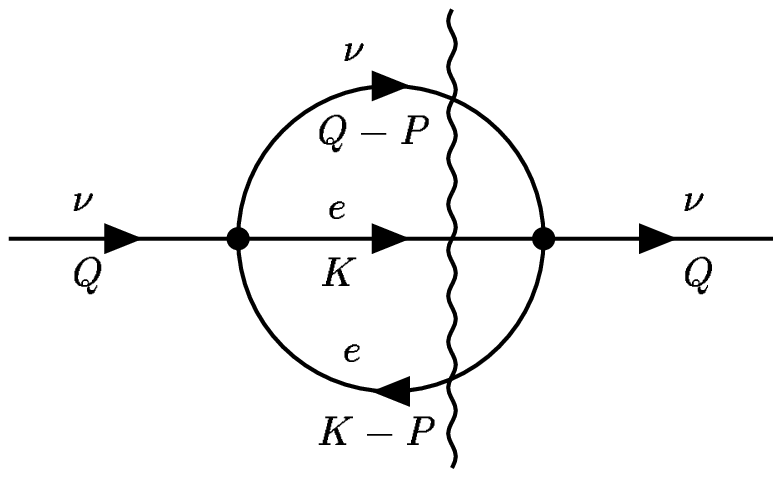}}
\end{picture}
  \caption{The two-loop diagram that contains all tree-level $\nu$-$e$
  processes in its imaginary part.}
\end{figure}

Let us start with the $\nu$-$e$-scattering process; the rules for
calculating $\Gamma$ are given in \cite{KobesS85} and correspond to a
cut setting-sun diagram (Fig.~1):
\begin{eqnarray}\label{ImSnue}
        \<\overline{\nu}|{\rm Im} \Sigma|\nu\>
        &=&-\frac{G_F^2}{4} \int \frac{d^4P\, d^4K}{(2\pi)^8}
        \rho(q_0,k_0,p_0;\mu_e,\mu_\nu) (2\pi)^3\nn
        &&\!\!\!\!\!\!\!\!
        \delta(K^2-m_e^2)\delta((K-P)^2-m_e^2)\delta((Q-P)^2)
        \<\overline{\nu}|\gamma^\mu(1-\gamma_5)(\slask Q-\slask P)
        \gamma^\nu(1-\gamma_5)|\nu\>\nn
        &&\!\!\!\!\!\!\!\!
        \tr[\gamma^\mu(g_V-g_A\gamma_5)(\slask K+m_e)
        \gamma^\nu(g_V-g_A\gamma_5)(\slask K-\slask P+m_e)]~~,
\end{eqnarray}
where the distribution functions for initial/final state factors enter
through
\begin{eqnarray}\label{rho}
        \rho&=&\biggl(\theta(q_0-p_0)-f_\nu(q_0-p_0)\biggr)
        \biggl(\theta(k_0)-f_e(k_0)\biggr)
        \biggl(-\theta(-k_0+p_0)+f_e(k_0-p_0)\biggr)\nn
        &\!\!\!\!\!\!+&\!\!\!\!\!\!\!\!
        \biggl(-\theta(-q_0+p_0)+f_\nu(q_0-p_0)\biggr)
        \biggl(-\theta(k_0)+f_e(k_0)\biggr)
        \biggl(-\theta(k_0-p_0)-f_e(k_0-p_0)\biggr),
\end{eqnarray}
where $f(p_0)=(\exp[\beta(|p_0|-\sign(p_0)\mu)+1)^{-1}$.  The general
rules in \cite{Weldon83,KobesS85} for calculating $\Gamma$ include all
possible processes that change the number of neutrinos in a certain
state, and the thermal factors for scattering/absorption/creation are
all automatically included. It is not difficult to separate out
$\Gabs$ and $\Gcre$ from $\Gamma$. Absorption corresponds to the first
term of $\rho$ in \eq{rho} and creation is the second one. It should
be noted that $\Gamma$ represents the integrated rate at which the
incoming neutrino changes state.  In a Boltzmann equation the
differential rate of scattering into another momentum would be needed,
and that is easily obtained by simply omitting the integration over
the neutrino inside the loop.  It is also easy to separate $\Gabs$ and
$\Gcre$ by noting that detailed balance requires
$\Gabs/\Gcre=(1-f_\nu)/f_\nu=\exp[\beta(q_0-\mu_\nu)]$ (for $q_0>0$)
in equilibrium, which is automatically satisfied with \eq{rho}.

So far the damping rate corresponds to the zero field case, but the
generalization to a background magnetic field is straightforward after
identifying the different factors in \eq{ImSnue}. The
$\delta$-functions come from the imaginary part of the usual electron
and neutrino propagators (or rather from the thermal part of the
real-time propagators). We now have to use the electron propagator in
an external field instead, and here it is convenient to use
Schwinger's formulation
\begin{eqnarray}\label{Schwprop}
        \frac{i(\slask P+m_e)}{P^2-m_e^2+i\epsilon}&\rightarrow&
        \int_0^\infty ds\frac{e^{ieBs\sigma_z}}{\cos(eBs)}
        \exp\biggl[is(P_\parallel^2+\frac{\tan(eBs)}{eBs}P_\perp^2-m_e^2
        +i\epsilon)\biggr]\nn
        &&\biggl(\slask P_\parallel+\frac{e^{-ieBs\sigma_z}}{\cos(eBs)}
        \slask P_\perp+m_e\biggr)~~,
\end{eqnarray}
where $a\cdot b_\parallel=a_0 b_b-a_z b_z$ and $a\cdot b_\perp=-a_x
b_x-a_y b_y$.  This expression is quite difficult to use in general
but the complicated parts, $\cos(eBs)$ and $\tan(eBs)/eBs$, have no
linear dependence on $eB$ and can therefore be approximated by 1 for
weak fields.  The $s$-integral can then be performed and after taking
the imaginary part we end up with the replacement rule
\begin{equation}\label{replace}
        (\slask P+m_e)\delta(P^2-m_e^2)\rightarrow
        (\slask P_\parallel+m_e) eB\sigma_z
        \sign(P^2-m_e^2)\delta((P^2-m_e^2)^2-(eB)^2)~~.
\end{equation}
To obtain this rule I have kept the full $eB$ dependence in the
exponentials in \eq{Schwprop} since they are necessary for the IR
convergence of the energy integrals. In the final result, after
performing all integrals, the damping rate has a very clear linear
dependence on $eB$ for weak~fields.

To compute the damping from absorption and creation of neutrinos in
URCA processes one can use the same formalism but with neutron and
proton propagators in the loop. In the present case absorption is
dominted by $\nu+n\rightarrow p+e$ and creation by the inverse
process. The nucleons are also essentially nonrelativistic which
simplifies the matrix elements. On the other hand, since the general
expression comes out for free and the trace is easily evaluated with a
symbolic algebraic program, we might as well keep the full
relativistic expression and all possible channels when we perform the
numerical integrals.

\subsection*{Results}

The physical conditions in a supernova vary considerably during the
short time of the explosion. I have as an illustration evaluated the
damping asymmetry for two more or less typical conditions. The first
case has a high electron chemical potential ($\mu_e=200$~MeV) and
neutrino energy $q_0=100$~MeV, and the other one has $\mu_e=50$~MeV
and $q_0=30$~MeV. In both cases I used $T=10$~MeV and $Y_e=0.3$. Local
thermal equilibrium was assumed and the proton chemical potential was
determined from charge neutrality. These two cases correspond to the
densities $2\EE{14}$~g/cm$^3$ and $4\EE{12}$~g/cm$^3$.  For simplicity
the incoming neutrino has its momentum parallel to the magnetic field
(the linear term in $eB$ is of course antisymmetric in $q_z$). Writing
the damping factor as $\Gamma=\Gamma^{(0)}+eB\,\Gamma^{(1)}$, I found
the relative asymmetry for URCA and $\nu$-$e$ 
processes to be ($B_{14}$
is the field strength in units of $10^{14}$~Gauss)
\begin{equation}\label{relG200}
        \abs{\frac{eB\,\Gamma^{(1)}_{\rm URCA}}{\Gamma^{(0)}_{\rm URCA}}}
        \simeq 3.6\EE{-6}\, B_{14}~,\quad
        \abs{\frac{eB\,\Gamma^{(1)}_{\nu-e}}{\Gamma^{(0)}_{\nu-e}}}
        \simeq 9.8\EE{-7}\, B_{14}~~,
\end{equation}
for $\mu_e=200$~MeV and
\begin{equation}\label{relG50}
        \abs{\frac{eB\,\Gamma^{(1)}_{\rm URCA}}{\Gamma^{(0)}_{\rm URCA}}}
        \simeq 1.0\EE{-5}\, B_{14}~,\quad
        \abs{\frac{eB\,\Gamma^{(1)}_{\nu-e}}{\Gamma^{(0)}_{\nu-e}}}
        \simeq 7.3\EE{-6}\, B_{14}~~,
\end{equation}
for $\mu_e=50$~MeV. These relative asymetries are as large as the ones
calculated in \cite{BeneshH97} and one can expect the nucleon
polarization to give a comparabel contribution. I therefore conclude,
based on the results in \cite{ReddyPL97}, that neutrino absorption is
an important process for the neutrino opacity (see Fig.~4 in
\cite{ReddyPL97}) and that the asymmetry in URCA processes needs to be
taken into account in any reliable estimate of the total neutrino
propagation asymmetry in a neutron star. However, it is also my
impression that the final word has not yet been said about the
influence from collective behavior of nucleons on the neutrino
propagation.

\subsection*{Acknowledgements}
I would like to thank D.~Grasso and P.~Ullio for collaboration and
the organizers for the invitation to Ringberg.

}\newpage{

\head{Cherenkov Radiation by Massless Neutrinos in a\\ Magnetic Field}
     {Ara N.~Ioannisian$^{1,2}$ and Georg G.~Raffelt$^2$}
     {$^1$Yerevan Physics Institute, Yerevan 375036, Armenia \\
      $^2$Max-Planck-Institut f\"ur Physik 
      (Werner-Heisenberg-Institut)\\ 
      F\"ohringer Ring 6, 80805 M\"unchen, Germany}

\subsection*{Abstract}

We discuss the Cherenkov process $\nu\to\nu\gamma$ in the presence of
a homogeneous magnetic field. The neutrinos are taken to be massless
with only standard-model couplings.  The magnetic field fulfills the
dual purpose of inducing an effective neutrino-photon vertex and of
modifying the photon dispersion relation such that the Cherenkov
condition $\omega<|{\bf k}|$ is fulfilled.  For a field strength
$B_{\rm crit}=m_e^2/e=4.41\times10^{13}~{\rm Gauss}$ and for $E=2m_e$
the Cherenkov rate is about $6\times10^{-11}~{\rm s}^{-1}$.

\vspace{0.5cm}

\noindent
In many astrophysical environments the absorption, emission, or
scattering of neutrinos occurs in dense media or in the presence of
strong magnetic fields \cite{ara1}. Of particular conceptual interest
are those reactions which have no counterpart in vacuum, notably the
decay $\gamma\to\bar\nu\nu$ and the Cherenkov process
$\nu\to\nu\gamma$. These reactions do not occur in vacuum because they
are kinematically forbidden and because neutrinos do not couple to
photons. In the presence of a medium or $B$-field, neutrinos acquire
an effective coupling to photons by virtue of intermediate charged
particles.  In addition, media or external fields modify the
dispersion relations of all particles so that phase space is opened
for neutrino-photon reactions of the type $1\to 2+3$.

If neutrinos are exactly massless as we will always assume, and if
medium-induced modifications of their dispersion relation can be
neglected, the Cherenkov decay $\nu\to\nu\gamma$ is kinematically
possible whenever the photon four momentum $k=(\omega,{\bf k})$ is
space-like, i.e.\ ${\bf k}^2-\omega^2>0$.  Often the dispersion
relation is expressed by $|{\bf k}|=n\omega$ in terms of the
refractive index~$n$. In this language the Cherenkov decay is
kinematically possible whenever $n>1$.

Around pulsars field strengths around the critical value $B_{\rm
crit}=m_e^2/e=4.41\times10^{13}~{\rm Gauss}$ are encountered.  The
Cherenkov condition is satisfied for significant ranges of photon
frequencies. In addition, the magnetic field itself causes an
effective $\nu$-$\gamma$-vertex by standard-model neutrino couplings
to virtual electrons and positrons. Therefore, we study the Cherenkov
effect entirely within the particle-physics standard model.  This
process has been calculated earlier in~\cite{ara2}, but we do not
agree with their results.

Our work is closely related to a recent series of papers~\cite{ara3}
who studied the neutrino radiative decay $\nu\to\nu'\gamma$ in the
presence of magnetic fields.  Our work is also related to the process
of photon splitting that may occur in magnetic fields as discussed,
for example, in Refs.~\cite{ara4,ara5}.

Photons couple to neutrinos by the amplitudes shown in Figs.~1(a) and
(b).  We limit our discussion to field strengths not very much larger
than $B_{\rm crit}=m_e^2/e$.  Therefore, we keep only electrons in the
loop.  Moreover, we are interested in neutrino energies very much
smaller than the $W$- and $Z$-boson masses, allowing us to use the
limit of infinitely heavy gauge bosons and thus an effective
four-fermion interaction [Fig.~1(c)].  The matrix element has the form
\begin{equation}
\label{m}
{\cal M}=-\frac{G_F}{\sqrt{2}\,e}Z\varepsilon_{\mu}
\bar{\nu}\gamma_{\nu}(1-\gamma_5)\nu\,
(g_V\Pi^{\mu \nu}-g_A\Pi_5^{\mu \nu}) ,
\end{equation} 
where $\varepsilon$ is the photon
polarization vector and $Z$ its wave-function renormalization
factor. For the physical circumstances of interest to us, the photon
refractive index will be very close to unity so that we will be able
to use the vacuum approximation $Z=1$. Further,  
$g_V=2\sin^2\theta_W+\frac{1}{2}$, 
$g_A=\frac{1}{2}$ for $\nu_e$,
$g_V=2\sin^2\theta_W-\frac{1}{2}$, and
$g_A=-\frac{1}{2}$ for $\nu_{\mu,\tau}$.

\begin{figure}[t]
\centering\leavevmode
\vbox{
\unitlength=0.8mm
\begin{picture}(60,25)
\put(8,15){\line(-1,1){8}}
\put(8,15){\line(-1,-1){8}}
\put(0,7){\vector(1,1){4}}
\put(8,15){\vector(-1,1){6}}
\multiput(9.5,15)(6,0){3}{\oval(3,3)[t]}
\multiput(12.5,15)(6,0){3}{\oval(3,3)[b]}
\put(31,15){\circle{10}}
\put(31,15){\circle{9}}
\multiput(37.5,15)(6,0){3}{\oval(3,3)[t]}
\multiput(40.5,15)(6,0){3}{\oval(3,3)[b]}
\put(0,10){\shortstack{{}$\nu$}}
\put(18,18){\shortstack{{$Z$}}}
\put(43,18){\shortstack{{$\gamma$}}}
\put(30,11){\shortstack{{e}}}
\put(40,5){\shortstack{{(a)}}}
\end{picture}
\hspace{0.5cm}
\unitlength=0.8mm
\begin{picture}(60,32)
\put(16,15){\line(-1,1){7.5}}
\put(16,15){\line(-1,-1){7.5}}
\put(15,15){\line(-1,1){7}}
\put(15,15){\line(-1,-1){7}}
\put(16,15){\line(-1,1){16}}
\put(16,15){\line(-1,-1){16}}
\put(1,0){\vector(1,1){6}}
\put(16,15){\vector(-1,1){10}}
\multiput(17.5,15)(6,0){3}{\oval(3,3)[t]}
\multiput(20.5,15)(6,0){3}{\oval(3,3)[b]}
\multiput(8.2,8.7)(0,6){3}{\oval(3,3)[l]}
\multiput(8.2,11.7)(0,6){2}{\oval(3,3)[r]}
\put(2,15){\shortstack{{}$W$}}
\put(0,3){\shortstack{{}$\nu$}}
\put(23,18){\shortstack{{$\gamma$}}}
\put(25,5){\shortstack{{(b)}}}
\end{picture}
\unitlength=0.8mm
\begin{picture}(60,25)
\put(8,15){\line(-1,1){8}}
\put(8,15){\line(-1,-1){8}}
\put(0,7){\vector(1,1){4}}
\put(8,15){\vector(-1,1){6}}
\put(13,15){\circle{10}}
\put(13,15){\circle{9}}
\multiput(19.5,15)(6,0){3}{\oval(3,3)[t]}
\multiput(22.5,15)(6,0){3}{\oval(3,3)[b]}
\put(0,10){\shortstack{{}$\nu$}}
\put(26,18){\shortstack{{$\gamma$}}}
\put(12,11){\shortstack{{e}}}
\put(25,5){\shortstack{{(c)}}}
\end{picture}
}
\smallskip
\caption[...]{Neutrino-photon coupling in an external magnetic field.
The double line represents the electron propagator in the presence of
a $B$-field. 
(a)~$Z$-$A$-mixing. (b)~Penguin diagram (only for $\nu_e$).
(c)~Effective coupling in the limit of infinite gauge-boson masses.
\label{Fig1}}
\end{figure}
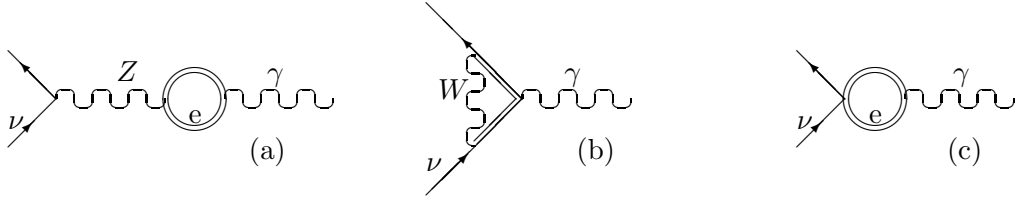

Following Refs.~\cite{ara4,ara6,ara7,ara8} we find
\begin{eqnarray}
\Pi^{\mu\nu}(k) &=& \frac{e^3B}{(4\pi)^2}
\Bigl[(g^{\mu \nu}k^2-k^{\mu}k^{\nu})N_0
-\,(g^{\mu \nu}_{\|}k^2_{\|}-k_{\|}^{\mu}k^{\nu}_{\|})N_{\|}+
(g^{\mu\nu}_{\bot}k^2_{\bot}-k^{\mu}_{\bot}k^{\nu}_{\bot})N_{\bot}
\Bigr], \nonumber\\
\Pi_5^{\mu \nu}(k) &=& \frac{e^3}{(4\pi)^2m_e^2}
\Bigl\{-C_\|\,k_{\|}^{\nu}(\widetilde{F} k)^{\mu}\
+ \ C_\bot\,\Bigl[k_{\bot}^{\nu}(k\widetilde{F})^{\mu}
+k_{\bot}^{\mu}(k\widetilde{F})^{\nu}-
k_{\bot}^2\widetilde{F}^{\mu \nu}\Bigr]\Bigr\},
\end{eqnarray} 
where $\widetilde{F}^{\mu \nu}= \frac{1}{2}\epsilon^{\mu \nu \rho
\sigma}F_{\rho \sigma}$ and $F_{12}=-F_{21}=B$. The $\|$ and $\bot$
decomposition of the metric is $g_\|={\rm diag}(-,0,0,+)$ and
$g_\bot=g-g_\|={\rm diag}(0,+,+,0)$ and $k$ is the four momentum of
the photon.  $N_0$, $N_{\bot}$,$N_{\|}$, $C_\bot$ and $C_\|$ are
functions of $B$, $k^2_{\|}$ and $k^2_{\bot}$.  They are real for
$\omega<2m_e$, i.e.\ below the pair-production threshold.

Four-momentum conservation constrains the photon emission angle to
have the value
\begin{equation}\label{emissionangle}
\cos \theta = \frac{1}{n} \
\left[1+(n^2-1)\frac{\omega}{2E}\right],
\end{equation}
where $\theta$ is the angle between the emitted photon and incoming
neutrino.  It turns out that for all situations of practical interest
we have $|n-1|\ll 1$ ~\cite{ara4,ara9}.  This reveals that the
outgoing photon propagates parallel to the original neutrino
direction.

It is easy to see that the parity-conserving part of the effective
vertex ($\Pi^{\mu \nu}$) is proportional to the small parameter
$(n-1)^2 \ll 1$ while the parity-violating part ($\Pi_5^{\mu \nu}$) is
{\it not\/}.  It is interesting to compare this finding with the
standard plasma decay process $\gamma\to\bar\nu\nu$ which is dominated
by the $\Pi^{\mu \nu}$. Therefore, in the approximation
$\sin^2\theta_W=\frac{1}{4}$ only the electron flavor contributes to
plasmon decay. Here the Cherenkov rate is equal for (anti)neutrinos of
all flavors.

We consider at first neutrino energies below the pair-production
threshold $E<2m_e$. For $\omega<2m_e$ the photon refractive index
always obeys the Cherenkov condition $n>1$ ~\cite{ara4,ara9}.
Further, it turns out that in the range $0<\omega< 2m_e$ the functions
$C_\|$,$C_\perp$ depend only weakly on $\omega$ so that they are well
approximated by their value at $\omega=0$.  For neutrinos which
propagate perpendicular to the magnetic field, the Cherenkov emission
rate can be written in the form
\begin{eqnarray}\label{finalresult}
\Gamma\ \approx \ \frac{4\alpha G_F^2E^5}{135(4\pi)^4}\,
\left(\frac{B}{B_{\rm crit}}\right)^2 h(B)\
= \ 
2.0\times10^{-9}~{\rm s}^{-1}~\left(\frac{E}{2m_e}\right)^5
\left(\frac{B}{B_{\rm crit}}\right)^2 h(B),
\end{eqnarray}
where 
\begin{equation}
h(B)= 
\cases{(4/25)\,(B/B_{\rm crit})^4&for $B\ll B_{\rm crit}$,\cr
1&for $B\gg B_{\rm crit}$.\cr}
\end{equation}
Turning next to the case $E>2m_e$ we note that in the presence of a
magnetic field the electron and positron wavefunctions are Landau
states so that the process $\nu\to\nu e^+e^-$ becomes kinematically
allowed. Therefore, neutrinos with such large energies will lose
energy primarily by pair production rather than by Cherenkov 
radiation; for recent calculations see~\cite{ara10}.

The strongest magnetic fields known in nature are near
pulsars. However, they have a spatial extent of only tens of
kilometers. Therefore, even if the field strength is as large as the
critical one, most neutrinos escaping from the pulsar or passing
through its magnetosphere will not emit Cherenkov photons. Thus, the
magnetosphere of a pulsar is quite transparent to neutrinos as one
might have expected.

\subsection*{Acknowledgments}
It is pleasure to thank the organizers of the Neutrino Workshop at
the Ringberg Castle for organizing a very interesting and enjoyable
workshop.

}\newpage{


\head{Photon Dispersion in a Supernova Core}
     {Alexander Kopf}
     {Max-Planck-Institut f\"ur Astrophysik\\ 
      Karl-Schwarzschild-Str.~1, 85748 Garching, Germany}

\noindent 
In astrophysical plasmas the dispersion relation of photons is usually
dominated by the electronic plasma effect.  It was recently
claimed~\cite{kopf.MS96}, however, that in a supernova (SN) core the
dominant contribution is caused by the photon interaction with the
nucleon magnetic moments.  This contribution can have the opposite
sign of the plasma term so that the photon four-momentum $K$ could
become space-like, allowing the Cherenkov processes $\gamma\nu \to
\nu$ and $\nu \to \gamma\nu$. Because of a numerical error the results
of Ref.~\cite{kopf.MS96} are far too large~\cite{kopf.Raffelt97}, but
the correct effect is still large enough to call for a closer
investigation.
 
Therefore, we will estimate the different possible contributions to
the refractive index, which is taken to be a real number,
\begin{equation}
n_{\rm refr}= \sqrt{\epsilon\mu} = k/\omega.
\end{equation}
It relates the wavenumber $k$ and frequency $\omega$ of an
electromagnetic excitation and is given in terms of the dielectric
permittivity $\epsilon$ and the magnetic permeability $\mu$
\cite{kopf.Jackson}.  We may also write the dispersion relation in the
form $\omega^2 - k^2 = m_{\rm eff}^2$, where $m_{\rm eff}^2$ can be
positive or negative.

We first assume that the medium constituents (protons, neutrons,
electrons) interact with each other only electromagnetically. This
leaves us with the usual plasma effect which is much larger for the
electrons than for the protons. It is approximately described by $
n_{\rm refr}^2 \approx 1 - \omega_p^2/\omega^2$, where $\omega_{p}^2 =
4\pi\alpha n_p/m_p$ for the protons (nondegenerate, nonrelativistic)
and $\omega_{p}^2 = (4\alpha/3\pi)\mu_e^2 $ for the electrons
(degenerate, relativistic) with $\mu_e$ the electron chemical
potential.

If the nuclei are not completely dissociated, nuclear bound-free
transitions might be significant.  However, at least for the simple
example of deuteron states this is not the case.  For a medium
consisting of deuterons alone we employ the usual relation between
refractive index and forward-scattering
amplitude~$f_0$~\cite{kopf.Sakurai} $ n_{\rm refr} = 1 +
(2\pi/\omega^2)\,n_d\,{\rm Re}f_0(\omega)$.  From the values for ${\rm
Re}f_0$ of Ref.~\cite{kopf.AFW88} we conclude that the contribution
from deuteron-like states never exceeds the single-proton plasma
effect and is thus negligible.

The main point of our work is to calculate the contribution from
correlated spins. We study the simplest model, a medium consisting
purely of neutrons. In the nonrelativistic case they do not respond to
an applied electric field so that $\epsilon = 1$. We write the
magnetic permeability $\mu = 1 + \chi$ \cite{kopf.Jackson} in terms of
the magnetic susceptibility $\chi = \chi' + i \chi''$ which implies
$n_{\rm refr}^2 - 1 = \chi'(\omega,k)$.  The fluctuation dissipation
theorem \cite{kopf.Forster} establishes a relationship between the
absorptive part $\chi''$ and spin fluctuations. The neutron spin
fluctuations can be described by the dynamical spin-density structure
function $S_{\sigma}(\omega,k)$ \cite{kopf.JKRS96}. We use a simple
model for its actual shape \cite{kopf.KR97} in terms of the spin
fluctuation rate $\Gamma_{\sigma}$ which can be estimated from the
nucleon interaction potential.  We calculate the refractive index (or
rather $\chi'$) from $\chi''$ via a Kramers-Kronig relation and
express it as an effective mass $m_{\rm eff}^2 = (1 - n_{\rm
refr}^2)\omega^2$.

\begin{figure}[ht]
\centerline{\epsfxsize=10cm\epsffile{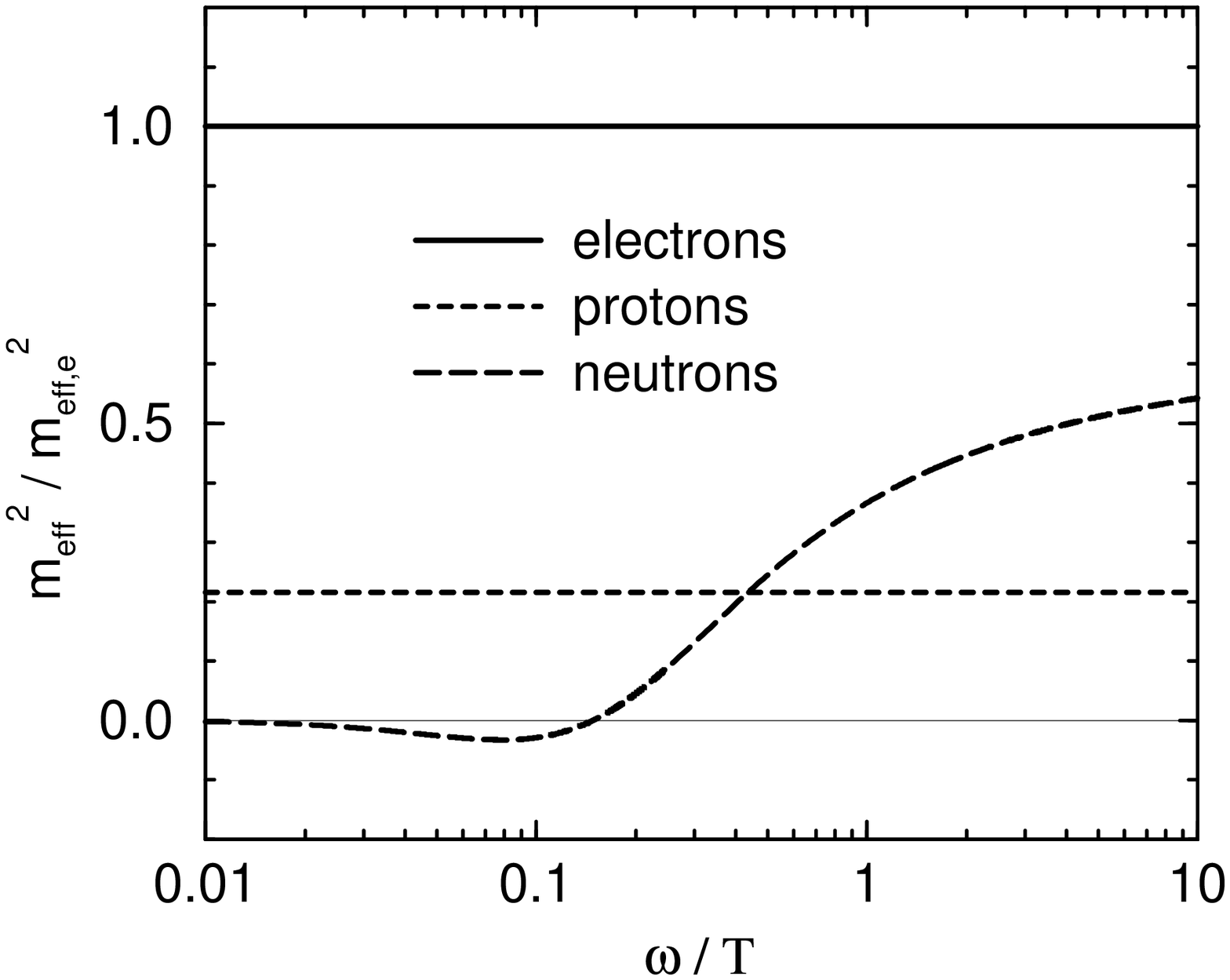}}
\caption{\label{fig:kopf} The contributions of electrons,
protons and correlated neutrons for $\rho = 7\times10^{14}\,{\rm
g\, cm^{-3}}$, $Y_e = 0.3$, $T = 40\,{\rm MeV}$ and $\Gamma_{\sigma} =
30\,T$.} 
\end{figure}

In Fig.~\ref{fig:kopf} we show the different contributions to $m_{\rm
eff}^2$ relative to the one by the electrons for typical conditions of
a SN core ($\rho = 7\times10^{14}\,{\rm g\, cm^{-3}}$, electron to
baryon ratio $Y_e = 0.3$, $T = 40\,{\rm MeV}$, $\Gamma_{\sigma} =
30\,T$). All the approximations made in our calculation work in the
same direction of overestimating the magnetic-moment contribution.

The contribution of the correlated spins has the opposite sign
relative to the plasma terms only in a small $\omega$ range and thus
cannot change the sign of the total $m_{\rm eff}^2$.  Thus the
refractive index is always less than 1 and Cherenkov effects remain
forbidden.

\subsection*{Acknowledgments}

This talk is based on work done in collaboration with G.~Raffelt
\cite{kopf.KR97}. I thank S.~Hardy for comments on the manuscript.

%
%

 } \newpage {\  }

\newpage{

\thispagestyle{empty}

\begin{flushright}
\Huge\bf
{\ }


Gamma-Ray\\
\bigskip
Bursts

\end{flushright}

\newpage

\thispagestyle{empty}

{\ }

\newpage

}\newpage {

\head{Gamma-Ray Burst Observations}
     {D.H.\ Hartmann$^1$, D.L.\ Band$^2$}
     {$^1$Department of Physics and Astronomy, Clemson University 
      Clemson, SC 29634-1911\\
      $^2$CASS, UC San Diego, La Jolla, CA 92093, USA}

\subsection*{Abstract}

The transients following GRB970228 and GRB970508 showed that these
(and probably all) GRBs are cosmological. However, the host galaxies
expected to be associated with these and other bursts are largely
absent, indicating that either bursts are further than expected or the
host galaxies are underluminous. This apparent discrepancy does not
invalidate the cosmological hypothesis, but instead host galaxy
observations can test more sophisticated models.

\subsection*{Absence of Expected Host Galaxies}

Observations of the optical transients (OTs) from GRB970228
\cite{jan97} and GRB970508 \cite{bond97} have finally provided the
smoking gun that bursts are cosmological.  In most cosmological models
bursts occur in host galaxies: are these galaxies present, and
conversely, what can we learn from them?  Underlying any confrontation
of theory and data must be a well defined model.  Here we show that
the host galaxy observations are not consistent with the expectations
of the simplest cosmological model, and that these observations can be
used to test more sophisticated models.

In the simplest (``minimal'') cosmological model the distance
scale is derived from the intensity distribution logN--logP assuming
bursts are standard candles which do not evolve in rate or intensity.
Bursts occur in normal galaxies at a rate proportional to a galaxy's
luminosity.  This model predicts the host galaxy distribution
for a given burst. Are the expected host galaxies present? 

For GRB970228 an underlying extended object was found \cite{jan97},
but its redshift and nature have not been established. If the observed
``fuzz'' is indeed a galaxy at $z\sim 1/4$, it is $\sim$5
magnitudes fainter than expected for a galaxy at this redshift. For
GRB970508 no obvious underlying galaxy was observed~\cite{pian97} and
the nearest extended objects have separations of several arcseconds,
but spectroscopy with the Keck telescopes \cite{metzger97} led to the
discovery of absorption and emission lines giving a GRB redshift of $z
\ge 0.835$. The {\it HST} magnitude limit $R_{\rm lim}\sim 25.5$
\cite{pian97} for a galaxy coincident with the transient again
suggests a host galaxy fainter than expected. Similar conclusions
follow from the inspection of IPN error boxes
\cite{BH98,schaefer92,schaefer97}, but see also
\cite{larson97,vrba97}. This absence of sufficiently bright host
galaxies is often called the ``no-host'' problem, which is a misnomer.
The point simply is that if galaxies such as the Milky Way provide the
hosts to most bursters, and if their redshifts are less than unity, as
predicted by the minimal model, we expect to find bright galaxies
inside a large fraction of the smallest IPN error boxes. 

To demonstrate this quantitatively, consider the apparent magnitude of
a typical host galaxy, which we assume has $M_*(B) = -20$
(approximately the absolute magnitude of an $L_*$ galaxy---see
discussion below). Using Peebles' notation \cite{peebles}, the
apparent magnitude is 
\begin{equation}
m = 42.38 + M + 5\ {\rm log}\left[y(z)(1+z)\right] + K(z) + E(z) + 
   A(\Omega,z) + \chi(z) \ \ ,
\end{equation}
where $K(z)$ is the usual K-correction, $E(z)$ corrects for the
possible evolution of the host galaxy's spectrum, $A$ is the sum of
Galactic foreground (position dependent) and intergalactic extinction,
and $\chi(z)$ represents any corrections that apply in hierarchical
galaxy formation scenarios, where galaxies are assembled through the
merger of star forming subunits. The commonly found term $5\log(h)$ is
already absorbed in eq.~(1). Neglecting potentially large corrections
from the $K$, $E$, $A$, and $\chi(z)$ terms, a host like the Milky Way
with $M\sim -20$ would have an apparent magnitude $m\sim 22$ for
redshifts of order unity. Several small IPN error boxes have no galaxy
of this magnitude or brighter. Our simplified treatment agrees with
Schaefer's conclusion\cite{schaefer92,schaefer97} that typical
galaxies at the calculated burst distance are absent from burst error
boxes. 

Thus bursts are further than predicted from the logN--logP
distribution without evolution, or they occur in underluminous
galaxies; an extreme limit of the latter alternative is that bursts do
not occur in galaxies. 

\subsection*{Host Galaxies as a Probe of Cosmological Models}

The search for host galaxies is a powerful test of cosmological burst
models.  From the two above mentioned OTs we conclude that GRBs are
cosmological, but the observations have not fixed the distance scale
quantitatively, nor have they determined the energy source. While the
x-ray, optical, and radio lightcurves (for GRB970508 only) are
consistent with the predictions of the basic ``fireball afterglow''
picture, the fireball's central engine could be the merger of a
neutron star binary, the collapse of a massive, rotating star, or the
jet produced by accretion onto a massive black hole residing at the
center of an otherwise normal galaxy.  

The host galaxies found within burst error boxes are a powerful
discriminant between different models for the burst energy source.
Note that the region within which a host would be acceptable
surrounding the sub-arcsecond localizations of an OT is effectively
the error box for the host galaxy. Almost all models assume that
bursts are associated with galaxies; the issue is the relationship
between the burst and the host.  In models such as the momentary
activation of a dormant massive black hole the burst rate per galaxy
is constant. On the other hand bursts are an endpoint of stellar
evolution in most models, and therefore to first order we expect the
burst rate per galaxy in these models to be proportional to the
galaxy's mass and thus luminosity.  These two model classes have
different host galaxy luminosity functions $\psi(M)$ with different
average values of $M$ (the absolute magnitude).  In the first case,
$\psi(M)$ is proportional to the normal galaxy luminosity function,
while in the second case $\psi(M)$ is proportional to the normal
galaxy luminosity function weighted by the luminosity $L\propto
10^{-0.4M}$.  We approximate the normal galaxy luminosity function
with the Schechter function: 
\begin{equation}
\Phi(M) = \kappa\ 10^{0.4(M_*-M)(\alpha+1)}\ \left[{\rm exp}\left(-10^{0.4
(M_*-M)}\right)\right]
\end{equation}
where $\kappa$ is the normalization, $\alpha$ is the slope of
the faint end, and $M_*$ is the absolute magnitude of an $L_*$ galaxy.
Here we use $\alpha=-1$. In the B band $M_*(B)= -19.53$, which
corresponds to $L_*(B) = 1.8\times 10^{10} L_\odot \ h_{75}^{-2} \sim
3\times 10^{11}\ L_{\odot}(B)\ h_{75}^{-2}$. In Figure~1 
we show the cumulative distributions for the host galaxy magnitudes
for the two model classes. As can be seen, the average host galaxy
magnitude (i.e., at 0.5) differs by $\sim1.75$ magnitudes. 

\begin{figure}[ht] 
\center
\epsfig{file=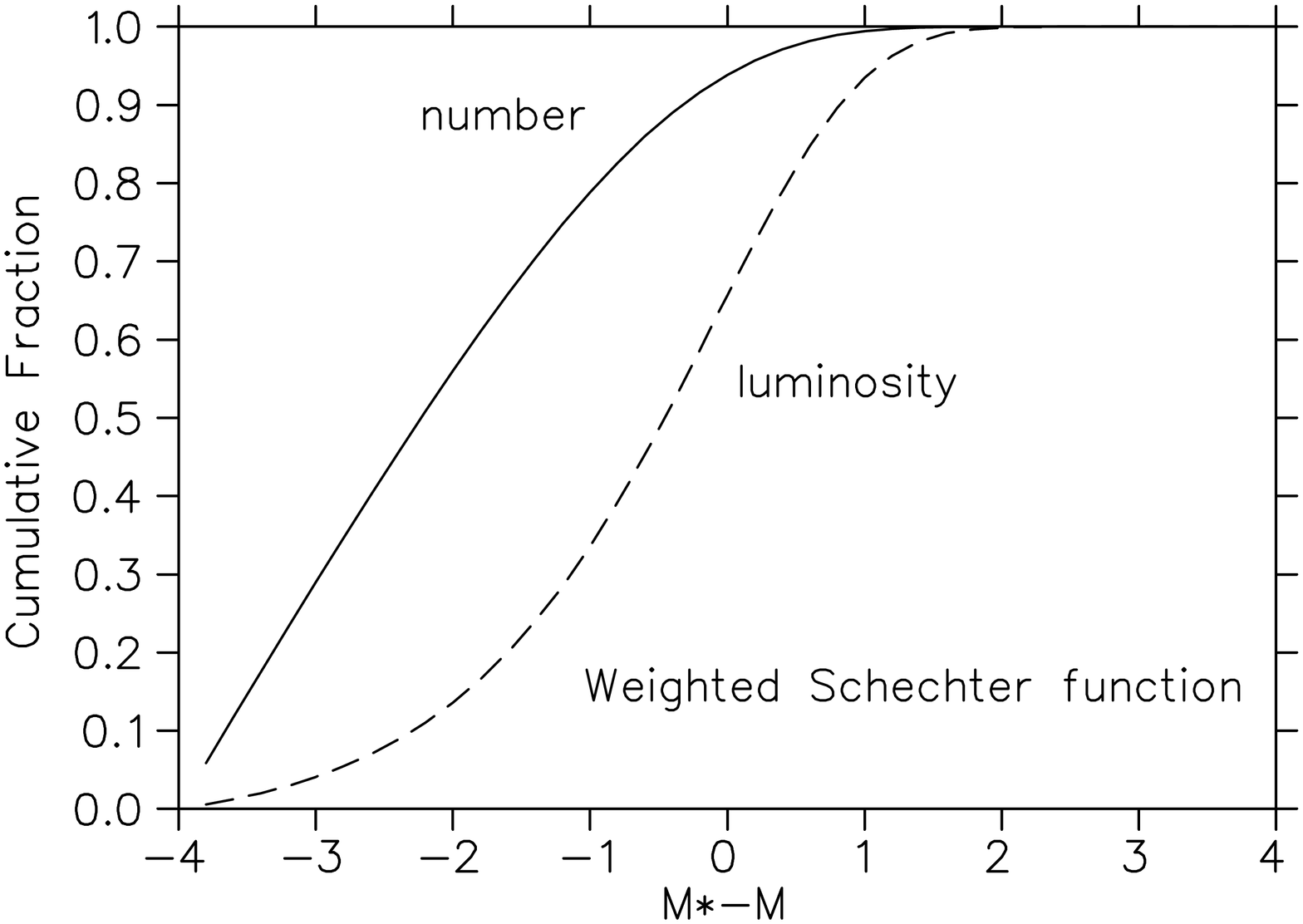, height=10cm, angle=90 }
\caption{The cumulative distribution of host galaxy magnitudes if
their luminosity function is weighted by number or luminosity. If GRBs
trace light, the 50$\%$ point is near $M_*$. If it is proportional to
galaxy number a typical host would be $\sim2$ magnitudes fainter.} 
\label{hartmann.fig1}
\end{figure}

However, we can make better predictions about the host galaxies in
cosmological models where bursts are a stellar endpoint.  In such
models, the burst rate should be a function of the star formation rate
(SFR).  If there is a substantial delay (e.g., of order a billion
years or more) between the GRB event and the star forming activity
that created the progenitor, then the burst rate integrates over a
galaxy's SFR, and we would not expect the host galaxy to display the
signatures of recent star formation.  Furthermore, if the progenitor
is given a large velocity, then it may travel a large distance from
the host galaxy before bursting, and it may become impossible to
associate a galaxy with the burst. 

In many models the burst occurs shortly after its progenitor star
forms (e.g., within a hundred million years or less).  We would then
expect that on average bursts would occur in galaxies showing evidence
of recent star formation.  The burst rate should be proportional to
the SFR, both for individual galaxies and for a given cosmological
epoch. 

In particular, the burst rate and the SFR should have the same
history, as was recently considered by several groups
\cite{vahe97,sahu97,totani97,wijers97}. Extensive redshift surveys and
data from the Hubble Deep Field have reliably determined the cosmic
star formation history to $z\sim5$
\cite{connolly97,ellis97,lilly97,madau96,madau97}. The data clearly
suggest a rapid increase in the comoving SFR density with increasing
redshift, SFR $\propto (1+z)^4$, reaching a peak rate (at $z\sim1.5$)
about 10--20 times higher than the present-day rate, and decreasing
slowly to the present value by $z\sim5$. This evolution function,
$\eta(z)$, enters the differential rate vs. (bolometric) peak flux 
\begin{equation}
\partial_P{R} \propto P^{-5/2}\ {\rm E}(z)^{-1}\ \eta(z)\ (1+z)^{-3}\
\left[(1+z)\partial_zy(z) + y(z)\right]^{-1} \ \ ,
\end{equation}
where E($z$) and $y(z)$ are defined in \cite{peebles}. For small
redshifts the logarithmic slope of this function is Euclidean, i.e.
$-$5/2.  The solid curve of Figure~2 
shows the effects of geometry (bending of logN--logP) and the dashed
curve demonstrates how $\eta(z)$ compensates for the geometry out to
the redshift at which the cosmic SFR peaks. At larger redshifts the
effects of geometry and decreasing SFR then combine and the slope
flattens quickly. Comparison with BATSE data suggests that this SFR
model deviates from the pseudo-Euclidean slope too abruptly. While
several studies \cite{sahu97,totani97,wijers97} report that the
observed SFR generates a brightness distribution consistent with BATSE
data, our findings support the different result of Petrosian $\&$
Lloyd \cite{vahe97}, who suggest that other evolutionary effects must
be present in addition to the density evolution described by
$\eta(z)$. While a good fit to the data requires a more sophisticated
model of source evolution, the basic message is likely to be the same:
the logN--logP distribution does not exclude GRB redshifts much
greater than unity. 

\begin{figure}
\center
\epsfig{file=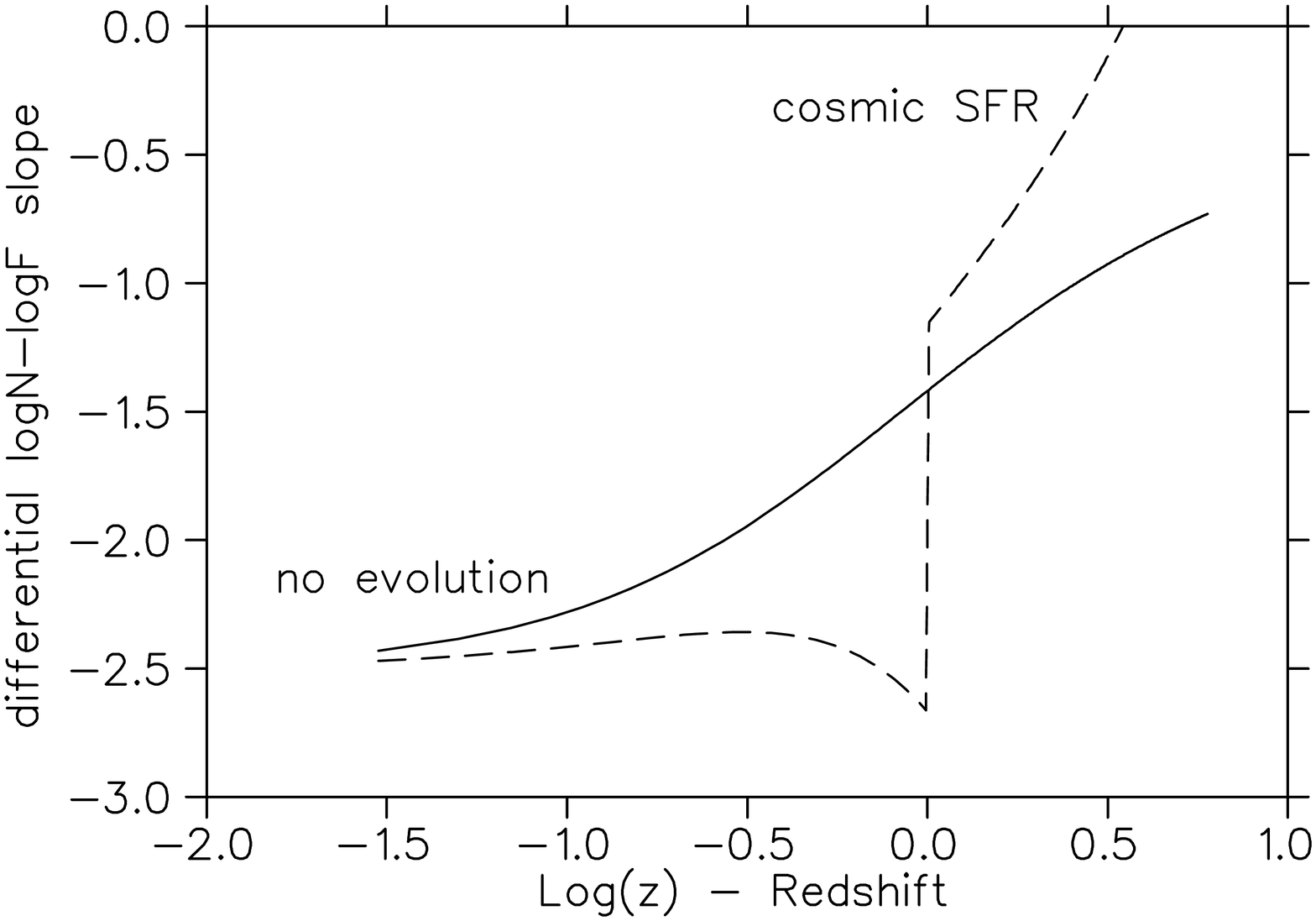, height=10cm, angle=90 }
\caption{The slope of the differential GRB brightness distribution.
Non-evolving sources (solid curve) quickly show significant deviation
from the Euclidean value $-$5/2 with increasing redshift. If the GRB
rate is proportional to the SFR (see text) the apparent Euclidean
slope extends to $z\sim 1$ (dashed curve). For greater $z$ the
geometry of the universe together with a now decreasing burst rate
cause the slope to deviate rapidly from $-$5/2.} 
\label{hartmann.fig2}
\end{figure}

Therefore, the absence of the host galaxies predicted by the
``minimal'' cosmological model does not call the cosmological origin
of bursts into question.  Instead, host galaxy observations will teach
us where bursts occur.

}\newpage{

\head{Gamma-Ray Bursts:\\ Models That Don't Work and Some that Might}
     {S.E.~Woosley$^{1,2}$ and A.~MacFadyen$^{1,2}$}
     {$^1$Max-Planck-Institut f\"ur Astrophysik, 85748 Garching, 
      Germany\\
      $^2$Astronomy Department, UC Santa Cruz (UCSC), USA}

\subsection*{Introduction}

Thirty years after their discovery and over 140 models later, we are
still struggling to understand the nature of gamma-ray bursts. There
has been considerable evolution in our views during the last few
years, especially the last six months, due largely to the recent
observational data summarized elsewhere in these proceedings by Dieter
Hartmann. A consensus, by no means universal, has developed that
gamma-ray bursts occur at cosmological distances, at least Gpc's, far
enough away for their distribution to be very isotropic, yet sampled
with sufficient sensitivity that the edge of the distribution is being
seen. This requires burst energies of over 10$^{51}$ erg (times a
beaming factor that could be $\ll$ 1) in an event that occurs about
1000 times annually, has median energy around 600 keV, and typical
duration from a small fraction of a second to hundreds of seconds
(bimodal with means 0.3 and 20 s). The large $\gamma \gamma$-opacity
above the pair threshold requires that a burst originate in a large
volume, perhaps 10$^{15}$ cm or more in radius, and the short duration
at this large size requires large relativistic factors, $\Gamma
\agt 100$. According to the current paradigm (e.g., Meszaros \&
Rees 1993), this is accomplished by having a small amount of matter,
$\alt 10^{-5}$~M$_{\scriptscriptstyle \odot}$, move with the
requisite $\Gamma$ and produce the gamma-rays either by internal
shocks or by interaction with the circum-source material, which may be
the interstellar medium.

The quest in current gamma may burst ``models''---there is still no
complete example of a natural event modeled thru to the production of
gamma-rays---has thus turned to an intermediate goal of delivering to
some small amount of matter the necessary tremendous energy to make it
move so close to the speed of light. There have been many
attempts. Most can be ruled out. 

\subsection*{Models that don't work}

\subsubsection*{a) All Forms of Accretion Induced Collapse 
in White Dwarf Stars}
In its simplest form (e.g., Dar et al.~1992), an unmagnetized white
dwarf collapses to a neutron star and produces a pair fireball by
neutrino annihilation above its surface. The problem is that the same
neutrinos blow a wind from the neutron star of about 
0.01~M$_{\scriptscriptstyle \odot}$~s$^{-1}$. 
The energy from neutrino annihilation is deposited deep
within this wind and adiabatic expansion leads to a very cool
photosphere. No gamma-rays are produced and the mass ejected has
$\Gamma$ much less than one.

A variation on this model is the rotating magnetic white dwarf that
collapses to a very energetic pulsar (Usov 1988). The pulsar converts
a portion of its rotational energy into gamma-rays and pairs---another
pair fireball model. The problem here is again the neutrino driven
wind, which since it has a much greater energy density than the
magnetic field, blows it away before any burst can be generated.

\subsubsection*{b) Phase Transitions in Neutron Stars} 
An example might be the transition of a neutron star to a strange star
after accreting a critical mass in a binary system. If the phase
transition results in a more compact core forming in less than a sound
crossing time, the bounce forms a shock that propagates to the edge of
the neutron star with energy that may be in excess of 10$^{51}$
erg. Steepening of the shock in the density gradient near neutron star
surface might concentrate this energy in a small amount of mass. The
problem here, besides the obvious issue of whether such supersonic
phase transitions ever occur, is that the shock hydrodynamics has been
calculated and shown inadequate to produce sufficient high energy
ejecta. Fryer \& Woosley (1998) show that, even in an optimal model,
less than 10$^{46}$ erg is concentrated in matter having $\Gamma \agt
40$. Most of the shock energy goes into heating and expanding the
neutron star and what remains, about 10$^{51}$ erg, ejects not
10$^{-5}$~M$_{\scriptscriptstyle \odot}$, but
0.01~M$_{\scriptscriptstyle \odot}$.  Most of this matter is at low
$\Gamma$.

\subsubsection*{c) Supermassive Black Holes} 
Such objects are thought to power AGN's including gamma-ray blazars;
why not gamma-ray bursts (Carter 1992)? One major problem is the long
orbit time near the event horizon, $\sim$5000~(M/10$^7$
M$_{\scriptscriptstyle \odot}$)~s, implies an event duration much
longer than observed in gamma-ray bursts. We presume that matter must
orbit the hole at least a few times before building up large magnetic
fields or other means of dissipating energy and forming jets.

\subsection*{Models That Might Work}

In general, the models that might work all involve accretion at a very
high rate into a newly formed black hole whose mass is between two and
a few hundred solar masses. The observational counterpart to neutron
star formation is a supernova. Perhaps a gamma-ray burst is the signal
that a black hole has been born. There are four ways of setting up the
requisite conditions: a) merging neutron stars forming a black hole
with a residual disk of a few hundredths of a solar mass; b) neutron
star, black hole mergers which might leave a disk of a few tenths
M$_{\scriptscriptstyle \odot}$; 
c) collapsars in which the core of a massive star collapses to
a black hole and accretes the rest of the star (several solar masses
endowed with sufficient rotation); and d) black hole, white dwarf
mergers. All of these events should occur chiefly where there are
stars, i.e., in galaxies. It would be difficult for binaries
consisting of compact stars and BH's to remain bound when kicked out
of the galaxy. WD-BH mergers might occur not only in ordinary
stellar systems, but in binaries formed in dense galactic nuclei
(Quinlan and Shapiro 1990). Collapsars in particular occur only in
star forming regions.

The accretion rate depends on the viscosity of the disk, or in the
case of the collapsar model upon the time required for the star to
collapse to the disk (446/$\rho^{1/2}$ s). Because the star has a
range of densities, accretion would continue at a diminishing rate for
a very long time after the gamma-ray burst perhaps providing an
enduring x-ray transient. Typically the accretion rate in all these
models starts at 
$\sim$~0.1~M$_{\scriptscriptstyle \odot}$~s$^{-1}$ 
and declines. For such
accretion rates one expects densities in the inner disk of around
10$^{12}$ g cm$^{-3}$ (Popham \& Gammie 1997; or simply seen by
equating $\pi R_s^2 h \rho v_{\rm drift} \sim \dot M$). At this
density the inner disk is marginally thick to neutrinos. Typical drift
velocities are $\sim$1000 km/s implying a residence time in the disk
of order 0.1 s (this number is viscosity ($\alpha$) dependent, and
uncertain), so the duration of bursts from merging compact objects
could be this short (longer burst time scales are allowable depending
on the circumstellar interaction, but none shorter than the cycle time
of the engine itself). A longer time scale would characterize the
collapsars where the disk is continually fed by stellar
collapse. Perhaps the long complex bursts (20 s mean) are due to
collapsars and the short bursts are merging compact objects.

For an accretion rate of 
0.1~M$_{\scriptscriptstyle \odot}$~s$^{-1}$, the energy dissipated,
mostly in neutrinos, is $1.8 \times 10^{53}$ F erg/s where F is 0.06
for Schwarzschild geometry (perhaps appropriate for merging compact
objects) and 0.4 for extreme Kerr geometry (perhaps appropriate for
collapsars). Coupled with their short time scale this implies that
merging compact objects may give bursts of much less total energy than
collapsars, but the efficiency for getting the jet out of a collapsar
may modulate this considerably (see below). In all models, a portion
of the energy goes into MHD instabilities in the disk that may result
in an energetic wind, or perhaps the rotational energy of the black hole
is tapped as in some models for AGN. But the simplest physical
solution for ejecting relativistic matter, if it works, is neutrino
annihilation along the rotational axis of the hole (Woosley 1993). The
efficiency for this is uncertain, but numerical simulations (see the
paper by Ruffert, these proceedings) suggest an efficiency for
converting neutrino luminosity into pair fireball energy of
$\sim$1\%. Thus one has a total energy for black holes merging with
neutron stars of order 10$^{50}$ erg (Ruffert et al.~1997) and a total
energy for collapsars that might approach 10$^{52}$ erg.

The chief uncertainty in all these models---aside from exactly how the
relativistic matter reconverts its streaming energy into 
gamma-rays---is 
the baryon loading. One expects this to be relatively small in
neutron star and black hole mergers, but still perhaps more than
10$^{-5}$~M$_{\scriptscriptstyle \odot}$, 
depending upon the mingling of the wind from the
disk, residual matter from the merger, and the relativistic
outflow. Detailed calculations remain to be done. 

The problem is much more severe in the collapsar model where the jet
that is formed has to penetrate the overlying star---or what is left
of it---a little like making the $\gamma$-ray burst at the center of
the sun. Still 10$^{52}$ erg is a lot of energy and we also expect
that the jet from collapsar models might be tightly beamed. This is
initially the case because the disk makes a transition from very thick
(geometrically) to thin at the point where neutrino losses start to occur
on an accretion time scale. This happens at a definite radius,
$\sim$100 km, because of the T$^9$ dependence of the neutrino rates
(Bob Popham, private communication). Calculations in progress by
M\"uller and colleagues at the MPA will show how effective this jet is
at tunneling thru the star. Our own recent 2D simulations suggest a
solid angle of only about 1\% for the jet. A lot of energy gets used
up ejecting all the mass above about 45 degrees from the equator. A
supernova is one consequence; a collapsar is not a ``failed
supernova'' after all, but this mass ejection at large angles is
non-relativistic. Along the axis the jet clears a path for more
energetic matter to follow, and since the burst continues for many
jet-stellar crossing times, $\Gamma$ rises. 
Only $\sim$0.01~M$_{\scriptscriptstyle \odot}$
initially lies in the path of the jet for the expected small opening
angle.

The energy of the ejecta and presumably the hardness of the transient
that is seen thereafter thus depends upon viewing angle. Along the
axis, high $\Gamma$ will overtake low $\Gamma$ material before the
observable event actually commences. The start of the burst occurs
when the jet sweeps up 1/$\Gamma$ times its rest mass, or about
10$^{-7}$~M$_{\scriptscriptstyle \odot}$. 
This happens after the jet has gone about 10$^{15}$
cm and interacted with the stellar wind ejected prior to the explosion
(about 10$^{-5}$~M$_{\scriptscriptstyle \odot}$~y$^{-1}$ assumed).

In summary we expect about 10$^{50}$ erg, $\alt$~0.1\% of the total
energy dissipated in the disk, to go into a jet with $\Gamma \agt$
100 and an opening angle of about 10 degrees. The event rate could be
10$^{-4}$ to 10$^{-3}$ y$^{-1}$ per bright galaxy.

Below we show pictures of our first 2D calculation of rotating stellar
collapse about 10 s after the collapse began (nb., no explosion,
contrary to Bodenheimer \& Woosley, 1983). A black hole and accretion
disk have clearly formed. The assumed angular momentum here was about
10$^{17}$ cm$^2$ s$^{-1}$ in the region that formed the disk. Stellar
evolution calculations suggest that a value of a few times 10$^{16}$
may be more appropriate to the stellar mantle, but larger values exist
farther out. This will merely delay the formation of the disk. The
figure on the right shows a first crude attempt at running a jet thru
this stellar model. The code employed lacked relativistic
hydrodynamics and the jet was artificially injected at the inner
boundary with a kinetic energy of $4 \times 10^{50}$ erg s$^{-1}$ 
and a speed of 2/3 c corresponding to a mass loss rate of 
0.001~M$_{\scriptscriptstyle \odot}$~s$^{-1}$. 
For now this is just a suggestive picture of what a more
realistic calculation might give. In a relativistic code, the same
energy 
injection with $\Gamma \sim 100$ would correspond to a (rest) mass
loss rate 100 times less, comparable to what one needs in a
$\gamma$-ray burst.

\begin{figure}[ht]
\vskip 0.25 in
\hbox{\hskip-0.8cm\epsfxsize=0.62\textwidth\epsffile{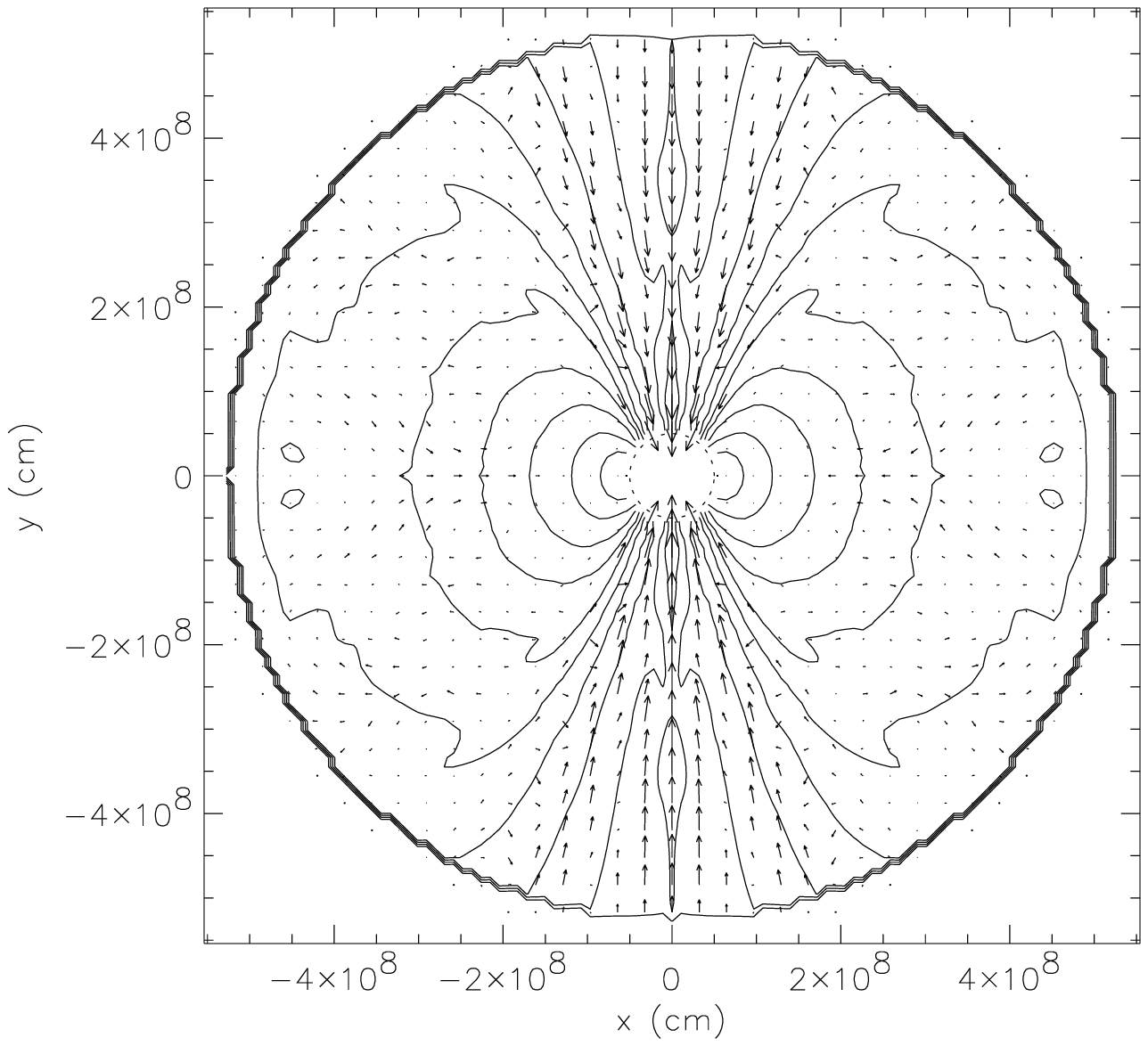}
 \hskip-0.9cm   \epsfxsize=0.48\textwidth\epsffile{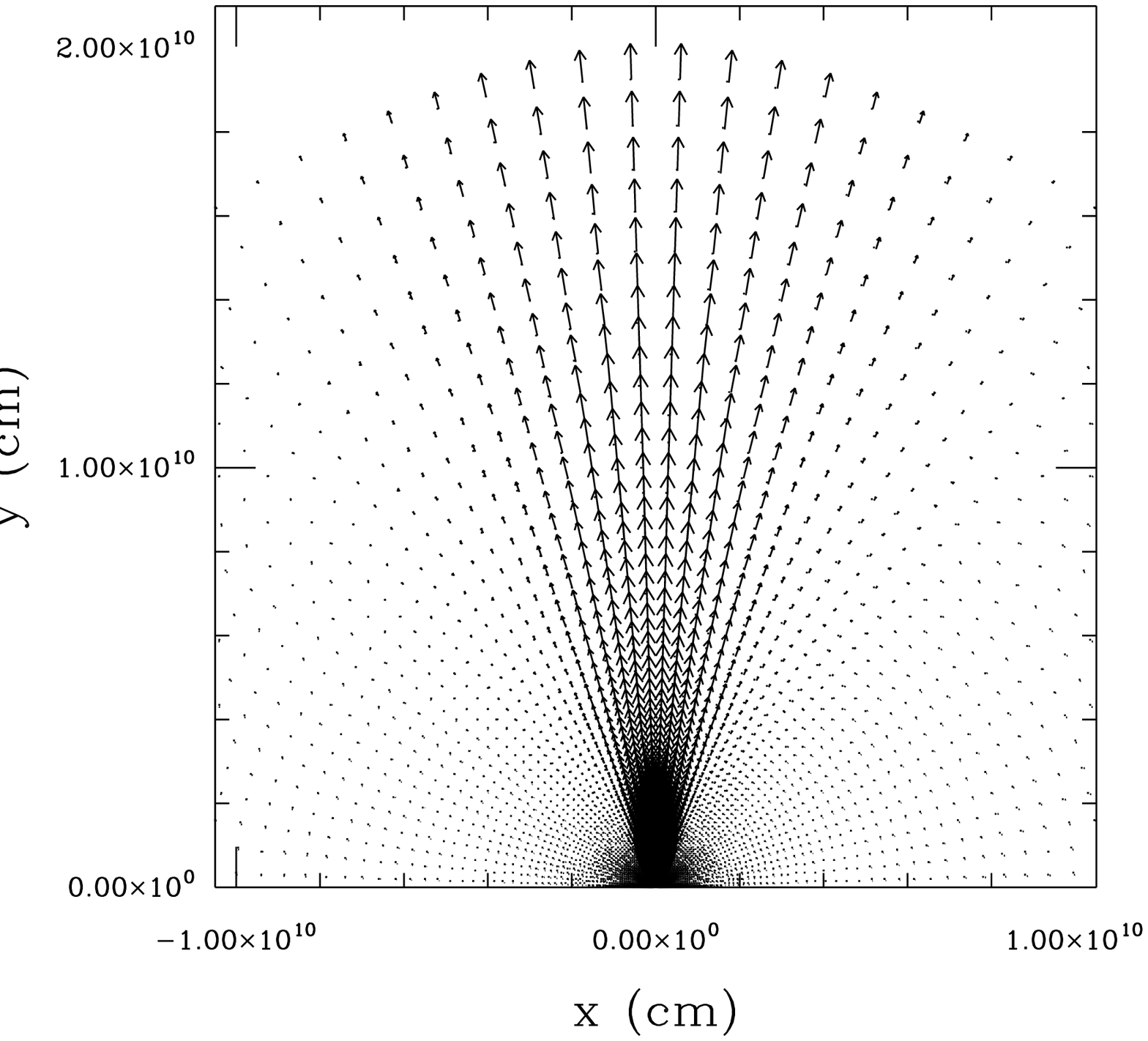}}
\caption{Left: Velocity field and density contours of a preliminary
rotating collapsar model at $t=10$~s after collapse.  The dashed
circle in the center represents a completely absorbing inner boundary
at $r=5 \times 10^7 $~cm.  The core has grown to 2.85
M$_{\scriptscriptstyle \odot}$. Right: Jet emerging from stellar
surface at $2 \times 10^{10}$~cm. The longest arrow is a velocity of
$1.7 \times 10^{10}$ cm s$^{-1}$. Clearly given enough momentum, the
jet can penetrate the star.}
\label{woosley.fig2}
\vskip 0.25 in
\end{figure}

The material that is ejected at lower $\Gamma$ at angles farther away
from the rotational axis may also be of considerable interest. For
intermediate values of $\Gamma$, say $\sim$1--10, one may still get a
bright circumstellar interaction, but because it is more massive and
must intercept a larger fraction of its rest mass before giving up its
energy, the display from this interaction may be delayed and
prolonged. We expect that this is the origin of the optical, x-ray,
and radio afterglows detected in the recent gamma-ray bursts of 022897
and 050897. As much or more energy could be in these displays than in
the $\gamma$-ray burst itself. The softer radiation would also be
beamed to a larger part of the sky suggesting the possibility that
there are many more optical, radio, and x-ray transients from
$\gamma$-ray burst sources than GRB's themselves.

Further details of our models and other issues discussed in this short
paper will appear in a paper to be submitted soon to the ApJ.

\subsection*{Acknowledgements}

We thank Alexander Heger, Thomas Janka, Ewald M\"uller, and Bob
Popham for informative conversations regarding presupernova evolution
including the effects of rotation, the neutrino physics of gamma-ray
burst models, multi-dimensional hydrodynamics, and black hole
accretion respectively. This work was supported by the NSF (AST
94-17161), NASA (NAG5-2843), and by a Humboldt Award at the MPA,
Garching.


}\newpage{


\head{Models of Coalescing Neutron Stars with Different Masses and 
Impact Parameters}
     {Maximilian Ruffert$\,^{1,2}$, Hans-Thomas Janka$\,^1$}
  {$^1$Max-Planck-Institut f\"ur Astrophysik, Postfach 1523, 
   85740 Garching, Germany \\
  $^2$Institute of Astronomy, Madingley Road, Cambridge CB3 0HA, U.K.}

\subsection*{Abstract}

We have simulated a variety of models, in order to shed light on the
accessible physical conditions during the mergers, 
on the amount of matter dynamically ejected during the merging,
on the timescales of mass accretion by the (forming) black hole,
on the conversion of energy into neutrino emission, on the 
amount of energy deposited by $\nu\bar\nu$-annihilation, and 
on the baryon loading of the created pair-plasma fireball.
To this end, we varied the masses and mass ratios
as well as the initial spins of the neutron stars, changed the
impact parameter to consider spiral-in orbits and direct collisions,
included a black hole (vacuum sphere) in our simulations, and
studied the dynamical evolution of the accretion torus around the
black hole formed after the neutron star merging until
a (nearly) stationary state was reached. While the neutrino 
emission during the dynamical phases of the mergings is definitely
too small to power gamma-ray bursts (GRBs), we find that the masses,
lifetimes, and neutrino luminosities of the accretion tori have
values that might explain short (${\cal O}$(0.1--1$\,{\rm s}$)) and
not too powerful ($\sim 10^{51}/(4\pi)\,{\rm erg/(s\cdot sterad}$)) 
gamma-ray bursts.
  
\subsection*{Summary of Numerical Procedures and Initial Conditions}

The hydrodynamical simulations were done with a code based on the
Piecewise Parabolic Method (PPM) developed by Colella \&
Woodward~\cite{ruffert.col84}.  The code is basically Newtonian, but
contains the terms necessary to describe gravitational wave emission
and the corresponding back-reaction on the hydrodynamical flow
(Blanchet et al.~\cite{ruffert.bla90}).  The terms corresponding to
the gravitational potential are implemented as source terms in the PPM
algorithm.  In order to describe the thermodynamics of the neutron
star matter, we use the equation of state (EOS) of Lattimer \&
Swesty~\cite{ruffert.lat91}.  Energy loss and changes of the electron
abundance due to the emission of neutrinos and antineutrinos are taken
into account by an elaborate ``neutrino leakage
scheme''~\cite{ruffert.ruf96}.  The energy source terms contain the
production of all types of neutrino pairs by thermal processes and of
electron neutrinos and antineutrinos also by lepton captures onto
baryons.  Matter is rendered optically thick to neutrinos due to the
main opacity producing reactions which are neutrino-nucleon scattering
and absorption of neutrinos onto baryons.  More detailed information
about the employed numerical procedures can be found in Ruffert et
al.~\cite{ruffert.ruf96}.  The following modifications compared to the
previously published simulations (\cite{ruffert.ruf96} and
\cite{ruffert.ruf97}) were made in addition.  (a) The models were
computed on multiply nested and refined grids to increase locally the
spatial resolution while at the same time computing a larger total
volume. Our grid handling is described in detail in Section~4 of
Ruffert~\cite{ruffert.ruf92}.  (b) An entropy equation instead of the
equation for the total specific energy (specific internal energy plus
specific kinetic energy) was used to calculate the temperature of the
gas in order to be able to determine low temperatures more accurately.
(c) The equation of state table was extended to $100\,{\rm
MeV}<T<0.01\,$MeV and
$2.9\times10^{15}\,$g/cm$^3>\rho>5\times10^7\,$g/cm$^3$ because in
extreme cases very high temperatures can be reached and the density of
ejected gas decreases to low values at large distances.

We started our simulations with two cool (temperatures initially some
$0.01\,{\rm MeV}$ to a few MeV) Newtonian neutron stars with baryonic
masses between 1.2~$M_\odot$ and 1.8~$M_\odot$ (depending on the
particular model), a radius of approximately 15~km, and an initial
center-to-center distance of 42--46~km.  The distributions of density
$\rho$ and electron fraction $Y_e$ were taken from one-dimensional
models of cold, deleptonized neutron stars in hydrostatic equilibrium.
We prescribed the orbital velocities of the coalescing neutron stars
according to the motions of point masses, as computed from the
quadrupole formula.  The tangential components of the velocities of
the neutron star centers correspond to Kepler velocities on circular
orbits, while the radial velocity components result from the emission
of gravitational waves that lead to the inspiral of the orbiting
bodies.  For an observer not corotating with the system, the neutron
stars were given either no spins or the two stars were assumed to be
tidally locked or to have spins opposite to the direction of the
orbital angular momentum~\cite{ruffert.ruf96}.

\begin{figure}[ht]
\centerline{ \epsfxsize=10.cm  \epsffile{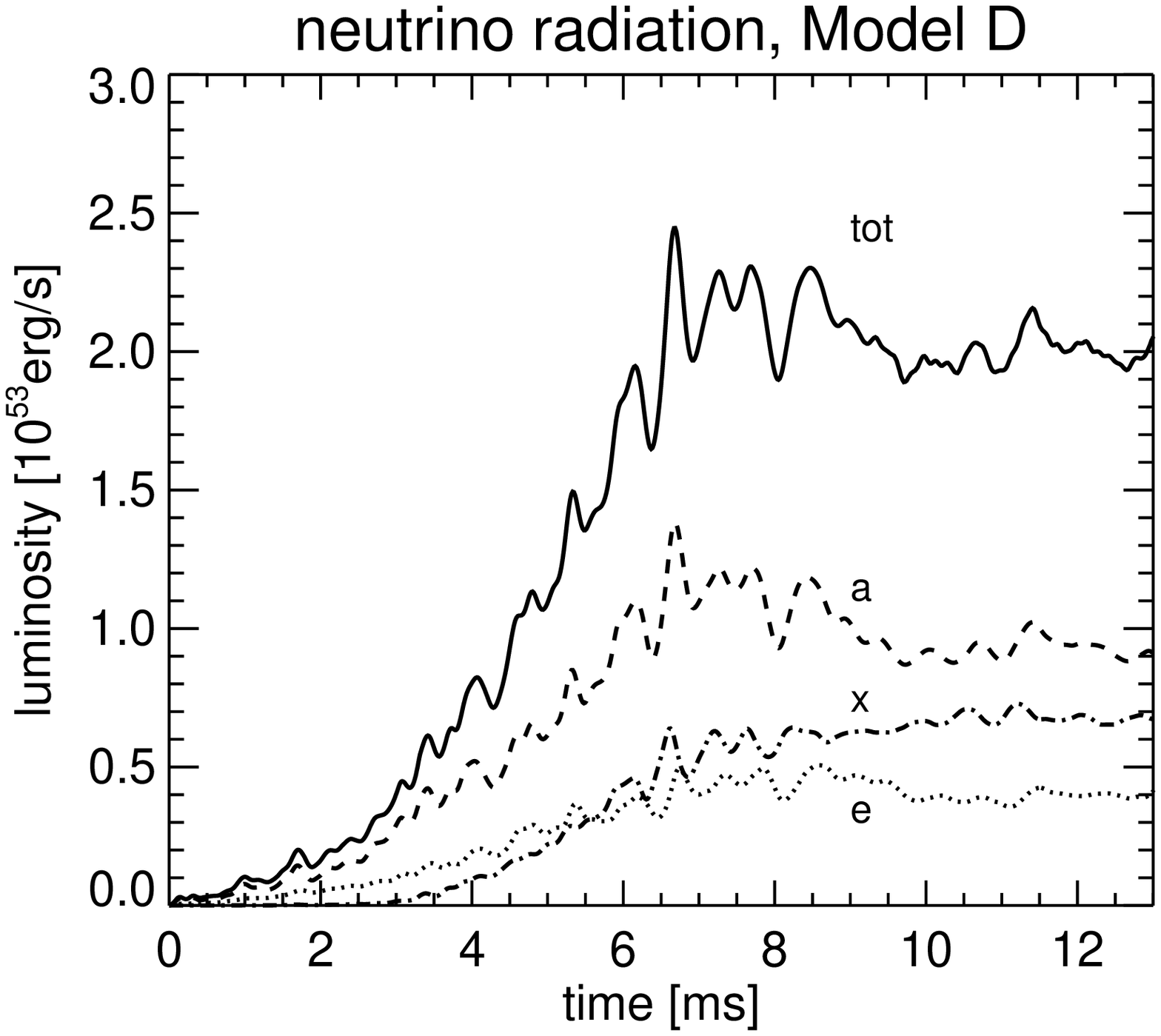} }
  \caption{ \label{F.mr.neutD} 
Luminosities of individual neutrino types ($\nu_e$, 
$\bar{\nu}_e$, and the sum of all $\nu_x$) and of the total of all
  neutrinos as functions of time for the merging of a $1.2M_\odot$ 
  and a $1.8M_\odot$ neutron star.}
\end{figure}

\begin{figure}[ht]
\centerline{
\begin{tabular}{cc}
  \epsfxsize=6.71cm  \epsffile{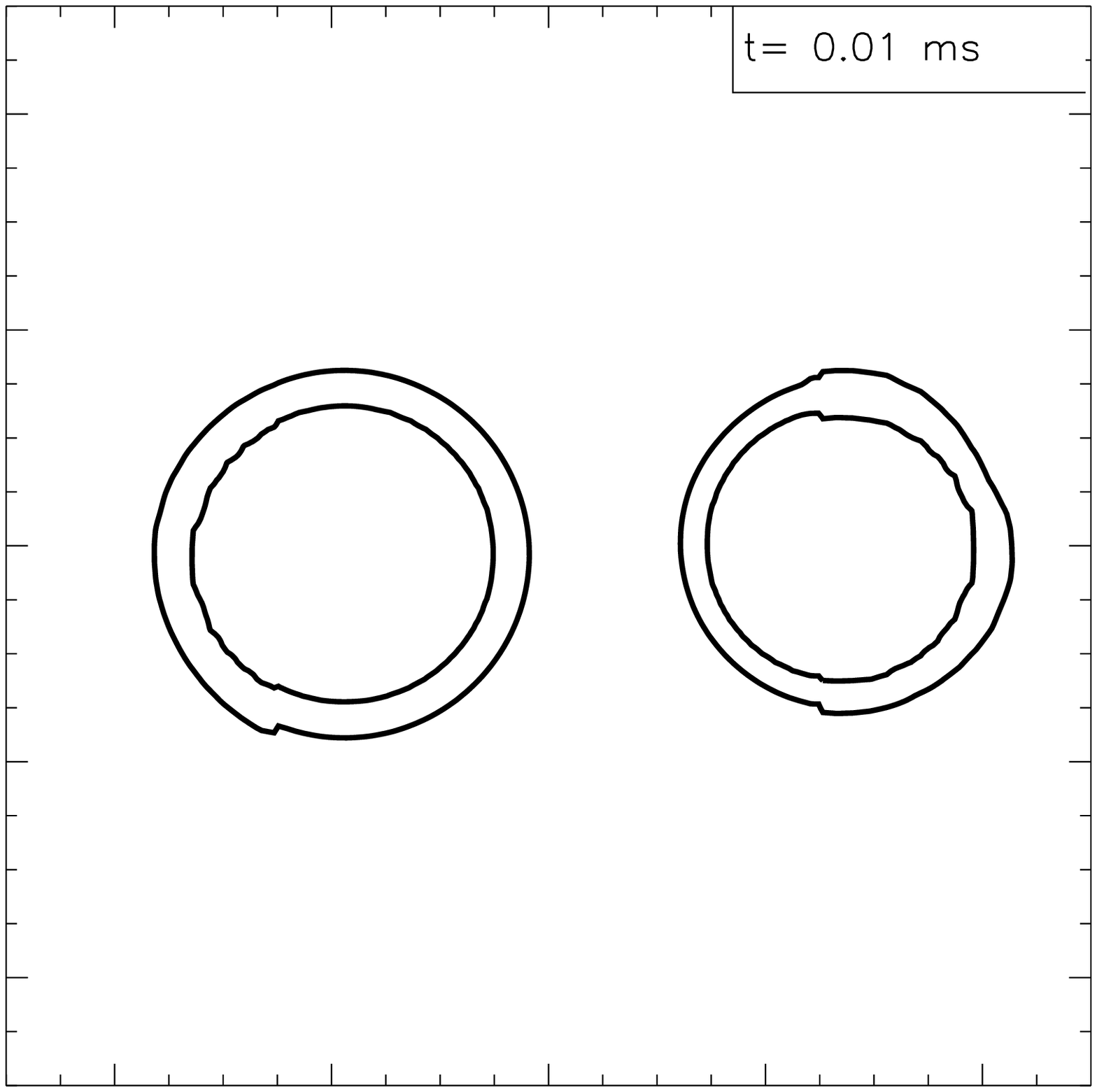} &
  \epsfxsize=6.71cm  \epsffile{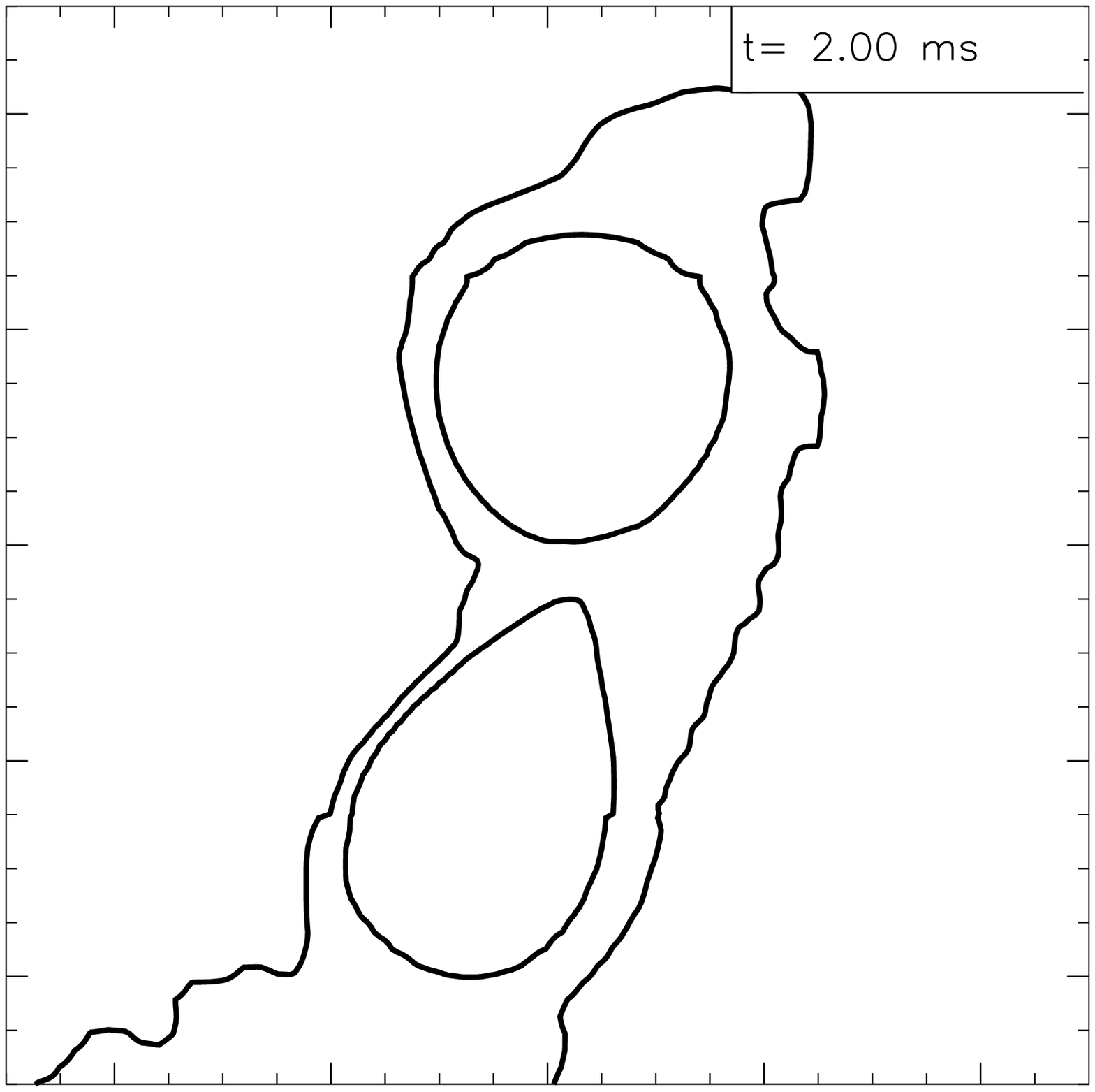} \\[-1.5ex]
  \epsfxsize=6.71cm  \epsffile{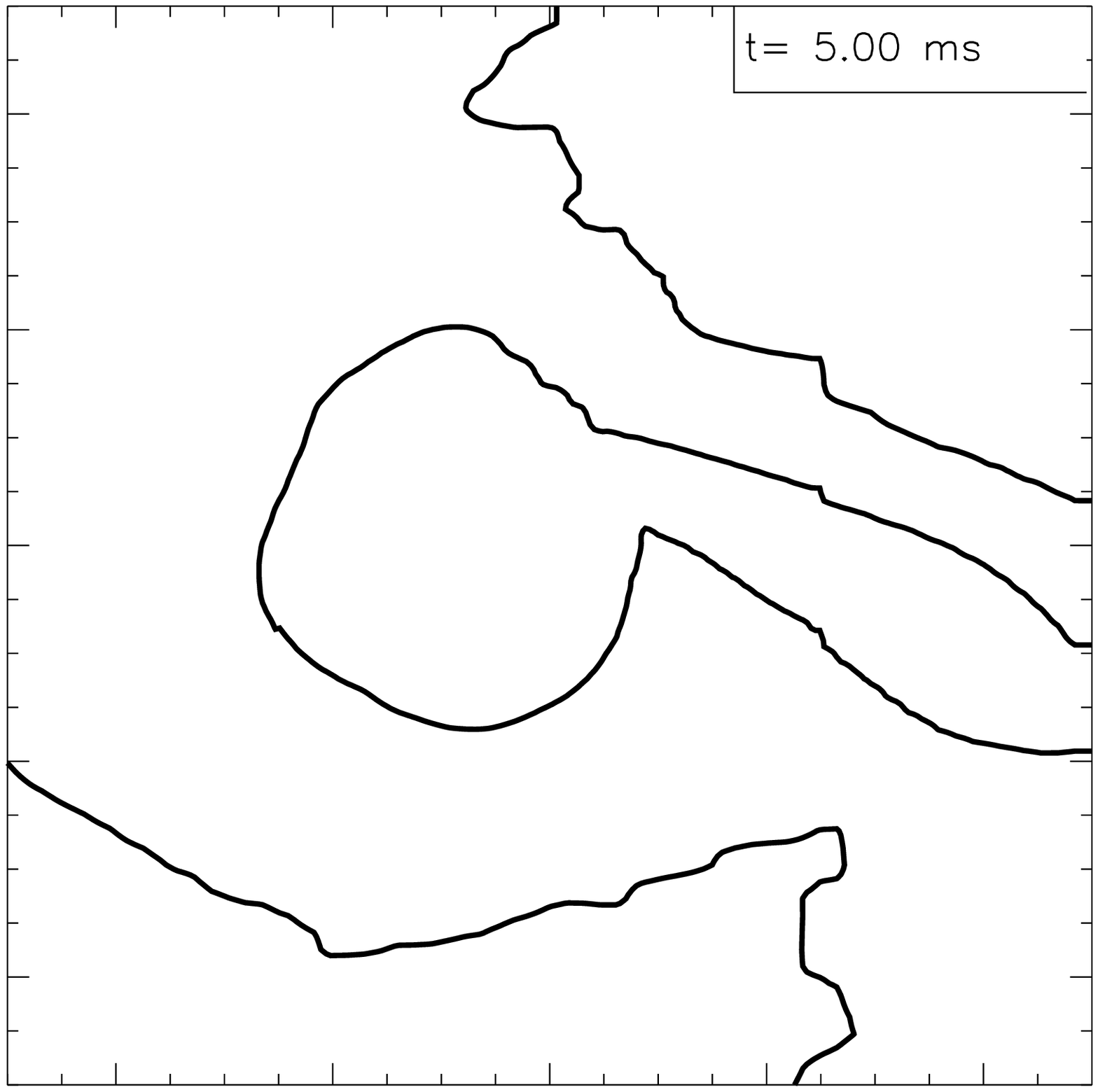} & 
  \epsfxsize=6.71cm  \epsffile{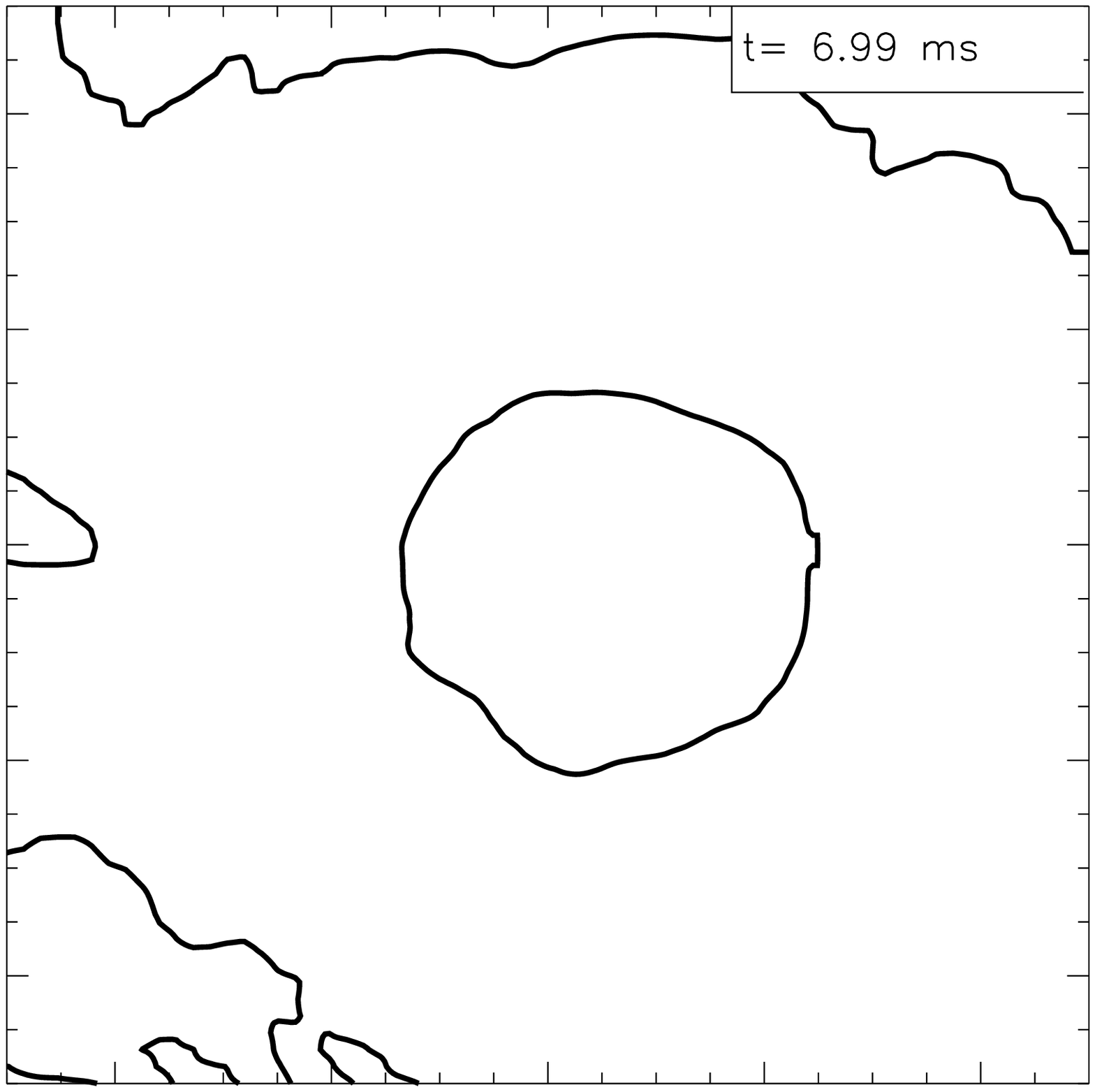} \\
\end{tabular}}
\caption{\label{F.mr.numres}  
Contour plots of the mass density distribution in the 
orbital plane
for the merging of a $1.2M_\odot$ and a $1.8M_\odot$ neutron star.
The outer density contour corresponds to $10^{10}$~g/cm$^3$,
the inner to $10^{14}$~g/cm$^3$.
In the top right corner of each panel one finds the time elapsed since
the beginning of the simulation.
}
\end{figure}

\subsection*{Some Results for Merging Neutron Stars}

From the long list of our models we present here some results only for
one exemplary case in more detail: the coalescence of a $1.8M_\odot$
primary with a $1.2M_\odot$ secondary (baryonic masses), both
initially corotating (solid-body like). The results of this model were
rather typical and not the most extreme in any direction.
 
The neutrino luminosities during the neutron star merger are shown in
Fig.~\ref{F.mr.neutD}.  During the phase of dynamical merging, a
constantly rising neutrino luminosity is produced.  It saturates after
about 7~ms, when the secondary has effectively been tidally shredded
and wrapped up along the surface of the primary.  This evolution of
the neutrino luminosity is very similar to what has been published for
equal-mass neutron stars (Ruffert et al.~\cite{ruffert.ruf97}),
although somewhat higher luminosities of 2--5$\times 10^{53}\,$erg/s
were found in the recent simulations, because the use of the larger
grid (out to about 200~km distance from the center of mass) allowed to
keep track of the mass flung to very large radii and then falling back
towards the center and contributing to the neutrino emission. Since
the geometry and neutrino luminosities are similar for equal-mass and
nonequal-mass neutron star mergers, also the energy deposition rate by
$\nu\bar\nu$-annihilation is of the same order. We calculated values
of several $10^{51}$~erg/s, about a factor of 10 larger than the ones
given in~\cite{ruffert.ruf97}, mainly because of the larger neutrino
luminosities of the more recent models. However, the dynamical phase
of the merging lasts only a few milliseconds before one must expect
the collapse to a black hole of the merged object with significantly
more than $2 M_\odot$.  Therefore the energy pumped into a
$e^{\pm}\gamma$ fireball during this phase hardly exceeds a few
$10^{49}\,$erg. Even worse, by far most of this energy is deposited in
the surface-near regions of the merged stars and will drive a mass
loss (``neutrino wind'') which will pollute the fireball with an
unacceptably large baryon load.

The neutrino emission from the coalescence of two neutron stars is
very different from the case of head-on collisions in which neutrinos
are emitted in two very short (about 1~millisecond) and extremely
luminous bursts reaching peak values of up to $4\times
10^{54}\,$erg/s~\cite{ruffert.ruf98}! This gigantic neutrino
luminosity leads to an energy deposition of about $10^{50}\,$erg by
$\nu\bar\nu$-annihilation within only a few milliseconds. However, the
surroundings of the collision site of the two neutron stars are filled
with more than $10^{-2}M_\odot$ of ejected matter and the maximum
values of the Lorentz-factor $\Gamma$ are only about $10^{-3}$, five
orders of magnitude lower than required for relativistic fireball
expansion that could produce a GRB.

Snapshots of the density contours in the orbital plane for the
considered $1.2M_\odot$-$1.8M_\odot$ merger can be seen in
Fig.~\ref{F.mr.numres}.  Initially (0ms $<$ $t$ $<$ 3ms) the secondary
is tidally elongated by the primary and a mass transfer is initiated.
The top left panel of Fig.~\ref{F.mr.numres} nicely shows how matter
is concentrated to flow through the L1-point onto the surface of the
primary. As more and more matter is taken away from the secondary it
becomes ever more elongated (bottom left panel).  Most of the matter
of the secondary finally ends up forming a rapidly rotating surface
layer of the primary, while a smaller part concentrates in an
additional, extended thick disk around the primary.  This matter has
enough angular momentum to stay in an accretion torus even after the
massive central body has most likely collapsed to a black hole.

\subsection*{Neutron Tori around Black Holes and GRBs}

If the central, massive body did not collapse into a black hole, it
would continue to radiate neutrinos with high luminosities, like a
massive, hot proto-neutron star in a supernova. But instead of
producing a relativistically expanding pair-plasma fireball, these
neutrinos will deposit their energy in the low-density matter of the
surface and thus will cause a mass flow known as neutrino-driven wind
(e.g.,~\cite{ruffert.bar92}). Most of this energy is consumed lifting
baryons in the strong gravitational potential and the expansion is
nonrelativistic.

This unfavorable situation is avoided if the central object collapses
into a black hole on a timescale of several milliseconds after the
merging of the neutron stars. We simulated the subsequent evolution by
replacing the matter inside a certain radius by a vacuum sphere (black
hole) of the same mass. The region along the system axis was found to
be evacuated on the free fall timescale of a few milliseconds as the
black hole sucks up the baryons. Thus an essentially baryon-free
funnel is produced where further baryon contamination is prevented by
centrifugal forces. This provides good conditions for the creation of
a clean $e^{\pm}\gamma$ fireball by $\nu\bar\nu$-annihilation.  Also
the thick disk closer to the equatorial plane loses matter into the
black hole until only gas with a specific angular momentum larger than
$j^\ast \cong 3R_{\rm s}v_{\rm k}(3R_{\rm s}) = \sqrt{6}GM/c$ ($R_{\rm
s} = 2GM/c^2$ is the Schwarzschild radius of the black hole with mass
$M$, $v_{\rm k}(3R_{\rm s})$ the Kepler velocity at the innermost
stable circular orbit at $3R_{\rm s}$) is left on orbits around the
black hole. We find torus masses of up to $M_{\rm t} \approx
0.2$--$0.3\,M_\odot$ at a time when a quasi-stationary state is
reached.  The temperatures in the tori are 3--10$\,$MeV, maximum
densities a few $10^{12}\,$g/cm$^3$.  Typical neutrino luminosities
during this phase are of the order of $L_{\nu}\approx 10^{53}\,$erg/s
(60\% $\bar\nu_e$, 35\% $\nu_e$).  With a maximum radiation efficiency
of $\varepsilon_{\rm r} \approx 0.057$ for relativistic disk accretion
onto a nonrotating black hole we calculate an accretion rate of $\dot
M_{\rm t}= L_{\nu}/(c^2\varepsilon_{\rm r})\sim 1\,M_\odot/{\rm s}$
and an accretion timescale of $t_{\rm acc} = M_{\rm t}/\dot M_{\rm t}
\sim 0.2$--0.3$\,$s. This is in very good agreement with the 
analytical estimates of Ruffert et al.~\cite{ruffert.ruf97}.

From our torus models we find that the energy deposition rate by
$\nu\bar\nu$-annihilation near the evacuated system axis is about
$\dot E_{\nu\bar\nu}\approx 5\times 10^{50}\,$erg/s. The corresponding
annihilation efficiency $\varepsilon_{\rm a}$ is of the order of $\dot
E_{\nu\bar\nu}/L_\nu \approx 5\times 10^{-3}$ which is rather large
because of the large $\nu_e$ and $\bar\nu_e$ luminosities and because
the torus geometry allows for a high reaction probability of neutrinos
with antineutrinos.  In order to account for an observed GRB
luminosity of $L_\gamma \sim 10^{51}\,$erg/s the required focussing of
the $\nu\bar\nu$-annihilation energy is moderate, $\delta\Omega/(4\pi)
= \dot E_{\nu\bar\nu}/(2L_\gamma)\sim 1/4$, corresponding to an
opening angle of about 60~degrees. This value does not seem
implausible for the geometry of a thick accretion torus.  Taking into
account general relativistic effects into the treatment of
$\nu\bar\nu$-annihilation reduces the above estimates only by about
10--50\%.

In summary, our numerical simulations show that stellar mass black
holes with accretion tori that form after the merging of neutron star
binaries have masses, lifetimes, and neutrino luminosities that might
provide enough energy by $\nu\bar\nu$-annihilation to account for
short and not too powerful GRBs. The longer bursts and very energetic
events, however, would require an alternative explanation, e.g.,
failed supernovae (or ``collapsars'')~\cite{ruffert.woo93} where a
stellar mass black hole could have a more than 10 times more massive
accretion torus than in the binary neutron star scenario.

\subsection*{Acknowledgments}
The calculations were performed at the Rechenzentrum Garching on an IBM~SP2.

}\newpage{


\head{Physical Processes Near Black Holes}
     {R.A.\ Sunyaev}
     {Max-Planck-Institut f\"ur Astrophysik\\ 
      Karl-Schwarzschild-Str.~1, 85740 Garching, Germany}

\noindent
The talk was devoted to the new information about black holes in the
Galaxy and in the nuclei of the galaxies:
\begin{itemize}
\item optical observations and the mass function of the invisible
   components of transient X-Ray binaries,
\item differences in X-Ray spectra of accreting neutron stars and black
   hole candidates \hfill\break
        (GRANAT and MIR-KVANT data),
\item superluminal radiosources in two galactic X-Ray transients,
\item quasiperiodic oscillations of X-Rays from GRS 1915+105 (RXTE data),
\item scaling in the radiojet lengths in galactic and extragalactic
   superluminal radiosources,
\item X-Ray fluorescent lines in SGR B2 molecular cloud (ASCA data) and
   information about the X-Ray luminosity of the Center of our Galaxy
   in the recent past,
\item scattering of X-Ray lines on neutral and molecular gas (the spectrum
   of recoiled photons and the Lyman gap).
\end{itemize}


}


\newpage {


\thispagestyle{empty}

\begin{flushright}
\Huge\bf
{\ }


High-Energy\\
\bigskip
Neutrinos\\

\end{flushright}

\newpage

\thispagestyle{empty}

{\ }

\newpage

 }\newpage {


\head{Astrophysical Sources of High-Energy Neutrinos}
     {K.~Mannheim}
     {Universit\"ats-Sternwarte, Geismarlandstra{\ss}e 11,\\
     D-37083 G\"ottingen, Germany (kmannhe@uni-sw.gwdg.de)}

\noindent
This contribution reviews currently discussed astrophysical sources of
high-energy ($>$TeV) neutrinos. Particle astrophysics with high-energy
neutrinos allows to study electroweak interactions, neutrino
properties, particle acceleration theories, cosmology, dark matter
candidates, and the enigmatic sources of cosmic
rays~\cite{mannheim.1}.  The detection of high-energy neutrinos is
greatly facilitated by the rising neutrino cross section and muon
range which cause the detection probablity for neutrinos to increase
with energy.  From an astrophysical point of view, the existence of
sources of high-energy neutrinos seems very likely, if not guaranteed,
for two strong reasons.

First of all, there are numerous cosmic synchrotron sources (radio,
optical, X-ray) for which simple arguments give electron energies in
the GeV--TeV range~\cite{mannheim.2}.  Since most electromagnetic
particle acceleration mechanisms predict accelerated baryons along
with the accelerated electrons, hadronic energy losses due to pion
production will lead to high-energy neutrino emission.  If the maximum
energies of particles emerging from cosmic accelerators are determined
by the balance between energy gains and energy losses, baryons must
reach much higher energies than electrons since their energy losses
are much weaker than those of the electrons {\em at the same energy}.
Therefore, pion energies may be expected to reach energies even above
TeV.  Neutrinos from pion decay will, of course, also be very
energetic, since each flavor carries $\sim 1/4$th of the decaying pion
energy.  Baryon acceleration could only be avoided in a plasma
exclusively composed of electrons and positrons.  Although pair
plasmas probably exist in pulsars, gamma-ray bursts, and active
galactic nuclei, baryon pollution (from the neutron star surface or
due to entrainment of environmental plasma) changes the picture quite
radically: As the pair plasma expands most of the internal energy is
quickly converted into kinetic energy of the polluting baryons which
then has to be tapped in a second stage dissipative process to give
rise to the observed radiation.  It has been widely discussed in the
context of pair fireball models for gamma-ray bursts~\cite{mannheim.3}
that the observed non-thermal gamma-ray spectra generally require
emission from a baryonic bulk flow at a large distance to the
energizing compact object rather than from a pair plasma close to the
compact object~itself.

\begin{figure}[b]
\vskip-0.4truecm
  \centerline{\epsfxsize=0.92\textwidth\epsffile{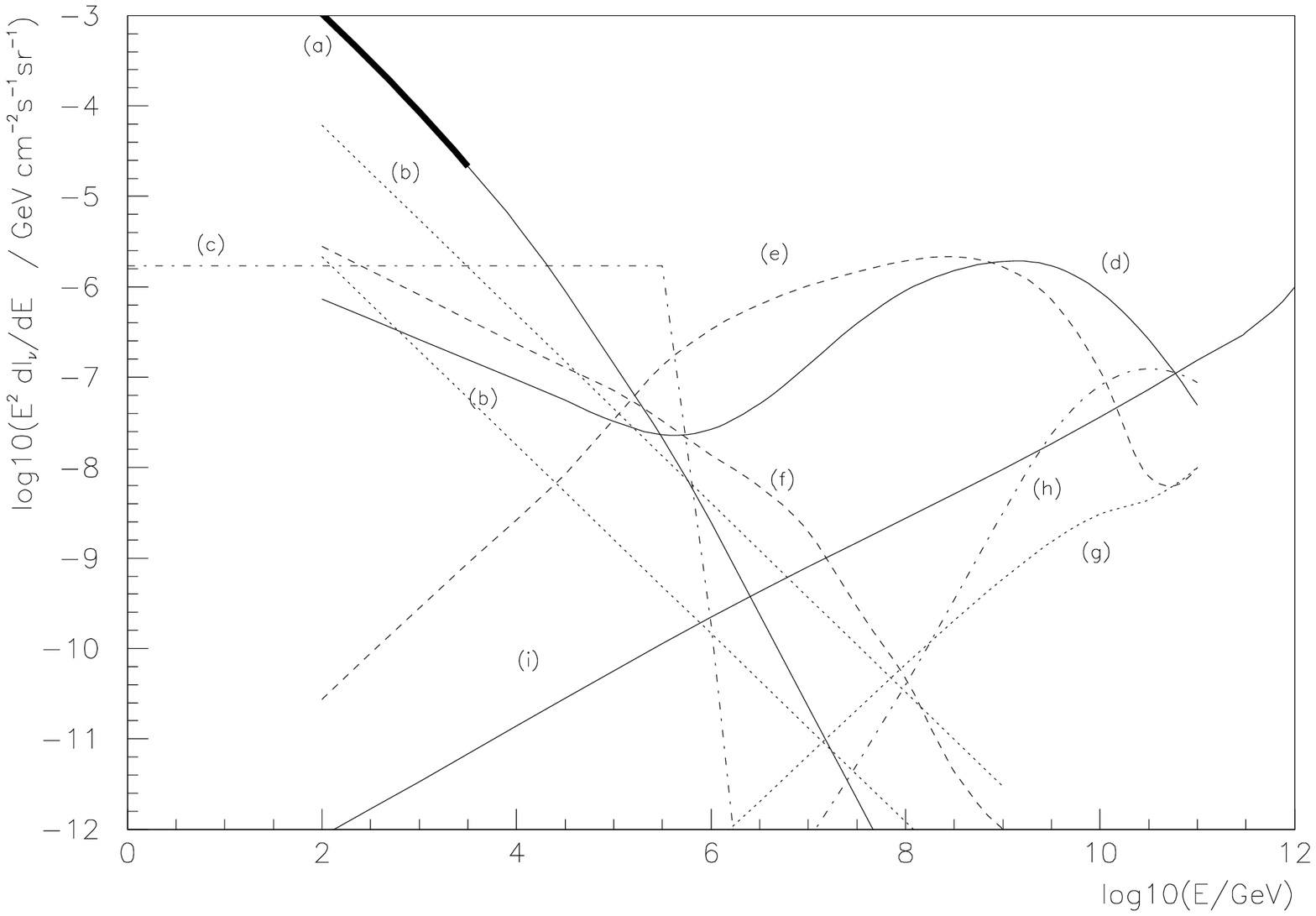}}
\vskip-9.2truecm
\caption[]{Overview of theoretical neutrino flux predictions as
compiled by R.J.~Protheroe during a workshop on high-energy neutrino
astrophysics in Aspen, 1996.  Labels refer to (a) atmospheric
neutrinos (the solid part indicates the part of the spectrum measured
by the Frejus experiment), (b) galactic disk in center and anti-center
directions (due to Domokos), (c) AGN~\cite{mannheim.11}, (d)
AGN~\cite{mannheim.8}, (e) AGN~\cite{mannheim.9}, (f) charm production
upper limit~\cite{mannheim.12}, (g) GRBs (due to Lee), (h) UHE CRs
(due to Stecker), (i) TDs~\cite{mannheim.10} (courtesy of W. Rhode).
\label{mannheim.fig}}
\end{figure}

Secondly, we know baryonic cosmic rays exist (up to energies of
$10^8$~TeV) and they must come from somewhere.  Supernova remnants are
still among the prime candidates for the acceleration of the cosmic
rays below $\sim 10^{3}$~TeV, they should therefore be visible as
gamma-ray and neutrino sources at comparable flux levels.  This flux
level is beyond reach for the first generation of high-energy neutrino
experiments. Giant molecular clouds could be somewhat more important,
since they provide most of the target mass in the vicinity of star
forming regions where supernovae occur.  Cosmic rays propagating
through the Milky Way produce pions in hadronic interactions with the
interstellar material and can be traced in gamma-rays.  A diffuse flux
of neutrinos at a flux comparable to the gamma-ray flux should be seen
from the Galactic plane~\cite{mannheim.4} corresponding to $\sim
150$~upward events above one TeV per year in a detector with an
effective volume of 1~km$^3$ (compared with $6.5\times 10^3$
atmospheric background events).  One can also speculate about
enshrouded sources which only show up in neutrinos but not in
gamma-rays.  However, the acceleration of particles to high energies
generally requires a low-density plasma.  The acceleration of cosmic
rays above $\sim 10^{18.5}$~TeV in sources belonging to the Milky Way
seems impossible.  Among the prime candidates for these ultra-high
energy cosmic rays are (radio and gamma-ray emitting) active galactic
nuclei~\cite{mannheim.5} and gamma ray bursts~\cite{mannheim.6}.  In
those sources cooling of accelerated protons (ions dissociate over
cosmological distances) is dominated by interactions with low-energy
synchrotron photons (from the accelerated electrons).  Due to
electromagnetic cascading most gamma-rays are shifted to below the TeV
range whereas the neutrinos remain unaffected from such reprocessing.
Hence the $\gamma/\nu$-ratio decreases with energy in spite of being a
constant function (from decay kinematics) at the production site (if
proton-matter collisions dominate the energy loss, the
$\gamma/\nu$-ratio in the TeV range may be constant and of order
unity~\cite{mannheim.7}.  The total electromagnetic power of the
gamma-ray emitting active galactic nuclei can explain the observed
extragalactic gamma-ray background and is of the same order as the
power in an extragalactic $E^{-2}$ differential cosmic ray spectrum
dominating above $10^{6.5}$~TeV.  Therefore, neutrino flux predictions
are generally bounded by $\sim
10^{-6}$~GeV~cm$^{-2}$~s$^{-1}$~st$^{-1}$ (Fig.~\ref{mannheim.fig}).
The models~\cite{mannheim.8,mannheim.9} yield $\sim 300$~upward events
above TeV per year per km$^3$.

A third, more speculative class of high-energy neutrino sources are
topological defects producing gauge bosons at the grand-unified-theory
scale of $\sim 10^{13}$~TeV~\cite{mannheim.10}.  The unstable gauge
bosons generate fragmentation jets with ultra-high energy protons,
gamma-rays, and neutrinos which may be responsible for the extended
air showers with primary energies above the so-called Greisen cutoff
at $10^{8}$~TeV above which protons from sources further way than
$\sim 30$~Mpc lose their energy in interactions with the microwave
background.

\subsection*{Acknowledgements}
Many thanks to all the organizers of this very stimulating workshop,
in particular to Wolfgang Hillebrandt and Georg Raffelt for inviting
me.

}\newpage{

\head{Atmospheric Muons and Neutrinos Above 1 TeV} 
  {P.~Gondolo} 
  {Max-Planck-Institut f\"ur Physik (Werner-Heisenberg-Institut)\\
  F\"ohringer Ring 6, 80805 M\"unchen, Germany}

\noindent 
The ultimate background to high energy neutrino astronomy are the
neutrinos produced in the interaction of cosmic rays with the Earth
atmosphere.
A cosmic ray nucleus penetrates on average $\approx 60$ g/cm$^2$ of
atmosphere before colliding with an air nucleus. The secondary mesons
produced in the collision can propagate an additional $\approx 100$
g/cm$^2$, but if their lifetime is short enough they may decay before
interacting with air. Their semileptonic decays then produce an
atmospheric neutrino flux.

The competition between interaction with air and decay is at the basis
of the energy and angular dependence of the atmospheric neutrino
fluxes.  Let us compare the mean free path of mesons in air to the
distance they travel in a mean lifetime.  In a cascade developing
vertically downwards, the primary interaction occurs at a height of
$\approx 20$ km, and the mean free path is $\approx 10$ km; in a
cascade developing horizontally, the primary interaction is $\approx
500$~km from the observer, and the mean free path is $\approx 100$
km. The distance travelled in a mean lifetime increases linearly with
the meson momentum: $d = p \tau / m $, where $p$, $\tau$, and $m$ are
the meson momentum, lifetime and mass. For $ p = 1 $ TeV, pions and
kaons can travel kilometers ($d_{\pi^+} = 55.9$ km, $d_{K^0_L} = 31.2$
km, $d_{K^+} = 7.52 $ km), but charmed mesons only centimeters
($d_{D^0} = 6.65 $ cm, $d_{D^+} = 1.70$ cm, $d_{D^+_s} = 7.1$ cm,
$d_{\Lambda^+_c} = 2.7 $ cm).

As a consequence: (1) vertical pions and kaons do not have time to
decay before interacting with air, while horizontal pions and kaons
decay readily: the subsequent ``conventiona'' neutrino flux is
stronger in the horizontal than in the vertical direction; (2) charmed
mesons decay promptly whether vertical or horizontal: the subsequent
``prompt'' neutrino flux is the same in the horizontal and in the
vertical direction; (3) as the energy increases, the mesons live
longer, and fewer and fewer pions and kaons have time to decay within
the available mean free path, while all charmed mesons have time to
decay:\footnote{At high enough energy ($\approx$ EeV), also charm
decays are fewer and fewer for an analogous reason.} the energy
spectrum of conventional neutrinos is steeper than that of prompt
neutrinos by one power of energy.

The different energy dependence means that beyond a certain energy,
the atmospheric neutrino flux is dominated by decays of charmed
mesons. The energy at which this happens depends on the relative
strength of charmed and non-charmed meson production in the primary
nucleus-nucleus collision.  Unfortunately, laboratory data do not
extend up to the relevant energies. One could either look for help
from measurements of the atmospheric muons produced in association
with the neutrinos or extrapolate laboratory data with the help of a
theoretical model for charm production.

In a search for muons, prompt and conventional components could be
separated by their different angular dependence. Several attempts have
been made with underground experiments and with air shower detectors.
In underground experiments, high energy muons, prompt muons in
particular, come only from directions close to the horizon, where
unfortunately there is an intense muon background generated by
low-energy atmospheric neutrinos in the rock surrounding the detector.
Subtraction of this neutrino-induced background is a source of big
systematic errors \cite{gondolo.systematics} which spoil the search
for prompt muons.  Air shower detectors search for horizontal showers
generated by a bremsstrahlung photon emitted by a high energy
muon. The AKENO array has seen none above 100 TeV, and has therefore
put only an upper limit to the prompt muon flux at high energy.

\begin{figure}[b]
  \centerline{\epsfxsize=0.92\textwidth\epsffile{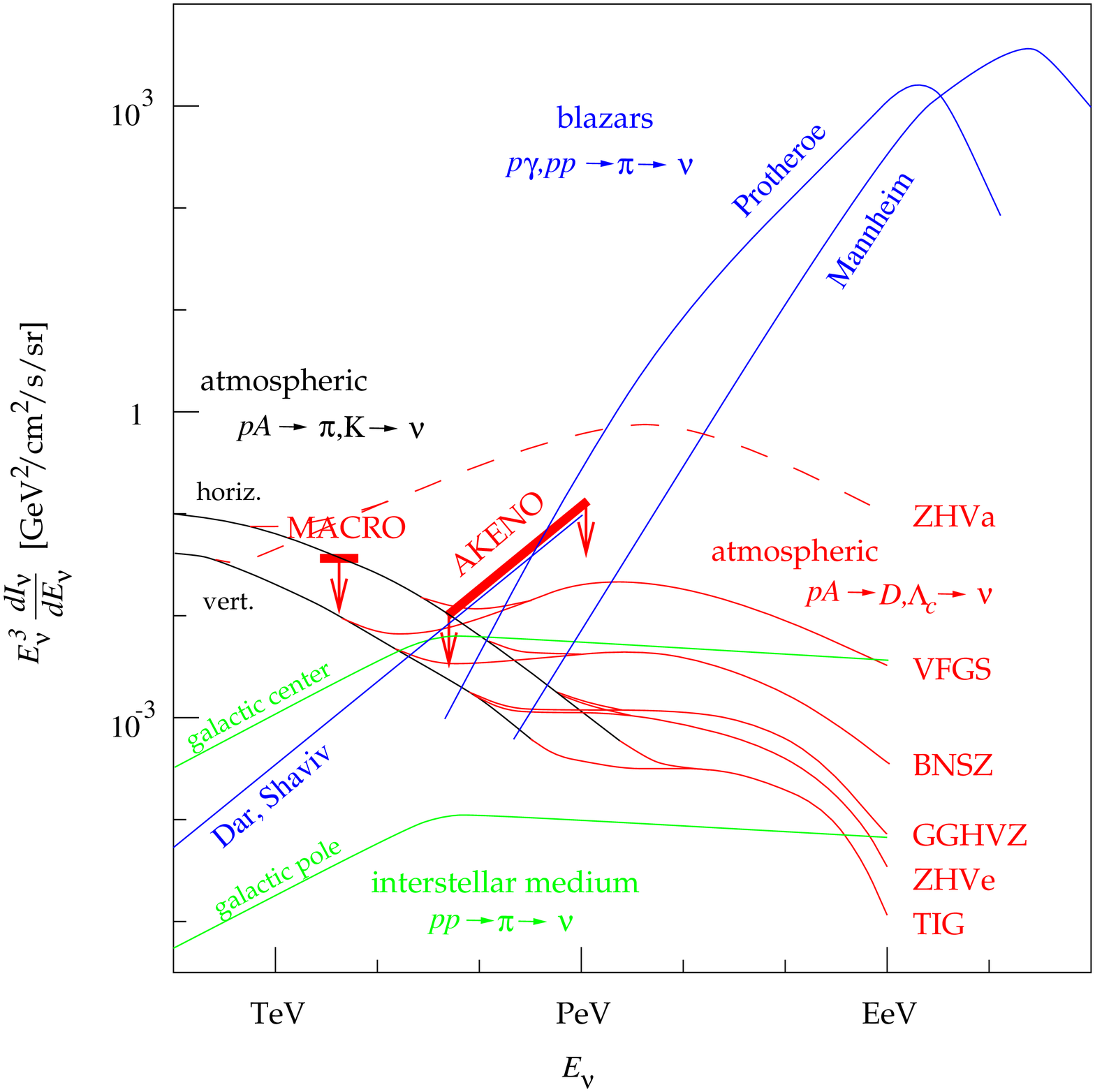}}
  \caption{Selected predictions for high energy neutrino fluxes.}
  \label{gondolo.fig1}
\end{figure}

Because of these experimental difficulties, we must resort to
theory. And here lie many uncertainties, because charm production
occurs half-way between the perturbative and non-perturbative regimes.
The theoretically preferred model, perturbative QCD, was thought to be
inadequate because to first order in $\alpha_s$ it could not account
for several aspects of the early data on open charm production, data
which by the way were in conflict with each other \cite{gondolo.ISR}.
So, even if a couple of QCD calculations had
appeared~\cite{gondolo.IK,gondolo.ZHV}, studies of atmospheric fluxes
have traditionally favored, besides semi-empirical parametrizations of
the cross section (see {\em e.g.}
\cite{gondolo.ZHV,gondolo.V,gondolo.Cast}), non-perturbative sources
of charm beyond basic QCD: for example, the dual parton model
\cite{gondolo.GGHVZ,gondolo.Batt}, based on Regge asymptotics, and
models incorporating the assumption of an intrinsic charm component in
the nucleon \cite{gondolo.VFGS,gondolo.BNSZ}.

Today, however, charm production experiments give a consistent set of
data, and are compatible with QCD calculations to next-to-leading
order \cite{gondolo.experiments}. This has removed the impetus to look
for unconventional mechanisms of charm production, and has motivated a
detailed study of atmospheric neutrinos within perturbative QCD
\cite{gondolo.TIG}.

In Fig.~\ref{gondolo.fig1} I have collected some of the expected
astrophysical neutrino signals and several predictions for the high
energy atmospheric background. The astrophysical signals shown are the
diffuse fluxes from blazars estimated by Dar and Shaviv
\cite{gondolo.DS}, Protheroe \cite{gondolo.P}, and Mannheim
\cite{gondolo.Mannheim}, and the neutrino emission from the galactic
interstellar medium as calculated by Domokos {\em et al.}
\cite{gondolo.DD}. The conventional atmospheric background is from
Lipari \cite{gondolo.Lipari}, and the prompt neutrino fluxes are from
Zas {\em et al.} (ZHVa for model A, ZHVe for model E)
\cite{gondolo.ZHV}, Volkova {\em et al.} (VFGS) \cite{gondolo.VFGS},
Bugaev {\em et al.} (BNSZ) \cite{gondolo.BNSZ}, Gonzalez-Garcia {\em
et al.} (GGHVZ) \cite{gondolo.GGHVZ}, and Thunman {\em et al.} (TIG)
\cite{gondolo.TIG}. Also indicated are the upper limits on the prompt
neutrino flux from MACRO \cite{gondolo.MACRO} and AKENO
\cite{gondolo.AKENO}. Although these limits exclude the highest
prediction ZHVa, the remaining uncertainty in the prompt neutrino
background is still a factor of 100.

Since the high-energy atmospheric neutrino background may be a
nuisance to the detection of astrophysical diffuse neutrino fluxes, it
may be of value to decrease the uncertainty in the predictions. The
best way would be a detection of the associated prompt muons, maybe
through larger air shower arrays. A second option would be to increase
the precision of the theoretical inputs, for example through better
measurement of the charm production cross section and the charm
semileptonic decay rates, which is under way/under project in high
statistics charm production experiments. At last, it may happen that
we have to wait for the neutrino telescopes themselves to know the
high energy atmospheric neutrino background.

}\newpage{


\def\nue{$\nu_e$}
\def\numu{$\nu_\mu$}

\head{Atmospheric Neutrinos in Super-Kamiokande}
     {Danuta Kie{\l}czewska$^{1,2}$
     (for the Super-Kamiokande Collaboration)}
     {$^1$The University of California, Irvine, California 92717, 
      USA\\
      $^2$Warsaw University, Warsaw, Poland}      

\subsection*{Abstract}

The measurements of atmospheric neutrino interactions during 326 days
of the Super-Kamio\-kande detector operation are reported.  The
measured ratio of muon and electron neutrinos has been found smaller
than expected from theoretical models. The angular distribution of
muon neutrinos is also inconsistent with expectations.

\subsection*{Introduction}

Atmospheric neutrinos are produced in a layer of about 15 km as a
result of hadronic cascades originating from interactions of cosmic
rays.  A number of models have been developed to calculate the fluxes
of atmospheric neutrinos \cite{kielczewska.honda,kielczewska.gaisser}
at energies below 5~GeV. The absolute neutrino fluxes are predicted
with uncertainties of 20\%.  However the ratio of
$\nu_{\mu}+\bar\nu_{\mu}$ to $\nu_{e}+\bar\nu_{e}$ fluxes is known to
better than 5\% and therefore experiments measure the double ratio $R
\equiv (\mu /e)_{\rm DATA}/(\mu /e)_{\rm MC}$, where $\mu/e$ denote
the ratio of $\mu$-like to $e$-like neutrino interactions
correspondingly in the data and simulated event samples.

The flavor composition of atmospheric neutrinos was studied in the
earlier underground experiments. The Kamiokande
\cite{kielczewska.hirata} and IMB \cite{kielczewska.imb} experiments,
using water Cherenkov detectors, and the Soudan
\cite{kielczewska.soudan} experiment, using iron calorimeters, have
reported the double ratio to be smaller than one. A dependence of $R$
on the zenith angle, and hence on the neutrino path length, was also
observed in Kamiokande~\cite{kielczewska.fukuda}. Those results have
been often interpreted by neutrino~oscillations.

Here the preliminary results of measurements of atmospheric neutrino
fluxes are presented for 20.1 kton-year exposure of the
Super-Kamiokande detector.

\subsection*{Super-Kamiokande Detector}

Super-Kamiokande is a 50 kton water detector located in the mine near
Kamioka town in Japan. It is situated at a mean overburden of 1000
meters below the peak of Mt.~Ikeno. Cherenkov photons emitted by
relativistic particles inside the inner cylindrical volume of water,
16.9 m in radius and 36.2 m high, are recorded by 11146
photomultiplier tubes (PMTs) of 50~cm diameter (40\% photocathode
coverage).  Each inner PMT signal provides the first photon
arrival-time in 1.2 $\mu s$ range at 0.3 ns resolution and the
collected charge at 0.2 pC resolution (equivalent to 0.1 p.e.).  The
outer layer of water, about 2.7 m thick, is instrumented with 1885
outward facing PMTs of 20 cm diameter with wavelength shifting
plates. The two detector regions are optically separated. The outer
detector tags events with incoming or exiting particles.
  
Water transparency is measured using a dye laser and is about 100 m at
wavelenth of 420 nm.  The neutrino interactions were collected inside
the fiducial volume 2 m from the inner detector walls, comprising 22.5
ktons of water.  The accuracy of the absolute energy calibration is
estimated to be $\pm2.4\%$ based on the detailed analysis of
cosmic-ray muons, muon-decay electrons and electrons from a linac. The
estimated momentum resolutions for electrons and muons are
$2.5\%/\sqrt{E({\rm GeV})}+0.5\%$ and $+3$\%, respectively.

\subsection*{Results}

The data reduction, single-ring selection and lepton identification
have been done by 2 groups independently and the results were found to
agree very well.  Three simulated samples have been compared with the
data. The simulations used 2 different flux calculations, 3 different
neutrino interaction models and 2 codes for particle and light
propagation in the detector. All the simulated samples were processed
by the same procedures as the data. The differences were used in
estimates of systematic errors.  The results discussed below come from
one data analysis and Monte Carlo simulations based on flux
calculations of Ref.~\cite{kielczewska.honda}.

Out of about 300 million triggers recorded during 320 days of detector
operation between May 1996 and June 1997, about 5000 events were
classified as fully-contained events, i.e. events with no significant
energy deposit outside of the inner detector.  Using the photon
time-arrival information recorded by PMTs, 2708 events with energy
$E_{\rm vis}>30$ MeV had vertices reconstructed in the fiducial
volume. The visible energy $E_{\rm vis}$ is defined as the energy of
an electron which would produce the same number of Cherenkov photons.

From Monte Carlo simulation it is estimated that 83.2\% of all
charged-current neutrino interactions in the fiducial volume are
retained in the current sample.  The remaining fraction of 16.8\%
consists mostly of events with exiting particles (9.3\%) or energy
lower than~30~MeV~(5.8\%).

To study the neutrino flavor composition we select a sample of
single-ring events.  Quasi-elastic interactions, for which the
information about neutrino flavor is provided by the charged lepton
identification, consist $75\pm3$\% of the sample.  Particle
identification is based on the pattern of \v{C}erenkov rings.
Diffuse, showering patterns associated with electromagnetic cascades
($e$-like events), are separated from sharper, non-showering rings
caused by stopping muons ($\mu$-like events).  The misidentification
probability for single-ring events is estimated to~be~$0.8\pm0.1$\%.

The results for events with $E_{\rm vis}<1.33$ GeV (Sub-GeV sample)
are presented in Table~\ref{kielczewska.table1}.  For single-track
events the electron and muon momenta are greater than 100 and 200
MeV/c, respectively. From these data one obtains: $R=0.63\pm0.03({\rm
stat})\pm0.05({\rm sys})$.  The systematic error of 8\% comes mostly
form uncertainty in the calculation of the $\nu_{\mu}/\nu_{e}$ ratio
(5\%), uncertainties in interaction cross sections and nuclear effects
in $^{16}$O (4.1\%), single-ring selection (4\%) and particle
identification (2\%).

\begin{table}[ht]
\vbox{
 \renewcommand{\arraystretch}{1.5}
 \newcommand{\lw}[1]{\smash{\lower2.ex\hbox{#1}}}
 \begin{center}
  \begin{tabular}{lrcrrrr} \hline\hline
     & \lw{Data}\hfil &&\multicolumn{4}{c}{Monte Carlo} \\
     \cline{4-7}
     & & &total &\nue\  CC & \numu\  CC & NC\\
     \hline
     single ring      &  1453 && 1551.5 & 539.7 & 927.8.0 & 84.0 \\
     {} $e$-like      &   718 &&  593.8 & 537.2 &   8.9 &  47.7 \\
     {} $\mu$-like    &   735 &&  957.7 &   2.5 & 918.9 &  36.3 \\
     multi ring       &   533 &&  596.6 & 148.4 & 241.3 & 206.9 \\
     \hline
     total            &  1986 && 2148.1 & 605.6 & 1076.5& 290.9 \\
     \hline\hline
  \end{tabular}
 \end{center}
\caption{Summary of the Sub-GeV sample compared with Monte Carlo
estimation, for 20.1 kton-years of Super-Kamiokande.
\label{kielczewska.table1}}
\bigskip
\renewcommand{\arraystretch}{1.5}
 \begin{center}
  \begin{tabular}{lrr} \hline\hline
     & Data\hfil &Monte Carlo\hfil \\
     \hline
     single ring      &  288 & 304.8 \\
     {} $e$-like      &  149 & 120.5 \\
     {} $\mu$-like    &  139 & 184.3 \\
     2 rings          &  207 & 221.0 \\
     3 rings          &  110 & 133.0 \\
     \hline
     total            &  605 & 658.8 \\
     \hline\hline
  \end{tabular}
 \end{center}
\caption{Summary of the Multi-GeV sample compared with Monte Carlo
estimation, for 20.1 kton-years of Super-Kamiokande. 
\label{kielczewska.table2}}
}
\end{table}

An independent signature of muon neutrino interactions is provided by
electrons from muon decays. The detection efficiency for muon decay is
estimated to be 0.80 for $\mu^{+}$ and 0.63 for $\mu^{+}$. The
fraction of $\mu$-like events with a decay electron is $67.5\pm1.7$\%
for the data and $68.2\pm1.0$\% for MC.  For $e$-like events the
corresponding fractions are $9.3\pm1.1$\% and $8.0\pm0.3$\%.

Angular distributions of the reconstructed track directions for
single-ring events have also been studied.
Figure~\ref{kielczewska.fig1} displays the distributions of zenith
angles for $\mu$-like and $e$-like events.  The double ratio $R$
dependence on the zenith angle is shown in
Figure~\ref{kielczewska.fig2}.

\begin{figure}[p]
\vbox{
\centerline{\epsfxsize=0.6\textwidth\epsffile{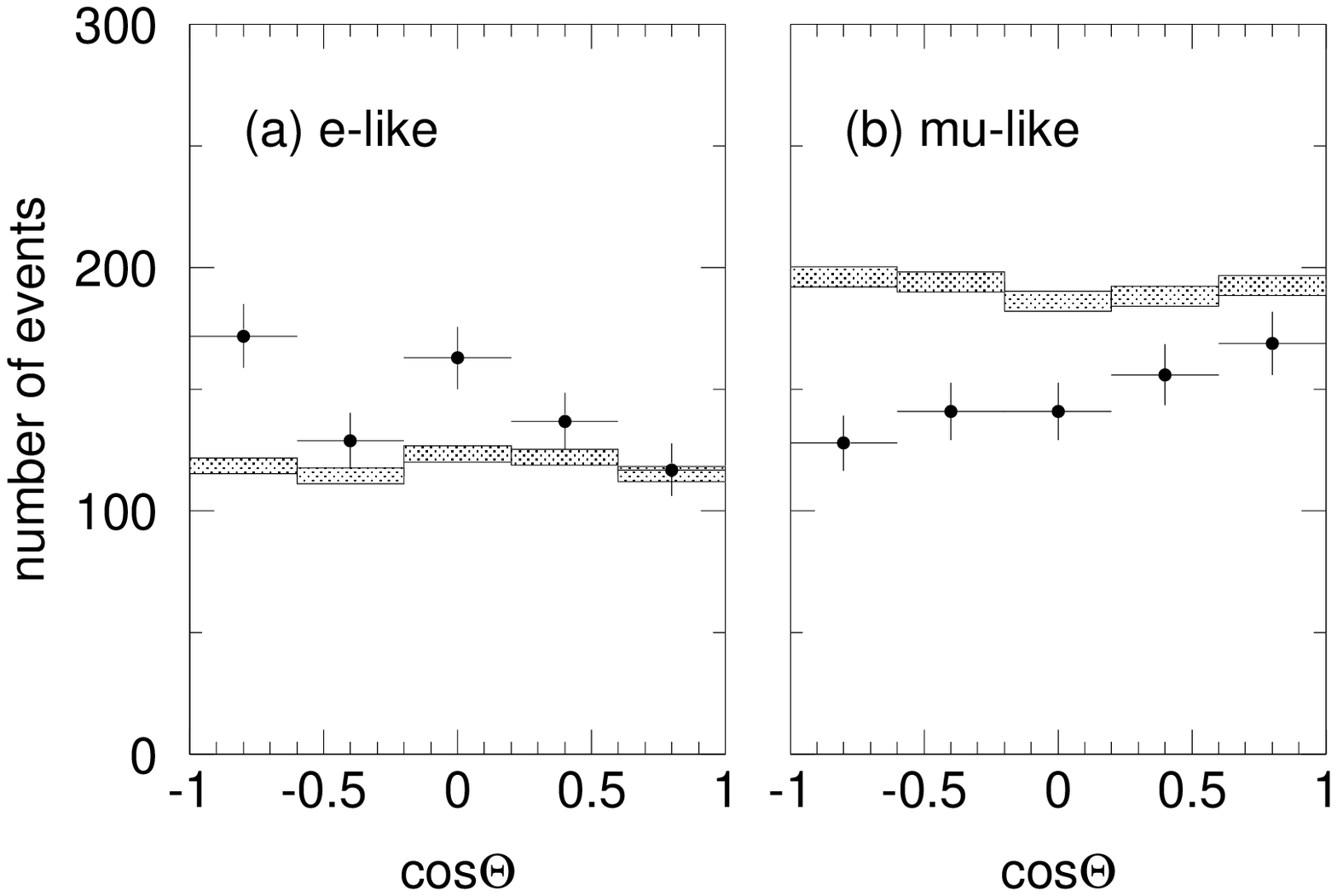}}
\caption{Zenith angle distributions for: (a) $e$-like events, (b)
$\mu$-like, 
where $\cos \Theta =1$ for down-going particles. Histograms
with hatched error bars show the MC prediction with its statistical
error.  Dotted histograms show the 25\% systematic uncertainty on the
absolute normalization, which is correlated between $\mu$-like and 
$e$-like events.\label{kielczewska.fig1}}
\bigskip\bigskip
\centerline{\epsfxsize=0.6\textwidth\epsffile{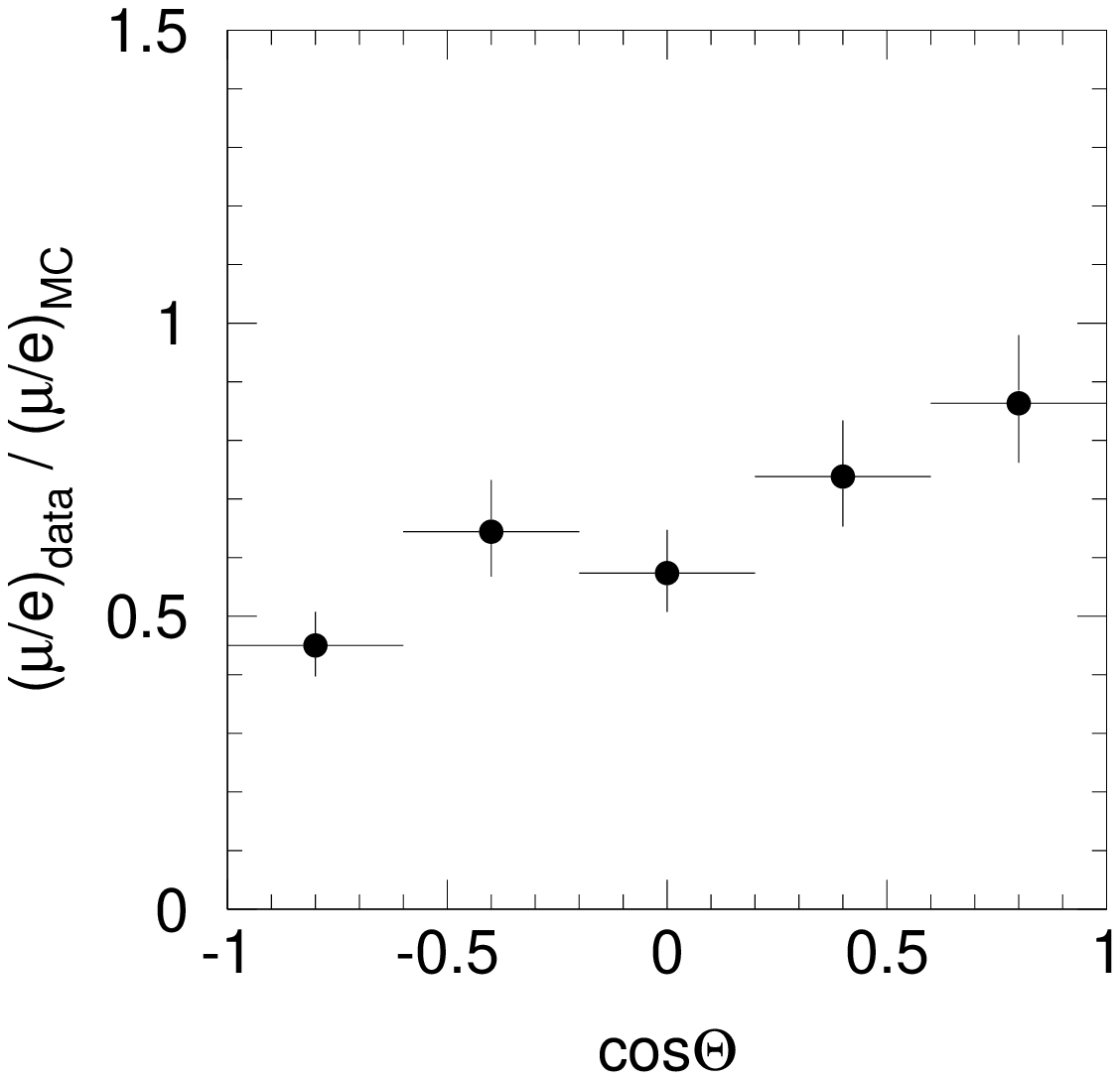}}
\caption{Zenith angle distribution of $R$.
\label{kielczewska.fig2}}
}
\end{figure}

Preliminary results of the analysis of the Multi-GeV sample of
events with $E_{\rm vis}>1.33$ GeV is shown in
Table~\ref{kielczewska.table2}.  The resulting double ratio is
$R=0.60\pm0.06({\rm stat})\pm0.07({\rm sys})$.

Finally the preliminary analysis of partially contained events, i.e.\
interactions with particles exiting the inner detector, provided 156
data and 210.6 MC events.  According to the simulations charged
current $\nu_{\mu}$ and $\bar\nu_{\mu}$ interactions constitute 97\%
of this sample.

\subsection*{Conclusions}

The deficit of muon neutrinos reported from earlier experiments have
been confirmed during the first year of data collection in
Super-Kamiokande.  The double rato values of $R=0.63\pm0.03({\rm
stat})\pm0.05({\rm sys})$ for the Sub-GeV sample and
$R=0.60\pm0.06({\rm stat})\pm0.07({\rm sys})$ for the Multi-GeV sample
are consistent with results from Kamiokande, IMB and Soudan.
 
The angular asymmetry observed with much smaller statistical weight in
Ref.~\cite{kielczewska.fukuda} is observed in the present sample for
both Sub-GeV and Multi-GeV cuts. Those results strongly suggest the
neutrino oscillation interpretation. However, more data are needed to
exclude any possibility of detector related biases.

\subsection*{Acknowledgements}

D.K.~is supported by a grant from the Polish Committee for Scientific
Research.  The Super-Kamiokande experiment was built and operated from
funding by the Japanese Ministry of Education, Science, Sports and
Culture, and the United States Department of Energy.  D.K.~has the
pleasure to thank organizers for their kind hospitality and excellent
organization of this meeting.

}\newpage{


\head{Neutrino Astronomy with AMANDA}
     {Ch.~Wiebusch$^1$ (for the AMANDA Collaboration$^2$)}
     {$^1$ DESY IfH Zeuthen, Platanenallee 6, D-15738 Zeuthen 
        (wiebusch@ifh.de)\\
      $^2$ http://amanda.berkeley.edu/ }

\subsection*{Introduction}

The AMANDA experiment plans the construction of a neutrino detector of
about $1$~km$^3$ size (ICECUBE).  Deep ($>2$~km) holes are melted into
the thick ice-cover of Antarctica at the South Pole.  Large (8 inch)
photomultiplier tubes, embedded in glass pressure housings, are
attached to an electro-optical cable (vertical separation of 10--20 m)
and lowered into the hole before re-freezing starts.  By deploying
several of these ``strings'' a 3 dimensional matrix of optical sensors
is achieved.

\begin{figure}[htp]
\hfill
  \parbox{0.67\textwidth}{%
        \epsfxsize=0.66\textwidth\epsffile{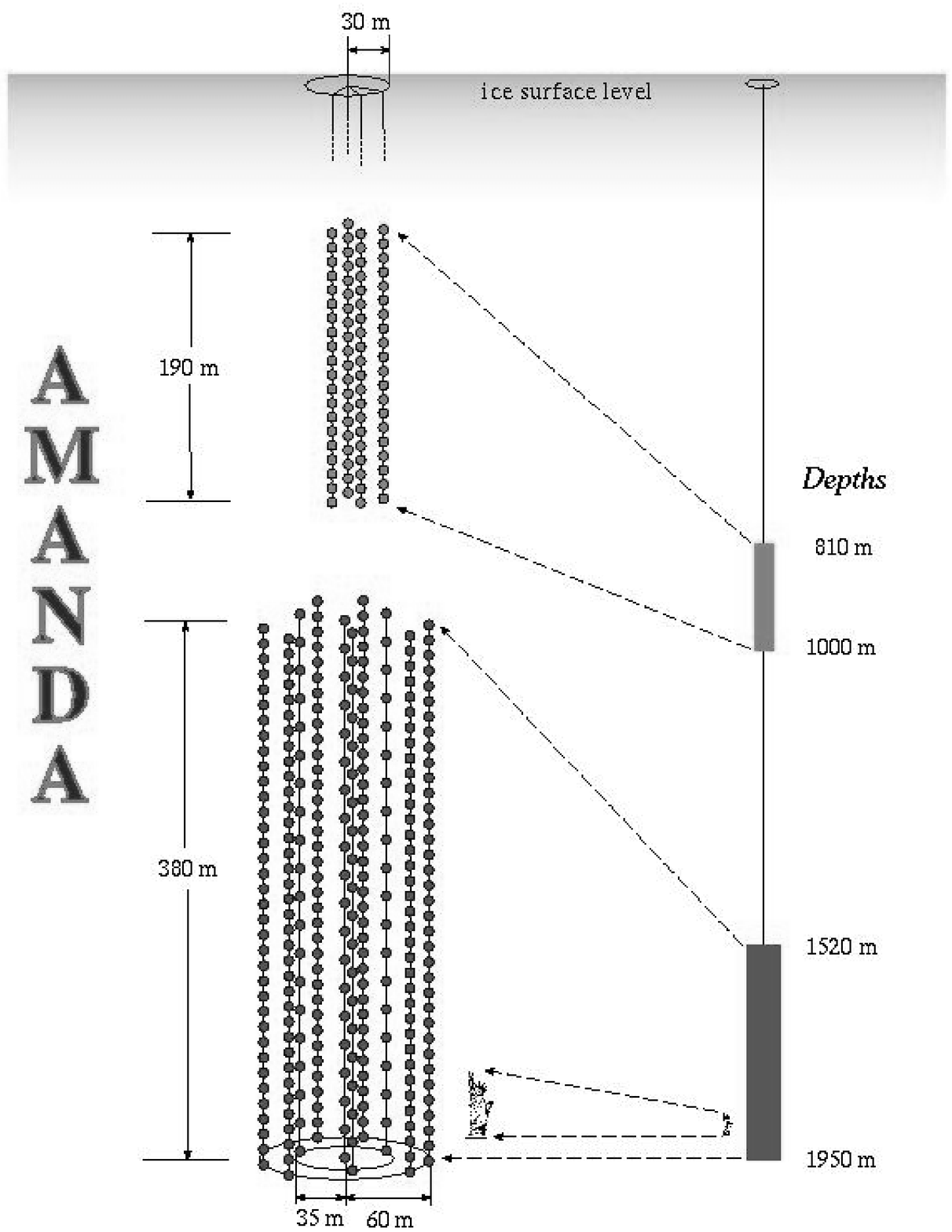}} 
  \parbox{4cm}{%
        \vspace*{3.9cm} 
        \vbox{\hbox{``AMANDA-A'' (1994)}}    
        \vspace*{3.2cm}
        \vbox{\hbox{``AMANDA-B'' (1997)}}}
\hfill
\caption{The AMANDA-A and AMANDA-B detectors.}
  \label{wiebusch.fig1}
\end{figure}

Key signatures for $\nu_\mu $ nucleon interactions below the detector
are upward going muon tracks traversing the detector.  $\nu_e$,
$\nu_\tau$ can be identified by large electromagnetic or hadronic
cascades not associated with a down-going muon track.  Muon tracks
emit \v{C}erenkov light with a cone of fixed angle. By measuring the
arrival time and number of photons the direction and energy of the
muon is reconstructed. The muon direction is pointing close into the
direction of the initial neutrino, thus enabling us to operate the
detector as a neutrino telescope.  The main background comes from
down-going atmospheric muons, which may be falsely reconstructed as
up-going tracks.  In case of showers the vertex of the interaction can
be reconstructed from the times of the spherically propagating
\v{C}erenkov photons.

The rate of neutrinos produced in the Earth's atmosphere is falling
off more rapidly with higher energies than expected for cosmic
sources. Therefore the AMANDA detector is optimised for muon detection
above energies of $1$~TeV.  However, muons are detected down to
energies of a few GeV with decreasing efficiency.

A sketch of the currently operating AMANDA-A (4 shallow strings) and
AMANDA-B (10~deep strings) detectors is shown in figure
\ref{wiebusch.fig1}.  In a next stage, AMANDA-II, 11 additional $1$~km
long strings are being installed around the AMANDA-B detector.  The
first three strings of 1.2~km length are being deployed\footnote{By
the deadline of this report all 3 have been successfully installed,
covering a depth from 1200m to 2400m.} during the winter season
97/98. The results of these strings will outline the cubic km scale
technology~\cite{wiebusch.4}.

Besides the main purpose---the search for point-sources of high energy
neutrinos (e.g.~from active galactic nuclei, AGN)---a large variety of
research topics are covered, like the measurement of the total
neutrino fluxes (from all AGN), neutrinos in coincidence with gamma
ray bursts, atmospheric neutrinos, or neutrinos from decays of exotic
dark matter.  Due to the low temperatures and clean environment the
noise rates of the photomultipliers are small.  A special DAQ system
continuously monitors these noise rates in different time windows,
allowing to search for bursts of low energy (MeV) neutrinos.  Already
in the present detector a supernova within our galaxy would yield a
statistically significant excess in the summed count-rate of all
sensors.  Another possible source of low energy neutrino bursts could
be due to gamma-ray bursts~\cite{wiebusch.1,wiebusch.2,wiebusch.4}.

\subsection*{The South Pole Site}

Besides an excellent local infrastructure and good transportation
logistics provided by the American Amundsen-Scott South-Pole station
the construction and long-term operation of a large neutrino telescope
in the deep Antarctic ice is suggested by a variety of
advantages~\cite{wiebusch.4}.

A continuous 3 month access (and year round maintenance) is possible.
Long good weather periods, 24~h daylight and large available space
allow complex deployment operations and preparations on the stable ice
surface.  This includes the use of heavy equipment and parallel work
e.g.~at two holes.

Relatively short cables are sufficient to connect the 2~km deep
strings to the surface data acquisition system.  While only simple and
robust components are buried deep into the ice, more vulnerable
electronic components are located at the surface, reducing the risks
of failure.  The short distance also allows to connect each PMT by its
own cable to the surface electronics.  The analog PMT signals are
transmitted to the surface via electrical and optical fibres. In case
of failure of an optical connector the electrical one still has
sufficient accuracy to serve as a backup.  The PMT anode current
drives an LED directly without using complex electronics.  Thus an
Amanda optical module consist of a bare PMT with an LED and a voltage
divider in a pressure housing. It is connected to a high voltage cable
and an optical cable via two connectors.

Though increasing the price per channel, the non-hierarchical approach
of individual cables provides great advantages, when operating in a
non-laboratory environment.  First, the risk of single-point failure,
e.g.~the failure of essential components is minimised---single PMTs
may fail but not larger parts of the detector as in the case of
several PMTs connected to one cable or ``digitisation module.''
Secondly, an evolutionary approach during the construction of AMANDA-B
allowed to improve certain components year by year, based on practical
experience.  Older installations are still operated, without the need
to redesign the detector. Despite of the fact that details of a later
km-size detector are not decided yet, it is obvious that AMANDA-B can
serve as a nucleus for the ICECUBE~\cite{wiebusch.5} detector.

The very stable cold temperatures in the deep ice result in low
noise-rates of PMTs and a high reliability.  No photomultiplier, which
was successfully deployed in 1994 or later, was lost since then. No
daily or annual changes of the environment are observed and the PMT
sensitivity is not degraded by effects such as sedimentation.

The South-Pole site is largely used by various astrophysical
facilities.  One interesting aspect are coincidences between AMANDA
and the surface air-shower detectors SPASE and GASP.  Shower cores,
which point from SPASE to AMANDA, give an important source of muons of
known direction.  These events are widely used for calibration and
verification of reconstruction algorithms.  Several hundred coincident
events per year originate from cosmic rays of energies above the knee
of the primary cosmic ray spectrum.  SPASE aims to use the information
e.g.~on the muon number, measured with AMANDA, to determine the cosmic
ray composition at high energies.

\subsection*{Results from Amanda-B}

\begin{figure}[htp]
   \centerline{\epsfxsize=0.9\textwidth\epsffile{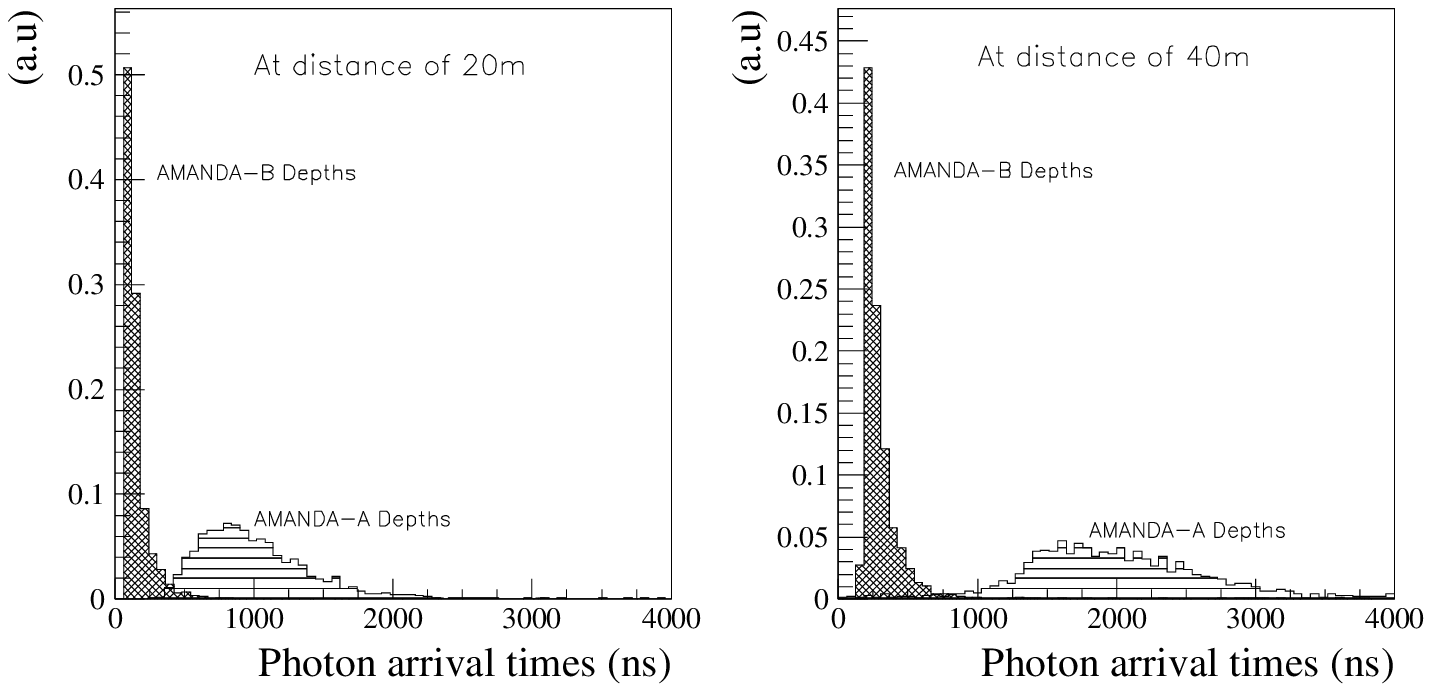}}
\caption{Ice properties measured at 1000 and 2000 m depth. The figures
show the distributions of photon arrival times relative to a laser
source, measured with PMTs at 20~m (left) and 40~m (right) distance to
the source.  Scattering due to residual air bubbles leads to a
significant increase of the photon propagation time (and a broadening
of the spectrum). This is a strong effect for AMANDA-A depths.  It is
much smaller at depths of 2~km (AMANDA-B).}  \label{wiebusch.fig2}
\end{figure}

As a major result from the analysis of laser light propagating through
ice it was found, that the optical absorption is extremely large, $>
200$~m at 800~m depths and about 100~m at 1800~m, considerably
exceeding the values for clearest natural water.  Residual bubbles at
shallow depths of AMANDA-A lead to a small scattering
length. Scattering of photons improves the calorimetric properties for
energy reconstruction but lead to a loss of time-information and thus
makes the reconstruction of track directions difficult.  At the larger
depths of AMANDA-B the scattering length is significantly improved
(see figure \ref{wiebusch.fig2}).  The challenge of track
reconstruction for the remaining level of scattering was solved. The
energy resolution seems to be superior to water experiments,
e.g.~BAIKAL.

\begin{figure}[htp]
   \centerline{
        \epsfysize=8.5cm\epsfxsize=0.7\textwidth\epsffile{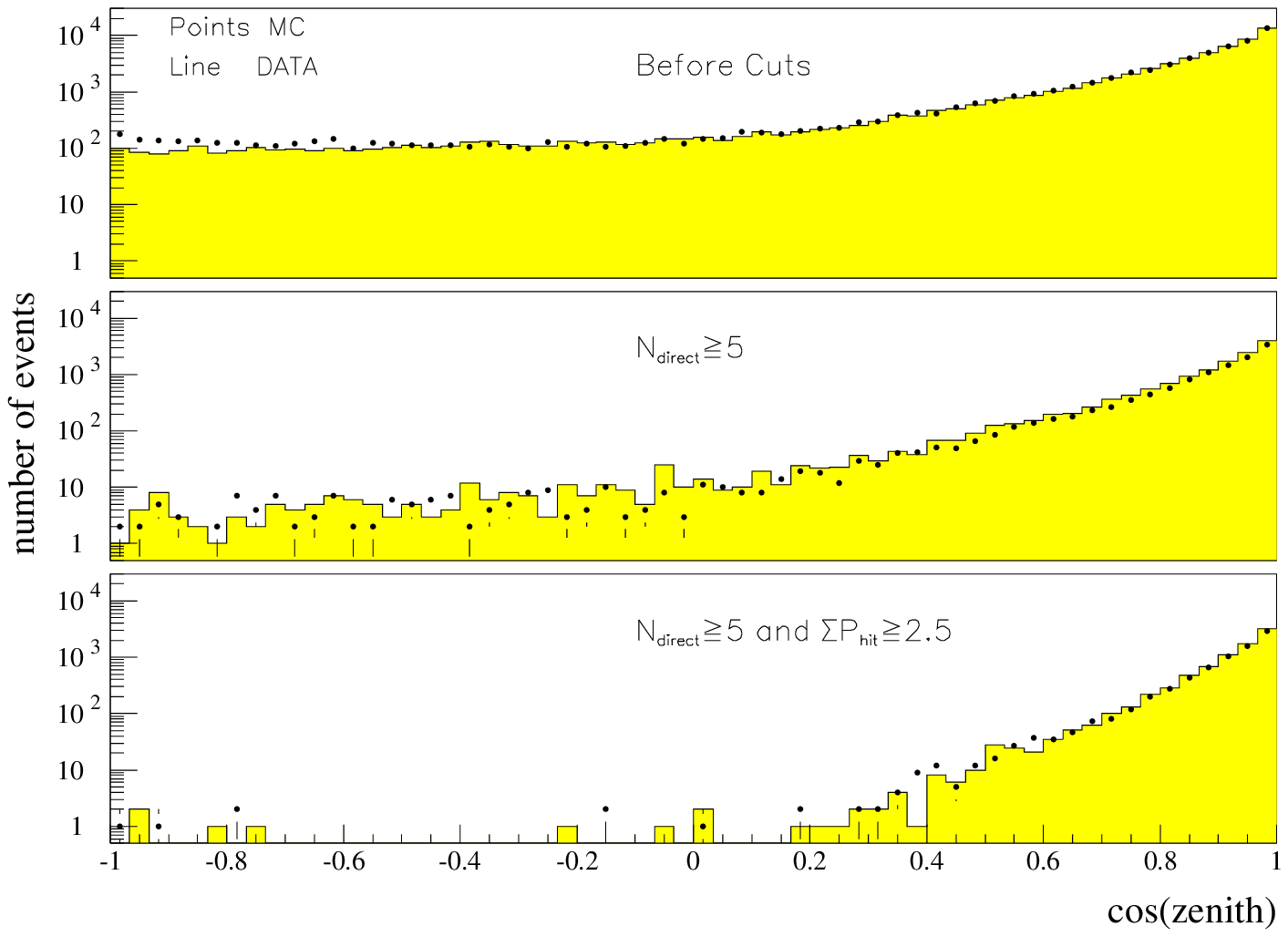}
        \hfill
        \epsfysize=9cm\epsfxsize=0.25\textwidth\epsffile{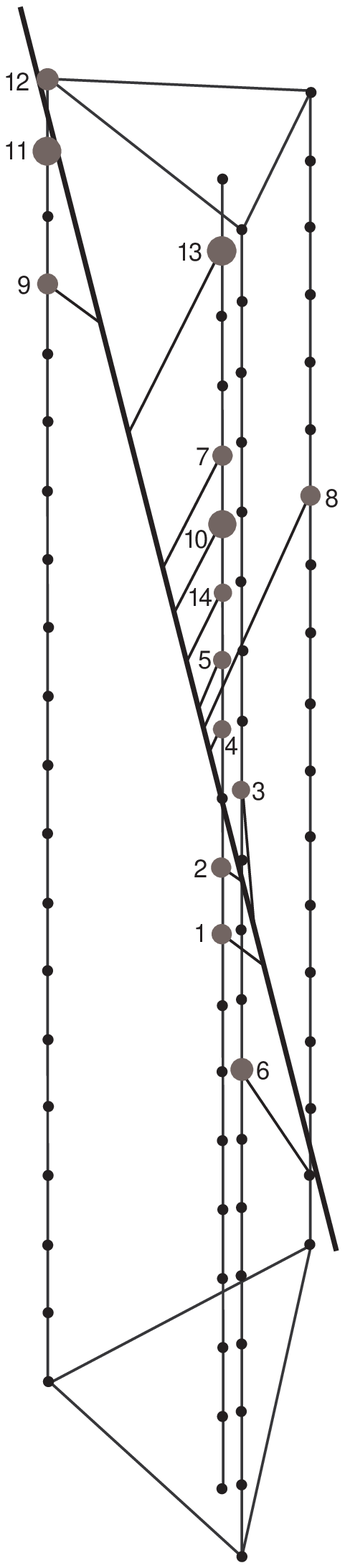}
        }
  \caption{Background rejection and search for $\nu $ candidates
          (see text).}
  \label{wiebusch.fig3}
\end{figure}

The data which was recorded in 1997 has been transferred from the Pole
and is currently being analysed. However, using the data taken in 1996
with only 4 installed strings, first results have been obtained.

As an example for neutrino search in AMANDA-B (4 strings), figure
\ref{wiebusch.fig3} shows the reconstructed zenith angle 
distribution of recorded muon events.  From top to bottom quality
criteria for filtering of up-going events are subsequently
tightened. Without quality criteria (top) a large fraction of events
are falsely reconstructed to originate from the lower
hemisphere. These events can be rejected in good agreement with the MC
expectation.  The right picture shows a neutrino candidate event
filtered out of the 4-string data. The event has a vertical length of
about 300~m and is reconstructed as up-going.

\begin{figure}[htp]
   \centerline{
        \epsfxsize=0.48\textwidth\epsffile{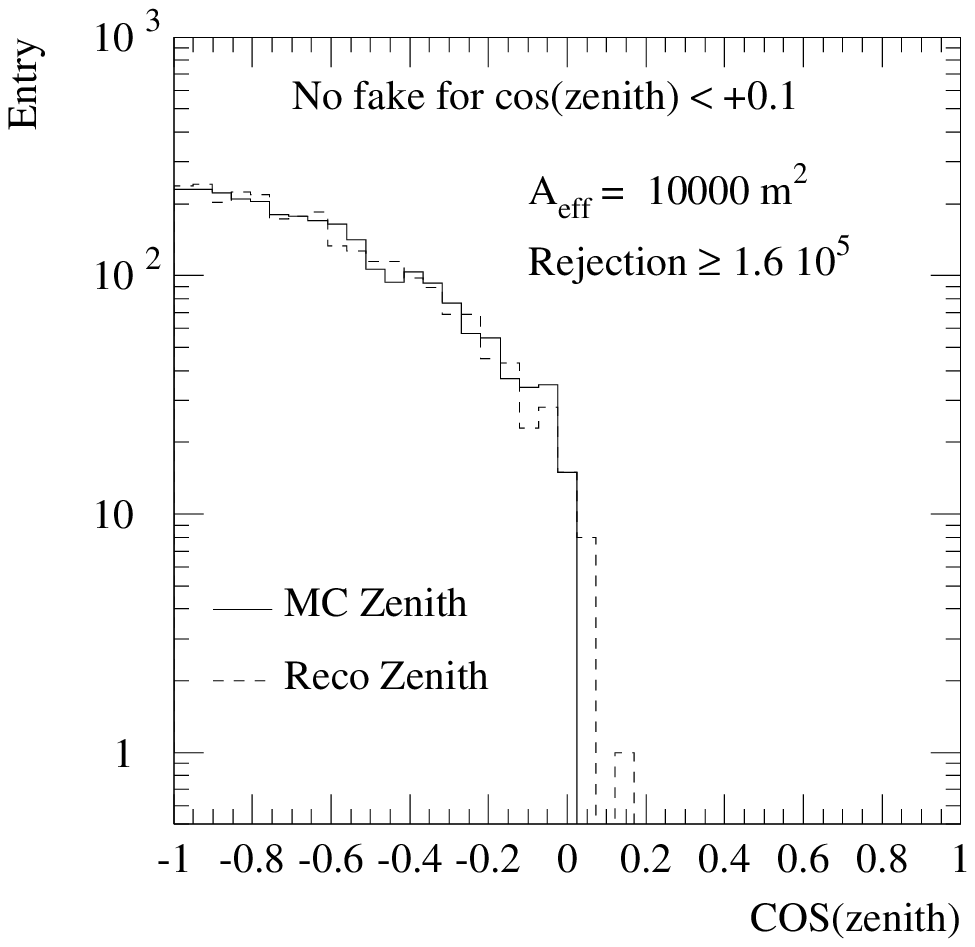}
        \hfill
        \epsfxsize=0.48\textwidth\epsffile{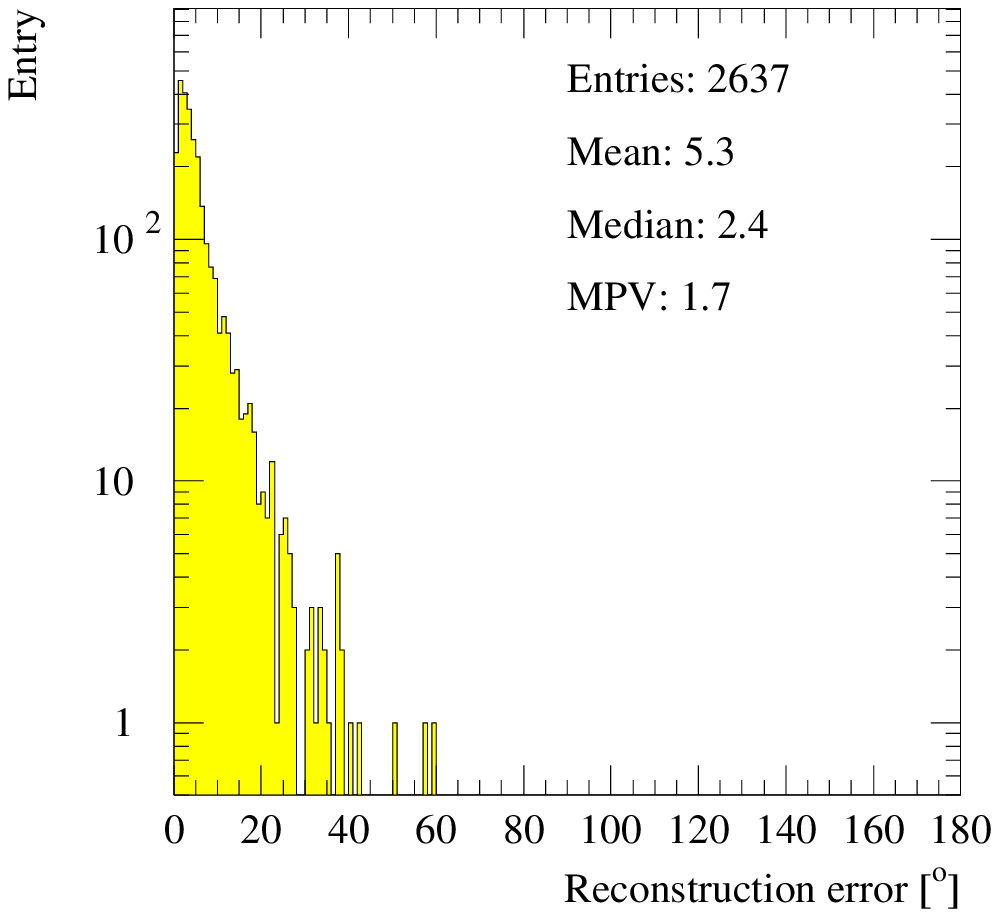}
        }
\caption{Performance expected for the 10 string  AMANDA-B detector.
The left figure shows the zenith angle distribution of up-going muon
  events passing all quality cuts. In this calculation minimum
  ionizing muons have been initially generated isotropically from the
  lower hemisphere. The difference in reconstructed and generated muon
  directions are shown on the right plot. A median accuracy of
  $2.4^\circ$ is achieved.} \label{wiebusch.fig4}
\end{figure}

The performance for the completed AMANDA-B detector (10 strings) is
expected to be significantly better compared to the 4-string
installation.  The data is currently being analysed.  Figure
\ref{wiebusch.fig4} shows results from a full Monte Carlo calculation.
With background suppression (S/N $> 10^5$) an effective area of the
order of $10 \cdot 10^3~$m$^2 $ and a median angular resolution of
$2.4^\circ $ is achieved. Several atmospheric neutrinos per day should
be detectable~\cite{wiebusch.3}.

The full AMANDA-II detector is aimed to be installed until the year
2000.  It should reach effective areas from $50 \cdot 10^3~$m$^2 $ to
$ 100 \cdot 10^3~$m$^2 $.  It is thus almost two orders of magnitude
more sensitive than current detectors.  The angular resolution is
expected to be of the order of~$1^\circ$~\cite{wiebusch.4}.

\subsection*{Towards ICECUBE}

First experiences with data of a km scale telescope have already been
gathered in 1996 with coincidences between the shallow AMANDA-A and
the deep AMANDA-B detector
\cite{wiebusch.1,wiebusch.2,wiebusch.4}. The purpose of
the current AMANDA-II installation campaign is to
monitor the depth profile of optical properties from 1200 m down to 
2500 m. The tests of a large variety of technical
issues will help to define a robust technology, which is suitable
for the construction of the large km$^3$ scale telescope.
Besides the tests of different PMTs, optical fibres and connectors 
also wavelength-shifters (to improve the UV sensitivity) and other
strategies for signal processing (Digitising optical modules)
are tested~\cite{wiebusch.4}.

A straw-man ICECUBE detector might consists out of 60 additional 1~km
long strings, each with 70 OMs. Based on the experience from AMANDA-B
and AMANDA-II the prices per channel may be estimated to 6~k\$. This
includes the optical module, cables, calibrations sources, surface
electronics (DAQ) and workshop labour.  Adding 5~M\$ for off-line
computing and additional detector components the bare detector price
would be of the order of 30~M\$~\cite{wiebusch.4}.

However, in order to operate the detector and perform the installation
within a reasonable time (5 years) additional logistic costs have to
be taken into account.  These are mainly the costs of modernisation
the drilling equipment, a new building, fuel, cargo and technical
personal. These support costs can be estimated to about
10~M\$~\cite{wiebusch.4}.

Adding the costs of AMANDA-B and AMANDA-II, the ICECUBE detector would
cost of the order of 50~M\$~\cite{wiebusch.4}.

It is planned to start the installation in the beginning of year 2001
and finish it by 2005. A workshop on the ICECUBE project will be held
in March 1998~\cite{wiebusch.5}.


}\newpage {


\head{High Energy Neutrino Astronomy with ANTARES}
     {M.E.~Moorhead (for the ANTARES Collaboration)}
     {University of Oxford, Particle and Nuclear Physics Laboratory,
       Keble Road, Oxford OX1 3RH, UK}

\noindent Neutrino astrophysics has already generated considerable
interest in both the particle physics and astrophysics communities,
through the detection of solar neutrinos, atmospheric neutrinos and
neutrinos from supernova 1987A. ANTARES~\cite{MEM.ANTARES} aims to
extend this scientific reach to higher energies (10~GeV--10 PeV) where
potential neutrino sources include Active Galactic Nuclei, Gamma Ray
Bursters, pulsars, X-ray binaries, young supernovae remnants and
neutralino annihilation in the center of the Sun or Earth. These
neutrino observations should provide complementary information that
might be difficult or even impossible to obtain through gamma or
cosmic ray observations. The observable energy range for neutrinos
extends well above 1--10 TeV where extra-galactic gamma rays are
severely attenuated over cosmic distances by their interactions with
infra red and cosmic microwave background photons.  Moreover, unlike
charged cosmic rays, neutrinos are not deflected by magnetic fields
and thus point back to their source.
               
There are several potential astrophysical sites for accelerating cosmic
rays up to the energies of 10$^{20}$ eV that have been observed. In general,
these sites will also produce significant fluxes of neutrinos and gamma rays
through the decays of charged and neutral pions, which are produced in
approximately equal numbers through soft interactions of the high energy
protons with other protons or photons present in the acceleration site. The
emitted fluxes of neutrinos and gamma rays from this mechanism would thus be
roughly equal, however the gamma rays may be severely attenuated at the source
by interactions with matter or photons, leaving the neutrinos as the best probe
of the physics. 

Several authors (see~\cite{MEM.Gaisser} and references therein) have
constructed models which account for the observed TeV gamma ray
emission of Active Galactic Nuclei (AGNs) in terms of proton
acceleration in the jets which emerge from the central black hole of
10$^6$--10$^{10}$ solar masses. These authors predict between 100 and
1000 events per year from all AGNs in a neutrino detector with 1
km$^2$ effective area. Another possible cosmic acceleration site is
the expanding fireball produced by coalescing neutron stars
immediately after merging. These events are sufficiently energetic and
numerous to explain the cosmic origin of the observed Gamma Ray Bursts
(GRBs). Waxman and Bahcall~\cite{MEM.Waxman} predict between 30--100
neutrino events per km$^2$ per year. Pulsed neutrino sources such as
GRBs are easier to detect since the background from atmospheric
neutrinos is greatly reduced by looking for time and direction
coincidences with gamma-ray observations of GRBs with satelite
experiments. They also offer the possibility of kinematic neutrino
mass tests through the relative times of the neutrino and gamma
signals at the Earth.  Potential galactic acceleration sites of cosmic
rays include pulsars, X-ray binaries and young supernova
remnants. Calculations show that for these Galactic sources, total
proton luminosities of a few percent of the Eddington limit could be
detected by a km$^2$ neutrino detector.

A very interesting signal for ANTARES arises if the non-baryonic dark
matter is primarily in the form of weakly interacting massive particles, the
favoured particle physics candidate being the lightest supersymmetric particle
(LSP), normally assumed to be the neutralino ($\chi$).  A relic neutralino
undergoing an elastic scatter while passing through a star or planet could be
gravitationally captured, the resulting accumulation of neutralinos continuing
until balanced by $\chi\chi$ annihilation. Final states from $\chi\chi$
annihilation include many particles which subsequently decay to final states
including energetic neutrinos which can be detected by ANTARES.  If we assume
that the dark matter in the halo of our galaxy consists of neutralinos, then a
large fraction of the currently allowed SUSY parameter space predicts neutrino
fluxes from $\chi\chi$ annihilation in the centres of the Sun and of the Earth
which would be detectable~\cite{MEM.Gaisser} with exposure times of around 
10$^5$ m$^2$ yr. 

High energy muon-neutrinos can be detected by observing long-range
muons produced by charged current neutrino-nucleon interactions in the
matter surrounding the detector. An array of photomultiplier tubes,
placed in an optically transparent medium such as deep ocean water or
antarctic ice, can be used to detect the \v{C}erenkov photons produced
along the muon tracks. Since \v{C}erenkov radiation is emitted in a
cone of fixed angle (42 degrees in water) the direction of the muon,
which preserves the neutrino direction, can be reconstructed from the
location and time of the PMT `hits'. Simulation studies of prototype
ANTARES detectors show that the angular resolution for reconstructed
muons could be better than 0.2$^{\circ}$. Moreover, for muons with
energies above 1--10 TeV, the average angle between the muon and the
parent neutrino is lower than 0.1$^{\circ}$. Above 1 TeV, the average
number of \v{C}erenkov photons produced per meter of muon track is
proportional to the muon's energy so that the number of detected
photons can be used to measure the muon energy, albeit with poor
resolution due to the stochastic nature of muon energy loss above
1~TeV.
               
Muons from cosmic ray interactions in the Earth's atmosphere produce a large
background of down-going muons which is many orders of magnitude greater than
any potential neutrino signal. Thus the aperture of the telescope is restricted
to the 2$\pi$ solid angle of up-going muon events where this background is
negligible, provided that the detector missreconstructs less than 1 in 10$^5$
(for a detector at 2.5 km depth) down-going events as up-going events. If this
is achieved then the dominant background source of up-going muons will be from
up-going atmospheric neutrinos which can interact, just like the cosmic
neutrinos, in the rock or water below the detector. This background is
isotropic and falls steeply with energy: angular resolution and energy
resolution will therefore play a critical role in determining the detector's
ability to find point sources of neutrinos above this background.  Electron and
tau flavor neutrinos can also be detected via the \v{C}erenkov photons produced
from the electromagnetic and hadronic showers from neutrino interactions inside
or near the detector. The pointing accuracy and event rates for these events
will be significantly less favourable than for muon events but the energy
resolution should  be better.
                                              
ANTARES is currently in an R\&D phase where the technology for
constructing a large-area neutrino detector in the mediterranean sea
is being developed through deployments at a test site which is 30 km
from the French coast near Toulon and at a depth of 2300~m. In 1998 a
first prototype string will be deployed and data from 8 PMTs
transmitted to shore by a 40 km long electro-optical cable which is
ready to be laid in place by France Telecom. By the end of 1999 one or
two fully-equipped strings and up to 100 PMTs will be deployed in an
array that will be directly scalable to the large-area detector.
Great emphasis is being placed on developing a reliable and
cost-effective deployment strategy by using well established
commercial technology from companies such as France Telecom and by
involving two Oceanographics institutes (IFREMER and Centre
d'Oceanologie de Marseille) directly in the collaboration.  These two
institutes have many years experience of deploying scientific
instruments at depths relevant to ANTARES.
               
In parallel with these activities, ANTARES has already started deploying
autonomous test strings at the Toulon site and at a site near Corsica. These
strings will be used to select the optimal site for the large-area detector by
measuring the relevant environmental parameters: background noise on PMTs due
to $^{40}$K decays and bioluminescence, the rate at which glass surfaces lose
transparency from sedimentation and biological growth and, most importantly,
the transparency of the water in terms of absorption and scattering lengths as
a function of wavelength. Twelve deployments of these strings have been carried
out so far, showing encouraging results that can be accessed from the ANTARES
web site~\cite{MEM.ANTARES}.

After 1999, when the R\&D phase is completed and the site for the large array
chosen, ANTARES aims to deploy an array with an effective area for up-going
muons of order 0.1 km$^2$ over a timescale of 3 years. Further development of
the array up to 1.0 km$^2$ area is anticipated but would depend on the results
obtained to that point. 

\subsection*{Acknowledgments}
I wish to thank Drs.\ Botton and Kajfasz for providing me with transparencies
for my talk and Dr.\ Moscoso for help in preparing the proceedings.


}\newpage {


\head{Ground-Based Observation of Gamma-Rays\\ (200 GeV--100 TeV)}
     {R.~Plaga}
     {Max-Planck-Institut f\"ur Physik
      (Werner-Heisenberg-Institut), \\ 
      80805 M\"unchen, Germany (plaga@hegra1.mppmu.mpg.de)}

\subsection*{Introduction}

The first attempts to detect very high energy (VHE) gamma rays from
cosmic point sources with ground based detectors go back to the early
1960s. After a long period in which there was only evidence from one
point source of VHE gamma-rays (the Crab nebula), in the last few
years technological advances in background suppression have led to the
discovery of six additional sources, three of them extragalactic.  The
study of these sources gives clues to the still very incompletely
understood acceleration mechanisms and sites of cosmic particles at
nonthermal energies.  Moreover, the raised sensitivities for the first
time allow to critically test predictions for theories of cosmic-ray
origin.  A very intense outburst of TeV gamma-rays observed in 1997
from the active galaxy Mkn 501 is an example for objects that emit a
major part of their total energy output in the high energy range. VHE
gamma-ray astronomy is thus becoming a part of main-stream
astrophysics, but there are also possibilities for investigations
which are unique to this energy range, particularly in cosmology.

\begin{figure}[ht]
\centerline{\epsfxsize=\textwidth\epsffile{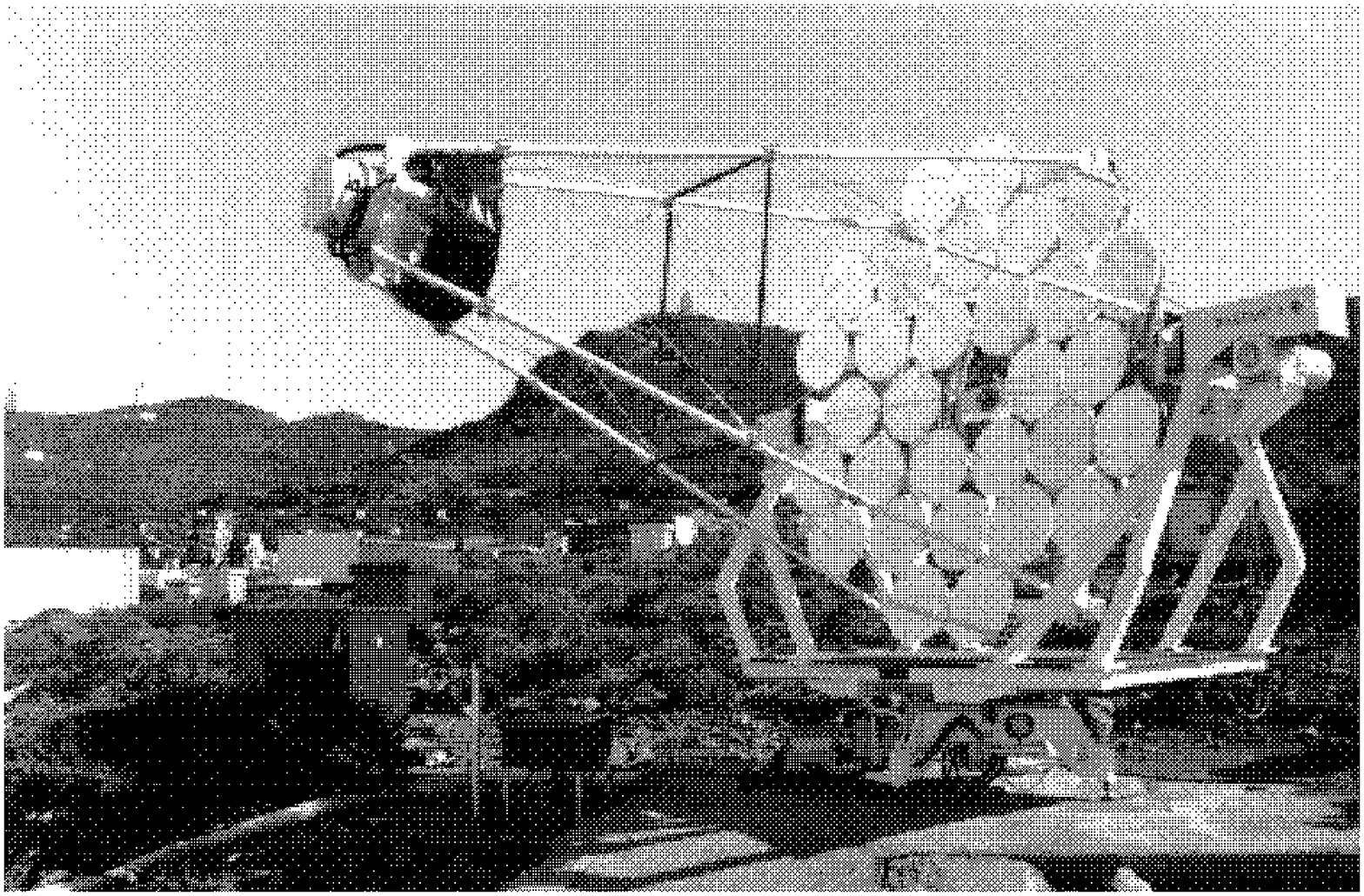}}
\caption{Typical \v{C}erenkov telescope (HEGRA telescope 2 at La
Palma). The main mirror consists of 30 submirrors with a total area of
8.5 m$^2$. The energy threshold of this telescope for $\gamma$-rays is
about 1.1 TeV.  In the background quadratic boxes from the HEGRA array
which operates at energies above 14 TeV are visible.}  \label{ring1}
\end{figure}

\subsection*{Status of Instrumentation}

In the energy range from 200 GeV to about 20 TeV the \v{C}erenkov
telescope technique to detect gamma-rays dominates presently. The
primary gamma ray develops an electromagnetic cascade which emits
\v{C}erenkov light. This light is detected in an optical telescope
with typical mirror areas from about 5 to 70 m$^2$.  The \v{C}erenkov
light is emitted over an area of about 20000 m$^2$ and each telescope
situated whithin this pool can detect the shower.  The resulting very
large detection area leads to very high count rates. For example, the
Whipple collaboration detected in one observing season (Spring 96)
about 5000 photons from the active galaxy Mkn~421 above
300~GeV~\cite{whipplep}.  This is about the same number as a satellite
experiment, EGRET at the Compton gamma-ray observatory, detects in one
year above 100 MeV, where the flux is typically a few thousand times
higher, from {\it all} active galaxies (about 50 active galaxies were
detected above 100 MeV)!

Until recently the sensitivities were seriously degraded by the fact
that the hadronic cosmic radiation produces a background which even
for the most intense sources is about a factor 100 higher than the
signal from photons.  The imaging technique, pioneered by the Whipple
collaboration~\cite{weekes} allows to reduce this background.  The
image of the developing shower is registered in an array of fast
photomulipliers which detect the few nanosecond long \v{C}erenkov
pulses.  The showers produced by hadrons are much more irregular in
shape than the ones from photons, and this allows to reject the
background showers with an efficiency reaching e.g.~98~\% for the
HEGRA telescope 1, a single telescope of modest size with a camera of
moderate angular resolution (127 pixels with a field of view of
$3^\circ$)~\cite{petry}; another telescope of the same experiment is
shown in Fig.~\ref{ring1}).

This technique is still in a state of evolution, the most advanced
camera presently in operation by the French CAT
collaboration~\cite{cat} has 546 pixels. Its high angular resolution
($0.12^\circ$) allows not only a further reduction in background but
also a low energy threshold and a deduction of the core position of
the shower which in turns allow a very good energy resolution.
Another crucial technological advance for \v{C}erenkov telescopes is
the exploitation of stereo imaging. The same shower is viewed by
several telescopes at the same time and its full three dimensional
structure can be reconstructed.  Besides advantages similar to a
higher angular resolution the measurement of shower parameters is
overdetermined and thus allows important consistency checks of the
technique~\cite{hofmann}.

\begin{figure}[htb]
\centerline{\epsfxsize=12.1cm\epsffile{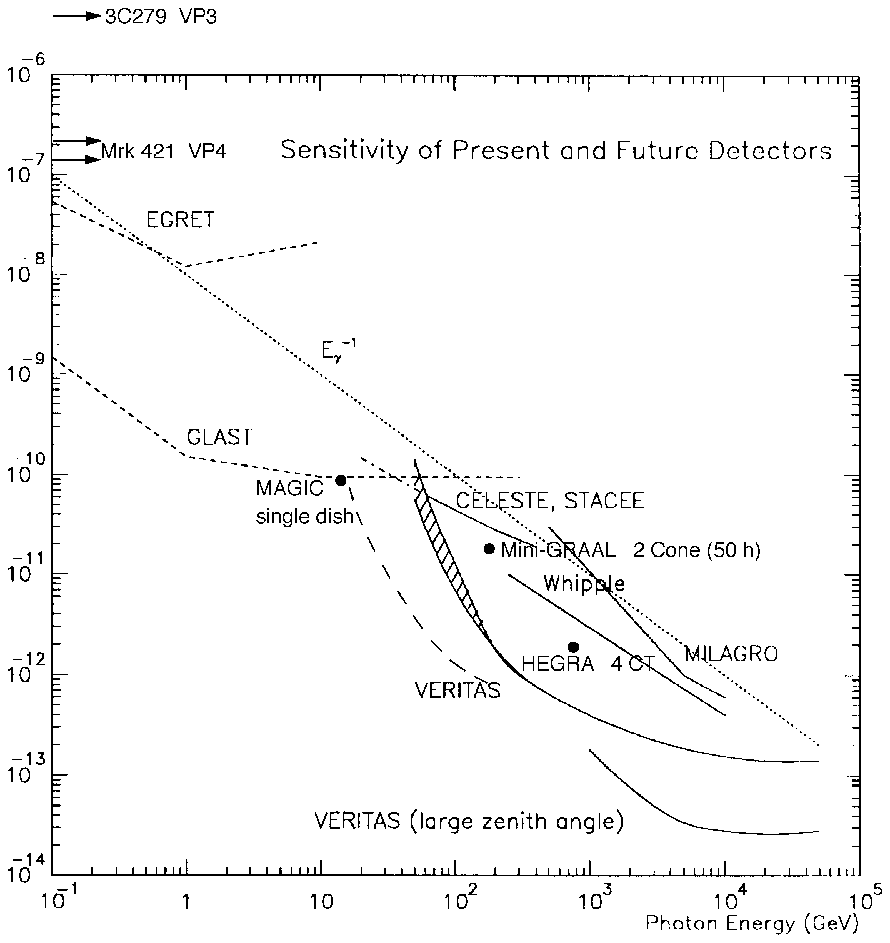}}
\caption{Sensitivities in units of photons cm$^{-2}$ sec$^{-1}$ of
various existing and planned detectors for high and very-high energy
gamma rays for various energies (full lines) or at the energy
threshold (big dot). The numbers are for a total measuring time of 50
hours for devices with a small field of view and 1 year for devices
with a field of view of 1 steradian or larger~[satellites: EGRET
(existing) and GLAST (planned) and the array MILAGRO (planned)].
WHIPPLE and HEGRA are existing telescopes, VERITAS and MAGIC proposed
large future devices. STACEE, CELESTE and GRAAL are detectors which
are currently being set up using the large mirror areas at solar power
plants. The dotted line gives the extrapolation of a source spectrum
from GeV to VHE energies (active galaxy Mkn~421). The arrows indicate
the intensity of this source during various viewing periods (``VP'')
of the EGRET satellite.}
\label{ring2}
\end{figure}

Array techniques where the lateral distribution of \v{C}erenkov light
and/or particles at ground level is sampled with high spatial
resolution are an important technological alternative to imaging
telescopes.  At higher energies (above 10 TeV in existing detectors)
they offer the distinct advantage of a field of view on the order of a
steradian, comparable to satellite detectors. The array technique will
also be exploited by using heliostat mirrors in existing solar-power
plants to reach low energy thresholds~\cite{solar}.

\newpage

\subsection*{Recent results}

This summer (1997) the Australian-Japanese CANGAROO collaboration
announced detection of radiation above 2 TeV from the remnant of the
supernova (SNR) of A.D.~1006~\cite{tanimori}.  The detected flux was
close to previous theoretical predictions~\cite{masti} on the basis of
the hypothesis that cosmic-ray electrons are accelerated by the shock
front of SNRs.  The accelerated electrons suffer inverse Compton
scattering on the cosmological 3~K background radiation, which are
then detected in the TeV range.  This is strong experimental support
for the idea that the Galactic cosmic-ray electrons (which account for
about 1 $\%$ of the total cosmic-ray flux at earth) are due to shock
wave acceleration in SNRs. Perhaps this brings us near to a solution
of the historic question of the origin of the main hadronic part of
cosmic rays: it had long been speculated that {\it all} cosmic rays
with energies below about 100 TeV are accelerated~in~SNRs.

\begin{figure}[b]
\centerline{\epsfxsize=12cm\epsffile{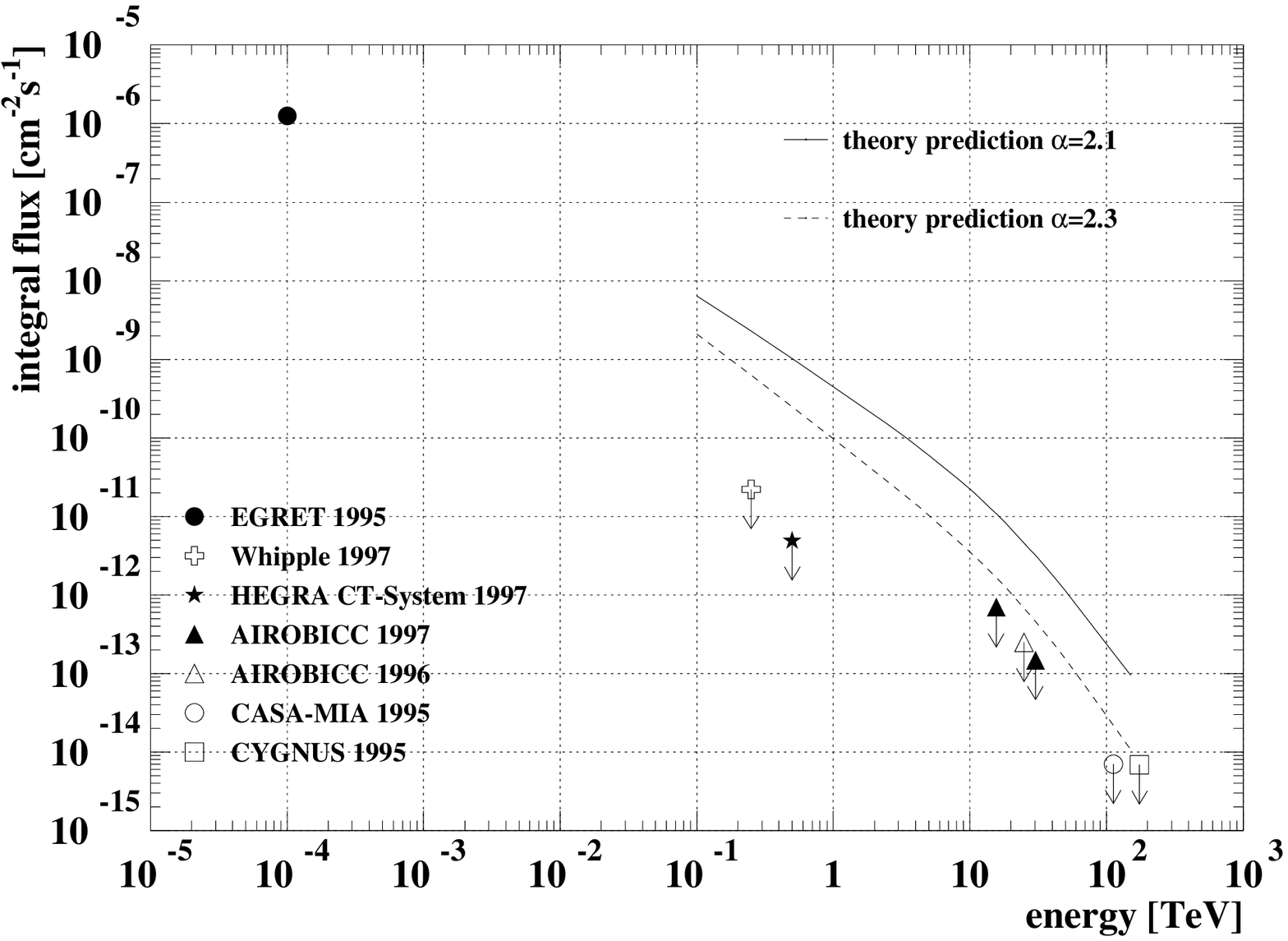}}
\caption{Upper limits on VHE radiation from the SNR G78.2+2.1. Whipple
and HEGRA are \v{C}erenkov telescopes, AIROBICC, CASA-MIA and CYGNUS
are arrays. It now seems likely that the detection of $\gamma$-rays
from a satellite above 100 MeV (EGRET) is due to a pulsar.}
\label{ring3}
\end{figure}

The VHE upper limits on the SNR G78.2+2.1 shown as an example in
Fig.~\ref{ring3} \cite{chris} present a problem for this scenario,
though.  This SNR lies near a dense molecular cloud which should act
as a ``target'' for the accelerated cosmic-rays, leading to the
prediction of a copious production of VHE gamma-rays via the reaction
p + p $\rightarrow$ $\pi_0$ $\rightarrow$ 2 $\gamma$.  It is seen that
the upper limits from various experiments shown in Fig.~\ref{ring3}
are below theoretical predictions in the mentioned
scenario~\cite{chris,voelk}.  This has recently led to a revival of
speculations on an extragalactical origin of the main part of {\it
hadronic} cosmic rays~\cite{plaga}.

\begin{figure}[ht]
\vskip9.5cm
\centering
\epsfig{file=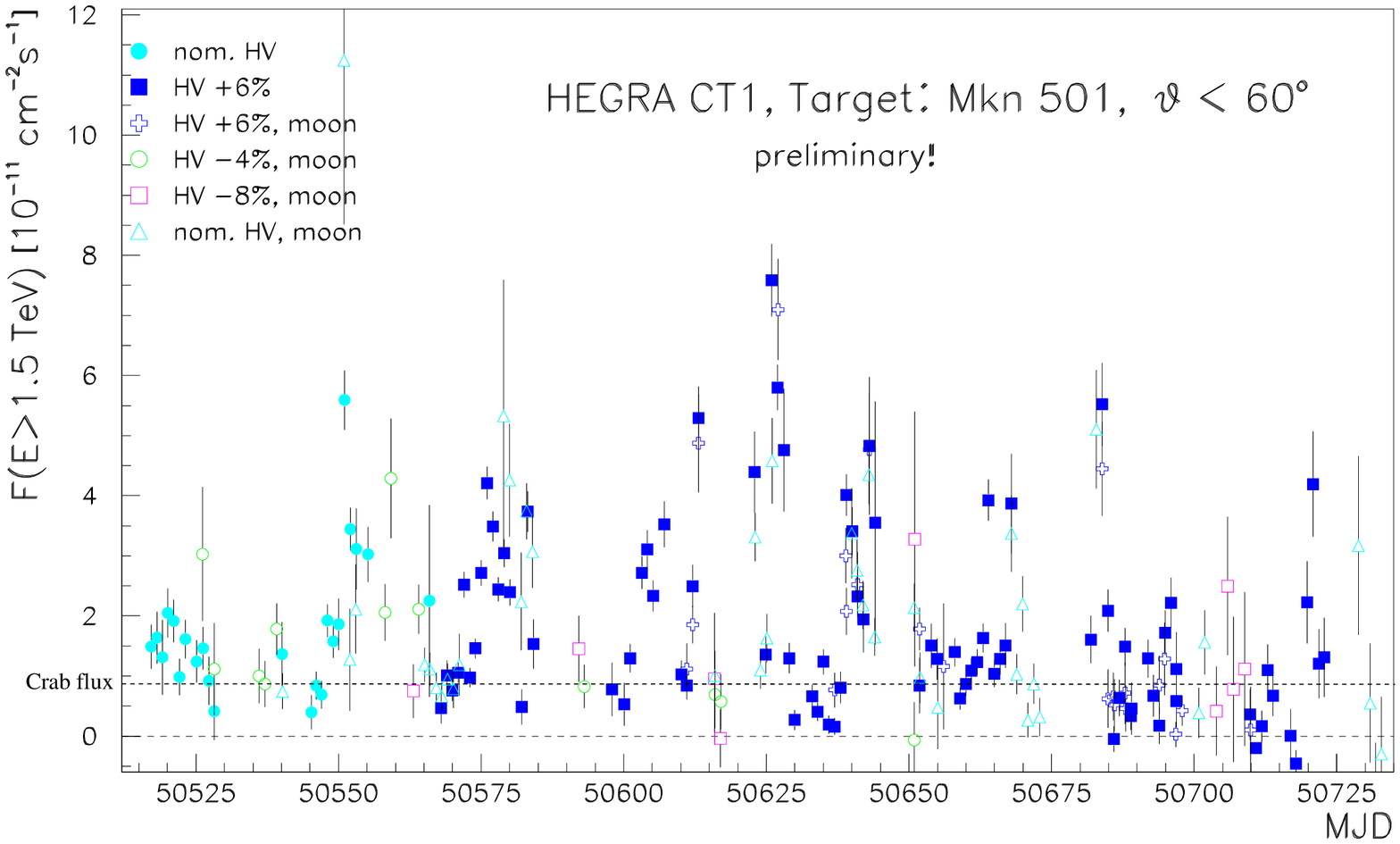,height=\textwidth,angle=270}
\caption{\label{ring4} Gamma-ray flux above 1.5 TeV from the 
direction of the active galaxy Mkn~501, determined with the 
HEGRA telescope 1. The various symbols are for different
modes of operation with and without ambient moon light.
The time axis is given in mean Julian days and extends from March 8
to October 13, 1997 (courtesy D.~Kranich).}
\end{figure} 

The active galaxy Mkn~501 at a distance of about 100 Mpc from earth
showed a spectacular outburst of TeV radiation 1997 which was observed
in detail by 5 different collaborations~\cite{protheroe}.  At the
highest activity level in April 1997 the flux was about 40 times
higher than during the discovery of the source in
1995. Figure~\ref{ring4} shows the light curve determined by the HEGRA
telescope 1.  This light curve is the most complete of all measured
ones because data were also taken during moonshine.  The outburst was
also observed in the X-ray and optical region and these
multiwavelength studies are especially valuable to discriminate
between various models for the origin of the TeV radiation and to find
the reasons for its fluctuation.  The measured VHE spectra already
allow interesting conclusions about the intensity of cosmological
infrared background radiation field~\cite{stecker}.  Interactions of
VHE photons with cosmological background fields will allow a variety
of cosmological studies unique to this wavelength range~\cite{cosmo}.

\subsection*{Projects for the Future}

VHE astrophysics seems to be in a similar state as X-ray astronomy in
the 1960s.  The next generation of telescopes which will combine
higher angular resolution, larger mirror area and stereoscopic
imaging~\cite{proj}.  It will probably allow to detect on the order of
hundreds of sources and lower the energy threshold thus closing the
present ``gap'' in detectable primary gamma energy between satellites
(up to 20 GeV presently) and ground based devices.

}\newpage {\  }


\newpage {


\thispagestyle{empty}

\begin{flushright}
\Huge\bf
{\ }


Cosmology

\end{flushright}

\newpage

\thispagestyle{empty}

{\ }

\newpage

}\newpage{


\head{Helium Absorption and Cosmic Reionization}
     {Craig J.~Hogan}
    {Departments of Physics and Astronomy, University of Washington\\ 
     Box 351580, Seattle, WA 98195, USA}

\noindent
In my talk at Ringberg I emphasized the current concordance of light
element abundances with the Big Bang predictions and with the observed
density of baryons. The substance of most of these remarks together
with further references can be found in recent papers
\cite{hogan.1,hogan.1a,hogan.2}.  Here I want to highlight one aspect
of this puzzle which has not been commented on very much, the
interesting current state of observations on singly ionized helium at
high redshift.

The gas in the emptier regions of the universe at the highest redshift
observed (almost 5) is already almost entirely ionized; the fraction
of hydrogen which is neutral is less than $10^{-4}$. It is small
enough that over most of redshift space, the absorption from hydrogen
has small optical depth.  HeII's higher ionization potential (54.4 eV
as opposed to 13.6 eV for hydrogen) means that its ionization is
delayed relative to hydrogen as such hard photons are relatively
rare. HeII is therefore more abundant in absolute terms than any other
species which makes it the best tool for studying matter in the
emptier regions of space---the voids between the density
concentrations.

Currently there are two quasars where published high resolution
HST/GHRS quasar absorption line spectra permit the separation of the
more diffuse component of the HeII resonance absorption from the
component in the concentrations of matter which appear as HI
Lyman-$\alpha$ forest clouds \cite{hogan.3,hogan.4}. Both
datasets imply an upper bound on the diffuse matter density at
$z\approx 3$ of less than $0.01 h_{70}^{-1.5}$, derived based on the
maximum permitted mean HeII Lyman-$\alpha$ optical depth allowed after
subtracting the minimal contribution from the detected HI
Lyman-$\alpha$ forest clouds, while adopting the hardest ionizing
spectrum allowed by the data.  This upper bound is important for
constraining ideas about the evolution of the baryons and especially
the possibility of a large repository of baryons in the voids. Both
datasets also agree that there is HeII absorption at nearly all
redshifts---that is, there is at least some matter everywhere, even
in the emptiest voids.

However the current published results do conflict in one important
respect. Hogan et al.~\cite{hogan.3} find in Q0302-003 that although
the HeII optical depth is rather high (greater than 1.3 everywhere),
there is also detectable quasar flux at all redshifts in between the
identified HI clouds, whereas Reimers et al.~\cite{hogan.4} find in
another quasar (HE2347-4342) large regions where the flux is
consistent with zero.  The first result suggests that there is rather
little HeII and that the helium is mostly doubly ionized HeIII already
(unless the voids are swept implausibly clean of gas), whereas the
second result suggests that large volumes of space are still mostly
HeII and that helium ionization is not yet complete.

Since the ionization is expected to occur in patchy domains around the
strongest sources of ionizing photons (indeed near 0302-003 itself the
sphere of influence is observed in the spectrum to about 4000 km/sec
in radius), it is possible that both interpretations are correct and
that the different directions are just different. This would be
interesting because the scale of the inhomogeneity in ionization
history would be much larger than expected, with likely consequences
for large scale structure in the galaxy distribution.  On the other
hand it could also be that one or both of the published spectra have
an incorrect zero level---a natural suspicion since they are taken
with the one-dimensional GHRS for which background subtraction
requires considerable modeling. If this is the case our data are more
suspect than those of Reimers et al.  since Q0302-003 is a fainter
quasar.

The reason I highlight this situation now is that it is likely to be
resolved soon as the new STIS detector is now being trained on both of
these targets and should with its better background subtraction
resolve this issue. By the time this note is published there should be
a new result which will give us a direct insight into the final stages
of cosmic reionization. It should also allow a much tighter
measurement of the range of densities allowed; at present the lower
limit on the density is about two orders of magnitude below the upper
limit, because of the uncertainty in modeling the ionization. In
addition, new higher resolution observations will allow a better
defined separation of diffuse and cloud components, and observations
of a third quasar (PKS 1935-692) should help to differentiate between
the accidental properties of particular sightlines and universal
properties of the intergalactic medium.

}\newpage{


\def\simge{\, {}^>_{\sim }\,}
\def\simle{\, {}^<_{\sim }\,}

\head{Non-Standard Big Bang Nucleosynthesis 
      Scenarios}
     {K.~Jedamzik}
     {Max-Planck-Institut f\"ur Astrophysik, Karl-Schwarzschild-Str.1,
     D-85748 Garching}

\subsection*{Abstract}

An overview of non-standard big bang nucleosynthesis (BBN) scenarios
is presented.  Trends and results of the light-element nucleosynthesis
in BBN scenarios with small-scale or large-scale inhomogeneity, the
presence of antimatter domains, stable or unstable massive neutrinos,
neutrino oscillations, neutrino degeneracy, or massive decaying
particles are summarized.

\subsection*{Introduction}

Light-element nucleosynthesis during the BBN epoch below cosmic
temperatures $T\approx 1\,$MeV contributes significantly to $^4$He,
$^3$He and $^7$Li abundances and likely, to all of the $^2$H abundance
observed throughout the universe.  BBN is a freeze-out process from
nuclear statistical equilibrium such that light-element abundance
yields are sensitively dependent on the cosmic conditions during the
BBN era as well as the properties of neutrinos governing the
freeze-out process from weak equilibrium.  Calculations of abundance
yields in a standard BBN scenario are performed under the assumptions
of a universe homogeneous in the baryon-to-photon ratio, with massless
neutrinos and vanishing neutrino chemical potentials, and in the
absence of massive decaying particles or other degrees of freedom.
For reviews on the physics of standard BBN see \cite{kj:skm93}.  I
presently summarize results and trends in theoretical calculations of
non-standard BBN scenarios where one of the above assumptions is
relaxed.  This summary is in no way intended to be complete in
discussing all possible modifications to a standard BBN scenario or in
referencing the thousands of articles on non-standard BBN.  I
apologize for including only key references due to the limited scope
of these proceedings and for presenting my personal view of the field
of non-standard BBN.  For two excellent reviews on non-standard BBN
the reader is referred to \cite{kj:mm93}.  The determination of
observationally inferred primordial abundance constraints, which
represents possibly the most important branch of the field of big bang
nucleosynthesis at present, will not be discussed here. In what
follows, abundance yields in non-standard BBN will either be given in
relation to abundance yields in standard BBN or in absolute values and
shall be understood as indicative of approximate trends.

\subsection*{Non-Standard BBN}

In the following a list of non-standard BBN scenarios, their
respective modifications to standard BBN (hereafter; SBBN), as well as
trends and results in these scenarios are given.

\subsubsection*{a) Inhomogeneity}

The baryon-to-photon ratio $\eta$ is the one free parameter in
SBBN. Any inhomogeneity in this quantity results in modified
nucleosynthesis yields which depend on the typical amplitude and
spatial seperation scale of inhomogeneities.  Substantial changes in
the abundance yields only result when $\delta\eta /\eta \simge 1$.

\vskip 0.1in
\noindent
{\em i) Inhomogeneity in the Baryon-to-Photon Ratio on Small Mass
Scales:}
\vskip 0.05in

\noindent
Fluctuations in $\eta$ which may arise on sub-horizon scales at
earlier cosmic epochs as, for example, possibly during a first-order
QCD transition or electroweak transition, result in a highly
nonstandard BBN scenario.  The nucleosynthesis in an environment with
$\eta$-fluctuations is characterized by coupled nuclear reactions and
hydrodymamic processes, such as baryon diffusion and late-time
expansion of high-density regions \cite{kj:ahs87}.  Fluctuations in
$\eta$ persist down to the onset of BBN provided the mass of an
individual high-density region exceeds $10^{-21}M_{\odot}$.  One of
the main features of such scenarios is the differential diffusion of
neutrons and protons leading to the existence of neutron- and
proton-rich environments. The trend in inhomogeneous BBN is the
overabundant production of $^4$He when compared to SBBN at the same
average $\eta$, nevertheless, there exists parameter space where less
$^4$He than in SBBN is synthesized.  Scenarios which don't overproduce
$^4$He typically have high ($^2$H/H) $\sim 1{-}2\times 10^{-4}$ and
high ($^7$Li/H) $\sim 10^{-9}{-}10^{-8}$. Such BBN may agree with
observational abundance constraints for fractional contributions of
the baryon density to the critical density $\Omega_b$ about 2--3 times
larger than in SBBN, but only in the seemingly unlikely advent of
efficient $^7$Li depletion in population II stars.

\vskip 0.1in
\noindent
{\it ii) Inhomogeneity in the Baryon-to-Photon Ratio on 
Large Mass Scales:}
\vskip 0.05in

\noindent
When the baryonic mass within a typical fluctuation exceeds ($M \simge
10^{-12}M_{\odot}$), baryon diffusion and hydrodymanic processes
during the BBN era are of no significance such that BBN abundance
yields may be given by an average over the SBBN abundance yields of
seperate regions at different $\eta$. For non-linear fluctuations
exceeding the post-recombination Jeans mass $M \simge 10^5M_{\odot}$,
which may exist in primordial isocurvature baryon (PIB) models for
structure formation, early collapse of high-density ($\eta$) regions
is anticipated \cite{kj:sm86}. The nucleosynthesis yields of
collapsing regions may be excluded from the primordial abundance
determination if either dark objects form or significant early star
formation in such high-density regions occurs.  If only low-density
regions contribute to the observable primordial abundances,
characteristic average abundance yields for scenarios designed to
possibly agree with observationally inferred primordial abundances
are: ($^2$H/H) $\sim 1{-}3\times 10^{-4}$, ($^{7}$Li/H) $\sim 5\times
10^{-10}{-}2\times 10^{-9}$, and $^4$He mass fraction $Y_p\approx
0.22{-}0.25$ at a total $\Omega_b \simle 0.2$ (i.e.~including possible
dark objects), larger than inferred from a SBBN scenario
\cite{kj:jf95,kj:cos95}.  One feature of such models is the prediction
of fairly large intrinsic spatial variations in the primordial
abundances, which may be observationally tested by ($^2$H/H)
determinations in Lyman-limit systems \cite{kj:jf95}.  These models
may only agree with observationally inferred abundance limits when
there are no fluctuations below $M \simle 10^5M_{\odot}$ and collapse
efficiencies of high-density regions are large.

\vskip 0.1in
\noindent
{\it iii) Matter/Antimatter Domains:}
\vskip 0.05in

\noindent
A distribution of small-scale matter/antimatter domains in
baryon-asymmetric universes (i.e.~where net $\eta \neq 0$) may result
from electroweak baryogenesis scenarios. If the baryon (antibaryon)
mass in individual domains is $\simge 10^{-21}M_{\odot}$ the BBN
process in such scenarios is characterized by differential diffusion
of neutrons (antineutrons) and protons (antiprotons) which causes a
preferential annihilation of antimatter on neutrons
\cite{kj:rj97}. When annihilation of antimatter occurs before
significant $^4$He synthesis ($T\simge 80$keV) but after weak
freeze-out ($T\simle 1\,$MeV) 
a modest to substantial reduction of $Y_p$
results.  When annihilations occur mainly after $^4$He synthesis the
dominant effect is significant production of $^3$He and $^2$H
\cite{kj:betal88}, with $^3$He/$^2$H ratios likely to be in conflict
with observational constraints.

\newpage

\subsubsection*{b) Non-standard Neutrino Properties}

The BBN process may be approximated as the incorporation of all
available neutrons into $^4$He nuclei at a temperature $T\approx
80\,$keV.  The neutron abundance at this temperature, and hence the
final $^4$He mass fraction $Y_p$, may be increased with respect to a
SBBN scenario due to an increased expansion rate of the universe
during the BBN era. An increased expansion rate raises the
neutron-to-proton ratio (hereafter; n/p) at weak freeze-out and
reduces the time for neutron decay to decrease the n/p ratio between
weak freeze-out and significant $^4$He synthesis.  We will refer to
this effect as the ``expansion rate effect.''  In addition,
the n/p ratio at $T\approx 80\,$keV may be either decreased or
increased by introducing additional electron- and/or anti-electron
neutrinos into the plasma.  In SBBN it is assumed that the left-handed
neutrino and right-handed antineutrino seas of three massless, stable
neutrino flavors $\nu_e$, $\nu_{\mu}$, and $\nu_{\tau}$ are populated
and that neutrino chemical potentials do vanish.  Modifications of
these assumptions usually result in either the expansion rate effect
or additional (anti) electron neutrinos, or both. The principal effect
of such modifications is to change the $^4$He abundance, and to a less
observationally significant degree the abundances of other
light-element isotopes.

\vskip 0.1in
\noindent
{\it i) Massive, Long-Lived $\tau$-Neutrinos:}
\vskip 0.05in

\noindent
Neutrinos are considered massless and long-lived in the context of BBN
for neutrino masses $m_{\nu}\simle 100\,$keV and lifetimes
$\tau_{\nu}\simge 10^3\,$s [see ii) below, however, for possible
photodisintegration]. A massive, long-lived $\tau$-neutrino leads to
the expansion rate effect since the contribution to the total energy
density from the rest mass of $\tau$-neutrinos continually increases
as the universe expands between weak freeze-out and $^4$He synthesis,
possibly even resulting in matter domination during the BBN era. BBN
with massive, long-lived $\tau$ neutrinos and for experimentally
allowed $\nu_{\tau}$-masses therefore results in increased $^4$He and
useful limits on the allowed mass of a long-lived $\tau$-neutrino have
been derived \cite{kj:ketal91}.

\vskip 0.1in
\noindent
{\it ii) Massive, Unstable $\tau$-Neutrinos:}
\vskip 0.05in

\noindent
The effects of decaying $\tau$ neutrinos on the light-element
nucleosythesis \cite{kj:ks82} sensitively depend on the decay
products. One distinguishes between (i) decay into sterile particles,
in particular, particles which interact neither weakly with nucleons
interchanging neutrons and protons nor electromagnetically with the
ambient plasma (e.g.~$\nu_{\tau}\mapsto \nu_{\mu} + \phi$, where
$\phi$ is a weakly interacting scalar), (ii) decay into sterile and
electromagnetically interacting particles (e.g.~$\nu_{\tau}\mapsto
\nu_{\mu} + \gamma$), and (iii) decay into (anti) electron neutrinos
and sterile particles (e.g.~$\nu_{\tau}\mapsto \nu_{e} + \phi$ )
\cite{kj:dgt94,kj:ketal94}.  For decay channel (i) and $\tau_{\nu}
\simge 1\,$s increased $^4$He mass fraction results due to the
expansion rate effect which, nevertheless, is weaker than for
long-lived $\tau$-neutrinos since the energy of the (massless) decay
products redshifts with the expansion of the universe.  In contrast,
for life times $\tau_{\nu_{\tau}} \simle 1\,$s and
$m_{\nu_{\tau}}\simge 10\,$MeV it is possible to reduce the $Y_p$
since effectively the distributions of only two neutrino flavors are
populated \cite{kj:kks97}.  Decay channel (ii) would have interesting
effects on BBN but is excluded by observations of supernova 1987A for
$\tau_{\nu_{\tau}} \simle 10^4\,$s.  For decay via channel (iii)
additional (anti) electron neutrinos are injected into the plasma and
their effect depends strongly on the time of injection \cite{kj:ts88}.
When injected early ($\tau_{\nu_{\tau}}\sim 1\,$s), the net result is
a reduction of $Y_p$ \cite{kj:h97} since weak freeze-out occurs at
lower temperatures.  In contrast, when injected late ($\tau_{\nu} \sim
10^2 {-}10^3\,$s) the resulting non-thermal electron neutrinos affect
a conversion of protons into neutrons, yielding higher $Y_p$ and/or
higher $^2$H depending on injection time.  It had been suggested that
a scenario with late-decaying $\nu_{\tau}$ via channel (iii) may
result in the relaxation of BBN bounds on $\Omega_b$ by a factor up to
ten \cite{kj:dgt94}, nevertheless, this possibility seems to be ruled
out now by the current upper laboratory limit on the
$\nu_{\tau}$-mass.  For life times $\tau_{\nu_{\tau}}\simge 10^3\,$s,
modifications of the light-element abundances after the BBN era by
photodisintegration of nuclei may result (cf. radiative decays).

\vskip 0.1in
\noindent
{\it iii) Neutrino Oscillations:}
\vskip 0.05in

\noindent
Neutrino oscillations may occur when at least one neutrino species has
non-vanishing mass and the weak neutrino interaction eigenstates are
not mass eigenstates of the Hamiltonian. One distinguishes between (i)
flavor-changing neutrino oscillations (e.g.~$\nu_{e}\leftrightarrow
\nu_{\mu}$) and (ii) active-sterile neutrino oscillations
(e.g.~$\nu_{e}\leftrightarrow \nu_{s}$).  Here $\nu_s$ may be either
the right-handed component of a $\nu_e$ ($\nu_{\mu}$, $\nu_{\tau}$)
Dirac neutrino, or a fourth family of neutrinos beyond the standard
model of electroweak interactions. In the absence of sterile neutrinos
and neutrino degeneracy, neutrino oscillations have negligible effect
on BBN due to the almost equal number densities of neutrino
flavors. When sterile neutrinos exist, neutrino oscillations may
result into the population of the sterile neutrino distribution,
increasing the energy density, and leading to the expansion rate
effect. The increased $Y_p$ has been used to infer limits in the plane
of the neutrino squared-mass difference and mixing angle
\cite{kj:bd91}.  In the presence of large initial lepton number
asymmetries (see neutrino degeneracy) and with sterile neutrinos, it
may be possible to reduce $Y_p$ somewhat independently from the
detailed initial conditions, through the dynamic generation of
electron (as well as $\mu$ and $\tau$) neutrino chemical potentials
\cite{kj:fv}.

\subsubsection*{c) Neutrino Degeneracy}

It is possible that the universe has net lepton number.  Positive net
cosmic lepton number manifests itself at low temperatures through an
excess of neutrinos over antineutrinos. If net lepton number in either
of the three families in the standard model is about ten orders of
magnitude larger than the net cosmic baryon number BBN abundance
yields are notably affected. Asymmetries between the $\nu_{\mu}$
($\nu_{\tau}$) and $\bar{\nu}_{\mu}$ ($\bar{\nu}_{\tau}$) number
densities result in the expansion rate effect only, whereas
asymmetries between the $\nu_e$ and $\bar{\nu}_e$ number densities
induce a change in the weak freeze-out (n/p) ratio as well.  Since the
expansion rate effect leads to increased $^4$He production $\nu_{\mu}$
($\nu_{\tau}$) degeneracy may be constrained.  However, one may find
combinations of $\nu_{\mu}$, $\nu_{\tau}$, {\it and} $\nu_e$ chemical
potentials which are consistent with observational abundance
constraints for $\Omega_b$ much larger than that inferred from SBBN
\cite{kj:by77}.  Nevertheless, such solutions not only require large
chemical potentials but also an asymmetry between the individual
chemical potentials of $\nu_e$ and $\nu_{\mu}$ ($\nu_{\tau}$).
Asymmetries between the different flavor degeneracies may be erased in
the presence of neutrino oscillations.

\subsubsection*{d) Massive Decaying Particles}

The out-of-equilibrium decay or annihilation of long-lived particles
($\tau \simge 0.1\,$s), 
such as light supersymmetric particles, as well
as non-thermal particle production by, for example, evaporating
primordial black holes or collapsing cosmic string loops, during, or
after, the BBN era may significantly alter the BBN nucleosynthesis
yields \cite{kj:detal78}.  The decays may be classified according to
if they are (i) radiative or (ii) hadronic, in particular, whether the
electromagnetic or strong interactions of the injected non-thermal
particles are most relevant.

\vskip 0.1in
\noindent
{\it i) Radiative decays:}
\vskip 0.05in

\noindent
Electromagnetically interacting particles (i.e.~$\gamma$, $e^{\pm}$)
thermalize quickly in the ambient photon-pair plasma at high
temperatures, such that they usually have little effect on BBN other
than some heating of the plasma. In contrast, if radiative decays
occur at lower temperatures after a conventional BBN epoch, they
result in a rapidly developing $\gamma$-$e^{\pm}$ cascade which only
subsides when individual $\gamma$-rays of the cascade do not have
enough energy to further pair-produce $e^{\pm}$ on the ambient cosmic
background radiation. The net result of this cascade is a less rapidly
developing $\gamma$-ray background whose properties only depend on the
total amount of energy released in electromagnetically interacting
particles and the epoch of decay.  At temperatures $T\simle 5\,$keV 
the
most energetic $\gamma$-rays in this background may photodissociate
deuterium, at temperatures $T\simle 0.5\,$keV 
the photodisintegration of
$^4$He becomes possible.  Radiative decays at $T\approx 5\,$keV may
result in a BBN scenario with low $^2$H {\it and} low $^4$He if an
ordinary SBBN scenario at low $\eta$ is followed by an epoch of
deuterium photodisintegration \cite{kj:hkm96}.  Radiative decays at
lower temperatures may produce signficant amounts of $^2$H and
$^3$He. Nevertheless, this process is unlikely to be the sole producer
of $^2$H due to resulting observationally disfavored $^3$He/$^2$H
ratio. The possible overproduction of $^3$He and $^2$H by the
photodisintegration of $^4$He has been used to place meaningful limits
on the amount of non-thermal, electromagnetically interacting energy
released into the cosmic background. These limits may actually be more
stringent than comparable limits from the distortion of the cosmic
microwave background radiation \cite{kj:setal96}.

\vskip 0.1in
\noindent
{\it ii) Hadronic decays:}
\vskip 0.05in

\noindent
The injection of hadrons into the plasma, such as $\pi^{\pm}$'s,
$\pi^0$'s and nucleons, may affect light-element nucleosythesis
during, or after, the BBN era and by the destruction {\it and}
production of nuclei. In general, possible scenarios and reactions are
numerous. If charged hadrons are produced with high energies below
cosmic temperature $T\simle$keV they may cause a cascade leading to
the possibility of photodisintegration of nuclei (see radiative
decays).  Through charge exchange reactions the release of about ten
pions per nucleon at $T\approx 1\,$MeV results in a significant
perturbation of weak freeze-out and increased $Y_p$ \cite{kj:rs88}.  A
fraction of only $\sim 10^{-3}$ antinucleons per nucleon may cause
overproduction of $^3$He and $^2$H through antinucleon-$^4$He
annihilations.  It is also conceivable that the decaying particle
carries baryon number such that the cosmic baryon number is created
{\it during} the BBN era.  A well-studied BBN scenario is the
injection of high-energy ($\sim 1\,$GeV) nucleons created in hadronic
jets which are produced by the decay of the parent particle
\cite{kj:detal88}.  If released after a conventional BBN era these
high-energy nucleons may spall pre-existing $^4$He, thereby producing
high-energy $^2$H, $^3$H, and $^3$He as well as neutrons. Such
energetic light nuclei may initiate an epoch of non-thermal
nucleosythesis.  It was found that the nucleosynthesis yields from a
BBN scenario with hadro-destruction/production and photodisintegration
may result in abundance yields independent of $\eta$ for a range in
total energy injection, and half-life of the decaying
particles. Nevertheless, such scenarios seem to produce $^6$Li in
conflict with observational constraints.

\subsubsection*{e) Other Modifications to BBN}

There are many other variants to a standard big bang nucleosythesis
scenario which have not been mentioned here.  These include
anisotropic expansion, variations of fundamental constants, theories
other than general relativity, magnetic fields during BBN,
superconducting cosmic strings during BBN, among others.  The
influence of many, but not all, of those scenarions on BBN is due to
the expansion rate effect. Studies of such variants is of importance
mainly due to the constraints they allow one to derive on the
evolution of the early universe.

}\newpage{


\head{Big Bang Nucleosynthesis With Small-Scale\\
     Matter-Antimatter Domains}
     {J.B.~Rehm and K.~Jedamzik}
     {Max-Planck-Institut f\"ur Astrophysik, Karl-Schwarzschild-Str.1,
     D-85748 Garching}

\noindent
Big Bang Nucleosynthesis (BBN) is one of the furthest back-reaching
cosmological probes available.  By means of comparing the predicted
and observationally inferred light element abundances the physical
conditions as early as a few seconds after the Big Bang can be
scrutinized~\cite{jr:MM:93}.  Nevertheless, presence of
matter-antimatter domains during BBN on length scales comparable in
size to the neutron diffusion length at temperatures between 10 and
10$^{-2}$~MeV (see e.g.~\cite{jr:JF:94}) has so far not been
investigated.  To our knowledge, the influence of antimatter on BBN
has only been studied \cite{jr:BBB:88} with respect to a homogeneous
injection of antiprotons into the primordial plasma after the end of
BBN ($T < 10$~keV).  The conclusion of such studies is that only a
small fraction of about 10$^{-3}$ antiprotons per proton is allowed to
be injected after the BBN epoch for the abundances of deuterium and
$^3$He produced in $\bar{p}\,{}^4{\rm He}$ reactions not to exceed the
observationally inferred values.

In this work \cite{jr:RJ:97} we discuss the BBN in a baryo-asymmetric
universe with a inhomogeneous distribution of matter-antimatter
domains.  For possible scenarios which may yield matter-antimatter
domains during an electroweak baryogenesis scenario see
e.g.~Refs.~\cite{jr:CPR:94,jr:GS:97a}.  As long as the diffusion
length of neutrons and antineutrons is shorter than the size of a
typical domain, matter and antimatter remain segregated. Note that
proton diffusion is suppressed due to Coulomb scattering.  Therefore,
annihilations may likely only proceed when (anti)neutrons diffuse out
of their domains. We identify three main scenarios: 1.~Annihilations
before weak freeze-out: No effect on the light-element abundances
because the $n/p$-ratio, which governs the abundances, is reset to the
equilibrium value by the fast weak interactions.  2.~Annihilations
between weak freeze-out ($\approx1$ MeV) and the onset of $^4$He
synthesis ($T\approx0.08$ MeV): The $n/p$-ratio is strongly affected.
Annihilation of antimatter mostly occurs on neutrons, since neutrons
can both be annihilated by antineutrons diffusing into the matter
region and diffuse themselves into the antimatter region to annihilate
there on antiprotons and antineutrons. Protons on the other hand are
confined to the matter region and can only be annihilated by diffusing
antineutrons.  By providing this effective sink for neutrons, the
$^4$He abundance may be substantially lower than in a standard BBN
scenario at the same net entropy-per-baryon, simply because there are
less neutrons left to build up $^4$He.  In extreme cases, when all
neutrons have annihilated on antimatter before the onset of $^4$He
synthesis, zero net $^4$He production results. This may well represent
the only known BBN scenario in a baryo-asymmetric universe which
precludes the production of any light elements.  3.~Annihilations well
after BBN: This scenario was already mentioned above.  Annihilations
of only small amounts of antimatter on $^4$He may easily overproduce
the cosmic deuterium and $^3$He abundances.

In summary, the figure illustrates that matter-antimatter domains
present at the BBN epoch may have dramatic effects on the net
synthesized $^4$He abundance. By comparison, the abundances of other
light isotopes are generally less affected. A more detailed analysis
of BBN with matter-antimatter domains will be presented in a
forthcoming work \cite{jr:RJ:97}.

\begin{figure}[ht]
  \centerline{\epsfxsize=0.6\textwidth\epsffile{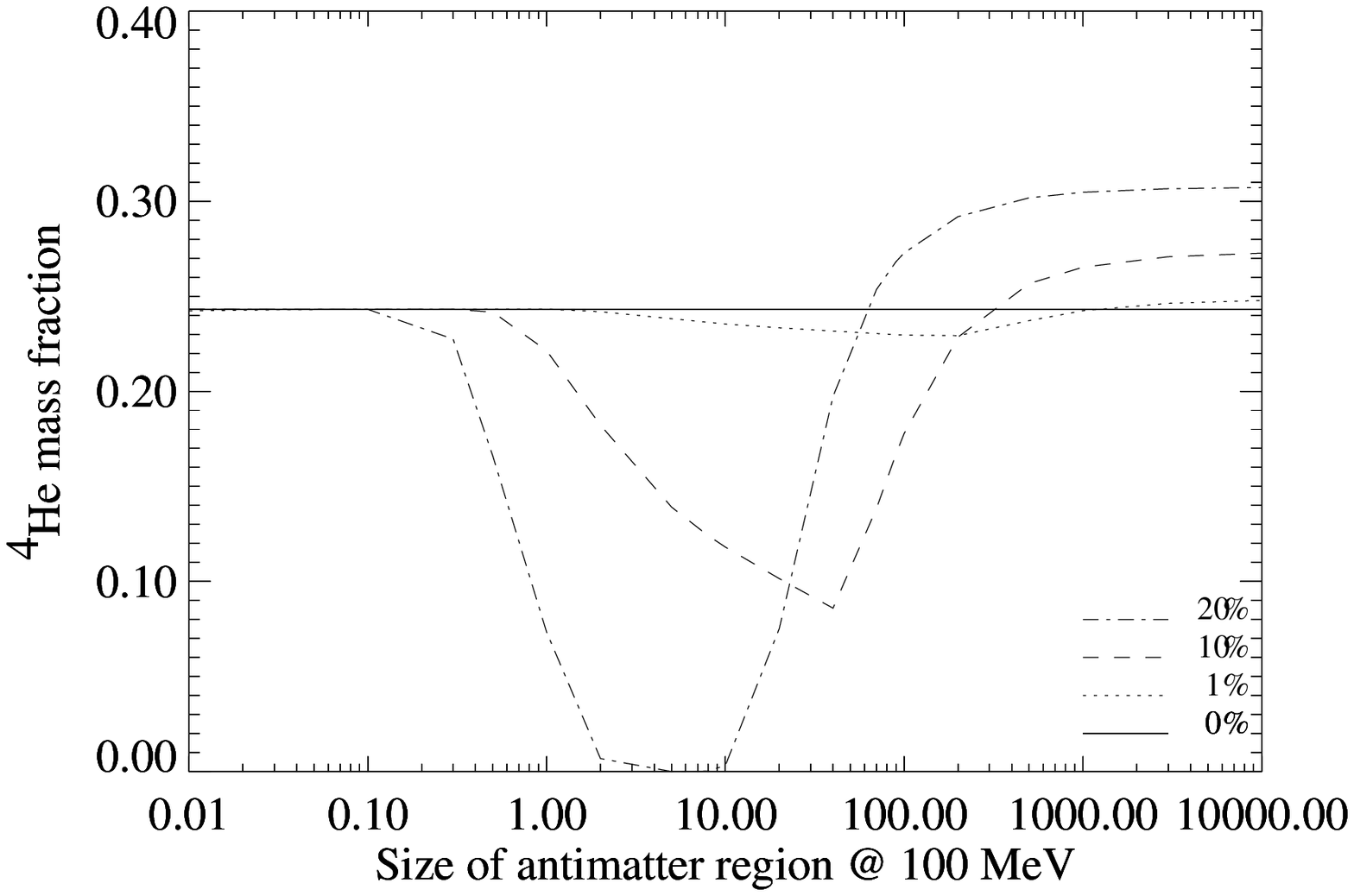}}
  \caption{Predicted $^4$He mass fraction as a function of the
  typical antimatter domain size.
  Length scales are given in proper units at a temperature
  of $T = 100$~MeV. The different lines give results for different
  amounts of antimatter, i.e. the ratio of $R = N_{\rm antimatter}/
  N_{\rm matter}$ takes values of 0, 1, 10, and 20\%, respectively.
  All models have the same net baryon density.}
\end{figure}

}\newpage{


\head{Neutrinos and Structure Formation in the Universe}
     {Matthias Bartelmann}
     {MPI f\"ur Astrophysik, P.O.\ Box 1523, D--85740 Garching,
      Germany}

\subsection*{Abstract}

I review the standard theory of structure formation in the Universe,
starting from the assumptions that (i)~the Universe is globally
described by a Friedmann-Lema{\^\i}tre model, (ii)~structure forms via
gravitational instability from primordial density fluctuations, and
(iii)~these primordial density fluctuations were Gaussian in nature. I
describe how density fluctuations grow in time, what the density
perturbation power spectrum looks like, and why hot dark matter like
neutrinos has a considerable impact on the process of structure
formation.

\subsection*{Introduction}

The simplest and most widely accepted theory of structure formation in
the Universe starts from the following three assumptions:

\begin{enumerate}
\item Globally, the Universe is well described by the
  Friedmann-Lema{\^\i}tre-Robertson-Walker (FLRW) model.
\item Structure formed via gravitational instability from primordial
  density fluctuations.
\item These primordial density fluctuations were Gaussian in nature.
\end{enumerate}

By the first assumption, the global dynamics of the Universe is
described in terms of four parameters, viz.\ the density of ordinary
and relativistic matter, the Hubble constant, and the cosmological
constant. The second assumption, together with the dynamics implied by
the first, specifies how structures grow in time. The third assumption
asserts that the statistics of the density fluctuation field is
completely specified by its mean and variance. By definition, the mean
density {\em contrast\/} is zero, and the variance is specified by the
{\em power spectrum\/},
\begin{equation}
  P(k) = \left\langle|\delta^2(\vec k)|\right\rangle\;,
\label{MBeq:1}
\end{equation}
where $\vec k$ is the wave vector of the density perturbation. The
power spectrum itself does not depend on the direction of $\vec k$
because of the (assumed) isotropy of the density fluctuation field.

These three basic assumptions provide the structure of this article. I
will first describe the global dynamics of FLRW models, then discuss
how density fluctuations grow in an expanding universe, and finally
describe the gross features of the density fluctuation power spectrum,
before I come to the conclusions.

\subsection*{Global Dynamics}
\label{MBsec:2}

The scale factor $a$ of a FLRW universe is governed by Friedmann's
equation,
\begin{equation}
  \left(\frac{\dot{a}}{a}\right)^2 = 
  \frac{8\pi G}{3}\,\rho - \frac{K\,c^2}{a^2}\;,
\label{MBeq:2}
\end{equation}
where I have ignored the cosmological constant for
simplicity. Neglecting the curvature term (as one can do at early
times, see below), and approximating $\dot{a}\sim a/t$,
eq.~(\ref{MBeq:2}) implies the familiar result that the expansion time
scale $t\propto\rho^{-1/2}$. The scale factor $a$ grows monotonically
with the cosmic time $t$ and can therefore be used as a time
variable. Boundary conditions for (\ref{MBeq:2}) are chosen such that
$a(t_0)=1$ at the present cosmic time $t_0$, and $t=0$ for $a=0$. The
present value of the Hubble parameter, $H=\dot{a}/a$, is called the
Hubble constant $H_0$. $K$~is the curvature of spatial hypersurfaces
of space-time, and $\rho$ is the total matter density. The quantity
\begin{equation}
  \rho_{\rm cr} \equiv \frac{3\,H_0^2}{8\pi G} \approx
  2\times10^{-29}\,h^2\,{\rm g\,cm^{-3}}
\label{MBeq:3}
\end{equation}
is called the critical density, for reasons that will become clear
below. The density parameter $\Omega_0$ is the current total matter
density $\rho(t_0)=\rho_0$ in units of $\rho_{\rm cr}$,
\begin{equation}
  \Omega_0 \equiv \frac{\rho_0}{\rho_{\rm cr}}\;.
\label{MBeq:4}
\end{equation}
The Hubble constant is commonly written as
\begin{equation}
  H_0 = 100\,h\,{\rm km\,s^{-1}\,Mpc^{-1}} \approx
  3.2\times10^{-18}\,h\,{\rm s^{-1}}\;,
\label{MBeq:5}
\end{equation}
where $0.5\le h\le1$ expresses our ignorance of $H_0$. Note that the
Hubble parameter has the dimension (time)$^{-1}$, so that $H^{-1}$
provides the natural time scale for the expansion of the
Universe. Evaluating Friedmann's equation (\ref{MBeq:2}) at time
$t_0$, it follows
\begin{equation}
  K = \left(\frac{c}{H_0}\right)^2\,(\Omega_0-1)\;.
\label{MBeq:6}
\end{equation}
In other words, a universe which contains the critical matter density
is spatially flat.

The first law of thermodynamics, ${\rm d}U+p{\rm d}V=0$, can be
written as
\begin{equation}
  \frac{{\rm d}(\rho c^2 a^3)}{{\rm d}t} + 
  p\frac{{\rm d}(a^3)}{{\rm d}t} = 0\;.
\label{MBeq:7}
\end{equation}
For ordinary matter, $\rho c^2\gg p\sim0$, hence $\rho\propto
a^{-3}$. For relativistic matter, $p=\rho c^2/3$, hence $\rho\propto
a^{-4}$. Therefore, the matter density changes with $a$ as
\begin{equation}
  \rho(a) = \rho_0\,a^{-n} = \rho_{\rm cr}\,\Omega_0\,a^{-n}\;,
\label{MBeq:8}
\end{equation}
where $n=3$ for ordinary matter (``dust''), and $n=4$ for relativistic
matter (``radiation'').

Summarising, we can cast Friedmann's equation into the form,
\begin{equation}
  H^2(a) = H_0^2\,
  \left[\Omega_0\,a^{-n} - (\Omega_0-1)\,a^{-2}\right]\;.
\label{MBeq:9}
\end{equation}
Since the universe expands, $a<1$ for $t<t_0$, and so the expansion
rate $H(a)$ was larger in the past. At very early times, $a\ll1$, the
first term in eq.~(\ref{MBeq:9}) dominates because $n\ge3$, and we can
write
\begin{equation}
  H(a) = H_0\,\Omega^{1/2}\,a^{-n/2}\;.
\label{MBeq:10}
\end{equation}
This is called the Einstein-de Sitter limit of Friedmann's equation.

\subsection*{Photon and Neutrino Backgrounds}

The Universe is filled with a sea of radiation. It has the most
perfect black body spectrum ever measured, with a temperature of
$T_{\gamma,0}=2.73\,{\rm K}$. Apart from local perturbations and
kinematic effects, this {\em Cosmic Microwave Background\/} (CMB) is
isotropic to about one part in $10^5$. It is a relic of the photons
that were produced in thermal equilibrium in the early Universe.
The energy density in the CMB is
\begin{equation}
  u_{\gamma,0} = 2\,\frac{\pi^2}{30}\,
  \frac{(kT_{\gamma,0})^4}{(\hbar c)^3}\;,
\label{MBeq:11}
\end{equation}
and the equivalent matter density is
$\rho_{\gamma,0}=u_{\gamma,0}/c^2$. We have seen before that the
matter density in radiation changes with the scale factor like
$a^{-4}$, hence
\begin{equation}
  T_{\gamma}(a) = a^{-1}\,T_{\gamma,0}\;.
\label{MBeq:12}
\end{equation}
Converting the matter density in the CMB to the density parameter as
in eq.~(\ref{MBeq:4}), we find
\begin{equation}
  \Omega_{\gamma,0} = \frac{u_{\gamma,0}}{c^2\rho_{\rm cr}} =
  2.4\times10^{-5}\,h^{-2}\;.
\label{MBeq:13}
\end{equation}
The energy density in radiation today is therefore about five orders
of magnitude less than that in ordinary matter.

Like photons, neutrinos were produced in thermal equilibrium in the
early universe. Because of their weak interaction, they decoupled from
the plasma at very early times when the temperature of the universe
was $kT\sim1\,{\rm MeV}$. Electrons and positrons remained in
equilibrium until the temperature dropped below their rest-mass
energy, $kT\sim0.5\,{\rm MeV}$. The consequent annihilation of
electron-positron pairs heated the photons, but not the neutrinos
since they had already decoupled. Therefore, the entropy of the
electron-positron pairs $S_{\rm e}$ was completely dumped into the
entropy of the photon sea $S_\gamma$. Hence,
\begin{equation}
  (S_{\rm e} + S_\gamma)_{\rm before} = (S_\gamma)_{\rm after}\;,
\label{MBeq:14}
\end{equation}
where `before' and `after' refer to the electron-positron annihilation
time. Ignoring constant factors, the entropy per particle species is
$S\propto gT^3$, where $g$ is the statistical weight of the
species. For bosons, $g=1$, and for fermions, $g=7/8$ per spin
state. Before the annihilation, we thus have the total statistical
weight $g_{\rm before}=4\cdot7/8+2=11/2$, while after the
annihilation, $g_{\rm after}=2$ because only photons remain. From
eq.~(\ref{MBeq:14}),
\begin{equation}
  \left(\frac{T_{\rm after}}{T_{\rm before}}\right)^3 =
  \frac{g_{\rm before}}{g_{\rm after}} = \frac{11}{4}\;.
\label{MBeq:15}
\end{equation}
After the electron-positron annihilation, the neutrino temperature is
therefore lower than the photon temperature by the factor
$(4/11)^{1/3}$. In particular, the neutrino temperature today is
\begin{equation}
  T_{\nu,0} = \left(\frac{4}{11}\right)^{1/3}\,T_{\gamma,0} =
  1.95\,{\rm K}\;.
\label{MBeq:16}
\end{equation}
Although the neutrinos are long out of thermal equilibrium, their
distribution function remained unchanged since they decoupled, except
that their temperature gradually dropped. We can therefore calculate
their energy density as for the photons, and convert this to the
equivalent cosmic density parameter. The result is
\begin{equation}
  \Omega_{\nu,0} = \frac{u_{\nu,0}}{c^2\rho_{\rm cr}} =
  2.8\times10^{-6}\,h^{-2}
\label{MBeq:17}
\end{equation}
per neutrino species. Likewise, their number density is given by the
Fermi distribution with temperature $T_{\nu,0}$,
\begin{equation}
  n_{\nu,0} = \frac{3\zeta(3)}{2\pi^2}\,
  \left(\frac{kT_{\nu,0}}{\hbar c}\right)^3 \approx
  113\,{\rm cm}^{-3}
\label{MBeq:18}
\end{equation}
per neutrino species. Suppose now there is one neutrino species which
has rest mass $m_\nu>0$, while the other species are massless. For the
dark matter to be dominated by neutrinos, $m_\nu$ needs to satisfy
\begin{equation}
  m_\nu\,n_{\nu,0} \stackrel{!}{=} \Omega_0\,\rho_{\rm cr}\;,
\label{MBeq:19}
\end{equation}
hence,
\begin{equation}
  m_\nu\,c^2 \simeq 95\,{\rm eV}\,(\Omega_0h^2)\;.
\label{MBeq:20}
\end{equation}
Adding the energy density of the two massless neutrino species to that
of the photons, the total equivalent mass density in relativistic
particles is
\begin{equation}
  \Omega_{\rm R,0} = 3.0\times10^{-5}\,h^{-2}\;.
\label{MBeq:21}
\end{equation}

\subsection*{The Horizon}

The size of causally connected regions of the Universe is called the
{\em horizon size\/}. It is given by the distance a photon can travel
in the time since the Big Bang. Since the appropriate time scale is
provided by the inverse Hubble parameter, the horizon size is $d'_{\rm
H}=c\,H(a)^{-1}$, and the {\em comoving\/} horizon size is
\begin{equation}
  d_{\rm H}(a) = \frac{c}{a\,H(a)} =
  \frac{c}{H_0}\,\Omega^{-1/2}\,a^{n/2-1}\;,
\label{MBeq:21a}
\end{equation}
where we have inserted the Einstein-de Sitter limit (\ref{MBeq:10}) of
Friedmann's equation. $c\,H_0^{-1}=3\,h^{-1}\,{\rm Gpc}$ is called the
{\em Hubble radius\/}.

The previous calculations show that the matter density today
is completely dominated by ordinary rather than relativistic
matter. But since the relativistic matter density grows faster with
decreasing scale factor $a$ than the ordinary matter density, there
had to be a time $a_{\rm eq}\ll1$ before which relativistic matter
dominated. The condition
\begin{equation}
  \Omega_{{\rm R},0}\,a_{\rm eq}^{-4} \stackrel{!}{=}
  \Omega_0\,a_{\rm eq}^{-3}
\label{MBeq:22}
\end{equation}
yields
\begin{equation}
  a_{\rm eq} = 3.0\times10^{-5}\,(\Omega_0h^2)^{-1}\;.
\label{MBeq:23}
\end{equation}
We shall see later that the horizon size at the time $a_{\rm eq}$
plays a very important r\^ole for structure formation. Under the
simplifying assumption that matter dominated completely for all $a\ge
a_{\rm eq}$ (i.e.\ ignoring the contribution from radiation to the
matter density), eq.~(\ref{MBeq:21a}) yields
\begin{equation}
  d_{\rm H}(a_{\rm eq}) =
  \frac{c}{H_0}\,\Omega_0^{-1/2}\,a_{\rm eq}^{1/2}
  \approx 13\,(\Omega_0\,h^2)^{-1}\,{\rm Mpc}\;.
\label{MBeq:24}
\end{equation}
The (photon) temperature at $a_{\rm eq}$ is
\begin{equation}
  kT_\gamma(a_{\rm eq}) = a_{\rm eq}^{-1}\,kT_{\gamma,0}
  \simeq 8\,{\rm eV}\;.
\label{MBeq:25}
\end{equation}
This is much lower than the rest-mass energy of a cosmologically
relevant neutrino. Hence, if there is a neutrino species with mass
given by eq.~(\ref{MBeq:20}), it becomes non-relativistic much earlier
than $a_{\rm eq}$.

\subsection*{Growth of Density Fluctuations}

\subsubsection*{a) Linear Growth}

Imagine a FLRW background model into which a spherical, homogeneous
region is embedded which has a slightly different density than the
surroundings. Because of the spherical symmetry, this sphere will
evolve like a universe of its own; in particular, it will obey
Friedmann's equation. Let indices 0 and 1 denote quantities within
the surrounding universe and the perturbed region, respectively. Then,
Friedmann's equation reads
\begin{eqnarray}
  H_1^2 + \frac{Kc^2}{a_1^2} &=& \frac{8\pi G}{3}\,\rho_1
  \nonumber\\
  H_0^2 &=& \frac{8\pi G}{3}\,\rho_0
\label{MBeq:26}
\end{eqnarray}
for the perturbation and for the surrounding universe,
respectively. We have ignored the curvature term in the second
equation because the background has $K=0$ at early times.

We now require that the perturbed region and the surrounding
universe expand at the same rate, hence
\begin{equation}
  H_0 \stackrel{!}{=} H_1\;.
\label{MBeq:27}
\end{equation}
Assuming that the density inside the perturbed region only
weakly deviates from that of the background, we can put $a_1\simeq
a_0$, and eqs.~(\ref{MBeq:26}) and (\ref{MBeq:27}) yield
\begin{equation}
  \delta \equiv \frac{\rho_1-\rho_0}{\rho_0} = \frac{3\,Kc^2}{8\pi
  G}\,\frac{1}{\rho_0a^2} \propto (\rho_0a^2)^{-1}\;.
\label{MBeq:28}
\end{equation}
We have seen before that $\rho_0\propto a^{-n}$, with $n=4$ before
$a_{\rm eq}$ and $n=3$ thereafter. It thus follows that
\begin{equation}
  \delta \propto \left\{\begin{array}{ll}
    a^2 & \hbox{before $a_{\rm eq}$,}\\
    a   & \hbox{after $a_{\rm eq}$.}\\
  \end{array}\right.
\label{MBeq:29}
\end{equation}
This heuristic result is confirmed by accurate, relativistic and
non-relativistic perturbation calculations.

\subsubsection*{b) Suppression of Growth}

A perturbation of wavelength $\lambda$ is said to ``enter the
horizon'' when $\lambda=d_{\rm H}$. If $\lambda<d_{\rm H}(a_{\rm
eq})$, the perturbation enters the horizon while radiation still
dominates. Until $a_{\rm eq}$, the expansion time scale, $t_{\rm
exp}$, is determined by the radiation density $\rho_{\rm R}$, and
therefore it is shorter than the collapse time scale of the dark
matter, $t_{\rm DM}$:
\begin{equation}
  t_{\rm exp} \sim (G\rho_{\rm R})^{-1/2} <
  (G\rho_{\rm DM})^{-1/2} \sim t_{\rm DM}\;.
\label{MBeq:30}
\end{equation}
In other words, the fast radiation-driven expansion prevents
dark-matter density perturbations from collapsing. Light can only
cross regions that are smaller than the horizon size. The suppression
of growth due to radiation is therefore restricted to scales smaller
than the horizon, and larger-scale perturbations remain
unaffected. This suggests that the horizon size at $a_{\rm eq}$,
$d_{\rm H}(a_{\rm eq})$, sets an important scale for structure growth.

\begin{figure}[ht]
  \centerline{\epsfxsize=0.6\textwidth\epsffile{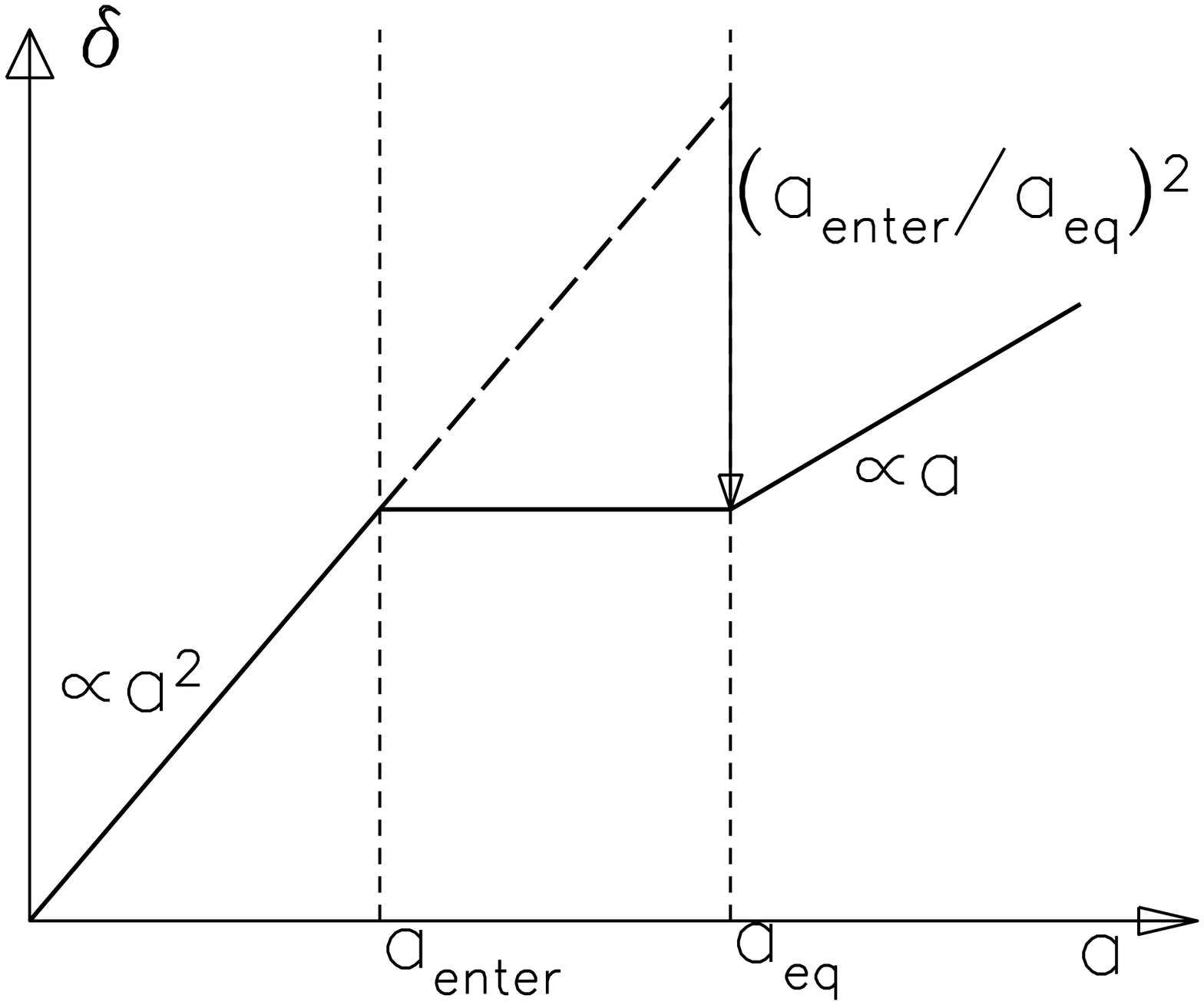}}
\caption{Sketch illustrating the suppression of structure growth
  during the radiation-dominated phase. The perturbation grows
  $\propto a^2$ before $a_{\rm eq}$, and $\propto a$ thereafter. If
  the perturbation is smaller than the horizon at $a_{\rm eq}$, it
  enters the horizon at $a_{\rm enter}<a_{\rm eq}$ while radiation is
  still dominating. The rapid radiation-driven expansion prevents the
  perturbation from growing further. Hence it stalls until $a_{\rm
  eq}$. By then, its amplitude is smaller by $f_{\rm sup}=(a_{\rm
  enter}/a_{\rm eq})^2$ than it would be without suppression.}
\label{MBfig:1}
\end{figure}

Figure~\ref{MBfig:1} illustrates the growth of a perturbation with
$\lambda<d_{\rm H}(a_{\rm eq})$, that is small enough to enter the
horizon at $a_{\rm enter}<a_{\rm eq}$. It can be read off from the
figure that such perturbations are suppressed by the factor
\begin{equation}
  f_{\rm sup} = \left(\frac{a_{\rm enter}}{a_{\rm eq}}\right)^2\;.
\label{MBeq:31}
\end{equation}
According to eq.~(\ref{MBeq:21a}), the comoving horizon size scales
with $a$ as $d_{\rm H}\propto a^{n/2-1}$. When a perturbation enters
the horizon, $\lambda=d_{\rm H}(a_{\rm enter})$, which yields $a_{\rm
enter}\propto\lambda^{2/(n-2)}=k^{-2/(n-2)}$, or $a_{\rm
enter}\propto\lambda=k^{-1}$ during the radiation-dominated
epoch. Thus we obtain from (\ref{MBeq:31})
\begin{equation}
  f_{\rm sup} = \left(\frac{\lambda}{d_{\rm H}(a_{\rm eq})}\right)^2
  = \left(\frac{k_0}{k}\right)^2\;,
\label{MBeq:32}
\end{equation}
where $k_0$ is the wave number corresponding to $d_{\rm H}(a_{\rm
eq})$.

\subsection*{The Perturbation Spectrum}

Consider now the primordial perturbation spectrum at very early times,
$P_{\rm i}(k)=|\delta^2_{\rm i}(k)|$. Since the density contrast grows
as $\delta\propto a^{n-2}$, the spectrum grows as $P(k)\propto
a^{2(n-2)}$. At $a_{\rm enter}$, the spectrum has therefore changed to
\begin{equation}
  P_{\rm enter}(k)=a_{\rm enter}^{2(n-2)}\,P_{\rm i}(k)\propto
  k^{-4}\,P_{\rm i}(k)\;.
\label{MBeq:33}
\end{equation}
It is commonly assumed that the total power of the density
fluctuations at $a_{\rm enter}$ should be scale-invariant. This
implies $k^3P_{\rm enter}(k)={\rm const.}$, or $P_{\rm
enter}(k)\propto k^{-3}$. Accordingly, the primordial spectrum has to
scale with $k$ as $P_{\rm i}(k)\propto k$. This {\em
scale-invariant\/} spectrum is called the Harrison-Zel'dovich-Peebles
spectrum. Combining that with the suppression of small-scale modes, we
arrive at
\begin{equation}
  P(k) \propto \left\{\begin{array}{ll}
    k & \hbox{for $k<k_0$,}\\
    k^{-3} & \hbox{for $k>k_0$.}\\
  \end{array}\right.
\label{MBeq:34}
\end{equation}

\subsection*{Damping by Free Streaming}

If the particles of the dominant dark-matter component are fast enough
(``hot''), perturbations can be washed out by free streaming. To see
this, consider the so-called {\em free-streaming length\/}
$\lambda_{\rm FS}$, that is the length that particles can cover within
the available time $t$. It is given by
\begin{equation}
  \lambda_{\rm FS}(t) = 
  a(t)\,\int_0^t\,\frac{v(t')}{a(t')}\,{\rm d}t'\;.
\label{MBeq:35}
\end{equation}
While particles are relativistic, $v=c$, and later $v\propto a^{-1}$
because the momentum is red-shifted. We thus introduce a new scale,
the scale factor $a_{\rm nr}$ where the dark-matter particles become
non-relativistic. It is defined by the condition $kT=a_{\rm
nr}^{-1}kT_{\gamma,0}\simeq m_\nu c^2$. Inserting
$T_{\gamma,0}=2.73\,{\rm K}$ and the neutrino mass $m_\nu$ from
eq.~(\ref{MBeq:20}), we find
\begin{equation}
  a_{\rm nr} = 4.7\times10^{-6}\,(\Omega_0h^2)^{-1}\;.
\label{MBeq:37}
\end{equation}
Evaluating eq.~(\ref{MBeq:35}), the {\em comoving\/} free-streaming
length turns out to be
\begin{equation}
  \lambda_{\rm FS} = \frac{2ct_{\rm nr}}{a_{\rm nr}}\,
  \left\{\begin{array}{ll}
    \frac{a}{a_{\rm nr}} &\hbox{for $a<a_{\rm nr}$,}\\
    \displaystyle
    1+\ln\left(\frac{a}{a_{\rm nr}}\right) & 
     \hbox{for $a_{\rm nr}<a<a_{\rm eq}$,}\\
    \displaystyle
    \frac{5}{2}+\ln\left(\frac{a_{\rm eq}}{a_{\rm nr}}\right) -
      \frac{3}{2}\left(\frac{a_{\rm eq}}{a}\right)^{1/2} &
      \hbox{for $a_{\rm eq}<a$,}\\
  \end{array}\right.
\label{MBeq:36}
\end{equation}
where
$\lambda_{\rm FS}$ is plotted in Fig.~\ref{MBfig:4} as a function of
$a$. This figure shows that $\lambda_{\rm FS}$ quickly reaches an
asymptotic value $\lambda_{\rm FS}^\infty$. For dark matter dominated
by neutrinos, this asymptotic free-streaming length is
\begin{equation}
  \lambda_{\rm FS}^\infty = 11.3\,{\rm Mpc}\,(\Omega_0h^2)^{-1}\;.
\label{MBeq:38}
\end{equation}

\begin{figure}[ht]
  \centerline{\epsfxsize=0.6\textwidth\epsffile{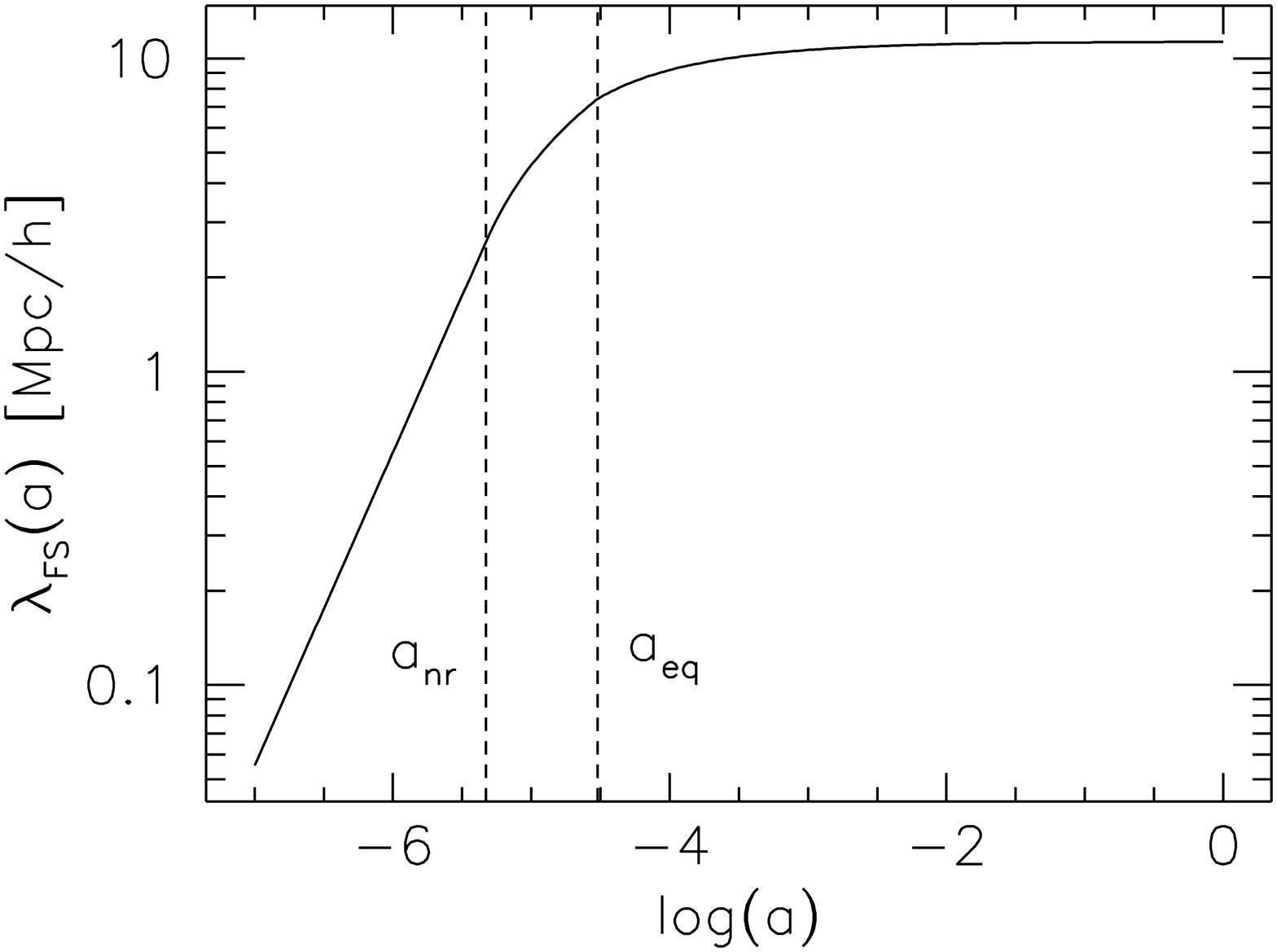}}
\caption{The comoving free-streaming length $\lambda_{\rm FS}(a)$ as a
  function of the scale factor $a$. $\lambda_{\rm FS}$ grows rapidly
  until $a_{\rm nr}$ and quickly approaches an asymptotic value
  $\lambda_{\rm FS}^\infty$ thereafter.}
\label{MBfig:4}
\end{figure}

\begin{figure}[ht]
  \centerline{\epsfxsize=0.45\textwidth\epsffile{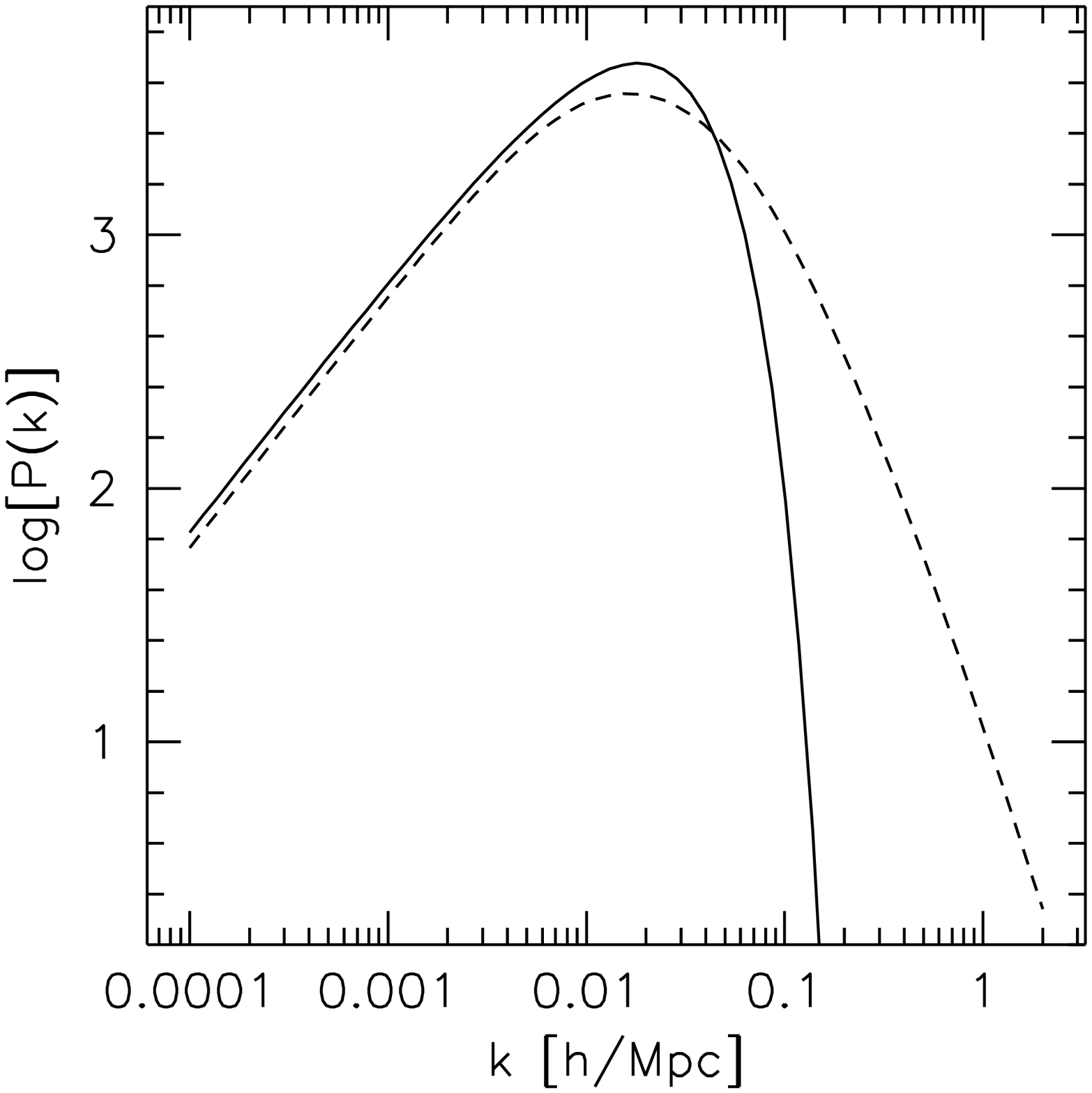}
              \epsfxsize=0.45\textwidth\epsffile{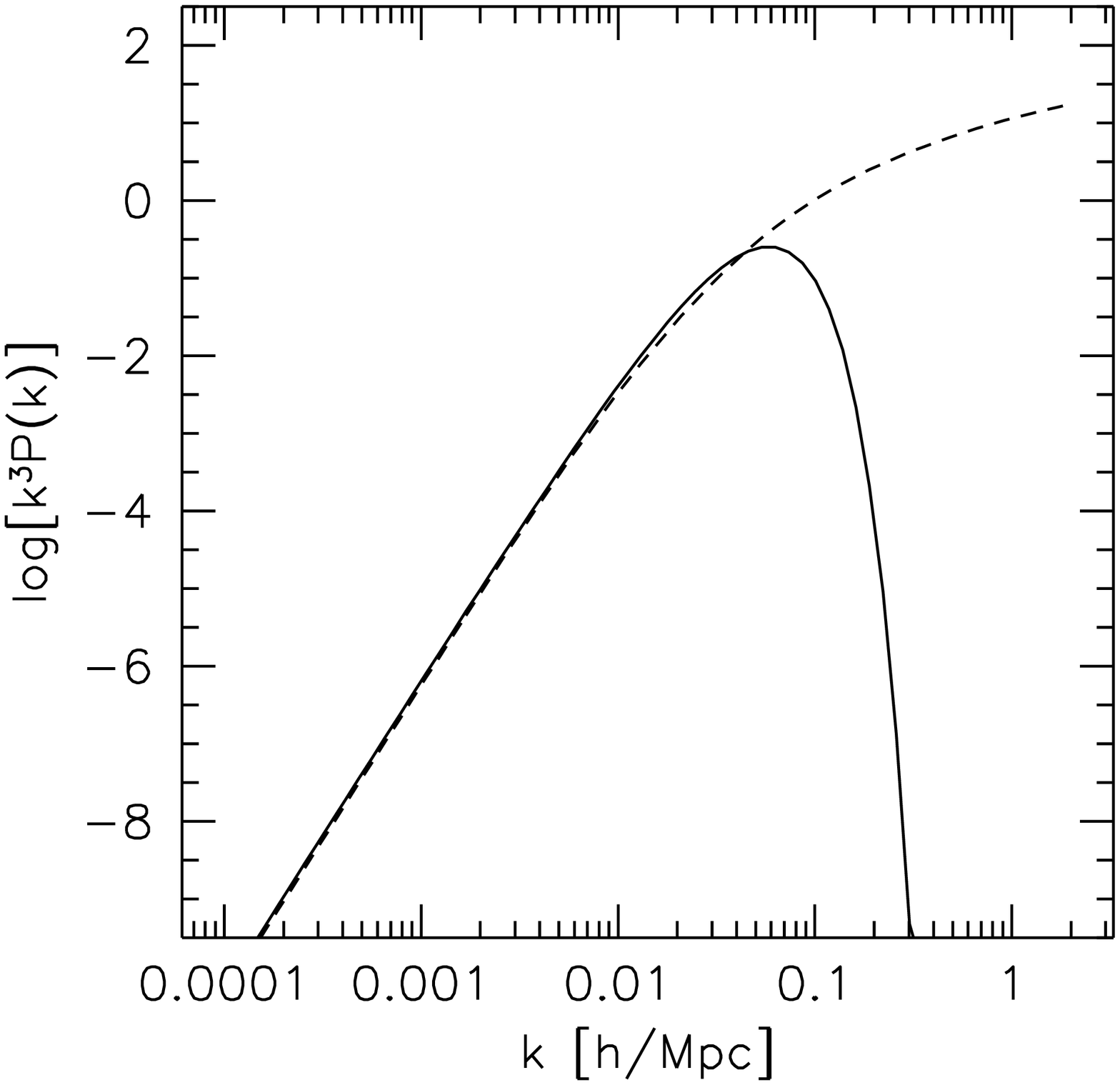}}
\caption{Left panel: Power spectrum $P(k)$ for cold and hot dark
  matter (dashed and solid lines, respectively). For cold dark matter,
  the spectrum falls off $\propto k^{-3}$ at large $k$, while free
  streaming of hot dark matter like neutrinos exponentially damps away
  perturbations smaller than the asymptotic free-streaming length
  $\lambda_{\rm FS}^\infty$. Right panel: Total power $k^3P(k)$ for
  cold and hot dark matter. While the power continues to increase
  monotonically for $k\to\infty$ for cold dark matter (solid curve),
  it reaches a peak at $k=k_0$ for hot dark matter and decreases
  rapidly on smaller scales.}
\label{MBfig:2}
\end{figure}

We therefore conclude that if neutrinos were to dominate the dark
matter, fluctuations with $\lambda\le\lambda_{\rm FS}^\infty$ would be
wiped out by free streaming. Such perturbations cannot keep the fast
neutrinos bound. Consequently, the power spectrum is exponentially cut
off at wave numbers $k\ge k_{\rm FS}=(\lambda_{\rm
FS}^\infty)^{-1}$. Since $k_{\rm FS}$ is only slightly larger than the
wave number where the power spectrum reaches its peak, $k_0$, the
neutrino-dominated power spectrum is cut off right above the
peak. This is illustrated in Fig.~\ref{MBfig:2} which
contrasts cold with hot dark matter spectra. For cold dark
matter (CDM), free streaming is unimportant, so the spectrum behaves
as described by eq.~(\ref{MBeq:34}). For hot dark matter (HDM), the
spectrum is cut off exponentially at large $k$. Consequently, the
total power of the density fluctuations, $k^3P(k)$, grows
monotonically with $k$ for CDM, while it peaks around $k_0$ for
HDM. Structure begins to form on scales where the power is
largest. For CDM, the smallest structures are formed first, and
structure formation proceeds towards larger scales. For HDM,
structures on scales of $\lambda\sim k_0^{-1}$ form first. They
correspond to objects somewhat more massive than galaxy
clusters. Smaller structures form later by fragmentation. Thus, free
streaming has the important consequence that the {\em bottom-up\/}
scenario of CDM is changed to the {\em top-down\/} scenario of
HDM. The formation of structures smaller than galaxy clusters, e.g.\
galaxies, is therefore strongly delayed in HDM
models. Figure~\ref{MBfig:3} illustrates this. It shows two
cosmological simulations starting from the same initial random density
perturbations, one with CDM and one with HDM. Three epochs of the
evolution at $z=2,1,0$ are shown. While significant structure on small
scales is already seen at $z=2$ in CDM, structure then just begins to
grow on much larger scales in HDM.

\begin{figure}
  \centerline{\epsfxsize=\textwidth\epsffile{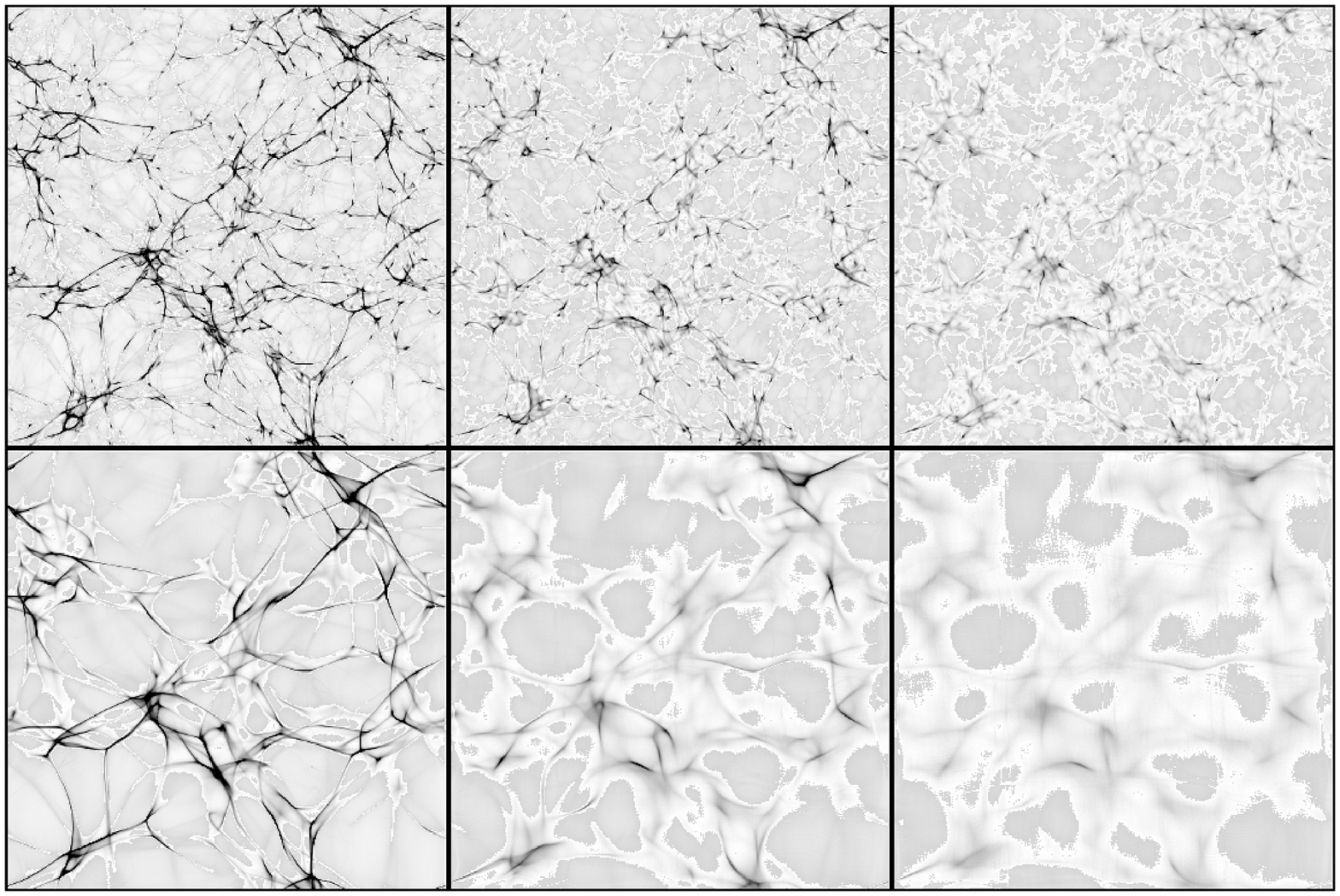}}
\caption{Growth of structure in CDM (top) and HDM (bottom) from the
  same initial random density perturbations; from left to right:
  $z=0$, $z=1$, and $z=2$. High-density regions are black.}
\label{MBfig:3}
\end{figure}

\subsection*{Summary}
\label{MBsec:8}

The main results of the preceding discussion can be summarised as
follows:

\begin{itemize}

\item The matter density of the Universe today is dominated by
  ordinary rather than relativistic matter. Going back in time, the
  radiation density grows faster than the matter density. The epoch
  when they equal each other plays an important r\^ole for structure
  formation. It is characterised by the scale factor
  $$
    a_{\rm eq} = 3.0\times10^{-5}\,(\Omega_0h^2)^{-1}\;.
  $$
\item Linear density perturbations grow $\propto a^2$ before $a_{\rm
  eq}$, and $\propto a$ thereafter.
\item Rapid radiation-driven expansion earlier than $a_{\rm eq}$
  suppresses the growth of modes which are smaller than the size of
  the horizon at $a_{\rm eq}$,
  $$
    d_{\rm H}(a_{\rm eq}) \sim 13\,(\Omega_0h^2)^{-1}\,{\rm Mpc}\;.
  $$
  Density perturbations with wave number $k>d_{\rm H}^{-1}(a_{\rm
  eq})=k_0$ are suppressed by the factor
  $$
    f_{\rm sup} = \left(\frac{k_0}{k}\right)^4\;.
  $$
  Together with the assumption of a scale-free primordial power
  spectrum, this yields the cold dark matter spectrum
  $$
    P(k) \propto \left\{\begin{array}{ll}
      k & (k<k_0), \\
      k^{-3} & (k>k_0). \\
    \end{array}\right.
  $$
\item Free streaming of hot dark-matter particles like neutrinos
  erases structures smaller than the asymptotic, comoving free
  streaming length
  $$
    \lambda_{\rm FS}^\infty \sim 11\,(\Omega_0h^2)^{-1}\,{\rm Mpc}\;.
  $$
  The hot dark matter spectrum therefore decreases exponentially for
  $k>(\lambda_{\rm FS}^\infty)^{-1}$.
\item This implies that structure formation in a neutrino-dominated
  universe proceeds according to the top-down scenario: small
  structures like galaxies form later than large structures like
  galaxy clusters, in conflict with observational results.
\end{itemize}

\noindent {\em Note:\/} There is a wealth of literature on the subject
briefly reviewed here. For further reading, and for references on
details, I can particularly recommend the following text-books:

}


\newpage {

\thispagestyle{empty}

\begin{flushright}
\Huge\bf
{\ }


Future\\
\bigskip
Prospects

\end{flushright}

\newpage

\thispagestyle{empty}

{\ }

\newpage

}\newpage{

\head{Neutrino Experiments with Cryogenic Detectors}
     {P.~Meunier$^{1,2}$}
     {$^1$Dipartimento di Fisica Universita' di Genova, 
      INFN Sezione di Genova, Genoa, Italy\\
      $^2$Max-Planck-Institut f\"ur Physik, F\"ohringer Ring 6, 
      80805 M\"unchen, Germany}

\noindent
In general an ideal cryogenic thermal detector consists of an absorber
for radiation or particles where the energy released has to be rapidly
converted into thermal energy, changing the detector temperature. The
absorber is coupled to a thermometer which responds to a change of
temperature by a change of resistance.  In the case of current biasing
of the thermistor, the temperature change is measured by observing the
voltage drop across it.  In order to have an increase of the detector
temperature which is not negligible (in the range of $\mu$K) for a
very small amount of energy release (100 eV), the detector heat
capacity has to be extremely small.  Since heat capacities decrease at
low temperature, the sensitivity is enhanced by running the detector
at very low temperatures ($\ll$ 1 K).  In the case of cryogenic
thermal detectors characterized by a small mass and a heat capacity of
the order of $10^{-12}\,{\rm J/K}$ at 0.1~K, it is theoretically
possible to obtain an energy threshold and an energy resolution of a
few~eV.  The intrinsic slowness of the developed low temperature
detectors (total pulse duration is about 100 ms) limits their
application only to low activity measurements.

A possible application of this kind of detector is the measurement of
the $\beta$ calorimetric spectrum with a high energy resolution.  The
investigation of the $\beta$ decay spectrum of particular isotopes,
such as $^{187}$Re and $^{167}$Ho can provide more stringent
kinematics limits on anti-neutrino and neutrino mass respectively.
The isotope $^{187}$Re is a $\beta$$^-$ emitter; it has the lowest
value of the end-point $Q$ among the known $\beta$$^-$ isotopes.  The
study of the $\beta$$^-$ spectrum of $^{187}$Re can be an alternative
to the tritium experiment to set a limit on the anti-neutrino mass.
The effect of a finite anti-neutrino mass should manifest itself as a
count deficit, in first approximation proportional to $m^{2}_{\nu}$,
near the spectrum end-point.  The isotope $^{187}$Re is naturally
occurring at the 62\% level and the element Re is a superconductor at
the detector working temperature of 0.08 K. Therefore it is possible
to use a crystal of natural Rhenium as a detector absorber to perform
the $^{187}$Re spectrum calorimetric measurement. As a thermistor an
heavily doped germanium thermistor has been used.

\begin{figure}[ht]
 \centerline{\epsfxsize=\textwidth\epsffile{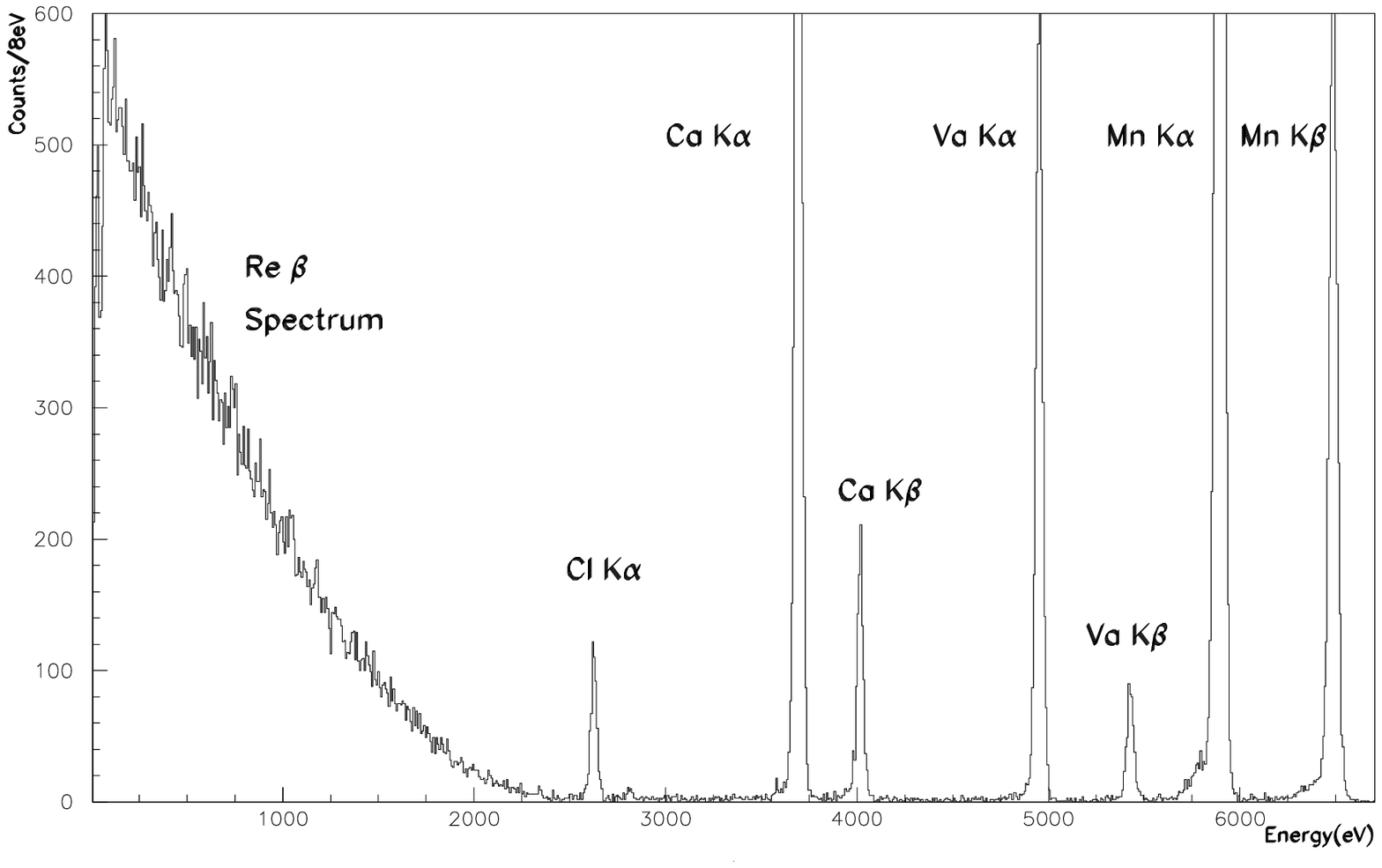}}
\caption{Calorimetric $\beta^{-}$ spectrum of $^{187}$Re plus 
a multi-lines X-ray fluorescence spectrum due to an external
calibration source (see text).}  \label{Meunier.fig1}
\end{figure}

In the preliminary measurement \cite{meunier.1} of the $^{187}$Re beta
spectrum performed in Genoa, an energy resolution of 30 eV FWHM has
been obtained. The end-point energy, obtained from the linear fit of
the Kurie Plot in the interval 112--2360 eV, is $E_{\rm end-point} =
(2482 \pm 12)\,$eV.  The detector energy calibration has been made
using a multi-lines fluorescence x-ray source; the K$_\alpha$ and
K$_\beta$ x-ray lines of Cl, Ca, Va, Mn have been detected, in the
energy range 100--7000~eV (see Fig.~1). Using this external source it
has been possible to investigate the detector linearity and the energy
dependence of the detector energy resolution.  A fit of the measured
energy lines versus the expected values with the function
$y=ax^2+bx+c$ has given the following result:
$a=(-0.7\pm0.7)E^{-4}\,{\rm eV}^{-1}$, $b=(1.000\pm0.005)$,
$c=(2.2\pm1.5)\,{\rm eV}$.  The rms value of the energy resolution has
been plotted against the energy lines, and a fit of this plot with the
function $y=Ax+B$ has given the following result:
$A=(0.9\pm2.5)\,E^{-4}$, $B=(15.3\pm1.4)\,\rm eV$.  It should be
possible to set a kinematics limit to the anti-neutrino mass by means
of this calorimetric method lower than 20 eV/c$^{2}$ in 50 days of
data taking. A high-statistics measurement of the $^{187}$Re spectrum
is under development.

\eject

The isotope $^{163}$Ho decays by electron capture (EC) with the
subsequent emission of an electron neutrino. The EC decay is followed
by the emission of Auger electrons and X-rays due to the electronic
cloud excitation of the daughter atom. For each occupied energy level
from which the electron capture is energetically allowed, there is a
line in the calorimetric energy spectrum. The effect of a finite
neutrino mass should manifest itself as a variation of the ratios of
the relative EC probabilities.  Measuring $N$ lines of the
calorimetric spectrum, it is possible to set $N-1$ simultaneous
constraints on the neutrino mass and the end-point value Q from the
relative ratios of the integral value of each spectral line.

Enclosing the $^{163}$Ho source in a suitable absorber of a
micro-calorimeter it has been possible to measure
\cite{meunier.2,meunier.3} four lines of the calorimetric spectrum
(see Fig.~2.a), but the statistics of this measurement is not high
enough to set a significant limit on the neutrino mass.  Considering a
zero value of the neutrino mass, the end-point energy has been
calculated by a fit procedure and it is turned out to be $E_{\rm
end-point} = (2800 \pm 50)\,$eV.  Taking into account this end-point
value it should be possible to set a kinematical limit to the neutrino
mass by means of this calorimetric method lower than 170 eV/c$^{2}$ in
50 days of data taking. Running more than one detector at the same
time it will be possible to further improve the achievable mass limit.
Although the limit on neutrino mass could not be competitive with the
achievable limit on electron anti-neutrino mass, we believe it is
meaningful to set independent limits on the masses of the electron
neutrino and anti-neutrino.

Another important application of this kind of detector is the
measurement of the solar neutrino flux in radiochemical solar neutrino
experiments.  In these experiments the integrated value of the solar
neutrino flux is measured counting the extracted isotope which is
produced in the target by neutrino capture.  The use of a detector
characterized by a low energy threshold and a good energy resolution
can improved the results obtained so far with proportional~counters.
 
A prototype of a new radiochemical lithium detector is now under
development at the Institute for Nuclear Research at
Moscow~\cite{meunier.4}.  The interest in this new experiment is
related to the fact that it could provide new data about the solar
neutrino flux at intermediate energy.  Even taking into account the
hypothesis of a strong attenuation of the neutrino flux of
intermediate energies, this experiment can measure the flux with an
accuracy of 12$\%$ in only one year of measurement time.

The full scale experiment will have a target mass of only 10 tons,
which is the absolutely minimal mass among all detectors of solar
neutrinos.  This experimental scale will be possible only if $^{7}$Be
can be detected with an efficiency close to 100$\%$.  The counting of
the $^{7}$Be isotope can be performed measuring the electron capture
decay of this isotope to $^{7}$Li, but in the 90$\%$ of the cases
$^{7}$Be decays into $^{7}$Li ground state with the subsequent
emission of a very low detectable energy, of the order of 100 eV.  The
only possible way to solve this problem appears to be the use of
cryogenic detectors for $^{7}$Be counting.  The detection of the
$^{7}$Be EC decay by means of cryogenic detector with an efficiency
close to 100\% has been successfully
realised~\cite{meunier.2,meunier.5}.

\begin{figure}[ht]
 \centerline{\epsfxsize=0.5\textwidth\epsffile{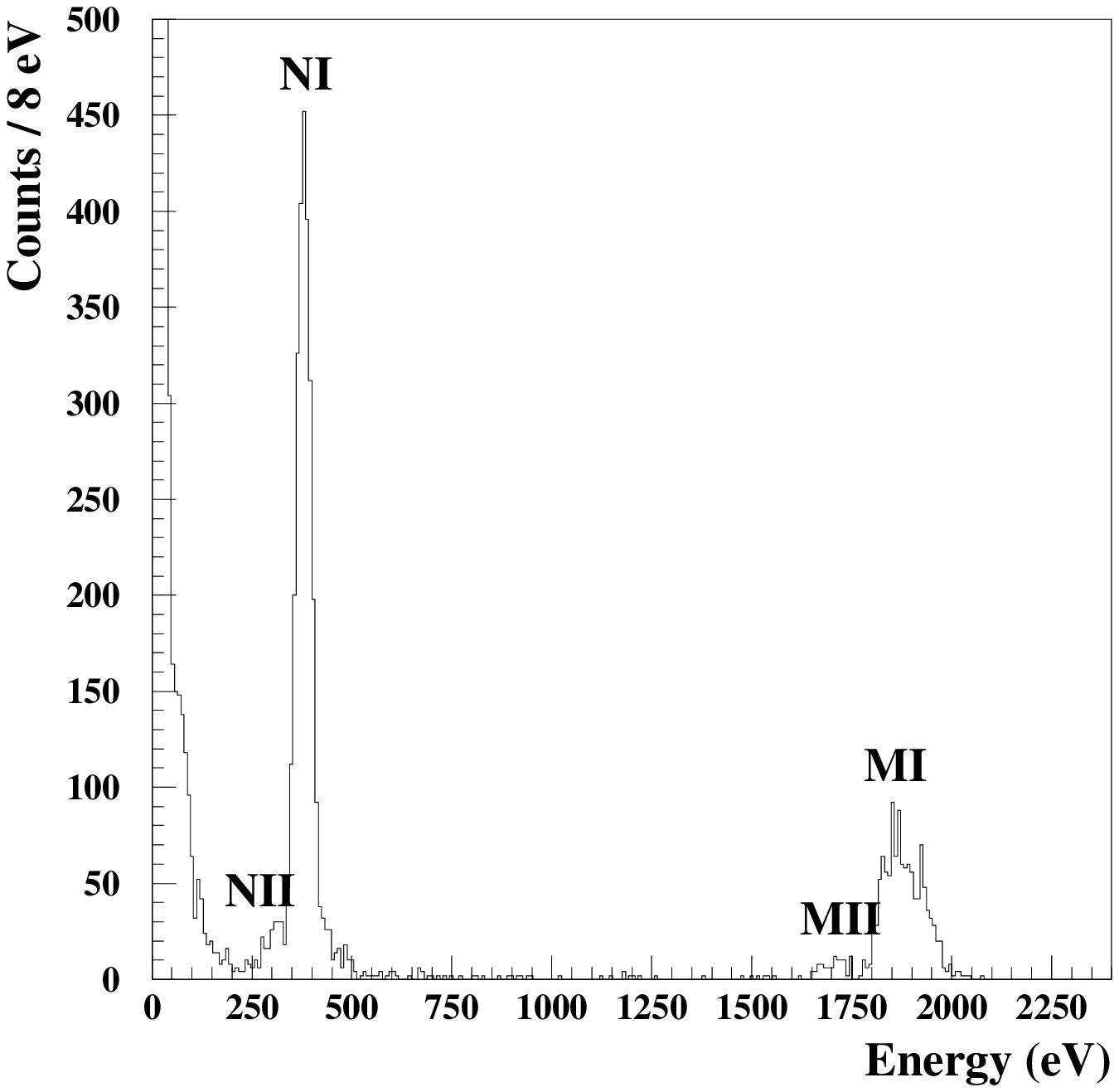}
             \epsfxsize=0.5\textwidth\epsffile{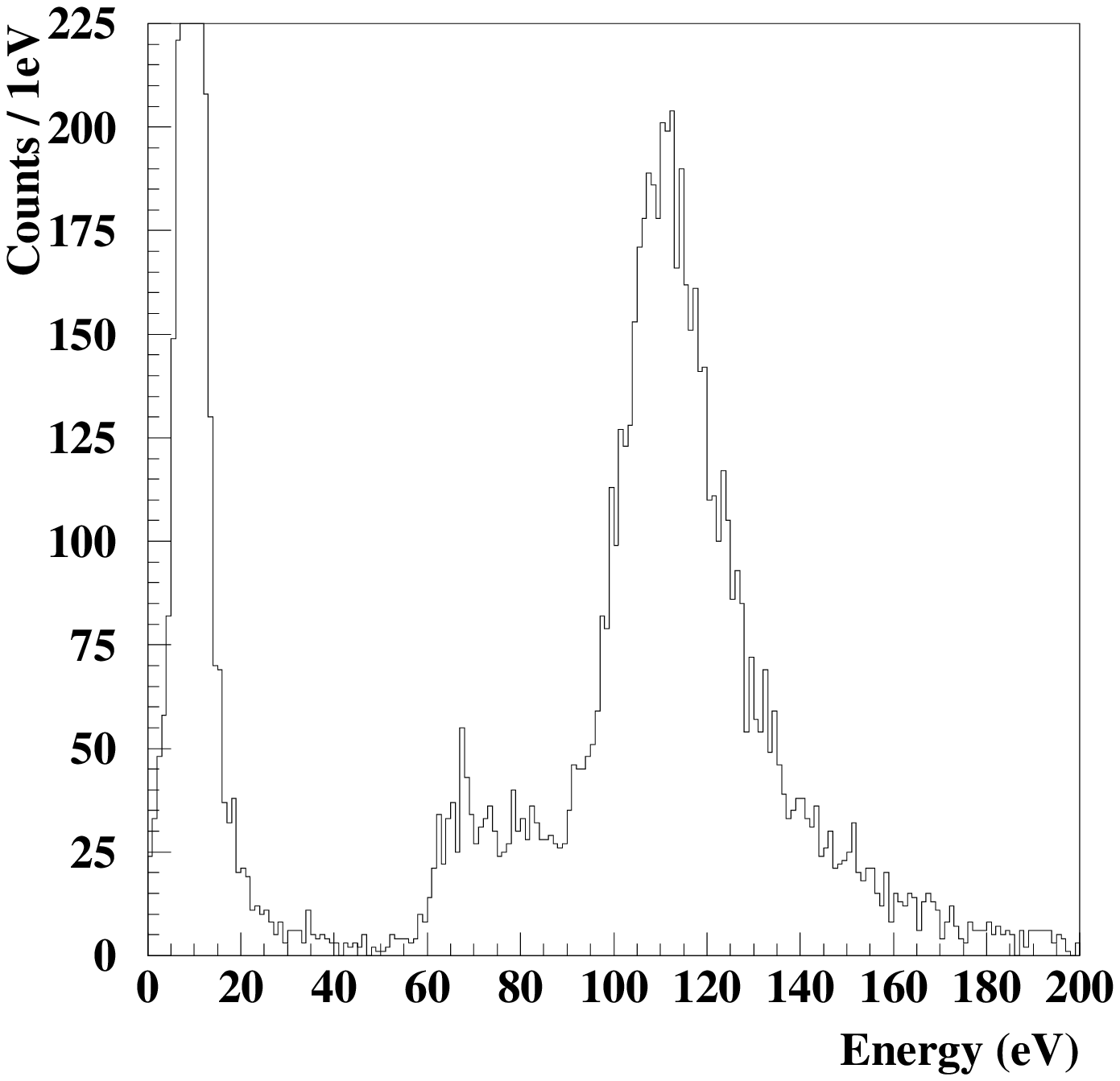}}
\caption{(a)~Calorimetric EC spectrum of a $^{163}$Ho source 
enclosed in a superconductive tin absorber. (b)~Calorimetric $^{7}$Be
EC spectrum.  The source has been produced by proton irradiation of a
BeO absorber.}
\label{Meunier.fig2-3}
\end{figure}

As an absorber a beryllium oxide crystal of mass 3 $\mu$g has been
used, as this is one of the possible forms in which Be can be
chemically extracted form the Li target.  The energy resolution has
been found to be 24 eV FWHM for the 112 eV $^{7}$Be line (see
Fig.~2.b), which would enable one to discriminate the background of
the counting system in the lithium Solar Neutrino Experiment.

}\newpage{


\head{OMNIS---A Galactic Supernova Observatory}
     {P.F.~Smith}
     {Rutherford Appleton Laboratory, Chilton, Oxfordshire, 
      OX11 0QX, UK}

\noindent 
A type II (or Ib) supernova explosion releases most of its energy as
neutrinos and antineutrinos with a time constant of a few seconds,
with all three neutrino types and their antiparticles produced in
comparable numbers.  The emission time profile from a typical model is
shown in Fig.~1 [1].  This provides a unique opportunity to discover
neutrino properties which are difficult or impossible to determine
using terrestrial neutrino sources.  In particular a non-zero neutrino
mass will alter the time profile arriving at the earth (Fig.~2),
allowing direct time-of-flight measurement of the mass of at least one
neutrino type.  In particular the most likely distance range for
Galactic supernovae ($\sim 2$--20~kpc) is ideal for time-of-flight
measurement of a ``cosmologically significant'' neutrino mass---i.e.~a
mass in the range 10--100~eV, for which it would form a major
component of the mass of the universe and could be a candidate for the
Galactic dark matter.

\begin{figure}[ht]
\centerline{\epsfxsize=8.2cm\epsffile{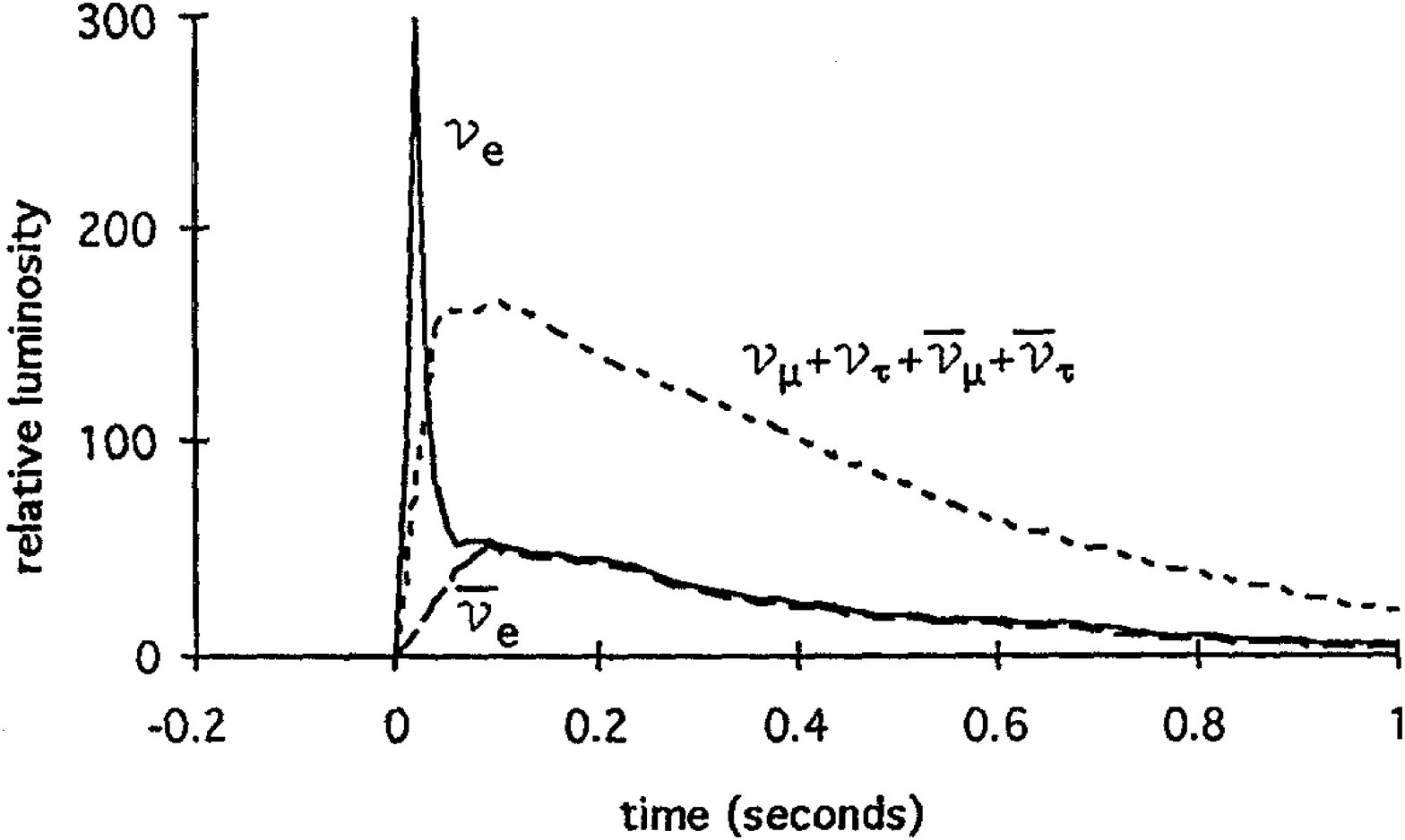}}
\caption{Time profiles of neutrinos emitted by supernova explosion. 
The mu and tau neutrinos have identical profiles when emitted.}
\label{smith1.fig}
\end{figure}

\begin{figure}[ht]
\centerline{\epsfxsize=8.2cm\epsffile{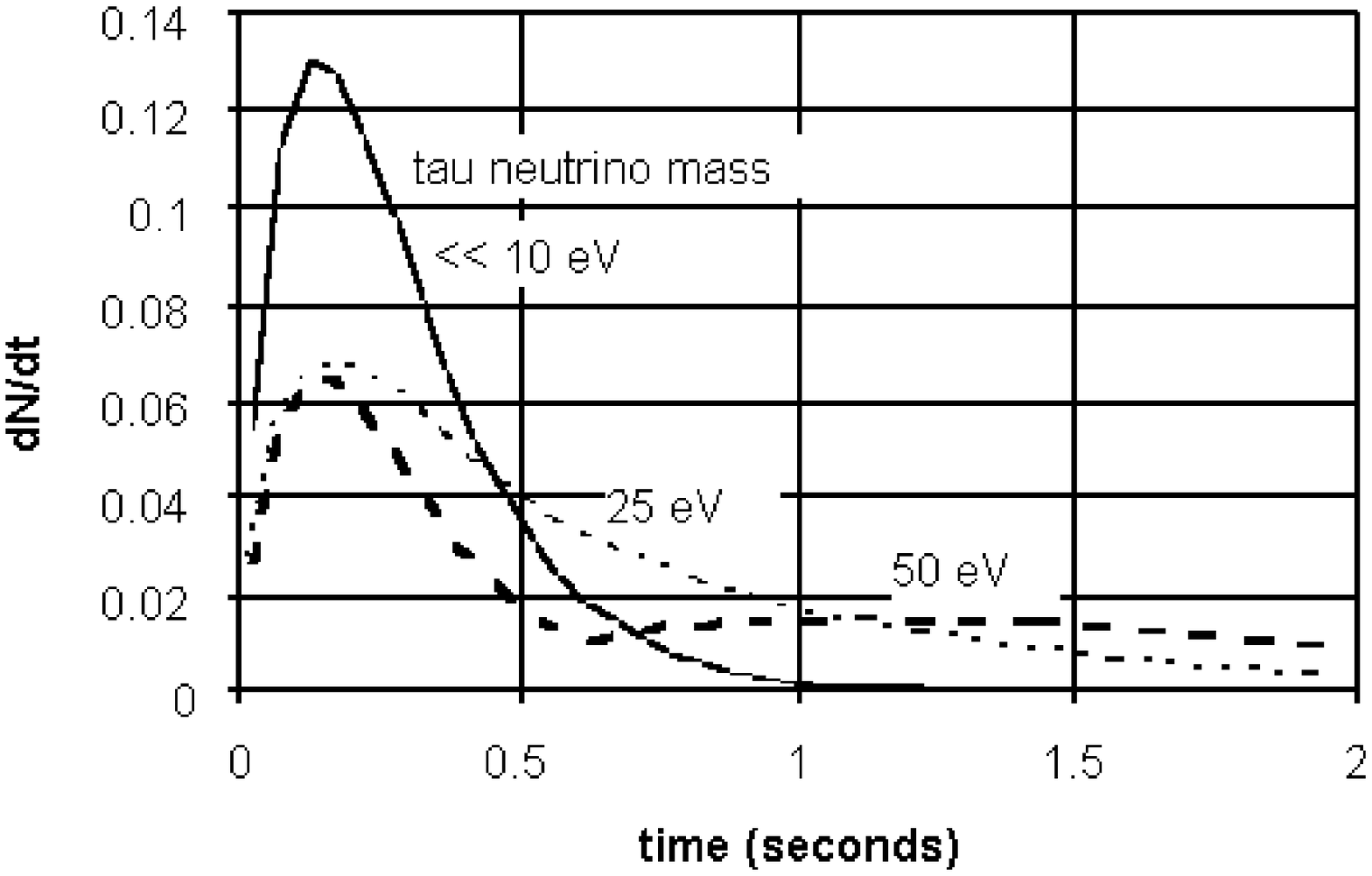}}
\caption{Effect of non-zero mass on mu/tau neutrino arrival time 
profile from distance 8~kpc.}
\label{smith2.fig}
\end{figure}

The frequency of Galactic supernovae is uncertain.  There is a
substantial discrepancy between the expectation of 1 per 30--100
years, from various astrophysical estimates, and the historical record
which indicates a higher frequency~[2, 3].  Figure~3 shows the
locations of known supernovae in our Galaxy during the past 1000
years.  Only those within 4~kpc of the sun are visible optically with
high efficiency.  A prediction of 1 in 50 years for the whole Galaxy
would imply an expectation of only 1 within 4~kpc of the sun, whereas
in fact 4 or 5 type II+Ib have been observed within this radius in the
past millennium.  This observation appears to exclude, with 90\%
confidence, an interval as large as 30 years or more, and is
consistent with a total Galactic rate of 1 in $15\pm5$ years.

\begin{figure}[ht]
\centerline{\epsfxsize=7.8cm\epsffile{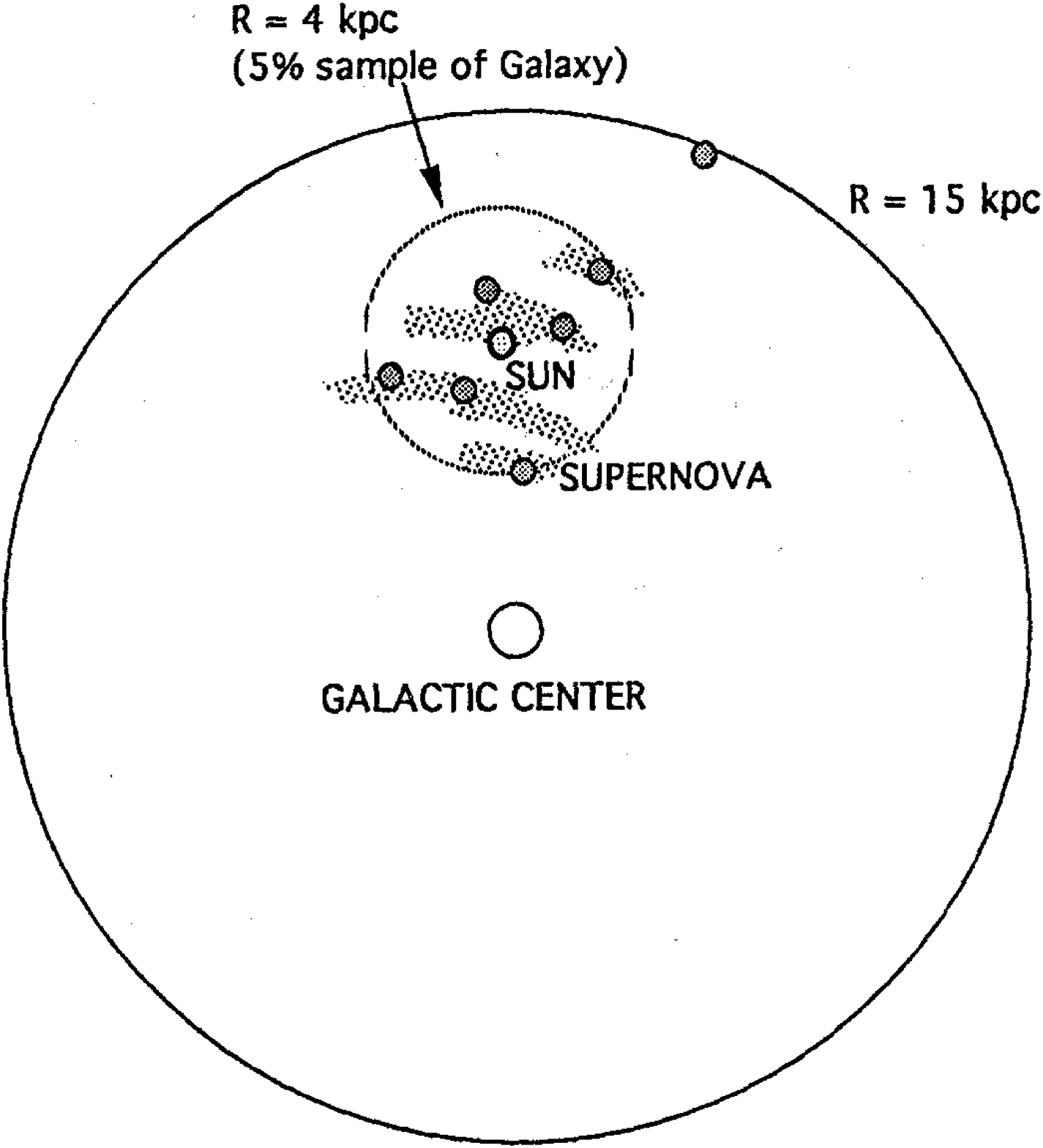}}
\caption{Galactic map of recorded supernova for past 1000 years,
showing that only the closest 5\% are visible optically. The much
larger number from the whole Galaxy would be observable as neutrino
bursts.}
\label{smith3.fig}
\end{figure}

With a time interval of this order, it becomes reasonable to set up an
underground detector array for such an event, run on a part-time basis
in parallel with other astrophysics and particle physics projects.
For a statistical precision sufficient to determine a neutrino mass in
the 10--100~eV range, we require about 2000 events (1000 each of mu
and tau neutrinos).  A number of existing world detectors are
sensitive to supernova neutrinos, in particular Super-Kamiokande, LVD,
MACRO, and SNO.  However, these detect principally the electron
antineutrino component through charged current interactions, and have
relatively low sensitivity to neutral current events~[4].  Numerical
estimates in Table~1 show that from an 8~kpc supernova
Super-Kamiokande would register $\sim 8000$ charged current events,
but only 1\% of that number from neutral currents.  We describe here a
low cost detection scheme which would be sensitive principally to the
mu/tau neutrino component. This would provide the first direct
observation of a cosmologically significant neutrino mass, and would
complement the electron antineutrino signal from other world detectors
to provide data on mixing between all three neutrino generations.
Information on the profile shape for all three neutrino types would be
also of considerable astrophysical interest.

The proposed detector array is referred to as OMNIS---an observatory
for multiflavour interactions from Galactic supernovae.  It has
evolved from the Supernova Neutrino Burst Observatory (SNBO) proposed
by D.B.~Cline et al., based on neutrino detection by neutral current
nuclear excitation in natural underground rock~[5]. The excited
nucleus would emit neutrons, which could be captured by counters
embedded in the rock.  The E2 dependence of the excitation cross
section makes the detector sensitive principally to the higher
temperature mu/tau neutrinos, producing a dominant neutral current
signal.  The evolution of OMNIS from SNBO is described in a more
detailed paper~[6].  There are two major improvements: (a)~neutron
collection efficiencies higher by an order of magnitude are obtained
by siting optimised detectors in open caverns, and (b)~additional
target materials are used, in particular Fe and Pb, which give higher
neutron production rates and with Pb also having a charged current
excitation route~[7, 8, 9]. Thus differences between lead, iron and
rock signals can be used to infer mixing between mu/tau and electron
neutrinos.

\begin{figure}[b]
\centerline{\epsfxsize=12cm\epsffile{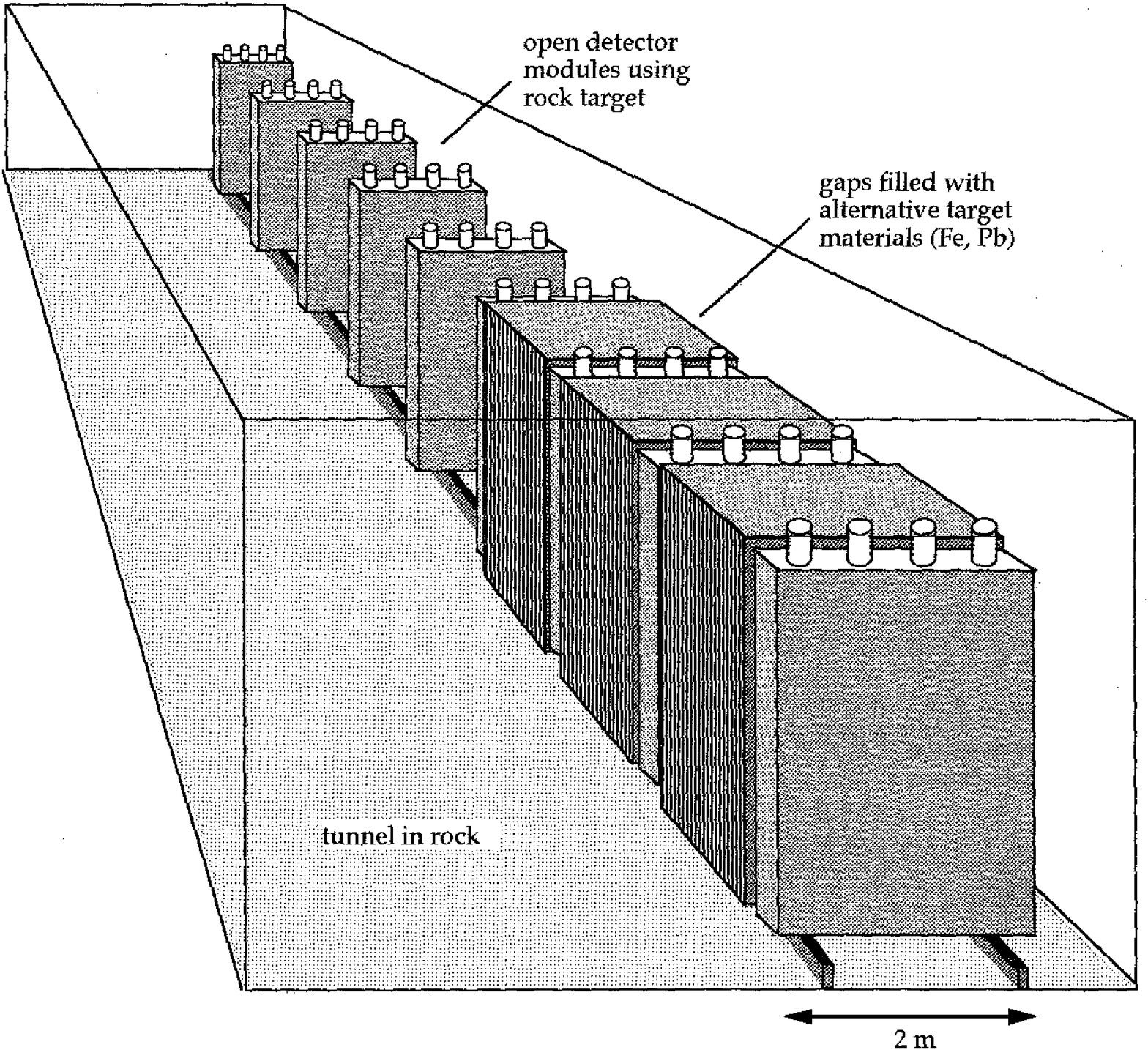}}
\caption{Possible arrangement of target materials and neutron
  detectors for OMNIS.}
\label{smith4.fig}
\end{figure}

A possible arrangement of detector/target modules for OMNIS is shown
in Fig.~4. Standard Gd or Li-loaded scintillator could be used for the
neutron detection, in modules of about 0.5--1 ton.  About 200--300
modules would be needed to observe 2000 events from an 8 kpc
supernova.  This demonstrates the uniquely high efficiency of the
detector array, giving it a unit cost much lower than other world
neutrino detectors.  An average of over 10 events per ton scintillator
could be achieved using indirect neutron detection compared with the
0.3 events per ton scintillator achieved with direct charged current
detection in MACRO.  The estimated neutron production rates in salt
rock, iron, and lead are 0.06, 0.15, 1 per ton for a supernova at
8~kpc.  In addition, higher collection efficiency is possible from Fe
and Pb due to lower absorption~[6], giving them an effectiveness in
the ratio $1:4:30$.  Thus lead is by far the most effective target to
maximise event numbers, but the design must include a sufficient
quantity of the other targets to provide information on mixing between
mu/tau and electron neutrinos, based on the fact that this would
enhance the charged current excitation in Pb but not in Fe or rock.

Table~1 shows the event numbers of different neutrino types from
principal world detectors, compared with those from the proposed OMNIS
detector.  Note that the neutron background from U/Th in the rock
would not be significant compared with the signal burst, and for sites
with depths $> 600$~m, the neutron background from muons would also be
an order of magnitude below the signal. Thus the project is
technically straightforward and based on conventional low-cost neutron
detection technology.  Neutron background is discussed in more detail
in~[6] where some discussion is also included on the enormously
greater background problem for extra-Galactic supernova detection.
For discrimination of supernova events from local neutron showers from
cosmic ray muons, it is desirable to split the array between several
underground caverns, and also between several world sites, also
permitting estimates of supernova direction by arrival time
differences.

\begin{table}[t]
\caption{Comparison of proposed OMNIS multi-target observatory with 
world detectors based on direct detection with water and scintillator
targets. Approximate event numbers for each neutrino type shown for
a supernova at 8 kpc.}
\label{smith1.tab}
\medskip
\centerline{
\begin{tabular}{lcccccc}
\hline
\noalign{\vskip2pt}
\hline
\noalign{\vskip4pt}
&Target  &Fiducial   &Target  &\multicolumn{3}{c}{Event Number}\\
&Material&Mass (tons)&Elements&$\nu_e$&$\bar\nu_e$&
$\nu_{\mu,\tau}+\bar\nu_{\mu,\tau}$\\
\noalign{\vskip4pt}
\hline
\noalign{\vskip4pt}
{\it Combined}           \\
{\it Target \& Detector:}\\
\quad Super-Kamiokande&H$_2$O    &32000&$p$, $e$, O&180&8300& 50\\
\quad LVD             &``CH$_2$''&1200 &$p$, $e$, C& 14& 540& 30\\
\quad MACRO           &``CH$_2$''&1000 &$p$, $e$, C&  8& 350& 25\\
\quad SNO             &H$_2$O    &1600 &$p$, $e$, O& 16& 520&  6\\
\quad SNO$^a$         &D$_2$O    &1000 &$d$, $e$, O&190& 180&300\\
\noalign{\vskip4pt}
{\it Separated}          \\
{\it Target \& Detector:}\\
\quad OMNIS              \\
\quad 200~t Scintillator &NaCl (rock)&``8000''&Na, Cl&10 &10&400\\
\quad + Natural Rock     &Fe         &  4000  &Fe    &10 &10&600\\
\quad + Installed Targets&Pb         &  4000  &Pb    &280&30&1000\\
\quad Direct Interaction&&&{\bf Total:}&{\bf300}&{\bf50}&{\bf2000}\\
\quad With Scintillator  &``CH$_2$'' &  200   &$p$, $e$, C&2&70&5\\
\noalign{\vskip4pt}
\hline
\noalign{\vskip4pt}
\multicolumn{7}{l}{$^a$Heavy water target not available indefinitely}
\end{tabular}
}
\end{table}

In conclusion the relatively low unit cost of the OMNIS scheme makes
it possible to propose a dedicated Galactic supernova neutrino
observatory, capable of observing the time profile of the mu and tau
neutrino component.  This would provide the first direct measurement
of a cosmologically-significant neutrino mass, and the use of several
different target elements, in conjunction with the large electron
antineutrino signal from SuperKamiokande would also provide data on
the presence or absence of neutrino mixing on the Galactic distance
scale.  It could be designed to run largely unattended, with minimal
maintenance, so would be a part-time activity for the participating
groups, alongside other underground experiments.  The current
collaboration on this project involves groups from RAL, Manchester,
Sheffield, Imperial College (for the UK Boulby Mine site) and UCLA,
UCSD, Ohio, LLNL, LANL, UT-Dallas, PNL, for the US Carlsbad site, both
sites in low activity salt rock and already containing suitable
infra-structure installed for other purposes.  First prototype modules
could now be set up in these sites and funding is being sought for
this.  New collaborators would be welcome to strengthen the case for
construction of the full array.

\subsection*{Acknowledgments}

I acknowledge clarifying discussions with D.B.~Cline, G.M.~Fuller,
J.R.~Wilson, B.~Cox, G.~McLaughlin, J.D.~Lewin, R.N.~Boyd, F.~Boehm,
J.~Kleinfeller, R.~Marshall, W.~Vernon, R.~Shirato, A.~Burrows, and
C.K.~Hargrove.

}\newpage{

\def\vallefig#1{{Fig.~\ref{#1}}}
\def\vallelsim{\raise0.3ex\hbox{$\;<$\kern-0.75em\raise-1.1ex\hbox{$\sim\;$}}}
\def\valleleq#1{{eq. (\ref{#1})}}
\def\vallene{\hbox{$\nu_e$ }}
\def\nm{\hbox{$\nu_\mu$ }}
\def\nt{\hbox{$\nu_\tau$ }}
\def\ns{\hbox{$\nu_{s}$ }}
\def\mne{\hbox{$m_{\nu_e}$ }}
\def\mnm{\hbox{$m_{\nu_\mu}$ }}
\def\mnt{\hbox{$m_{\nu_\tau}$ }}
\def\ap#1#2#3{           {\it Ann. Phys. (NY) }{\bf #1} (19#2) #3}
\def\arnps#1#2#3{        {\it Ann. Rev. Nucl. Part. Sci. }{\bf #1} (19#2) #3}
\def\cnpp#1#2#3{        {\it Comm. Nucl. Part. Phys. }{\bf #1} (19#2) #3}
\def\apj#1#2#3{          {\it Astrophys. J. }{\bf #1} (19#2) #3}
\def\app#1#2#3{          {\it Astropart. Phys. }{\bf #1} (19#2) #3}
\def\asr#1#2#3{          {\it Astrophys. Space Rev. }{\bf #1} (19#2) #3}
\def\ass#1#2#3{          {\it Astrophys. Space Sci. }{\bf #1} (19#2) #3}
\def\aa#1#2#3{          {\it Astron. \& Astrophys.  }{\bf #1} (19#2) #3}
\def\apjl#1#2#3{         {\it Astrophys. J. Lett. }{\bf #1} (19#2) #3}
\def\ap#1#2#3{         {\it Astropart. Phys. }{\bf #1} (19#2) #3}
\def\ass#1#2#3{          {\it Astrophys. Space Sci. }{\bf #1} (19#2) #3}
\def\jel#1#2#3{         {\it Journal Europhys. Lett. }{\bf #1} (19#2) #3}
\def\ib#1#2#3{           {\it ibid. }{\bf #1} (19#2) #3}
\def\nat#1#2#3{          {\it Nature }{\bf #1} (19#2) #3}
\def\nps#1#2#3{        {\it Nucl. Phys. B (Proc. Suppl.) }{\bf #1} (19#2) #3} 
\def\np#1#2#3{           {\it Nucl. Phys. }{\bf #1} (19#2) #3}
\def\pl#1#2#3{           {\it Phys. Lett. }{\bf #1} (19#2) #3}
\def\pr#1#2#3{           {\it Phys. Rev. }{\bf #1} (19#2) #3}
\def\prep#1#2#3{         {\it Phys. Rep. }{\bf #1} (19#2) #3}
\def\prl#1#2#3{          {\it Phys. Rev. Lett. }{\bf #1} (19#2) #3}
\def\pw#1#2#3{          {\it Particle World }{\bf #1} (19#2) #3}
\def\ptp#1#2#3{          {\it Prog. Theor. Phys. }{\bf #1} (19#2) #3}
\def\jppnp#1#2#3{         {\it J. Prog. Part. Nucl. Phys. }{\bf #1} (19#2) #3}
\def\cpc#1#2#3{         {\it Comp. Phys. Commun. }{\bf #1} (19#2) #3}
\def\rpp#1#2#3{         {\it Rep. on Prog. in Phys. }{\bf #1} (19#2) #3}
\def\ptps#1#2#3{         {\it Prog. Theor. Phys. Suppl. }{\bf #1} (19#2) #3}
\def\rmp#1#2#3{          {\it Rev. Mod. Phys. }{\bf #1} (19#2) #3}
\def\zp#1#2#3{           {\it Zeit. fur Physik }{\bf #1} (19#2) #3}
\def\fp#1#2#3{           {\it Fortschr. Phys. }{\bf #1} (19#2) #3}
\def\Zp#1#2#3{           {\it Z. Physik }{\bf #1} (19#2) #3}
\def\Sci#1#2#3{          {\it Science }{\bf #1} (19#2) #3}
\def\n.c.#1#2#3{         {\it Nuovo Cim. }{\bf #1} (19#2) #3}
\def\r.n.c.#1#2#3{       {\it Riv. del Nuovo Cim. }{\bf #1} (19#2) #3}
\def\sjnp#1#2#3{         {\it Sov. J. Nucl. Phys. }{\bf #1} (19#2) #3}
\def\yf#1#2#3{           {\it Yad. Fiz. }{\bf #1} (19#2) #3}
\def\zetf#1#2#3{         {\it Z. Eksp. Teor. Fiz. }{\bf #1} (19#2) #3}
\def\zetfpr#1#2#3{    {\it Z. Eksp. Teor. Fiz. Pisma. Red. }{\bf #1} (19#2) #3}
\def\jetp#1#2#3{         {\it JETP }{\bf #1} (19#2) #3}
\def\mpl#1#2#3{          {\it Mod. Phys. Lett. }{\bf #1} (19#2) #3}
\def\ufn#1#2#3{          {\it Usp. Fiz. Naut. }{\bf #1} (19#2) #3}
\def\sp#1#2#3{           {\it Sov. Phys.-Usp.}{\bf #1} (19#2) #3}
\def\ppnp#1#2#3{           {\it Prog. Part. Nucl. Phys. }{\bf #1} (19#2) #3}
\def\cnpp#1#2#3{           {\it Comm. Nucl. Part. Phys. }{\bf #1} (19#2) #3}
\def\ijmp#1#2#3{           {\it Int. J. Mod. Phys. }{\bf #1} (19#2) #3}
\def\ic#1#2#3{           {\it Investigaci\'on y Ciencia }{\bf #1} (19#2) #3}
\def\tp{these proceedings}
\def\pc{private communication}
\def\opc{\hbox{{\sl op. cit.} }}
\def\ip{in preparation}

\head{Neutrinos in Astrophysics}
     {Jos\'e W. F. Valle }
       {Instituto de F\'{\i}sica Corpuscular---C.S.I.C. \\
Departament de F\'{\i}sica Te\`orica, Universitat de
 Val\`encia,            
46100 Burjassot, Val\`encia, Spain      
 (http://neutrinos.uv.es)
}

\subsection*{Theory of Neutrino Mass}
\noindent
Neutrinos are the only apparently massless electrically neutral
fermions in the Standard Model (SM).  Nobody knows why they are so
special when compared with the other fermions. The masslessness of
neutrinos is imposed in an {\sl ad hoc} fashion, not dictated by an
underlying {\sl principle} such as gauge invariance. Moreover,
massless neutrinos may be in conflict with present data on solar and
atmospheric neutrino observations, as well as cosmological data on the
amplitude of primordial density fluctuations \cite{valle.1}.  The
latter suggest the need for hot dark matter in the Universe. On the
other hand, if massive, neutrinos would pose another puzzle, namely,
{\sl why are their masses so much smaller than those of the charged
fermions}? The key to the answer may lie in the fact that neutrinos
could be Majorana fermions. Such fermions are the most fundamental
ones. If neutrinos are majorana particles the suppression of their
mass could be related to the feebleness of lepton number
violation. This, in turn, would point towards physics beyond the SM.

Lepton number, or $B-L$ 
can be part of the gauge symmetry \cite{valle.2}
or, alternatively, can be a spontaneously broken global symmetry. In
the latter case there is a physical pseudoscalar Goldstone boson
generically called majoron \cite{valle.3}.  One can construct an enormous
class of such models \cite{valle.4} which may have important implications
not only in astrophysics and cosmology but also in particle physics
\cite{valle.5}.

Many mechanisms exist to explain the small masses of neutrinos as a
result of the violation of lepton number.  The first is the seesaw
mechanism, which is based on the existence of some relatively large
mass scale. However, neutrinos could acquire their mass
radiatively. In this case the smallness of the mass holds even if all
the new particles required to generate the neutrino masses are light
and therefore accessible to present experiments \cite{valle.6}.  The
seesaw and the radiative mechanisms of neutrino mass generation may be
combined. Supersymmetry with broken R-parity provides a very elegant
mechanism for the origin of neutrino mass \cite{valle.7} in which the
tau neutrino $\nu_{\tau}$ acquires a mass due to the mixing between
neutrinos and neutralinos. This happens in a way similar to the seesaw
mechanism, in which the large mass scale is now replaced by a
supersymmetry breaking scale characterizing the neutralino sector. In
supergravity models with universal soft breaking scalar masses, this
{\sl effective} neutralino-neutrino mixing is induced only radiatively
\cite{valle.7}. As a result the Majorana $\nu_{\tau}$ mass is naturally
suppressed, even though there is no large mass scale present. From
this point of view, the mechanism is a {\sl hybrid} between the
see-saw idea and the radiative mechanism. Despite the small neutrino
mass, many of the corresponding R-parity violating effects can be
sizeable.  An obvious example is the fact that the lightest neutralino
decay will typically decay inside the detector, unlike the case of the
minimal supersymmetric model.

Other than the seesaw scheme, none of the above models requires a
large mass scale. In all of them one can implement the spontaneous
violation of the lepton number symmetry leading to neutrino masses
that {\sl vanish} as the lepton-number scale goes to zero, in contrast
to the see-saw model. Such low-scale models are very attractive and
lead to a richer phenomenology, as the extra particles required have
masses at scales that could be accessible to present experiments. One
remarkable example is the possibility of 
invisibly decaying Higgs bosons~\cite{valle.5}.

The large diversity of neutrino mass schemes and the lack of a theory
for the Yukawa couplings imply that present theory is not capable of
predicting the scale of neutrino masses any better than it can fix the
masses of the other fermions, such as the muon: one should turn to
observations as means for constraining their properties.  Direct
laboratory observations leave a lot of room for massive neutrinos,
except beta decay experiments which are more restrictive.  A
complementary approach to the problem of neutrino mass tries to get
information from astrophysics and cosmology. It constitutes a
promising interdisciplinary field, now often-called {\sl astroparticle
physics}.  In this talk I will briefly mention recent work devoted to
the constraints that one can place on neutrino properties from
cosmological and astrophysical observations.

\subsection*{Limits from Cosmology }

The first cosmological bound on neutrino masses follows from avoiding
the overabundance of relic neutrinos \cite{valle.8}
\begin{equation} 
\label{RHO1}
\sum m_{\nu_i} \vallelsim 92 \: \Omega_{\nu} h^2 \: {\rm eV}\:, 
\end{equation} 
where $\Omega_{\nu} h^2 \leq 1$ and the sum runs over all stable
species of isodoublet neutrinos with mass less than $O(1 \:
{\rm MeV})$. 
Here $\Omega_{\nu}=\rho_{\nu}/\rho_c$, where $\rho_{\nu}$ is
the neutrino contribution to the total density and $\rho_c$ is the
critical density.  The factor $h^2$ measures the uncertainty in the
present value of the Hubble parameter, $0.4 \leq h \leq 1$, and
$\Omega_{\nu} h^2$ is smaller than 1.  For the $\nu_{\mu}$ and
$\nu_{\tau}$ this bound is much more stringent than the laboratory
limits.

Apart from the experimental interest, a heavy tau neutrino (say in the
MeV range) could also be interesting from the point of view of
structure formation \cite{valle.9}.  According to \valleleq{RHO1} such
neutrino can not exist if it has only the interactions prescribed by
the SM. However a heavy tau neutrino is theoretically viable if in
models with spontaneous violation of total lepton number
\cite{valle.3} since these contain new interactions of neutrinos 
majorons which may cause neutrinos to decay into a lighter neutrino
plus a majoron, for example \cite{valle.4},
\begin{equation}
\label{NUJ}
\nu_\tau \to \nu_\mu + J \:\: .
\end{equation}
or have sizeable annihilations to these majorons,
\begin{equation}
\label{nunuJJ}
\nu_\tau + \nu_\tau \to J + J \:\: .
\end{equation}
The possible existence of fast decay and/or annihilation channels
could eliminate relic neutrinos and therefore allow them to have
higher masses, as long as the lifetime is short enough to allow for an
adequate red-shift of the heavy neutrino decay products. These 2-body
decays can be much faster than the visible modes, such as radiative
decays of the type $\nu' \to \nu + \gamma$. Moreover, the majoron
decays are almost unconstrained by astrophysics and cosmology (for a
detailed discussion see ref.~\cite{valle.8}).

A general method to determine the majoron emission decay rates of
neutrinos was first given in ref.~\cite{valle.10}. The resulting decay
rates are rather model-dependent and will not be discussed here.
Explicit neutrino decay lifetime estimates are given, e.g.~in
ref.~\cite{valle.4,valle.11}.  The conclusion is that there are many
ways to make neutrinos sufficiently short-lived and that all mass
values consistent with laboratory experiments are cosmologically
acceptable.

Stronger limits on neutrino masses, lifetimes and/or annihilation
cross sections arise from cosmological Big-Bang
Nucleosynthesis. Recent data on the primordial deuterium abundance
\cite{valle.12} have stimulated a lot of work on the subject
\cite{valle.13}.  If a massive \nt is stable on the nucleosynthesis
time scale, ($\nu_\tau$ lifetime longer than $\sim 100$ sec), it can
lead to an excessive amount of primordial helium due to their large
contribution to the total energy density. This bound can be expressed
through an effective number of massless neutrino species ($N_\nu$). If
$N_\nu < 3.4{-}3.6$, 
one can rule out $\nu_\tau$ masses above 0.5~MeV
\cite{valle.14}.  If we take $N_\nu < 4$ the \mnt limit loosens
accordingly. However it has recently been argued that non-equilibrium
effects from the light neutrinos arising from the annihilations of the
heavy \nt's make the constraint a bit stronger in the large \mnt
region \cite{valle.15}. In practice, all $\nu_\tau$ masses on the few MeV
range are ruled out.  One can show, however that in the presence of
\nt annihilations the nucleosynthesis \mnt bound is substantially
weakened or eliminated \cite{valle.16}.  In \vallefig{valle.fig1} we
give the effective number of massless neutrinos equivalent to the
contribution of a massive \nt majoron model with different values of
the coupling $g$ between $\nu_\tau$'s and $J$'s, expressed in units of
$10^{-5}$. For comparison, the dashed line corresponds to the SM $g=0$
case. One sees that for a fixed $N_\nu^{\rm max}$, a wide range of tau
neutrino masses is allowed for large enough values of~$g$. No \nt
masses below the LEP limit can be ruled out, as long as $g$ exceeds a
few times $10^{-4}$.
\begin{figure}[t]
\centerline{\protect\hbox{\psfig{file=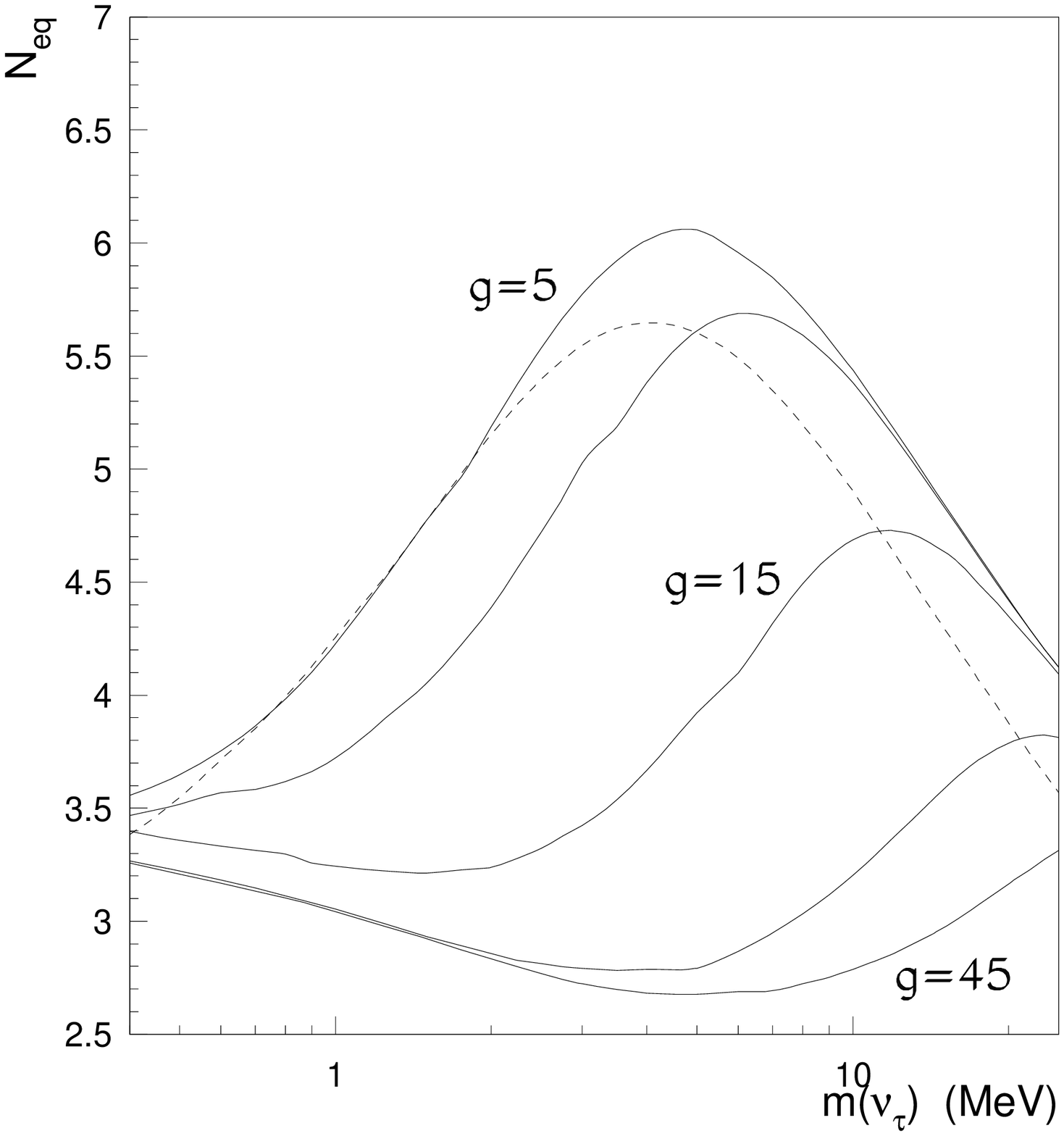,height=6cm,width=8cm}}}
\caption{A heavy \nt annihilating to majorons can lower
the equivalent massless-neutrino number in nucleosynthesis.}  
\label{valle.fig1} 
\end{figure} 
One can express the above results in the $m_{\nu_\tau}$-$g$-plane, as
shown in \vallefig{valle.fig2}.  
\begin{figure}[t]
\centerline{
\psfig{file=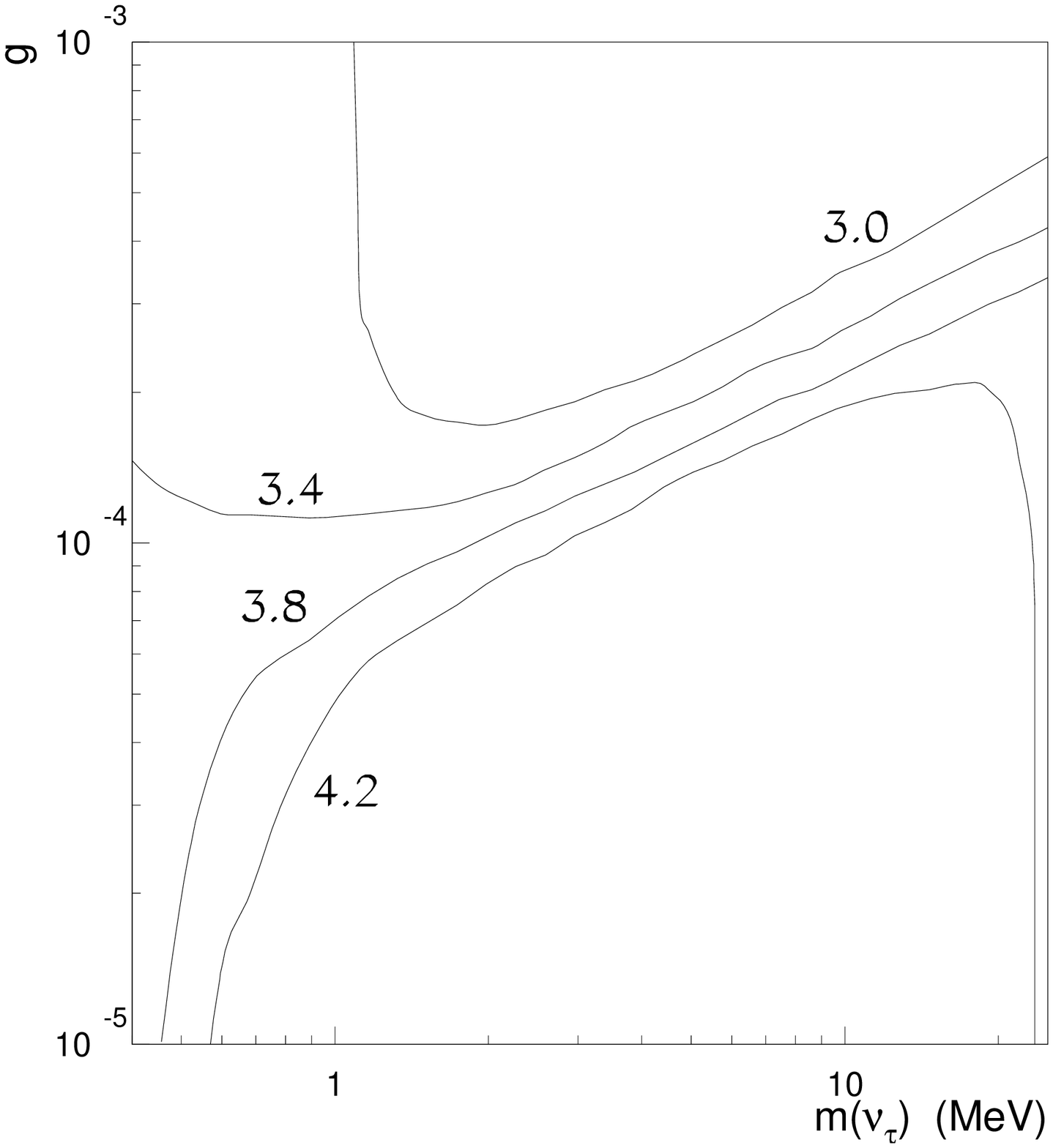,height=6cm,width=8cm}}
\caption{The region above each curve is allowed for the
corresponding $N_{\rm eq}^{\rm max}$.}  
\vglue -.5cm 
\label{valle.fig2}
\end{figure} 
Moreover the required values of $g(m_{\nu_\tau})$ are reasonable in
many majoron models \cite{valle.4,valle.17}.  Similar depletion in
massive \nt relic abundance also happens if the \nt is unstable on the
nucleosynthesis time scale \cite{valle.18} as will happen in many
majoron models.

\subsection*{Limits from Astrophysics  }

There are a variety of limits on neutrino parameters that follow from
astrophysics, e.g.~from the SN1987A observations, as well as from
supernova theory, including supernova dynamics \cite{valle.19} and
from nucleosynthesis in supernovae \cite{valle.20}. Here I briefly
discuss three recent examples of how supernova physics constrains
neutrino parameters.

It has been noted a long time ago that, in some circumstances, {\sl
massless} neutrinos may be {\sl mixed} in the leptonic charged current
\cite{valle.21}. Conventional neutrino oscillation searches in vacuo
are insensitive to this mixing. However,  such neutrinos may
resonantly convert in the dense medium of a supernova
\cite{valle.22}. The observation of the energy spectrum of
the SN1987A $\bar{\nu}_e$'s \cite{valle.23} may be used to provide very
stringent constraints on {\sl massless} neutrino mixing angles, as
seen in \vallefig{valle.fig3}.  The regions to the right of the solid curves are
forbidden, those to the left are allowed. Massless neutrino mixing may
also have important implications for $r$-process nucleosynthesis in
the supernova \cite{valle.20}. For details see ref.~\cite{valle.22}.
\begin{figure}[t]
\centerline{\protect\hbox{
\psfig{file=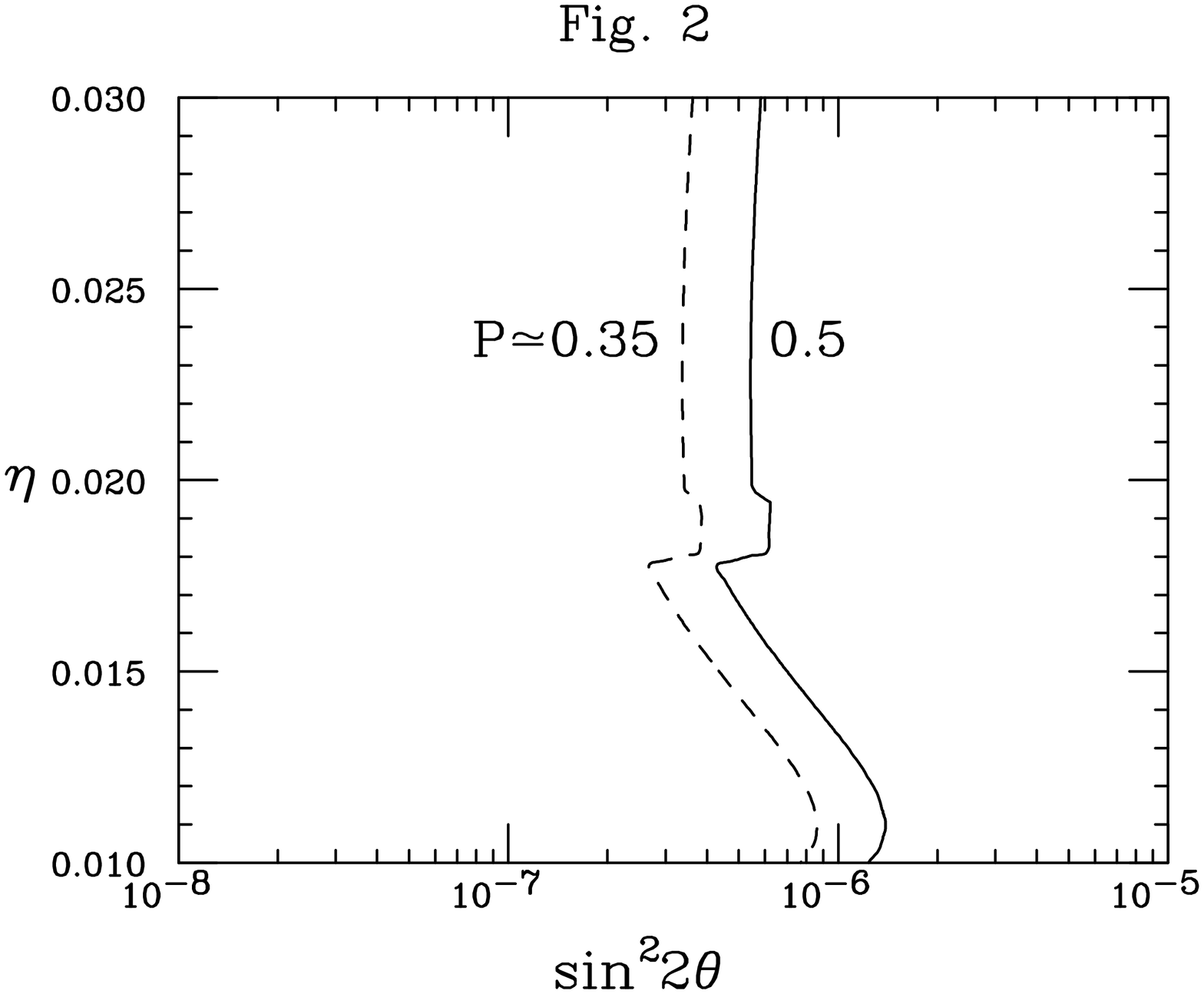,height=6.25cm,width=8cm,angle=90}
}}
\vglue -8.0cm
\hglue 7.5cm
\psfig{file=valle4.eps,height=.6cm,width=3cm,angle=0}
\vglue 7.0cm
\caption{SN1987A bounds on massless neutrino mixing. }
\label{valle.fig3}
\end{figure}

Another illustration of how supernova restricts neutrino properties
has been recently considered in ref.~\cite{valle.24}. In this paper
flavour changing neutral current (FCNC) neutrino interactions were
considered.  These could arise in supersymmetric models with $R$
parity violation. These interactions may induce resonant neutrino
conversions in a dense supernova medium, both in the massless and
massive case. The restrictions that follow from the observed
$\bar\nu_e$ energy spectra from SN~1987A and the supernova $r$-process
nucleosynthesis provide constraints on $R$ parity violation, which are
much more stringent than those obtained from the laboratory. In
\vallefig{valle.fig4} we display the SN~1987A constraints on explicit
$R$-parity-violating FCNCs in the presence of non-zero neutrino masses
in the hot dark matter eV range.  
\begin{figure}[t]
\centerline{\protect\hbox{
\psfig{file=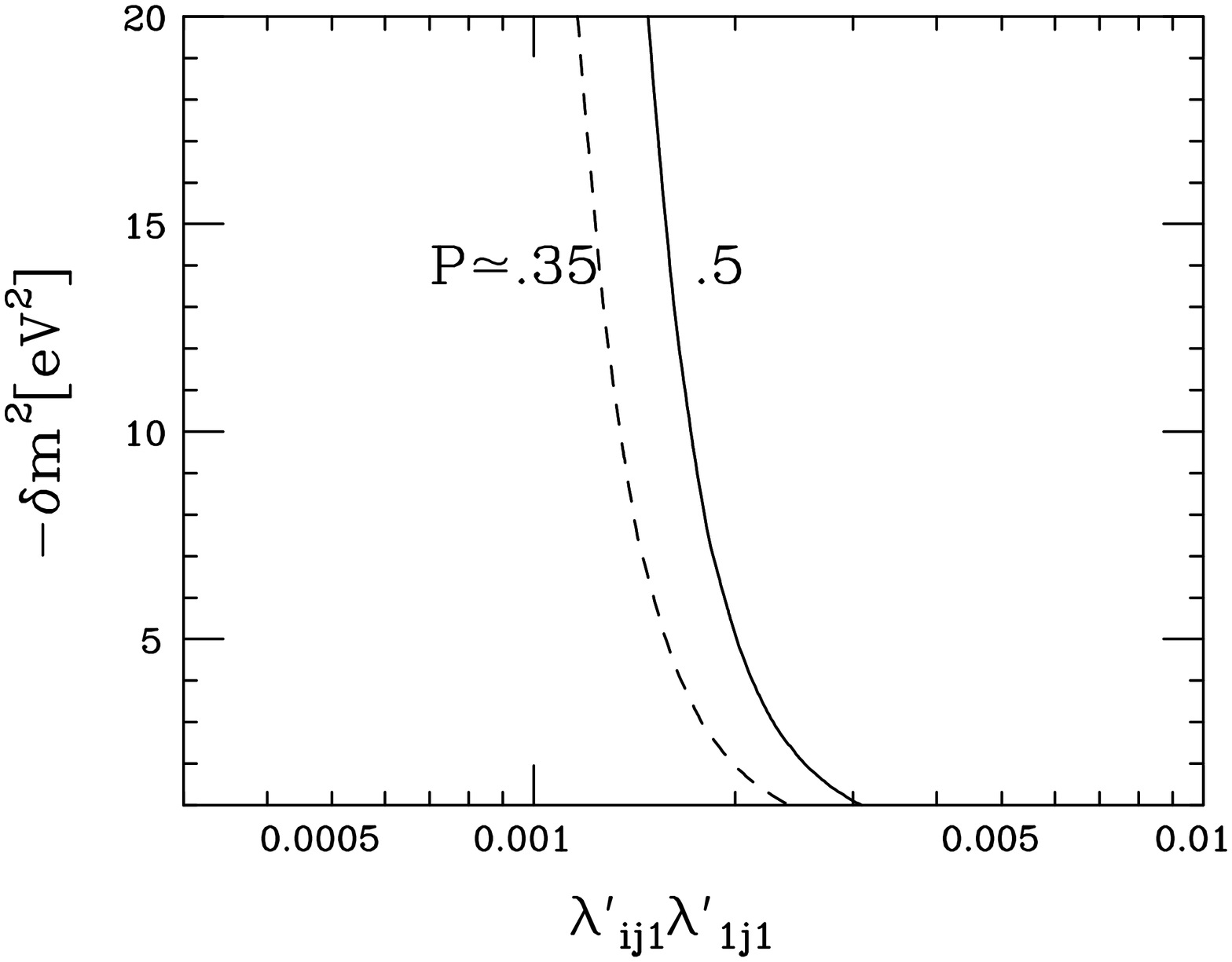,height=5cm,width=8cm,angle=90}
}}
\vglue -.7cm
\caption{SN~1987A bounds on  FCNC neutrino interactions. }
\label{valle.fig4}
\end{figure}
In \vallefig{valle.fig5} we display the sensitivity of the r-process
argument to the presence of FCNC neutrino interactions.
\begin{figure}[t]
\centerline{\protect\hbox{
\psfig{file=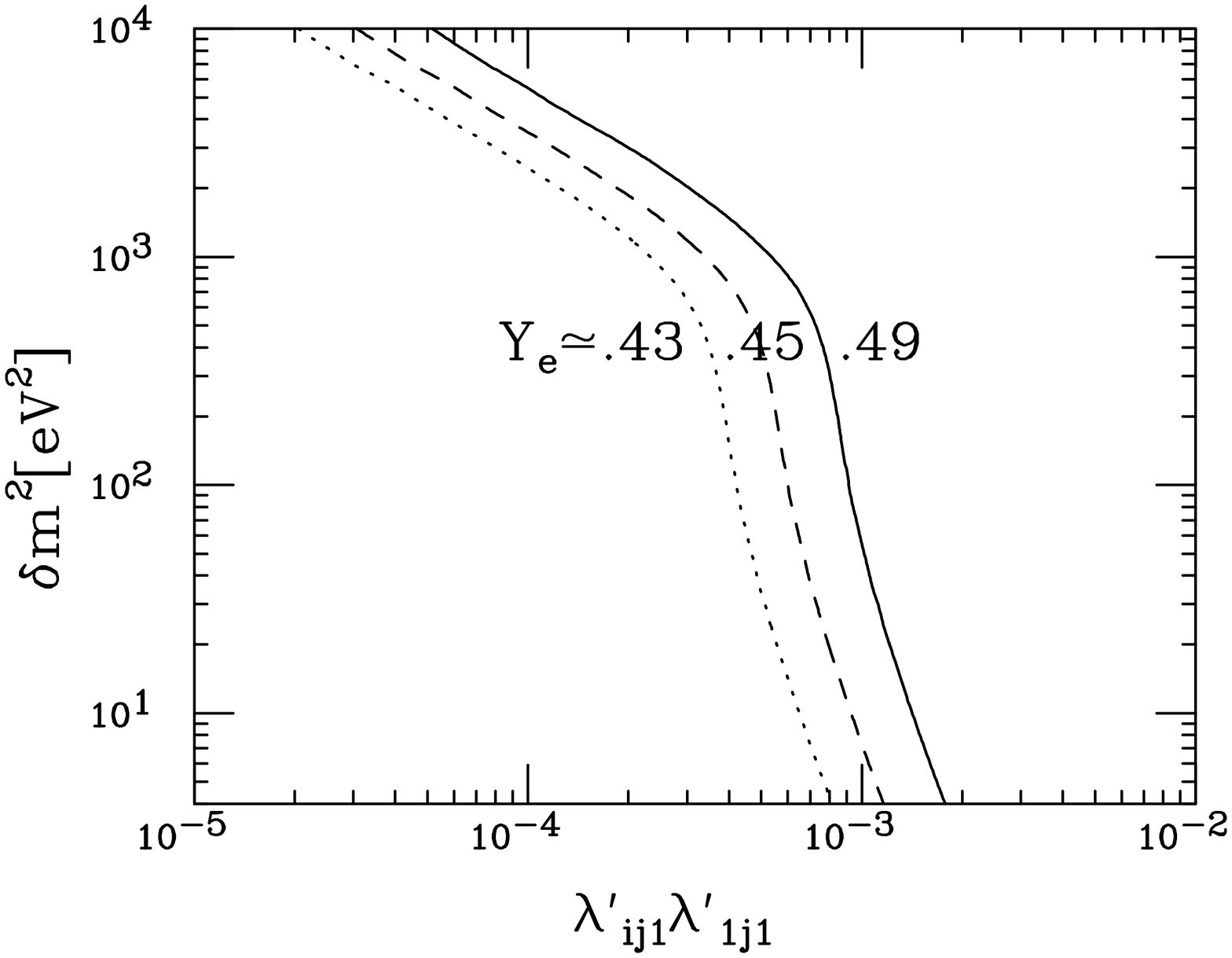,height=5cm,width=8cm,angle=90}
}}
\vglue -.7cm
\caption{Supernovae and FCNC neutrino interactions. R-process. }
\label{valle.fig5}
\end{figure}
Taken altogether, \vallefig{valle.fig4} and \vallefig{valle.fig5} disfavour a
leptoquark interpretation of the recent HERA anomaly.
\begin{figure}
\centerline{\protect\hbox{\psfig{file=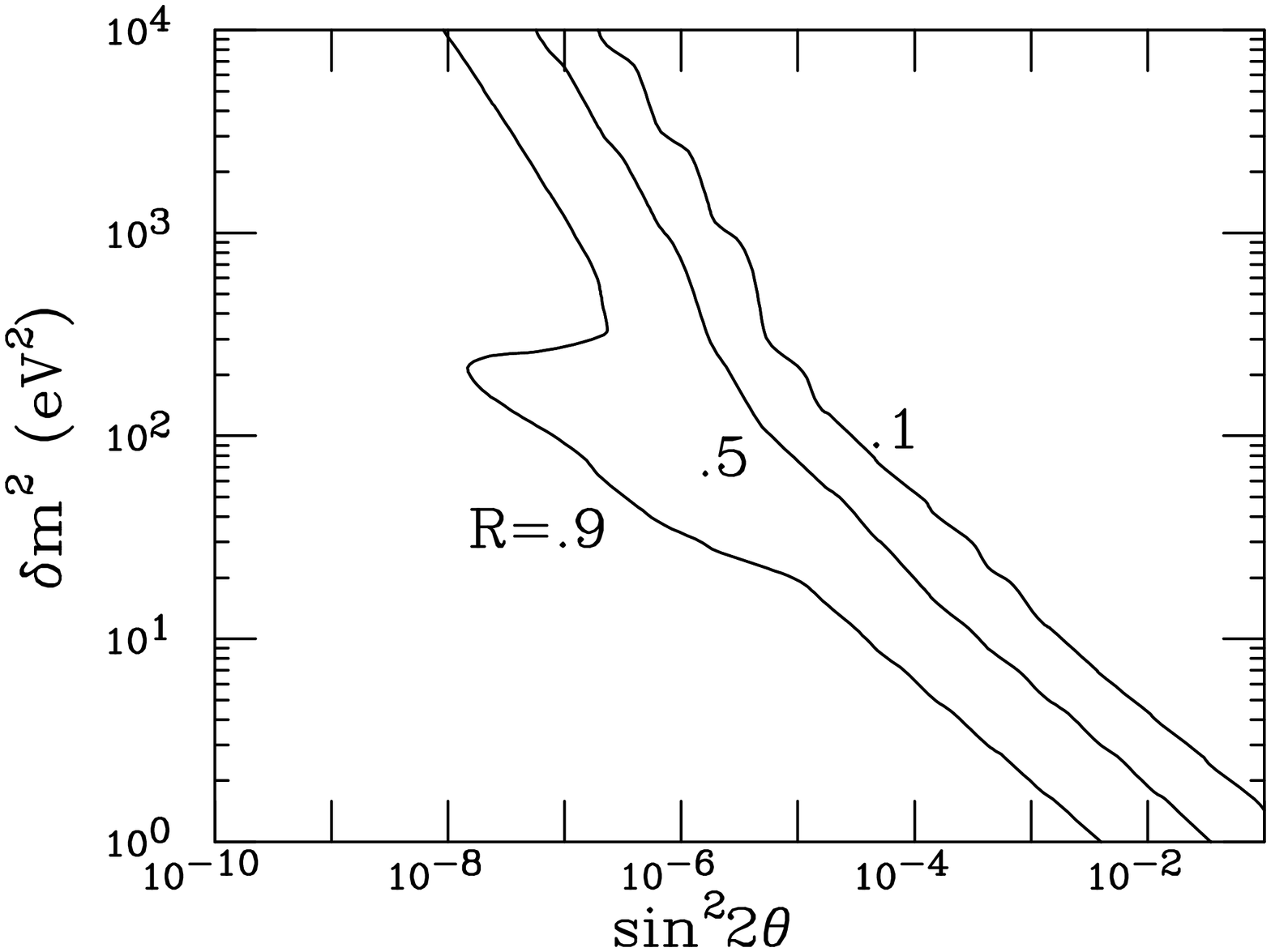, 
        height=5cm,width=8cm,angle=90}}}
\vglue -0.3cm
\caption{Supernovae shock reheating argument and $\nu_e-\nu_s$ neutrino 
conversions.  }
\label{valle.fig6}
\end{figure}

As a final example of how astrophysics can constrain neutrino
properties we consider the case of resonant $\nu_e \to\nu_s$ and
$\bar{\nu}_e\to\bar{\nu}_s$ conversions in supernovae, where $\nu_s$
is a {\it sterile} neutrino \cite{valle.25}, which we assume to be in
the hot dark matter mass range.  The implications of such a scenario
for the supernova shock re-heating, the detected $\bar\nu_e$ signal
from SN~1987A and for the $r$-process nucleosynthesis hypothesis have
been recently analysed \cite{valle.25}. In \vallefig{valle.fig6}, taken
from ref.~\cite{valle.25}, we illustrate the sensitivity of the
supernova shock re-heating argument to active-sterile neutrino mixing
and mass difference.  Notice that for the case of $r$-process
nucleosynthesis there is an allowed region for which the $r$-process
can be enhanced.

\section*{Acknowledgements}
Supported by DGICYT grant PB95-1077 and in part by EEC under the TMR
contract ERBFMRX-CT96-0090.  It is a pleasure to thank the organizers
for a very friendly atmosphere.


}\newpage{

\head{Some Neutrino Events of the 21st Century}
     {Leo Stodolsky}
     {Max-Planck-Institut f\"ur Physik, F\"ohringer Ring 6, 
      80805 M\"unchen, Germany}
 
\noindent
For this short concluding talk about the future of neutrino physics, I
would like to give the only talk of the meeting which has a really
good chance of being 100 percent wrong.  Let us imagine ourselves---or
our grandchildren---here a hundred years from now. We're reviewing
neutrino occurences of the last century.  Sticking our necks way out,
here are some guesses as to what happened.

\subsection*{Present--2015: Masses and Mixings }

In this period, we may suppose the various masses and mixings of the
neutrinos will be elucidated---or at least the larger ones where the
oscillation length is not greater than the earth-sun distance and the
mixing angle is not microscopic. I recall that a 30~eV neutrinos once
was the natural candidate for dark matter; so this point, that is if
there are neutrino masses in the eV range, is important in telling us
if neutrinos play any role in cosmology following the big bang.

\subsection*{2000--2025: Point Sources}

In this period the existence of high energy point sources of neutrinos
should be established, by very large detectors in the water or in the
ice.  If the recent spectacular observations~\cite{leo.1} of point
sources in high energy photons have a hadronic mechanism as their
origin, then there should be an equal number of similarly energetic
neutrinos.  That is, if we have a hadronic accelerator making pions
then the photons come from $\pi^0$ decay and the neutrinos from
$\pi^{\pm}$'s, giving $\nu/\gamma \sim 1$. On the other hand the
photons could be originating from something non-hadronic, such has
high energy electrons. Thus the neutrino observations will tell us
about the nature of the production mechanisms in these high energy
sources.

\subsection*{2010: Coherent Scattering in the Lab}

The coherent scattering of neutrinos is an old favorite topics. Due to
the (neutron number)$^2$ factor in the cross section, such high rates
result~\cite{leo.2} that one can imagine smallish neutrino detectors
for many interesting applications. This process is totally
uncontroversial in the Standard Model. But it still would be nice to
see it in the laboratory, which should be possible with cryogenic
detectors early in the century.

\subsection*{2023: Next Nearby Supernova}

All neutrino folk want to know when the next Milky Way supernova is
going to be. Let's put our money on 2023. By then several big
detectors should be operating smoothly, allowing a full exploitation,
of the thousands of events they will observe, both in neutral and
charged current modes.

\subsection*{2030: Neutrino Thermometer}

Now we're getting to some harder stuff, real challanges for the next
generation of experimentalists. As you know, the $^7{\rm Be}$
neutrinos from the sun are essentially monochromatic, coming from an
electron capture process.  However the ${\rm Be}$ nuclei in the center
of the sun are moving around due to the high temperature, giving a
spread and shift of the line. A calculation~\cite{leo.3} gives the
shift as 1.29~keV. A measurement, performed perhaps by comparing with
a terrestrial source, would give the temperature in the center of the
sun, an amusing use of neutrinos as a thermometer.

\subsection*{2040: Neutrino Geology}

Around this time we see the active development of neutrino geology.
The various radioactive processes in the earth give rise to neutrinos
(actually anti-neutrinos, except for the neutrino ``line'' from
electron capture in $^{40}{\rm K}$). Measurement of these terrestrial
neutrinos will give a direct snapshot of the energy production in the
earth and allow us to answer many interesting questions of geophysics
and the thermal history of the earth. Since the coherent
superconducting detector provides a light and portable instrument,
local and ``tomographic'' studies will be possible, as well as the
investigation of the planets and their moons. I set this relatively
late since the background from solar neutrinos and probably also
nuclear reactors must be well understood to see this signal (see
Fig.~11 of Ref.~[2]).

\subsection*{2050: Extra-Galactic Neutrino Burst Observatory}

Now we're getting to the big stuff: the extra-galactic neutrino
bursts. If we had a detector that could see the neutrinos from stellar
collapse in nearby clusters of galaxies we wouldn't have to wait
decades for the next event. With a thousand galaxies in the Virgo
cluster at 10~Mpc, we will be having them every few weeks and
supernova neutrino observations will become a systematic affair. This
will permit the study of many interesting points, such as flavor
``echos'' due to mixing, tests of CP for neutrinos and so
forth~\cite{leo.4}. The enormous distance involved means various
time-of-flight effects due to neutrino masses are greatly magnified
(in~fact there is a danger this could become too much of a good thing
since too much spreading of the pulse could make it disappear into the
background~\cite{leo.5}).  Such a detector, or perhaps we should say
observatory, is very ambitious, but not inconceivable. With the
coherent scattering process and cryogenic detection to see the small
recoils, about a megaton of cold material could suffice~\cite{leo.5}.
An alternative idea is the ``magic mountain'' or OMNIS~\cite{leo.7}
where neutral-current induced neutrons from natural Ca are the
signal. The joker in these proposals, as usual, is background.

\subsection*{2050: Neutrino Technology}

It took the laser about 50 years to get from the lab to the checkout
counter at the supermarket. Since neutrinos are more difficult than
photons, let's give them a hundred years from their discovery to get
into the economy. About mid-century, then, the light and portable
detector will allow us to monitor nuclear power stations from the
outside and to make geological investigations for minerals and
petroleum. Here again an understanding of the background is
essential. I'm not so sure about the sometimes mentioned neutrino
telecommunication channel, because I don't know what the transmitter
is supposed to be.

\subsection*{2099: Relic Neutrinos}

And last but not least, we have the observation of the big bang
neutrinos. This at the end of the century since I don't know how this
is going to be done. These relic neutrinos are a simple consequence of
big bang nucleosynthesis and should exist just as the relic microwave
photons do, with about the same density: $100/{\rm cm}^3$ for each
species. While their existence is not at all controversial, it would
be very satisfying to observe them and study their properties. The
great difficulty is their low energy, (actually momentum, which is the
correct thing to redshift if they are massive~\cite{leo.11}) which is
only millivolts. Previous schemes for low energy neutrino detection
involving some kind of energy transfer would seem to be unworkable,
leading one to consider coherent effects like forces and
torque~\cite{leo.11}. Two schemes have survived various criticisms. In
one, a net helicity in space because of an excess of one kind of
neutrino (not a feature of big bang cosmology) would exercise a torque
on an object with spin, in another a sponge-like object with pores on
about the wavelength of the neutrinos would get random kicks from
scattering, leading to a hopefully detectable force. A variant of this
would be to observe myriads of little grains of this size and wait til
one of them hops~\cite{leo.9}. Or, on another tack, there is the
possibility that extremely high energy neutrinos, near the cosmic ray
cutoff, hitting the background neutrinos and making a Z boson, lead to
an observable characteristic cascade~\cite{leo.8}.  None of these
ideas seem particularly practical, but then 2099 is a long way off and
maybe our grandchildren won't be so dumb $\ldots$
    
\subsection*{Comments}  
Some people have objected that I didn't include any results from
accelerator physics like observation of right-handed currents or new
heavy neutrinos.  True, but I don't know what these may be. I stress
that I have interpreted the word ``neutrino'' narrowly, as the
familiar known, almost massless, object.

Others have suggested I am much too conservative and are willing to
bet that it's all going to happen more and faster. That would be
great, let's hope so. Indeed, concerning the terrestrial neutrinos the
Borexino group~\cite{leo.10} says they are seriously considering
looking for an anti-neutrino signal, which could be detectable
according to estimates of Raghavan.


}


\newpage {\  }
\newpage{

\thispagestyle{empty}

\begin{flushright}
{\Huge \bf

{\ }

Workshop Program \\
\bigskip
and\\
\bigskip
List\\
\bigskip
of\\
\bigskip
Participants
}
\end{flushright}

\vskip 3truecm plus 3truemm minus 3truemm
\section*{Monday, Oct.~20: Solar Neutrinos}

\talk{10:45 -- 11:00}
        {F. v. Feilitzsch: Welcome}
\medskip
\leftline{{\bf Morning Session} (Chairperson: W.\ Hillebrandt)}
\talk{11:05 -- 11:50}
        {M. Stix: Solar Models}
\talk{12:00 -- 12:20}
        {H. Schlattl: Garching Solar Model -  Present Status}
\talk{12:30 -- 14:00}
        {\it{Lunch}}

\medskip
\leftline{{\bf Afternoon Session A} (Chairperson: F.\ von Feilitzsch)}
\talk{14:00 -- 14:30}
        {Y. Fukuda: Solar Neutrino Observation with Superkamiokande}
\talk{14:40 -- 15:10}
        {M. Altmann: Radiochemical Gallium Solar Neutrino 
         Experiments}
\talk{16:00 -- 16:30}
        {\it{Coffee break}}

\medskip
\leftline{{\bf Afternoon Session B} (Chairperson: R.\ L.\ M\"o\ss bauer)}
\talk{16:30 -- 17:00}
        {L. Oberauer: BOREXINO}
\talk{17:10 -- 17:40}
        {M. Moorhead: SNO}
\talk{17:50 -- 18:20}
        {M. Junker: Measurements of Low Energy Nuclear Cross Sections}
\talk{18:30 -- 20:00}
        {\it{Dinner}}

\medskip
\leftline{{\bf After Dinner Session} (Chairperson: A.\ Dar)}
\talk{20:00 -- 21:00}
        {G. Fiorentini: Solar Neutrinos - Where We Are and What Is Next?}

\vfill\eject

\section*{Tuesday, Oct.~21: Supernova Neutrinos}

\leftline{{\bf Morning Session A} (Chairperson: R.\ Bender)}
\talk{09:00 -- 09:30}
        {W. Hillebrandt: Phenomenology of Supernova Explosions}
\talk{09:40 -- 10:10}
        {H.-T. Janka: Supernova Explosion Mechanism}
\talk{10:20 -- 10:50}
        {\it{Coffee break}}

\medskip
\leftline{{\bf Morning Session B} (Chairperson: G.\ B\"orner)}
\talk{10:50 -- 11:20}
        {W. Keil: Convection in newly born neutron stars}

\talk{11:30 -- 12:00}
        {K. Sato: Cosmic Background Neutrinos}
\talk{12:30 -- 14:00}
        {\it{Lunch}}

\medskip
\leftline{{\bf Afternoon Session A} (Chairperson: S.\ Woosley)}
\talk{14:00 -- 14:30}
        {G.G. Raffelt: Neutrino Opacities}
\talk{14:40 -- 14:55}
        {S.J. Hardy: Neutrino Plasma-Wave Interactions}
\talk{15:00 -- 15:30}
        {P. Elmfors: Neutrino Propagation in a Magnetized Plasma}
\talk{15:40 -- 15:55}
        {A.N. Ioannisian: Neutrino Cherenkov Radiation}
\talk{16:00 -- 16:30}
        {\it{Coffee break}}

\medskip
\leftline{{\bf Afternoon Session B} (Chairperson: C.\ Hogan)}
\talk{16:30 -- 16:45}
        {A. Kopf: Photon Dispersion in a SN Core}
\talk{16:50 -- 17:20}
        {A. Lyne: Pulsar Proper Motions}
\talk{17:30 -- 18:00}
        {B. Leibundgut: Supernova Rates}
\talk{18:30}
        {\it{Dinner}}
%

\vskip 8truemm  minus 1truemm
\section*{Wednesday, Oct.~22: Gamma-Ray Bursts}

\leftline{{\bf Morning Session A} (Chairperson: H.-Th.\ Janka)}
\talk{09:00 -- 09:30}
        {D.H. Hartmann: Observations of Gamma-Ray Burst Sources}
\talk{09:40 -- 10:10}
        {S.E. Woosley: Theoretical Models}
\talk{10:20 -- 10:50}
        {\it{Coffee break}}

\medskip
\leftline{{\bf Morning Session B} (Chairperson: K.\ Sato)}
\talk{10:50 -- 11:20}
        {M. Ruffert: Merging Neutron Stars}
\talk{11:30 -- 12:00}
        {R.A. Sunyaev: Physical Processes Near Black Holes}
\talk{12:30 -- 14:00}
        {\it{Lunch}}

\smallskip
\talk{\bf Afternoon}
        {\it{Excursion}}
\talk{18:30 -- 20:00}
        {\it{Dinner}}

\medskip
\leftline{{\bf After Dinner Session} (Chairperson: K.\ Lande)}
\talk{20:00 -- 21:00}
        {R.L. M\"o\ss bauer: History of Neutrino Physics}
%

\vskip 1truecm plus 2truemm minus 2truemm
\section*{Thursday, Oct.~23: High Energy Neutrinos / Cosmology}

\leftline{{\bf Morning Session A} (Chairperson: E.\ Zas)}
\talk{09:00 -- 09:30}
        {K. Mannheim: Astrophysical Sources for High Energy Neutrinos}
\talk{09:40 -- 10:10}
        {C. Wiebusch: Neutrino Astronomy with AMANDA}
\talk{10:20 -- 10:50}
        {\it{Coffee break}}

\medskip
\leftline{{\bf Morning Session B} (Chairperson: L.\ Stodolsky)}
\talk{10:50 -- 11:20}
        {M. Moorhead: Status of ANTARES}
\talk{11:30 -- 12:00}
        {R. Plaga: Ground-Based Observations Gamma-Rays (200 GeV - 100 TeV)}
\talk{12:30 -- 14:00}
        {\it{Lunch}}

\medskip
\leftline{{\bf Afternoon Session A} (Chairperson: C.\ Hagner)}
\talk{14:00 -- 14:30}
        {D. Kie{\l}czewska: Superkamiokande - Atmospheric Neutrino Observations}
\talk{14:40 -- 15:10}
        {P. Gondolo: Atmospheric Muon and Neutrino Fluxes Above 1 TeV}
\talk{15:20 -- 15:50}
        {K. Jedamzik: Big Bang Nucleosynthesis - Theory}  
\talk{16:00 -- 16:30}
        {\it{Coffee break}}

\medskip
\leftline{{\bf Afternoon Session B} (Chairperson: M.\ Stix)}
\talk{16:30 -- 17:00}
        {C. Hogan: BBN - Observed Abundances}
\talk{17:10 -- 17:25}
        {J. Rehm: BBN - With Antimatter}
\talk{17:30 -- 18:00}
        {M. Bartelmann: Formation of Structure}
\talk{18:30 -- 20:00}
        {\it{Dinner}}

\medskip
\leftline{{\bf After Dinner Session} (Chairperson: D.\ Hartmann)}
\talk{20:00 -- 21:00}
        {A. Dar: What Killed the Dinosaurs?}
%

\vskip 1truecm plus 2truemm minus 2truemm
\section*{Friday, Oct.~24: The Future Prospects}

\leftline{{\bf Morning Session A} (Chairperson: L.\ Oberauer)}
\talk{09:00 -- 09:30}
        {P. Meunier: Neutrino Experiments with Cryogenic Detectors}
\talk{09:40 -- 10:10}
        {P. Smith: OMNIS - An Observatory for Galactic Supernovae}
\talk{10:20 -- 10:50}
        {\it{Coffee break}}

\medskip
\leftline{{\bf Morning Session B} (Chairperson: M.\ Lindner)}
\talk{10:50 -- 11:20}
        {J. Valle: Neutrinos and New Physics}
\talk{11:30 -- 12:00}
        {L. Stodolsky: Neutrinos in the 21st Century}
\talk{12:15}
        {\it{Lunch and Departure}}

\vfill\eject

\participant{Michael Altmann}
        {\TUM  
         {\tt email: altmann@e15.physik.tu-muenchen.de}
        }
\participant{Godehard Angloher}
        {\TUM  
         {\tt email: angloher@e15.physik.tu-muenchen.de}
        }
\participant{Matthias Bartelmann}
        {\MPA  
         {\tt email: msb@mpa-garching.mpg.de}
        }
\participant{Ludwig Beck}
        {\TUMS  
         {\tt email: lbeck@physik.tu-muenchen.de}
        }
\participant{Claus Beisbart}
        {\LMU  
         {\tt email: beisbart@stat.physik.lmu-muenchen.de}
        }
\participant{Ralf Bender}
        {\USM  
         {\tt email: bender@usm.uni-muenchen.de}
        }
\participant{Sergej Blinnikov}
        {Institute of Theoretical and Experimental Physics (ITEP)\\  
        B.~Cheremushkinskaya 25\\
        117259 Moskow, GUS\\
         {\tt email: blinn@sai.msu.su}
        }
\participant{Gerhard B\"orner}
        {\MPA  
         {\tt email: grb@mpa-garching.mpg.de}
        }
\participant{Thomas Buchert}
        {\LMU  
         {\tt email: buchert@stat.physik.uni-muenchen.de}
        }
\participant{Arnon Dar}
        {Department of Physics and Space Research Institute\\
         TECHNION---Israel Institute of Technology\\
         Haifa 32000, Israel\\ 
         {\tt email: arnon@physics.technion.ac.il}
        }
\participant{Maria Depner}
        {\MPA  
         {\tt email: maria@mpa-garching.mpg.de}
        }
\participant{Karin Dick}
        {\TUML  
         {\tt email: karin.dick@physik.tu-muenchen.de ???}
        }
\participant{Per Elmfors}
        {Stockholm University \\
        Fysikum, Box 6730 \\
         113 85 Stockholm, Sweden \\  
         {\tt email: elmfors@physto.se}
        }
\participant{Thomas Erben}
        {\MPA  
         {\tt email: erben@mpa-garching.mpg.de}
        }
\participant{Thomas Faestermann}
        {\TUMK
         {\tt email: Thomas.Faestermann@physik.tu-muenchen.de}
        }

\participant{Franz von Feilitzsch}
        {\TUM  
         {\tt email: feilitzsch@e15.physik.tu-muenchen.de}
        }

\participant{Gianni Fiorentini}
        {Dipartimento di Fisica dell'Universit\'a di Ferrara e INFN \\
        44100 Ferrara, Italy \\
         {\tt email: fiorentini@ferrara.infn.it}
        }

\participant{Chris Fryer}
        {Board Studies Astronomy and Astrophysics\\
         University of California at Santa Cruz\\
        Santa Cruz CA 95064, USA 
        }

\participant{Yoshiyuki Fukuda}
        {Institute for Cosmic Ray Research \\
         University of Tokyo\\
         Tokyo, Japan  \\
         {\tt email: fukuda@sukai07.icrr.u-tokyo.ac.jp}
        }

\participant{Narine Ghazarian}
        {\MPP  
         {\tt email: ara@mppmu.mpg.de}
        }

\participant{Paolo Gondolo}
        {\MPP  
         {\tt email: gondolo@mppmu.mpg.de}
        }

\participant{Caren Hagner}
        {\TUM  
         {\tt email: hagner@e15.physik.tu-muenchen.de}
        }

\participant{Steven Hardy}
        {\MPA  
         {\tt email: stephen@mpa-garching.mpg.de}
        }

\participant{Dieter Hartmann}
        {Department of Physics and Astronomy \\
        Clemson University \\
        Clemson, SC 29634-1911, USA\\
         {\tt email: hartmann@grb.phys.clemson.edu}
        }

\participant{Roger von Hentig}
        {\TUM  
         {\tt email: hentig@e15.physik.tu-muenchen.de}
        }

\participant{Wolfgang Hillebrandt}
        {\MPA  
         {\tt email: wfh@mpa-garching.mpg.de}
        }

\participant{Craig Hogan}
        {Departments of Physics and Astronomy \\
         University of Washington\\
         Box 351580 \\
          Seattle, WA 98195, USA \\
         {\tt email: hogan@centaurus.astro.washington.edu}
        }

\participant{Ara Ioannisian}
        {\MPP  
         {\tt email: ara@mppmu.mpg.de}
        }

\participant{Hans-Thomas Janka}
        {\MPA  
         {\tt email: thj@mpa-garching.mpg.de}
        }

\participant{Karsten Jedamzik}
        {\MPA  
         {\tt email: jedamzik@mpa-garching.mpg.de}\\
         {\tt email: karsten@igpp.llnl.gov}
        }

\participant{Matthias Junker}
        {Laboratori Nazionali Gran Sasso \\
         Assergi (AQ), Italy\\
         {\tt email: junker@lngs.infn.it}
        }

\participant{Wolfgang Keil}
        {\MPA  
         {\tt email: wfk@mpa-garching.mpg.de}
        }

\participant{Danka Kie{\l}czewska}
        {The University of California \\
          Irvine, CA 92717, USA \\
         {\tt email: danka@hepxvt.ps.uci.edu}
        }

\participant{Alexander Kopf}
        {\MPA  
         {\tt email: kopf@mpa-garching.mpg.de}
        }

\participant{Gunther Korschinek}
        {\TUM  
         {\tt email: korschin@physik.tu-muenchen.de}
        }

\participant{Kenneth Lande}
        {Department of Physics     \\
         University of Pennsylvania\\
         Philadelphia, PA 19104, USA \\
         {\tt email: klande@sas.upenn.edu}
        }

\participant{Bruno Leibundgut}
        {European Southern Observatory (ESO) \\
         Karl-Schwarzschild-Stra\ss e~2 \\
         85748 Garching, Germany\\
         {\tt email: bleibund@eso.org}
        }

\participant{Manfred Lindner}
        {\TUML  
         {\tt email: lindner@physik.tu-muenchen.de}
        }

\participant{Andrew Lyne}
        {University of Manchester\\
          Jodrell Bank \\
          Macclesfield, Cheshire SK11 9DL, UK \\
         {\tt email: a.lyne@jb.man.ac.uk}
        }

\participant{Andrew MacFadyen}
         {Board Studies Astronomy and Astrophysics \\
         University of California at Santa Cruz\\
         Santa Cruz CA 95064, USA
        }

\participant{Karl Mannheim}
        {Universit\"ats-Sternwarte \\
         Geismarlandstra\ss e~11 \\
          37083 G\"ottingen, Germany \\
         {\tt email: kmannhe@uni-sw.gwdg.de}
        }

\participant{Patrizia Meunier}
        {\MPP  
         {\tt email: meunier@vms.mppmu.mpg.de}
        }

\participant{Martin Moorhead}
        {University of Oxford \\
         Particle and Nuclear Physics Laboratory\\
          Keble Road \\
         Oxford OX1 3RH, UK \\
         {\tt email: m.moorhead1@physics.oxford.ac.uk}
        }

\participant{Rudolf L.\ M\"o\ss bauer}
        {\TUM  
         {\tt email: bvbellen@e15.physik.tu-muenchen.de}
        }

\participant{Marianne Neff}
        {\TUM  
         {\tt email: neff@e15.physik.tu-muenchen.de}
        }

\participant{Lothar Oberauer}
        {\TUM  
         {\tt email: oberauer@e15.physik.tu-muenchen.de}
        }

\participant{Axel Pichlmaier}
        {\TUMS  
         {\tt email: apichlma@physik.tu-muenchen.de}
        }

\participant{Rainer Plaga}
        {\MPP  
         {\tt email: plaga@hegra1.mppmu.mpg.de}
        }

\participant{Georg Raffelt}
        {\MPP  
         {\tt email: raffelt@mppmu.mpg.de}
        }

\participant{Jan Rehm}
        {\MPA  
         {\tt email: jan@mpa-garching.mpg.de}
        }

\participant{Maximilian Ruffert}
        {Institute of Astronomy \\
         Madingley Road \\
         Cambridge CB3 0HA, U.K.\\ 
         {\tt email: mruffert@ast.cam.ac.uk}
        }

\participant{Marisa Sarsa}
        {\TUM  
         {\tt email: sarsa@e15.physik.tu-muenchen.de}
        }

\participant{Katsuhiko Sato}
        {Department of Physics, School of Science \\
         University of Tokyo\\
         7-3-1 Hongo, Bunkyo-ku, Tokyo 113, Japan \\
         {\tt email: sato@utaphp1.phys.s.u-tokyo.ac.jp}
        }

\participant{Helmut Schlattl}
        {\MPA  
         {\tt email: schlattl@mpa-garching.mpg.de}
        }

\participant{Jens Schmalzing}
        {\MPA 
         {\tt email: jensen@mpa-garching.mpg.de}\\
        and\\
        \LMU 
         {\tt email: jens@stat.physik.uni-muenchen.de}
        }

\participant{Herbert Schmid}
        {\MPP  
         {\tt email: schmid@mppmu.mpg.de}
        }

\participant{Hans Schnagl}
        {\TUM  
         {\tt email: schnagl@e15.physik.tu-muenchen.de}
        }

\participant{Stefan Sch\"onert}
        {\TUM  
         {\tt email: schoenert@lngs.infn.it}
        }

\participant{Stella Seitz}
        {\MPA  
         {\tt email: stella@mpa-garching.mpg.de}
        }

\participant{Peter Smith}
        {Rutherford Appleton Laboratory \\
         Chilton, Oxfordshire, OX11 0QX,  UK \\
         {\tt email: p.f.smith@rl.ac.uk}
        }

\participant{Torsten Soldner}
        {\TUMS 
         {\tt email: tsoldner@physik.tu-muenchen.de}
        }

\participant{Michael Stix}
        {Kiepenheuer-Institut f\"ur Sonnenphysik \\
         Freiburg, Germany \\
         {\tt email: stix@kis.uni-freiburg.de}
        }

\participant{Leo Stodolsky}
        {\MPP  
         {\tt email: les@mppmu.mpg.de}
        }

\participant{Rashid Sunyaev}
        {\MPA  
         {\tt email: sunyaev@mpa-garching.mpg.de}
        }

\participant{Jose Valle}
        { Department of Physics \\
          University of Valencia \\
         Valencia, Spain \\
         {\tt email: valle@flamenco.ific.uv.es}
        }

\participant{Achim Weiss}
        {\MPA  
         {\tt email: weiss@mpa-garching.mpg.de}
        }

\participant{Christopher Wiebusch}
        {Institut f\"ur Hochenergiephysik  \\
         DESY-Zeuthen \\
         Berlin, Germany \\
         {\tt email: wiebusch@ifh.de}
        }

\participant{Stan Woosley}
        {Lick Observatory \\
         University of California at Santa Cruz\\
         Santa Cruz CA 95064, USA \\
         {\tt email: woosley@lick.ucsc.edu}\\
         {\tt email: woosley@ucolick.org}
        }
\participant{Shoichi Yamada}
        {\MPA  
         {\tt email: shoichi@mpa-garching.mpg.de}
        }
\participant{Enrique Zas}
        {\MPP  
        and\\
        Universidad de Santiago de Compostela, Spain\\
         {\tt email: zas@gaes.usc.es}
        }
\vfill

}
                   
\newpage {

\thispagestyle{empty}
\begin{flushright}
\Huge\bf
{\ }


Our\\
\bigskip
Sonderforschungsbereich (SFB)\\
\bigskip
``Astro-Teilchen-Physik''\\
\bigskip
and its\\
\bigskip
Divisions\\
\end{flushright}

\vfill
\participant{\bf Speaker:}
        {Prof. Dr. Franz von Feilitzsch \\
         \TUM
         Phone: ++49 +89 289 12511 \\
         email: {\tt feilitzsch@e15.physik.tu-muenchen.de}
        }
\participant{\bf Deputy Speaker:}
        {Prof. Dr. Wolfgang Hillebrandt \\
         \MPA
         Phone: ++49 +89 3299 3200 \\
         email: {\tt wfh@mpa-garching.mpg.de}
        }

\participant{\bf Secretary:}
        {N.N. \\
         \TUM
         Phone: ++49 +89 289 12503 \\
         Fax: ++49 +89 289 12680 \\
         email: {\tt depner@e15.physik.tu-muenchen.de}
        }


\eject

\section*{A Experimental Particle Physics}
\subsection*{A1 Spectroscopy of Solar Neutrinos}
\subsubsection*{A1-1 Borexino}
\subsubsection*{A1-2 Development of Cryogenic Detectors for GNO}
\participant{Principal Investigator}
        {Prof. Dr. Franz von Feilitzsch \\
         \TUM
         Phone: ++49 +89 289 12511 \\
         email: {\tt feilitzsch@e15.physik.tu-muenchen.de}
        }
%
\subsection*{A2 Experiments on the Decay of the Free
                Neutron}
\participant{Principal Investigator}
        {Prof. Dr. Klaus Schreckenbach \\
         \TUMS
         Phone: ++49 +89 289 12183 \\
         email: {\tt kschreck@physik.tu-muenchen.de}
        }
%
\subsection*{A3 Development and Application of Cryogenic Detectors for Dark
        Matter Detection via elastic WIMP-nucleus scattering }
\participant{Principal Investigator}
        {Prof. Dr. Franz von Feilitzsch \\
         \TUM
         Phone: ++49 +89 289 12511 \\
         email: {\tt feilitzsch@e15.physik.tu-muenchen.de}
        }
%
\subsection*{A4 Weak Decays of Highly Ionized Nuclei}
\participant{Principal Investigator}
        {Prof. Dr. Paul Kienle \\
         \TUMK
         Phone: ++49 +89 289 12421 \\
         email: {\tt paul.kienle@physik.tu-muenchen.de}
        }
%

\newpage
\section*{B Theoretical Particle Physics and Statistical Methods}
\subsection*{B1 Extensions of the Standard Model in View of Applications to
Astrophysics and Cosmology}
\participant{Principal Investigator}
        {Prof. Dr. Manfred Lindner \\
         \TUML
         Phone: ++49 +89 289 12196 \\
         email: {\tt lindner@physik.tu-muenchen.de}
        }
%
\subsection*{B2 String Cosmology}
\participant{Principal Investigator}
        {Prof. Dr. Stefan Theisen \\
         \LMU
         Phone: ++49 +89 2394 4375 \\
         email: {\tt theisen@mppmu.mpg.de}
        }
\subsection*{B3 Morphology of Cosmological Structures}
\participant{Principal Investigator}
        {Prof. Dr. Herbert Wagner \\
         \LMU
         Phone: ++49 +89 2394 4537 \\
         email: {\tt wagner@stat.physik.uni-muenchen.de}
        }
%
\subsection*{B4 Electroweak Symmetry Breaking in View of
Astrophysics and Cosmology}
\participant{Principal Investigator}
        {Prof. Dr. Manfred Lindner \\
         \TUML
         Phone: ++49 +89 289 12196 \\
         email: {\tt lindner@physik.tu-muenchen.de}
        }
%

\newpage
\section*{C Astrophysics and Cosmology}
\subsection*{C1 Neutrinos in Astrophysics and Cosmology}
\subsubsection*{C1-1 Neutrinos in Dense Media and Strong Fields}
\subsubsection*{C1-2 Stars and Weakly Interacting Particles}
\subsubsection*{C1-3 Supernovae and Compact Stars}
\subsubsection*{C1-4 Primordial Nucleosynthesis}
\participant{Principal Investigator}
        {Prof. Dr. Wolfgang Hillebrandt \\
         \MPA
         Phone: ++49 +89 3299 3200 \\
         email: {\tt wfh@mpa-garching.mpg.de}
        }
%
\subsection*{C2 Cosmology and Dark Matter: Theoretical Models}
\subsubsection*{C2-1 Dark Matter and Galaxy Distribution}
\subsubsection*{C2-2 Models of Galaxy Evolution}
\subsubsection*{C2-3 Microwave Background}
\participant{Principal Investigator}
        {Prof. Dr. Gerhard B\"orner \\
         \MPA
         Phone: ++49 +89 3299 3250 \\
         email: {\tt grb@mpa-garching.mpg.de}
        }
%
\subsection*{C3 Cosmology and Dark Matter: Astronomical Observations}
\participant{Principal Investigator}
        {Prof. Dr. Ralf Bender \\
         \USM
         Phone: ++49 +89 9220 9426 \\
         email: {\tt bender@usm.uni-muenchen.de}
        }
%
\subsection*{C4 Cosmology and Dark Matter: Gravitational Lensing}
\participant{Principal Investigator}
        {Prof. Dr. Ralf Bender \\
         \USM
         Phone: ++49 +89 9220 9426 \\
         email: {\tt bender@usm.uni-muenchen.de}
        }

\vfill

}                    

\end{document}